Study of the

Experimental Probe of Inflationary Cosmology - Intermediate Mission

for NASA's Einstein Inflation Probe

4 June 2009

Members of the EPIC-IM Study Team

| | | | |
|---|---|---|---|
| Abdullah Aljabri | JPL | Brian Keating | UC San Diego |
| Alex Amblard | UC Irvine | Chao-Lin Kuo | Stanford University |
| Daniel Bauman | Harvard University | Adrian Lee | UC Berkeley |
| Marc Betoule | IAS, France | Andrew Lange | Caltech/JPL |
| James Bock (PI) | JPL/Caltech | Charles Lawrence | JPL |
| Talso Chui | JPL | Steve Meyer | U. Chicago |
| Loris Colombo | USC | Nate Miller | UC San Diego |
| Asantha Cooray | UC Irvine | Hien Nguyen | JPL |
| Dustin Crumb | ATK Space | Elena Pierpaoli | USC |
| Peter Day | JPL | Nicolas Ponthieu | IAS, France |
| Clive Dickenson | JPL | Jean-Loup Puget | IAS, France |
| Darren Dowell | JPL/Caltech | Jeff Raab | NGAS |
| Mark Dragovan | JPL | Paul Richards | UC Berkeley |
| Sunil Golwala | Caltech | Celeste Satter | JPL |
| Krzysztof Gorski | JPL/Caltech | Mike Seiffert | JPL |
| Shaul Hanany | U. Minnesota | Meir Shimon | UCSD |
| Warren Holmes | JPL | Huan Tran | UC Berkeley/SSL |
| Kent Irwin | NIST | Brett Williams | JPL |
| Brad Johnson | UC Berkeley | Jonas Zmuidzinas | Caltech/JPL |

*EPIC-IM would like to thank the following people for their help in assembling and reviewing the material presented in the report: Scott Dodelson, Jo Dunkley, Aurelien Fraisse, Peg Frerking, Paul Goldsmith, Andrew Lange, Gary Parks, Paul Richards, John Ruhl, John Vaillancourt, Mike Werner and Matias Zaldarriaga*

# TABLE OF CONTENTS

## Summary

Measurements of Cosmic Microwave Background (CMB) anisotropy have served as the best experimental probe of the early universe to date. The CMB arises when the universe was a mere 380,000 years old, a small fraction of its present age of 13.7 billion years. Before this time the universe was ionized, and free electrons interacted strongly with photons to produce a thermal radiation spectrum. As the universe expanded, protons and electrons combined and the universe became transparent to electromagnetic radiation. The CMB, the thermal radiation from this epoch, can be used to probe the physical state of the universe at the time of recombination. The CMB also interacts, albeit weakly, with matter after the time of last scattering, via free electrons in the intergalactic medium or in inter-cluster gas, and through gravitational interaction with matter. Cosmological theory has been remarkably successful in predicting the tiny temperature and polarization anisotropies in CMB observations, and relating these observations to parameterized cosmology.

The theory of inflation, inspired in part by the extreme isotropy of the CMB, is now a cornerstone of modern cosmology. Inflation has passed a series of rigorous experimental tests, once again driven by increasingly precise measurements of the CMB anisotropy, both in temperature and polarization. Predictions of inflation have now been verified in the scale-invariant spectrum on large angular scales, the nearly perfectly flat geometry from the apparent angular scale of the acoustic peaks in the spectrum, the adiabatic and Gaussian nature of anisotropies, and the anti-correlation of temperature and polarization on scales outside of the causal horizon. Evidence is now starting to emerge for a small departure from scale invariance as expected in inflationary models. While evidence for inflation now seems inescapable, we still do not understand the physical mechanism or energy scale behind inflation.

Measurements of CMB polarization are emerging from their infancy into a powerful scientific observable. CMB temperature anisotropy measurements, based on the WMAP (Wilkinson Microwave Anisotropy Probe) satellite experiment, have now achieved the cosmic variance limit, based on the number of independent sky patches available in our view of the universe, up to an angular scale of $\ell \sim 500$. Measurements of temperature anisotropy will be completed by Planck, which will reach the cosmic variance limit into the Silk damping tail. Polarization measurements, which have now probed the scalar polarization spectrum with modest signal to noise, are improving rapidly both on ground-based and balloon-borne platforms, and in the upcoming Planck mission. Planck will measure the entire sky in polarization in 7 frequency bands at significantly better sensitivity than WMAP. While Planck will provide a wealth of cosmology, as well as information on Galactic foregrounds, it will not have the sensitivity to either extract all the cosmological information encoded in polarization, nor carry out a definitive measurement of the inflationary B-mode polarization spectrum.

The EPIC (Experimental Probe of Inflationary Cosmology) study team has investigated CMB polarization mission concepts based on scan-modulated bolometers. Bolometers offer the possibility of the highest possible sensitivity and coverage in multiple frequency bands from 30 to 300 GHz (or wider if necessary) to remove Galactic foregrounds. Large-format arrays of bolometers are needed to provide a large improvement in sensitivity over the single-element bolometers flying on Planck.

In a previous report [1] we studied a high-TRL mission configuration with six 30-cm apertures using existing NTD bolometer technology in a liquid helium cryostat, termed EPIC-LC (low-cost), as well as a large configuration with a 3 m telescope termed EPIC-CS (comprehensive science). In this report we study a mission with a 1.4 m aperture, termed the EPIC-Intermediate Mission, or EPIC-IM, because the aperture is intermediate between the 30 cm and 3 m cases studied previously. EPIC-IM's increased aperture allows access to a broader science case than the small EPIC-LC mission. In addition to the search for inflationary gravitational waves, the increased aperture allows us to mine the scalar polarization and shear polarization signals down to cosmological limits, so that we extract virtually all the cosmological information available from the CMB. In addition, a modest number of channels operating at higher frequencies allows for an all-sky measurement of polarized Galactic dust, which will provide a rich dataset for Galactic science related to magnetic fields. Using a combination of a large sensitivity focal plane with a new optical design, and an efficient 4 K mechanical cooler, EPIC-IM realizes higher sensitivity than EPIC-CS, but with a large mass savings.

# 1. Science

The high quality CMB data from sub-orbital and ground-based experiments [1-8] and now the Wilkinson Microwave Anisotropy Probe (WMAP) [9] have facilitated our current understanding of the Universe's geometry and composition, as well as the spectrum of primordial perturbations. While the data indicate that the global geometry of the Universe is spatially flat, we do not yet understand the physical nature of the dark matter and dark energy that form the dominant contributions to the energy density of the Universe. Furthermore, the physics behind the initial perturbations that were laid down at very high energies is unknown and is unlikely to ever be studied with a terrestrial particle accelerator.

Inflation, the prevailing paradigm related to the origin of density perturbations [10], posits that a nearly exponential expansion stretched space in the first moments after the Big Bang and promoted microscopic quantum fluctuations to perturbations on cosmological scales. Inflation makes detailed predictions for key statistical features of these fluctuations. These predictions have begun to be tested by a range of cosmological observations like the study of anisotropies in the CMB temperature and polarization and the distribution of galaxies on the sky. Observations of the CMB have been particularly instrumental in testing inflation, confirming all of the following generic predictions: 1) a nearly *flat* geometry; 2) a nearly *scale-invariant* spectrum of fluctuations, but with increasing evidence for a small departure from exact scale-invariance [14]; 3) *adiabatic* fluctuations; 4) nearly perfectly *Gaussian* fluctuations [15-17]; and 5) *super-horizon* fluctuations. The physics of inflation remains mysterious, but the theory generically predicts a stochastic background of gravitational waves – ripples in space-time that travel at the speed of light. The amplitude and the shape of the gravitational wave spectrum contain unique information about the physics of the early Universe that is not captured by the observed density perturbations.

The Experimental Probe of Inflationary Cosmology (EPIC) will pursue the CMB polarization signature associated with the inflationary gravitational wave signal. A detection of the primordial gravitational wave background would be a truly spectacular achievement and will not only establish inflation as the source of cosmological perturbations, but also allow a way to connect inflationary models to fundamental physics at a specific energy scale [10]. In addition to the inflationary gravitational wave signal, the polarization maps produced by the EPIC will be powerful new tools for cosmology, enabling us to precisely study neutrino masses with CMB lensing, to extract all information contained in CMB polarization about reionization, and to study the distribution of dust in our Galaxy.

## 1.1 Inflationary Gravitational Wave Background

The fundamental microscopic origin of inflation is still a mystery. What is the inflaton field that drove the expansion? How did the inflationary energy vary in time? And why did the universe start in a high energy state? What is the fundamental physics responsible for inflation? Is this new physics related to Grand Unified Theories? These are some of the many questions which remain unanswered. The challenge to explain the physics of inflation is considerable. Inflation is believed to have occurred at an enormous energy scale (maybe as high as $10^{16}$ GeV), far out of reach of terrestrial particle accelerators. Observations of the cosmic microwave background are therefore likely to remain our only experimental probe of the physics that shaped the earliest moments of the universe.

CMB polarization experiments are the next logical step to determine the ultra-high-energy physics responsible for inflation. In addition to density fluctuations inflation predicts a cosmological background of stochastic gravitational waves, produced through quantum-mechanical excitations of the gravitational field [22-23]. The theory predicts that the amplitude of the gravitational-wave background depends only on the universal expansion rate -- or equivalently on the cosmological energy density, the age of the Universe, or the height of the inflaton potential -- during inflation. The signal is insensitive to other details of inflation like the precise shape of the inflationary potential. If detected, the gravitational wave signal therefore directly determines the energy scale at which inflation occurred. This information is the single most important clue scientists would like to get about the physical origin of inflation.

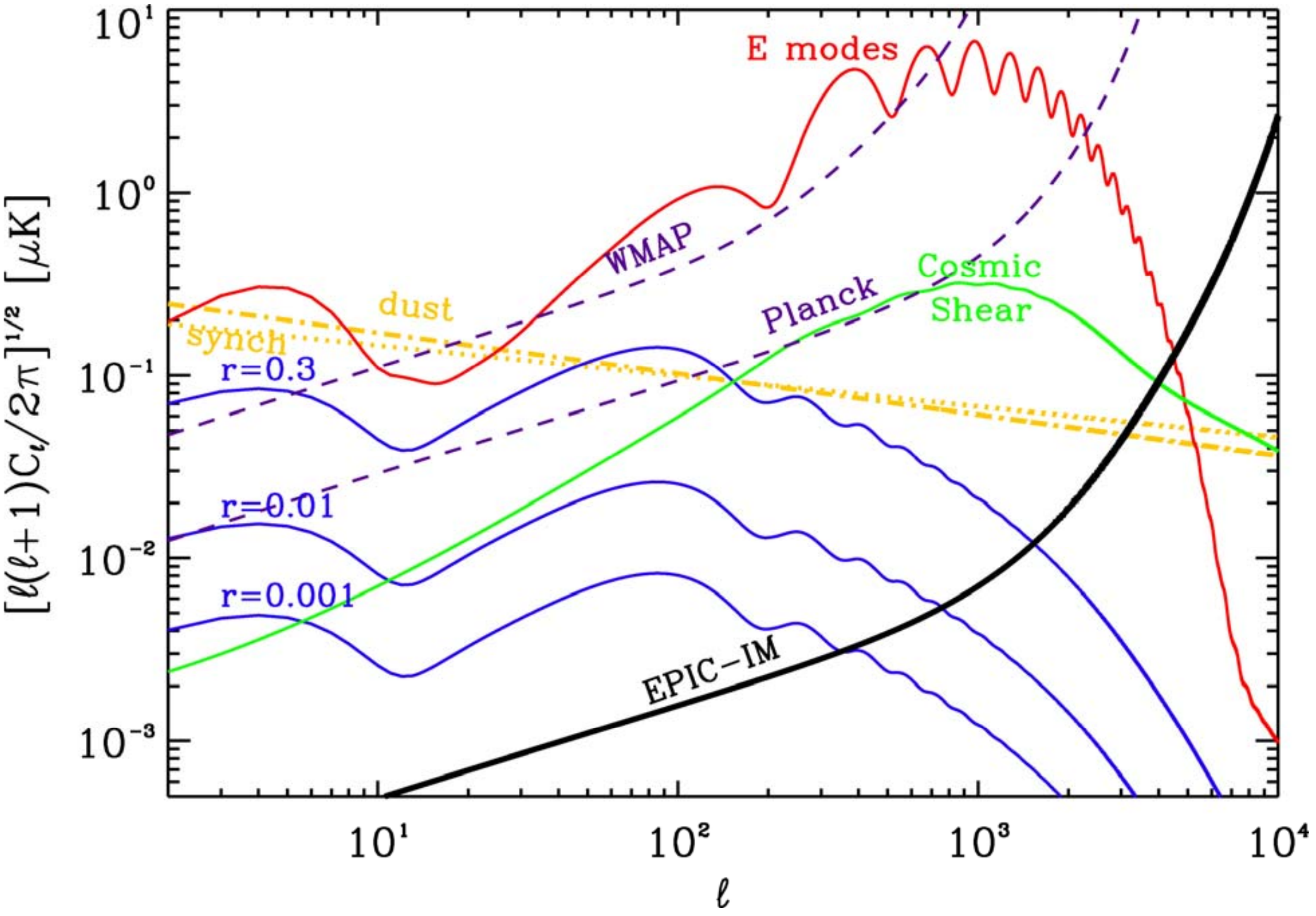

Fig. 1.1 The sensitivity of EPIC-IM, WMAP and Planck to CMB polarization anisotropy. E-mode polarization from scalar perturbations is shown in red; B-mode polarization from tensor perturbations is shown in blue for r = 0.3 and r = 0.01; and B-mode polarization produced by lensing of the E-mode polarization is shown in green. The science goal of EPIC is to measure the inflationary B-mode spectrum the level of r = 0.01 for the entire 2 < l < 200 multipole range after foreground subtraction. Expected B-mode foreground power spectra for polarized dust (orange dash-dotted) and synchrotron (orange dotted) at 70 GHz are determined by power-law models fits to the foreground power in a combination of WMAP 23 GHz polarization maps [11], low frequency radio maps [12], and 100 micron dust map for |b| > 20° [13] for a 65 % sky cut. The band-combined instrumental sensitivity of EPIC is given assuming the 4 K telescope option and 4 years of observations. WMAP assumes an 8-year mission life; Planck assumes 1.2 years at goal sensitivities for HFI. Note that the sensitivity curves show band-combined sensitivities to $C_\ell$ in broad $\Delta l/l = 0.3$ bins in order to compare the full raw statistical power of the three experiments in the same manner. The final sensitivity to r after foreground removal will naturally be reduced.

If the energy scale turns out to be around $10^{16}$ GeV, then inflation was most likely associated with processes involved in the unification of the strong, weak, and electromagnetic interactions. If the energy scale is lower, then inflation may be involved with the breaking of supersymmetry [24]. Finally, while there are sources of gravitational waves within the horizon leading to waves of sub-horizon wavelength, such as due to cosmic strings or massive black hole binaries [25], a primordial phenomenon such as inflation is the only mechanism to produce gravitational waves with super-horizon wavelengths [26]. If inflationary gravitational waves are detected this is universally accepted to be the smoking-gun signature of inflation.

The vector-like properties of the polarization allow it to be decomposed into two orthogonal modes: curl ("B-mode") and curl-free grad ("E-mode") components [27-28]. The "curl" component is only sourced by inflationary gravitational waves as primordial density perturbations do not have a handedness[1]. This provides a unique signature of inflationary gravitational waves. The curl modes also do not correlate with either the temperature or the gradient modes, providing a way to test for certain systematic effects. The power spectrum for the curl component of polarization due inflationary gravitational waves is shown in Fig. 1.1.

The amplitude of inflationary gravitational waves (*tensor* modes) is typically normalized relative to the known amplitude of density fluctuations (*scalar* modes) and quantified by the *tensor-to-scalar ratio* r.[2] With a detection of the tensor perturbations, two key properties of inflation can be directly established simply based on the value of the tensor-to-scalar ratio:

a) The energy scale of inflation via

$$V^{1/4} = 1.06 \times 10^{15}\,\mathrm{GeV}\left(\frac{r}{0.01}\right)^{1/4}$$

A tensor amplitude of order r~0.01 would demonstrate conclusively that inflation occurred at a tremendously high energy scale, comparable to that of Grand Unified Theories (GUTs). It is difficult to overstate the impact of such a result for the high-energy physics community, which to date has only two indirect clues about physics at this scale: the apparent unification of gauge couplings, and experimental lower bounds on the proton lifetime. A B-mode detection would therefore turn the early universe into a probe for ultra-high energy physics at energies entirely inaccessible to conventional terrestrial experimentation.

b) A tensor-to-scalar ratio bigger than 0.01 also correlates with a super-Planckian field variation between the time when CMB fluctuations exited the horizon during inflation and the end of inflation [29]. As explained in Baumann et al. [10], this would provide important information about certain properties of the ultraviolet completion of quantum gravity, since a natural explanation of the flatness of the inflaton potential over a super-Planckian range requires certain assumptions about the symmetries of the inflationary action and their validity near the Planck scale. An upper limit of r < 0.01 would also be very informative as it would rule out all large-field models of inflation, while leaving a well-defined class of inflationary mechanisms to consider (Fig. 1.2).

---

[1] The vector modes are also expected to produce polarization B-mode component, but primordial perturbations in the form of vector modes with sufficient amplitude to be of interest are not expected in the inflationary scenarios.

[2] We define the tensor-to-scalar ratio as the ratio of tensor and scalar power spectra at a certain wave number $k_0$: $r=P_t(k_0)/P_s(k_0)$

It is essential to recognize that CMB polarization experiments have almost unique potential to provide these two clues about physics at the highest scales. Finally, alternatives to inflation [30-33] almost universally predict an unobservably low gravitational wave amplitude and would hence be ruled out by a B-mode detection.

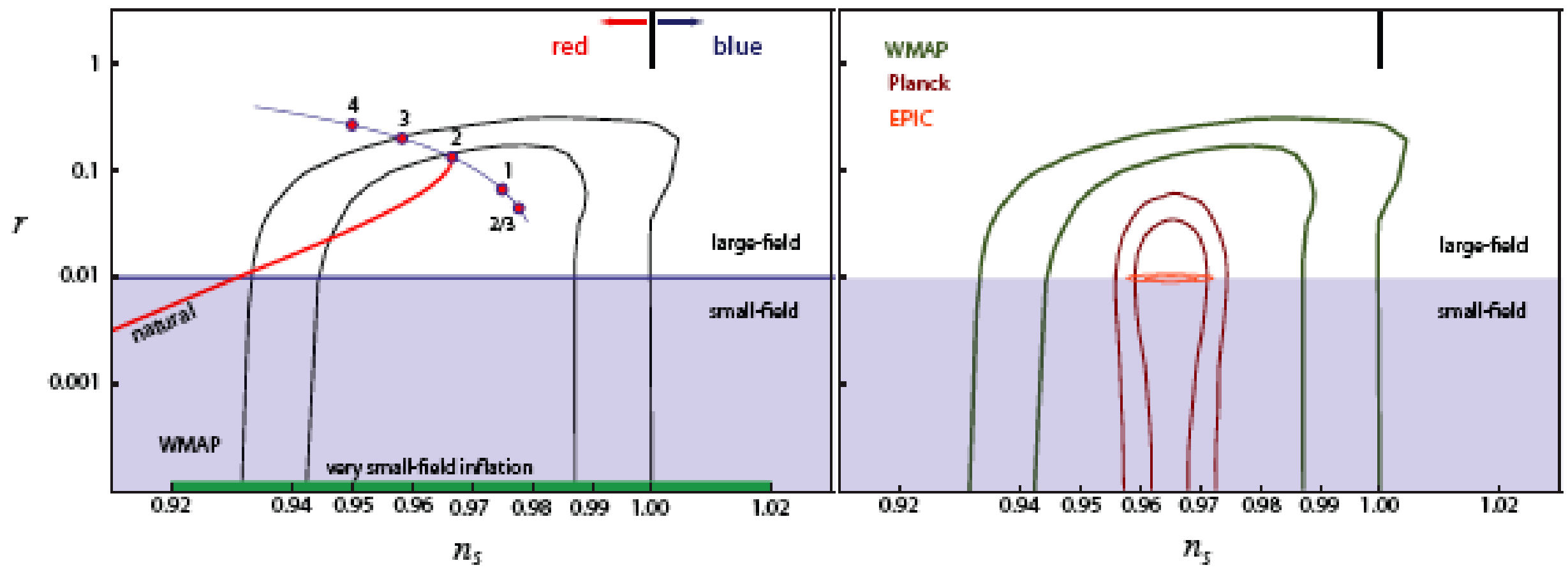

Fig. 1.2. Constraining inflationary models in the $n_s$-r plane. The left panel shows the theoretical expectations for a few representative models of slow-roll inflation, while the right panel shows the constraints derivable from data. The slow-roll models shown in the left panel are: *chaotic inflation*: $\lambda_p\phi^p$ for general *p* (thin solid line) and for *p*=4, 3, 2, 1, 2/3 (red circles); models with *p*=2, *p*=1 and *p*=2/3 have recently been obtained in string theory, *natural inflation*: $V_0$ [1-cos($\phi/\mu$)] (solid line); *very small-field inflation*: models of inflation with a very small tensor amplitude, r << $10^{-4}$ (green bar); examples of such models in string theory include brane inflation, Kahler inflation, and racetrack inflation. The right panel shows WMAP 5-year constraints [15] and projections for Planck [35] and EPIC-IM. Realistic forecasts for EPIC-IM specifications with a detailed model and removal of foregrounds show that gravitational waves can be detected at $\geq 7\sigma$ when r~0.01 [10]. This shows that EPIC-IM is a powerful instrument to test this crucial regime of the inflationary parameter space.

## 1.2 Further Constraints on Inflationary Physics

A detection of inflationary gravitational waves would be nothing short of revolutionary. However, even absent a B-mode detection, precision measurements of CMB polarization contain vital additional information on the physics of the inflationary era. Here we list some of the main observables that will be extracted from the data (more details and references may be found in [10]):

Deviations from scale-invariance: Inflation predicts that the primordial fluctuation spectra are nearly, but not exactly, scale-invariant. Any deviation from perfect scale-invariance is a powerful discriminator for different inflationary mechanisms (e.g. for slow-roll inflation the scale dependence reflects the shape of the inflationary potential). Fig. 1.3 shows the predictions for the scalar tilt $n_s$ from a few representative inflationary models. (Here, $n_s$=1 corresponds to perfect scale-invariance, $n_s$<1 (>1) corresponds to more (less) power on large scales). Five-year WMAP data have now provided the first evidence for a red ($n_s$<1) spectrum, even when the tensor-to-scalar ratio is taken to be a free parameter [15]. Planck [35] in combination with present and future small-scale CMB and LSS experiments will probe the scale dependence of the scalar spectrum in much more detail. The data from EPIC will further add to constraining important regions of the inflationary parameter space, with standard parameters of the LCDM cosmological model with tensors determined at an accuracy within a few percent of the cosmic variance limit of CMB temperature and anisotropy data (Table 1.1 and Fig. 1.3).

Non-Gaussianity: Non-Gaussianity is a measure of the interactions of the inflaton. During slow-roll inflation, the inflaton self-interactions are necessarily small and the fluctuations are very nearly Gaussian [36]. However, in multi-field inflation or inflationary models with non-trivial kinetic effects (non-trivial sound speed) the non-Gaussianity is often large [37] containing crucial information about the inflationary action. Many of these theoretical structures have recently been motivated by developments of inflationary model building in string theory. Moreover, alternatives to inflation often predict large non-Gaussianity [38] and are hence also tested.

**Table 1.1.** Parameter errors expected from EPIC, WMAP 5-year results and Planck forecasts

| Parameter | 5-year WMAP[15] | σ(Planck) | σ(EPIC) |
|---|---|---|---|
| $n_s$ | $0.963 \pm 0.015$[(a)] | $3.5 \times 10^{-3}$ | $1.6 \times 10^{-3}$ |
| $100\Omega_b h^2$ | $2.273 \pm 0.062$ | $1.4 \times 10^{-2}$ | $5.9 \times 10^{-3}$ |
| $\Omega_c h^2$ | $0.1099 \pm 0.0062$ | $1.2 \times 10^{-3}$ | $7.1 \times 10^{-4}$ |
| $\tau$ | $0.087 \pm 0.017$ | $4.5 \times 10^{-3}$ | $2.5 \times 10^{-3}$ |
| $A_s$ | $(2.41 \pm 0.11) \times 10^{-9}$ | $4.4 \times 10^{-11}$ | $1.1 \times 10^{-11}$ |
| R | $< 0.43$ $(2\sigma)$ | 0.011[(b)] | $5.4 \times 10^{-4}$ [(b)] |
| $H_0$ | $71.9 \pm 2.6$ | 0.62 | 0.11 |
| $f_{NL}$ (Local) | $38 \pm 21$[17] | ~5 | 2 |
| $\alpha_s$ | $-0.037 \pm 0.028$ | $5.2 \times 10^{-3}$ | $3.6 \times 10^{-3}$ |
| $f_{NL}$ (Equil.) | $73 \pm 101$ | 26 | 13 |
| $\Omega_k$ | $-0.0179 < \Omega_k < 0.0066$ $(2\sigma)$ | $4.5 \times 10^{-3}$ | $6 \times 10^{-4}$ |
| $\alpha_{-1}$ (curvaton) | $< 4.1 \times 10^{-3}$ $(2\sigma)$ | $1.2 \times 10^{-4}$ | $3.5 \times 10^{-5}$ |
| $\alpha_0$ (axion) | $< 7.2 \times 10^{-2}$ $(2\sigma)$ | $2.5 \times 10^{-2}$ | $6.6 \times 10^{-3}$ |
| $\Sigma m_\nu$ | $< 0.67$eV $(2\sigma)$[(c)] | $< 0.12$eV[(d)] | $< 0.047$eV[(d)] |

Notes: WMAP results are from Ref. [15], except for $f_{NL}$ (local). Planck and EPIC results make use of the foreground model described in Ref. [34] and described in Chapter 2. The parameter errors listed in the Table assume a LCDM cosmological model with no running in the scalar spectral index and a tensor spectral index based on the single-field slow-roll consistency relation, $r = -8n_t$. If the tensor spectral index is taken to be a free parameter, with EPIC, $\sigma_r = 0.0025$ (when r=0.01) and $\sigma(n_t) = 0.13$. (a) WMAP mean likelihood determination assuming no tensors with $r = 0$. (b) We assume r=0.01 when quoting expected errors on the tensor-to-scalar ratio for Planck and EPIC. (c) WMAP+BAO+SN [12]. (d) Using cosmic shear information in CMB and with a general dark energy equation of state.

The primary signature of non-Gaussian correlations is a non-zero three-point function. In Fourier space this relates to the bispectrum whose momentum dependence contains a large amount of information on the physical process generating the non-Gaussianity. Multi-field non-Gaussianity can in this way be distinguished from non-Gaussianity due to kinetic effects. Thus, if detected, non-Gaussianity of primordial perturbations would provide a unique avenue for studying the ultra-high-energy physics responsible for inflation or even testing alternative ideas for the dynamics of the early Universe.

WMAP 5-year data lead to a non-Gaussianity measurement with a value for the non-Gausianity parameter $f_{NL}$ in the local model of 38 ± 21 [17]. While consistent with zero, such a measurement could also be argued as a hint of non-Gaussianity. High resolution and high sensitivity temperature and polarization anisotropy data from EPIC will improve uncertainty on the non-Gaussianity parameter by a factor of 10 down to $f_{NL}$ of 2, reaching the interesting level where a non-Gaussian signal is expected to be present even under certain slow-roll inflationary models (Table 1.1).

Isocurvature fluctuations: Isocurvature density perturbations are a clean signature of multi-field models of inflation [10]. In single-field inflation the fluctuations are necessarily adiabatic, but in multi-field models a significant non-adiabatic or isocurvature component can arise. At late times we would observe this as variations in the relative density between different matter components. CMB E-mode polarization gives important constraints on primordial isocurvature fluctuations.

Defects, curvature and anisotropy: In addition to testing the physics during inflation EPIC has the potential to provide information on pre- and post-inflationary physics: e.g. *i*) defects like cosmic strings produced after inflation create a characteristic B-mode signature; *ii*) a remnant curvature and large-scale anisotropy from pre-inflationary initial conditions leaves signatures in the CMB temperature and polarization maps [10].

### 1.3 The Scientific Promise of Precision CMB Polarimetry

EPIC will improve upon Planck's raw sensitivity by a factor of 25 to 60. In addition to the measurements we review briefly here, EPIC will open a huge discovery space for breakthroughs we at present cannot even anticipate.

#### *1.3.1 E-mode polarization*

EPIC will extract all of the information encoded in the CMB surface-of-last-scattering, and achieve a cosmic-variance limited measurement of the E-mode polarization down to extremely small scales limited only by the beam resolution. A measurement of the scalar power spectrum will enable new tests of the physics of recombination, probes for exotic phenomena [39], and the best constraint possible with CMB anisotropies on the isocurvature amplitude (Table 1.1).

Cosmological Birefringence and Rotation of Polarization Vectors: Recent studies have shown the importance of looking for conversions between E and B-modes during the propagation of CMB photons from the last scattering surface due to, for example, a non-standard coupling of a scalar field to electromagnetism through a parity violating mechanism [40]. Another possibility is a different dispersion relation for the left and right-circularly polarized modes, similar to Faraday rotation. Initial attempts have been made to constrain an isotropic rotation of CMB polarization, through cross-correlations of parity violating spectra of TB and EB, with existing data from BOOMERanG, WMAP, and QUAD. WMAP 5-year data lead to the constraint that the rotation angle is between –5.9 and 2.4 degrees at the 95% confidence level [15]. Planck data allow will constrain the rotational angle with a 1$\sigma$ uncertainty of 0.1 degrees, while EPIC TB and EB correlations will improve this result significantly with an uncertainty at the level of $10^{-3}$ degrees [41]. Instead of an isotropic rotation, some mechanisms may lead to an anisotropic rotation, due to for example inhomogeneities in the coupling. With Planck, an anisotropic rotation can be constrained down to 0.01 $deg^2$ [42], while with EPIC this can be improved down to the level of $10^{-5}$ $deg^2$ [43].

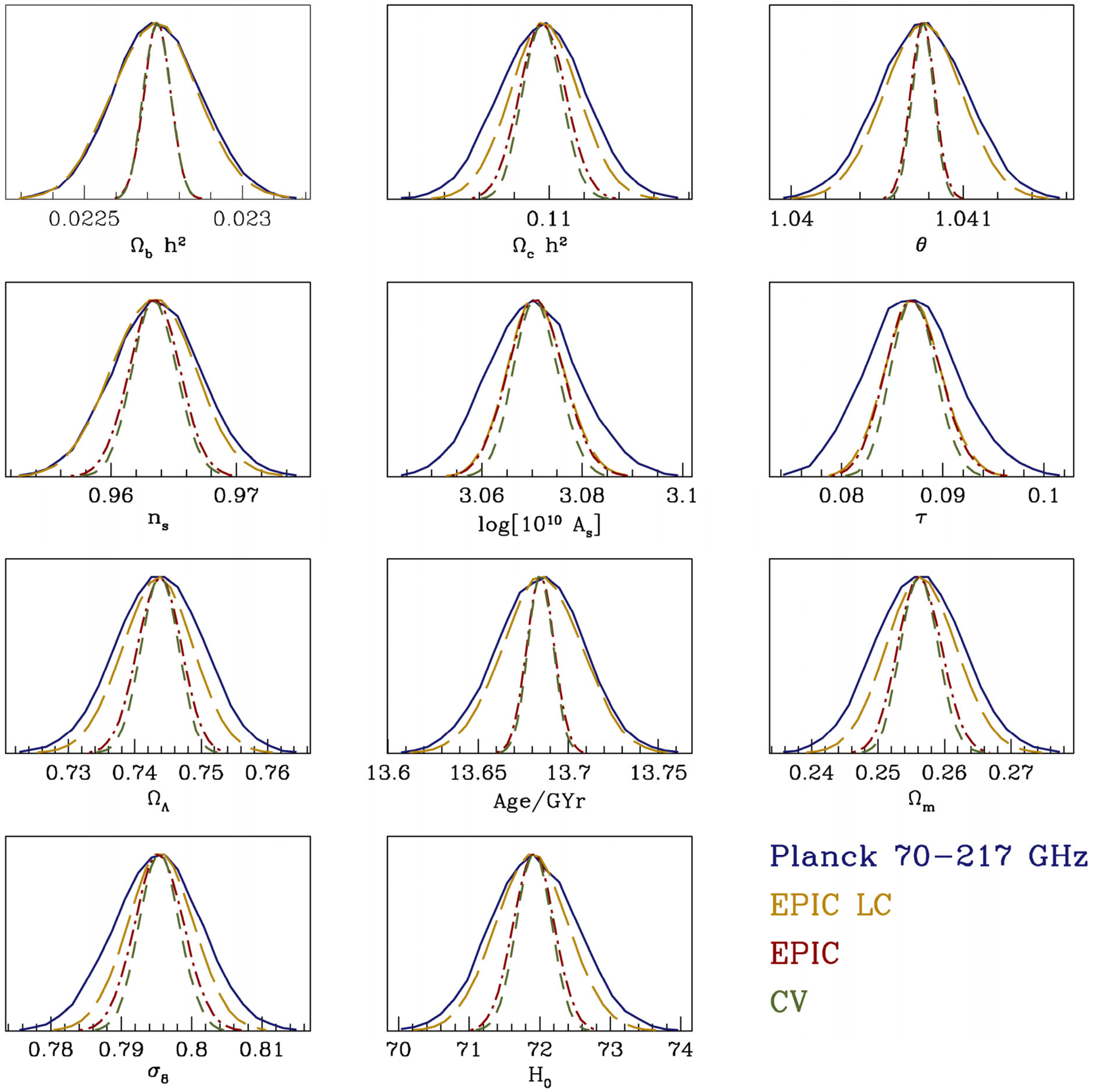

Fig. 1.3. Marginalized distributions of various cosmological parameters showing the expected uncertainty from Planck (using channels between 70 and 217 GHz) [35], EPIC *Low Cost* version [18], and EPIC-IM. For reference, we also calculate the uncertainty in a cosmic variance limited experiment out to $\ell$ = 2500 using all information in CMB temperature and polarization. The mocks analyzed for this exercise are centered on the WMAP 5-year ΛCDM best-fit cosmology with r = 0.01. The analysis assumes $f_{sky}$=0.8 and makes use of isotropic instrumental noise for each of Planck and two versions of EPIC. While EPIC-LC improves over Planck uncertainties for cosmological parameters listed above, new version of EPIC presented in this report reach the cosmic variance limit of all the parameters listed above within 5 % of the error.

<u>Sub-horizon and secondary tensors</u>: In addition to the primordial and super-horizon tensor amplitude detectable through CMB B-mode polarization, an energy density in a background of stochastic gravitational waves with frequencies greater than $10^{-15}$ Hz present during decoupling of the CMB can be probed through modifications to the expansion rate and growth of structure as seen by CMB temperature and E-mode anisotropies [44]. While with WMAP data, the 2σ limit for the energy density of gravitational waves $\Omega_{GW}h^2$ is at $6.9 \times 10^{-6}$ EPIC will improve this limit by roughly an order of magnitude down to $6 \times 10^{-7}$ at the same confidence level.

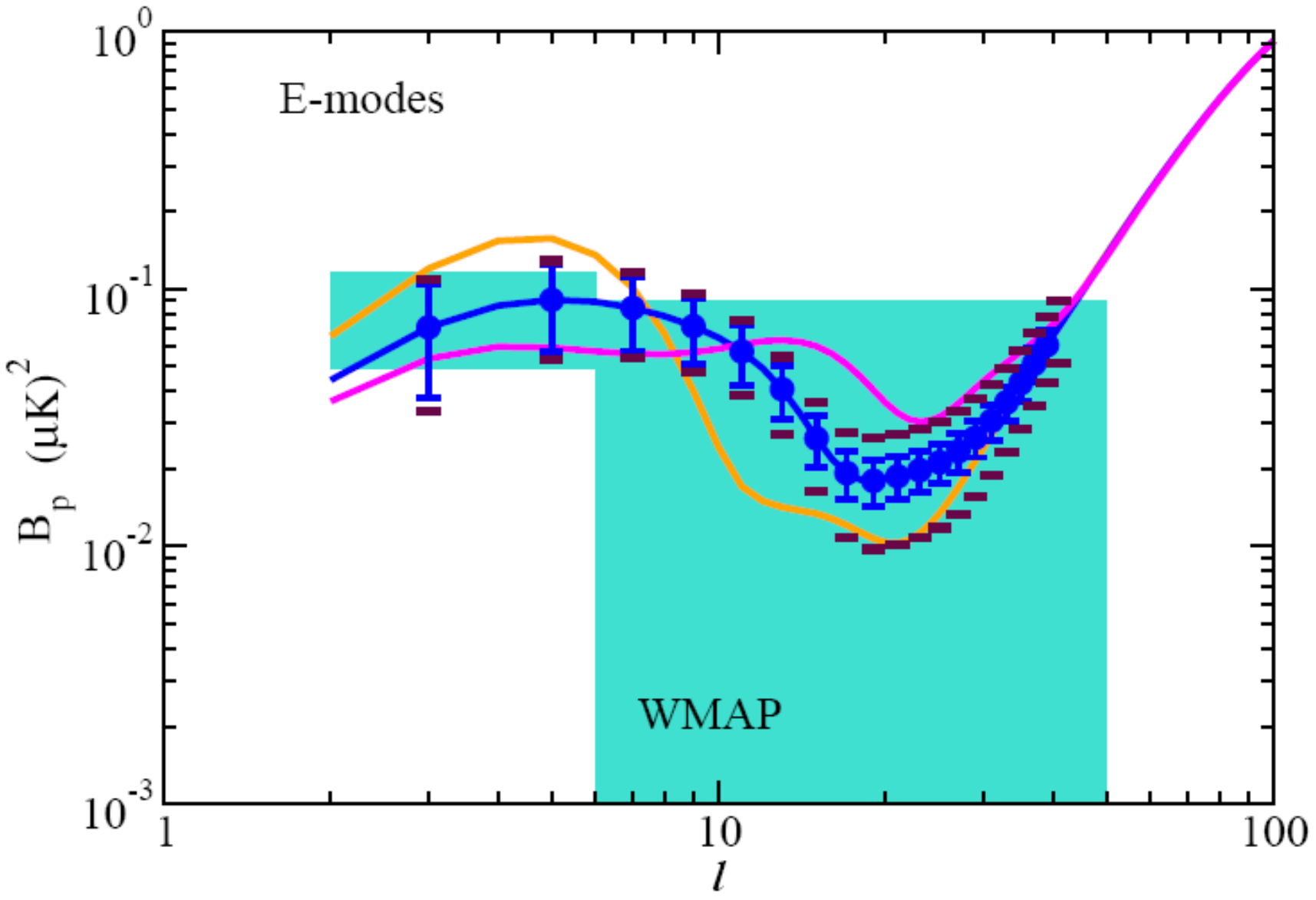

Fig. 1.4 The low-multipole (l< 100) region of the CMB polarization E-mode angular power spectrum with a bump related to reionization such that the ionization fraction of electrons has variations at two redshifts with complete reionization at a redshift of 6.3 consistent with SDSS [48]. Between a redshift of 6.3 and $z_{ri}$, the ionization fraction has a varying value such that the total optical depth is still normalized to 0.1 consistent with WMAP 5-year data [15]. Between 10 < l < 40, from bottom, middle, and top curves are for $z_{ri}$ = 13, 30, and 50, respectively. The error bars show the cosmic variance limited errors in the E-mode spectrum possible with EPIC, while extended errors marked by horizontal lines show the errors expected from Planck. With the cosmic variance limited measurements available with EPIC, one can establish additional details of the reionization process beyond the integrated optical depth [46-47].

### *1.3.2 Reionization*

The recent WMAP report of a large-angle polarization excess has indicated the possibility of reionization at a redshift of 11.0 ± 1.4 at the 68% confidence level [15]. There is still a large uncertainty, however, on the exact reionization history of the Universe. None of the ground-based experiments will improve this result to the limit allowed by foregrounds as observations will have limited sky coverage. The reionization signal in polarization is best studied with all-sky observations. EPIC will extract all information on reionization encapsulated in CMB at the cosmic-variance limit of the E-mode spectrum. The measurements will have enough precision to address questions such as whether the Universe reionized once or twice [45] (Fig. 1.4). Moreover, a model-independent study of reionization signature at large-scale CMB polarization has shown that five principle components of the reionization history with redshift can be achieved with cosmic-variance limited data [46]. Using TE cross-correlation and E-mode correlations, EPIC will conduct the definitive study on reionization with CMB polarization [47].

*1.3.3 Cosmic Shear*

Sum of the Neutrino masses: The tens of arcminute and finer angular scale CMB temperature and polarization anisotropy provide a unique probe of the integrated mass distribution along the line of sight through lensing modification of the anisotropy structure [49-50]. This secondary lensing signal can be studied through higher-order statistics leading to a direct measurement of the integrated mass power spectrum [51-53]. In combination the power spectrum provides a measure of the neutrino mass since massive neutrinos affect the formation of small-scale structure [54-55]. In Fig. 1.5 we summarize our results related to neutrino masses. While existing cosmological studies limit the sum neutrino masses to be below about 0.28 eV (95% CL) [56], a combination of CMB lensing studies with Planck can be used to probe a sum of the neutrino masses down to 0.15 eV (95% CL). For the LCDM cosmological model, EPIC reaches a sum of the neutrino masses of 0.042 eV and if the equation of state of dark energy is allowed to vary this constraint is at 0.047eV at the same 95% confidence level.

These cosmological results expected from EPIC can be put in the context of neutrino experiments motivated by the particle physics side. The existing neutrino oscillation experimental data fix the difference of neutrino mass squared between two states and for solar and atmospheric neutrinos the mass squared differences are $2.5\text{x}10^{-3}$ $eV^2$ [57] and $8\text{x}10^{-5}$ $eV^2$ [58], respectively. When combined, these estimates of mass-squared differences lead to two potential mass hierarchies shown in the inset of Fig.1.5 [60].

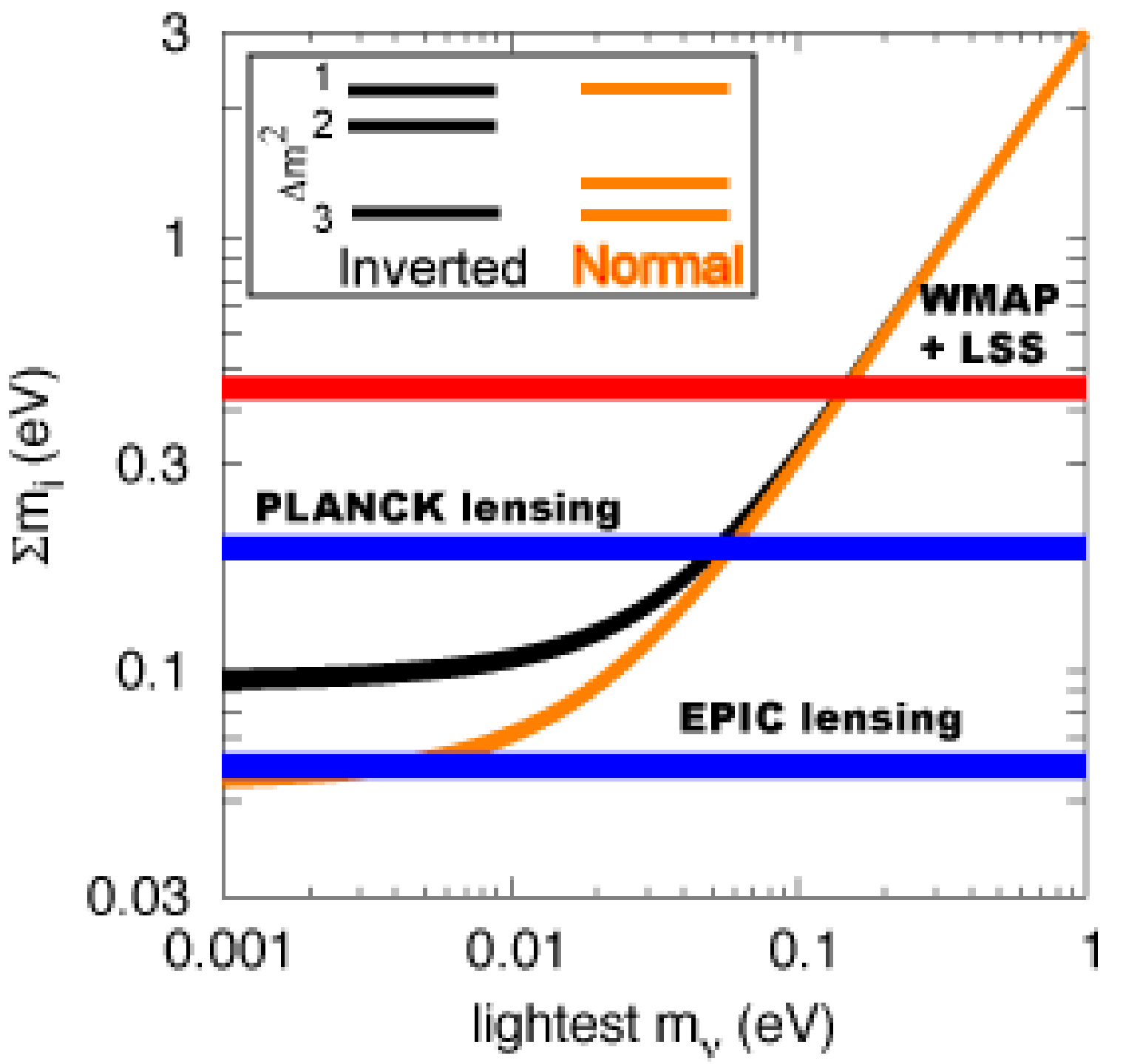

Fig. 1.5 The sum of neutrino masses as a function of the lightest neutrino mass by making use of atmospheric and solar neutrino oscillation data [57-58]. The two lines show the relation between sum of the neutrino masses and the mass of the lightest neutrino for the two possible mass hierarchies [59]. The horizontal lines show the limits reached with existing data (top line from [15]) and 95% confidence level limits reachable with EPIC either in terms of the two low-resolution and high-resolution versions. This figure was adapted from Ref. [60].

In the case of the inverted mass hierarchy, the sum of three neutrino masses is expected to be at minimum about 0.1 eV. Such a neutrino mass sum is detected at more than 5σ with lensing measurements in EPIC. While EPIC lensing measurements allow us to distinguish between the normal and inverted mass hierarchies at significance greater than 5σ, a total mass measurement below 0.05 eV also allows us to establish the full mass spectrum.[3]

Cosmological Modifications to General Relativity: Beyond the sum of neutrino masses, cosmic shear in CMB can also be used for several other cosmological tests. For example, there is a possibility that the observed cosmic acceleration results from a new theory of gravity at cosmological length scales. While a compelling underlying theory is still lacking in the community, CMB polarization data can be used for a consistency test related to General Relativity [61-62]. The gravitational potentials are equal in the presence of non-relativistic stress-energy under GR, but alternate theories of gravity make no such guarantee and a slip between the two is expected such that $\phi \neq \psi$ in the presence of non-relativistic stress-energy.

To perform a test of General Relativity at cosmological length scales, we make use of the Post-Parameterized General Relativity description for cosmological perturbations with $\psi=[1+\varpi(z)]\phi$ [63]. Here, when $\varpi(z)=0$ we recover the standard calculation for perturbations under General Relativity. With EPIC we can make use of both the large-scale ISW effect in temperature and cosmic shear in CMB polarization to constrain $\varpi(z)$. In the case of Planck data alone, the constraint is weak but Planck combined with a half-sky galaxy shear survey constrain $\varpi(z)$ with an error of 0.13 at the 95% confidence level. On the other hand, the integrated potential power spectrum extracted from CMB lensing information in EPIC polarization data, when combined with the EPIC temperature anisotropy spectrum, can constrain $\varpi(z)$ to a similar accuracy again at the 95% confidence level [64]. While competitive, when compared to the combination of Planck and a large area galaxy shear survey, the test with EPIC alone is also expected to be clean given that the test makes use of a single dataset.

Early Dark Energy Density: As CMB lensing integrates the project potential out to the last scattering surface, it is more sensitive to an early dark energy component in the Universe at redshifts of ~3 and above than its sensitivity to dark energy at late times. In the ΛCDM cosmological model, the fractional energy density contribution from the dark energy (in this case the cosmological constant) is about $10^{-9}$ at redshifts close to the last scattering surface. Alternative models of dark energy can contribute at percent level [65], especially if the dark energy were to trace the energy density of the dominant component of the Universe at a given epoch. Unfortunately, any changes to the sound horizon as probed by CMB will remain degenerate with changes to other cosmological parameters. CMB lensing, however, provides an independent way to constrain the presence of such an early dark energy component. In fact, unlike other cosmological probes that study the density and evolution of dark energy at redshifts less than 2, cosmic shear of CMB is one of the few avenues to study the presence of an early dark energy component in the Universe. Estimates making use of the accuracy of lensing measurements with EPIC suggest that an early dark energy component can be measured down to a fractional energy density of 0.002.

[3] In comparison the neutrino-less double beta decay experiments planned in a deep underground laboratory with source masses at the 1-ton scale are able to measure a different parameter set of the neutrino mixing matrix and reach a sensitivity to the sum of neutrino masses around 0.1 eV.

*1.3.4 Secondary anisotropy and polarization*

The resolution provided by EPIC will open up temperature and polarization maps for a variety of studies related to secondary polarization signals from the large-scale structure. In the case of temperature maps alone, EPIC will improve cluster detection through the Sunyaev-Zel'dovich (SZ) effect relative to the cluster catalog in Planck given the improvement in noise by at least an order of magnitude. This will allow detection of clusters with total mass above a few times $10^{14}$ solar mass over essentially 70% or more of the sky unobstructed by the Galaxy

While the homogenous reionization signal peaks at large angular scales, density or ionization fraction modulation of reionization will lead to an additional polarization signal at small angular scales. Moreover, scattering of the temperature anisotropy quadrupole by electrons in galaxy clusters will generate another polarization signal. The cluster locations can be identified based on SZ detections in the temperature map and the cluster polarization detection can be optimized through known locations and depths of the SZ signal.

By averaging over large samples of clusters, one can determine the CMB quadrupole projected at various cluster locations [66]. The evolution of the mean cluster polarization with redshift reflects the growth of the quadrupole, from the integrated Sachs-Wolfe effect, and this depends on dark energy properties [67]. This measurement will enable a unique measurement of the dark energy equation of state.

In addition to non-Gaussianity associated with primary anisotropy, a large number of secondary effects in CMB data will generate non-Gaussian signals, especially at small angular scales that will be probed with EPIC [68]. These signals provide information related to the growth of structures as well as cosmology and astrophysics during the reionization era and later.

*1.3.5 Galactic Science*

Dust polarization maps, which have best sensitivity and resolution in the highest frequency channels of EPIC, are of great interest to astronomers who study the Galactic interstellar medium and magnetic field. Along with gravity and gas pressure, the interstellar magnetic field is one of the three major forces acting on interstellar gas. Although key to understanding interstellar medium physics, our current data on Galactic magnetic fields (from Zeeman splitting, synchrotron polarization, Faraday rotation, dust absorption and emission polarization) is quite limited. Sparse measurements of the line-of-sight magnetic field strength indicate that the energy density of static magnetic fields is significant and even dominant in some cases (Fig. 1.6). Widespread, arcminute-resolution surveys of linear polarization with EPIC will revolutionize our understanding of the interstellar magnetic field on 0.03 – 3000 pc length scales.

Theoretical and phenomenological tools for precision application of dust polarization mapping are emerging. Our understanding of the underlying physics of grain alignment has made accelerating progress [70]. By the time EPIC flies, grain alignment theory will have been subjected to more stringent observational tests – in particular, the effects of environment – via polarization surveys at optical through millimeter wavelengths with modest sensitivity gains and greater coverage compared to present-day results. An important conclusion of existing observational work is that dust grain alignment is pervasive, from diffuse gas [71] to dense cores, with and without embedded stars [72-73].

Significant theoretical effort has gone into understanding the applicability and accuracy of estimating magnetic field strengths from the fine structure in the magnetic field, first applied to diffuse gas by Chandrasekhar and Fermi [74] and Davis [75]. The basic conclusion that B ~

$\rho^{1/2}$ $\Delta v/\Delta\theta$ still holds for dense, more complex clouds[4], and a numerical prefactor of order unity can be estimated from MHD simulations with good accuracy [76-79].

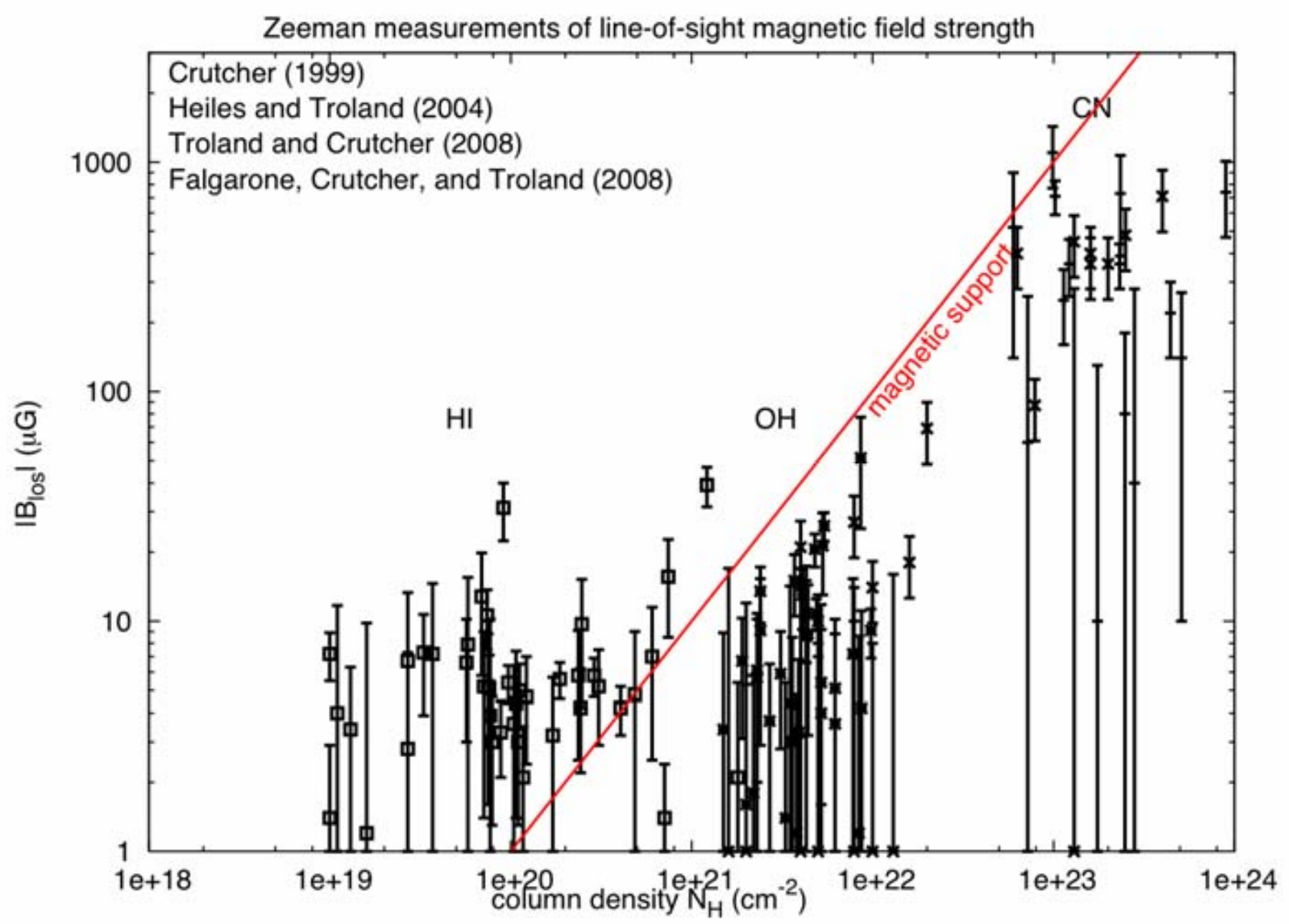

Fig. 1.6. Compilation of measurements of magnetic field strength along various sight lines through the Galactic interstellar medium, ranging from diffuse HI gas to dense molecular cores (based on [69]). The red line shows the case of magnetic criticality; clouds with measurements on or above the line are supported against gravitational collapse on the basis of static magnetic fields alone. Magneto-hydrodynamic waves provide additional support, not captured in the plot. EPIC is sensitive to magnetic field structures in gas at column densities of $N_H = 10^{19}$ $cm^{-2}$ and above, which includes the full range over which typical diffuse clouds apparently lose their magnetic support and form cores and stars. Using the Chandrasekhar-Fermi technique, EPIC can infer magnetic field strengths in as many as one million fields over the sky (see text below).

EPIC will provide the first all-sky polarization survey allowing us to trace magnetic fields from diffuse HI clouds into dense molecular cores (Fig. 1.7); it is the role of magnetic fields in the accumulation of gas clouds and star-forming cores which is the primary high-frequency science goal for EPIC. We will apply two basic approaches. The first relates the magnetic field geometry to the density and/or velocity structure. Basic questions include: Does the magnetic field direction correlate with the direction of disk- and filament-like features in the ISM? How does the mass-to-magnetic-flux ratio change from the center to the edge of clouds? Which models for molecular cloud formation most resemble the density, velocity, and magnetic field structures that we observe? The second approach utilizes the Chandrasekhar-Fermi method to infer magnetic field strengths in essentially all resolved clouds in the Galaxy. Interesting questions include: Can we verify the overall trend of magnetic field strength vs. gas density, as deduced from the radio Zeeman measurements (Fig. 1.6)? Given our much larger sample, we will study evolutionary effects. For example, is the degree of magnetic support of a cloud correlated with the presence or absence of protostars?

[4] The measured quantities are the gas density $\rho$, the velocity dispersion $\Delta v$, and the dispersion in polarization angle $\Delta\theta$.

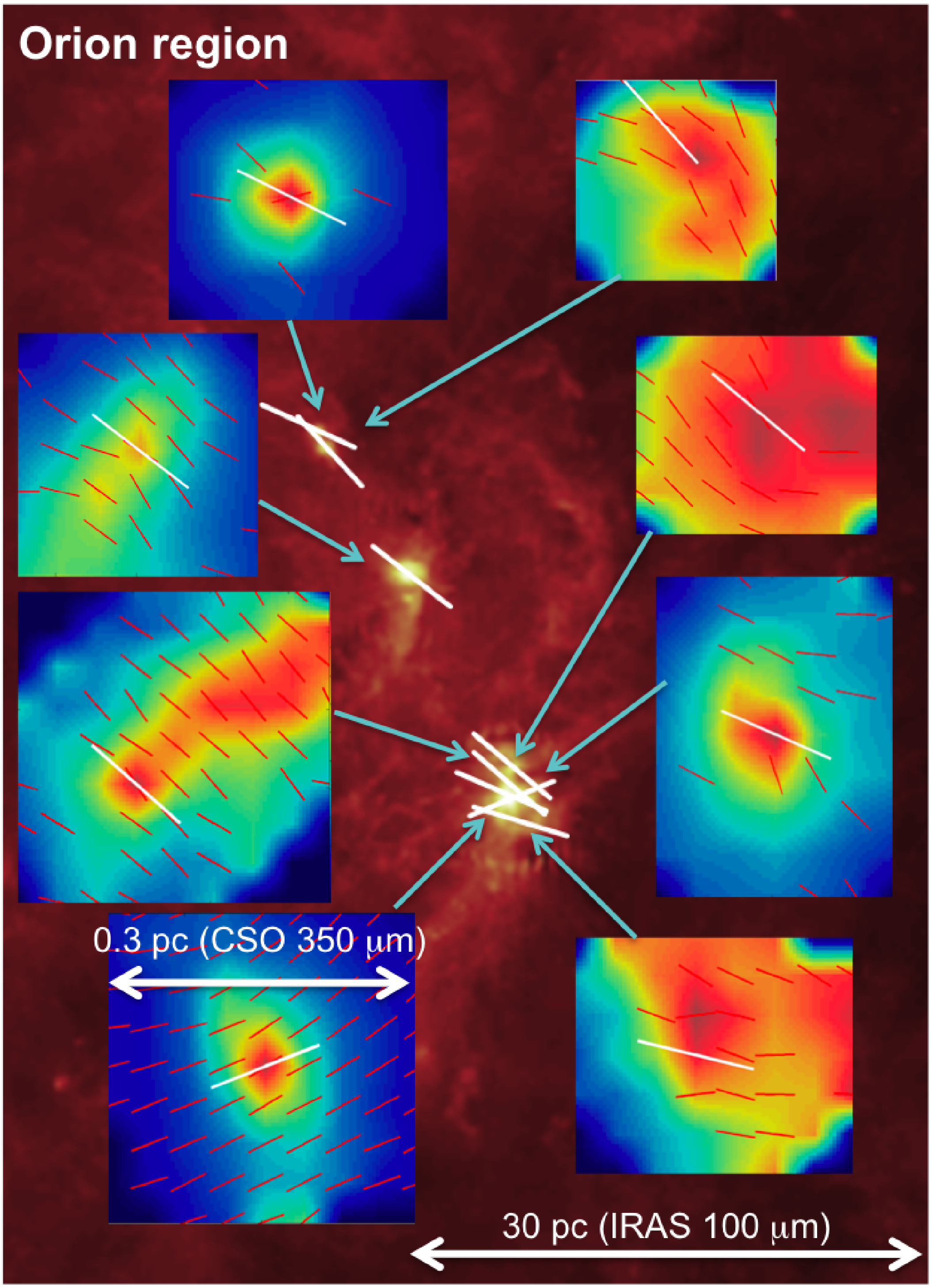

Fig. 1.7. Magnetic field direction maps of eight dense cores in the Orion region[83]. The red vectors show ground-based submillimeter measurements, which have been averaged for each core to produce the white vectors. Those results are superposed on an IRAS image showing the cores as well as the diffuse emission which can not be mapped well from the ground. The cores have a preferred (not random) magnetic field orientation, which also matches the average direction of optical polarization of surrounding lines of sight, implying a large-scale process organizing this entire region. Only sensitive space polarimetry with arcminute resolution (such as with EPIC) will reveal the full structure of the magnetic field in interstellar complexes like Orion.

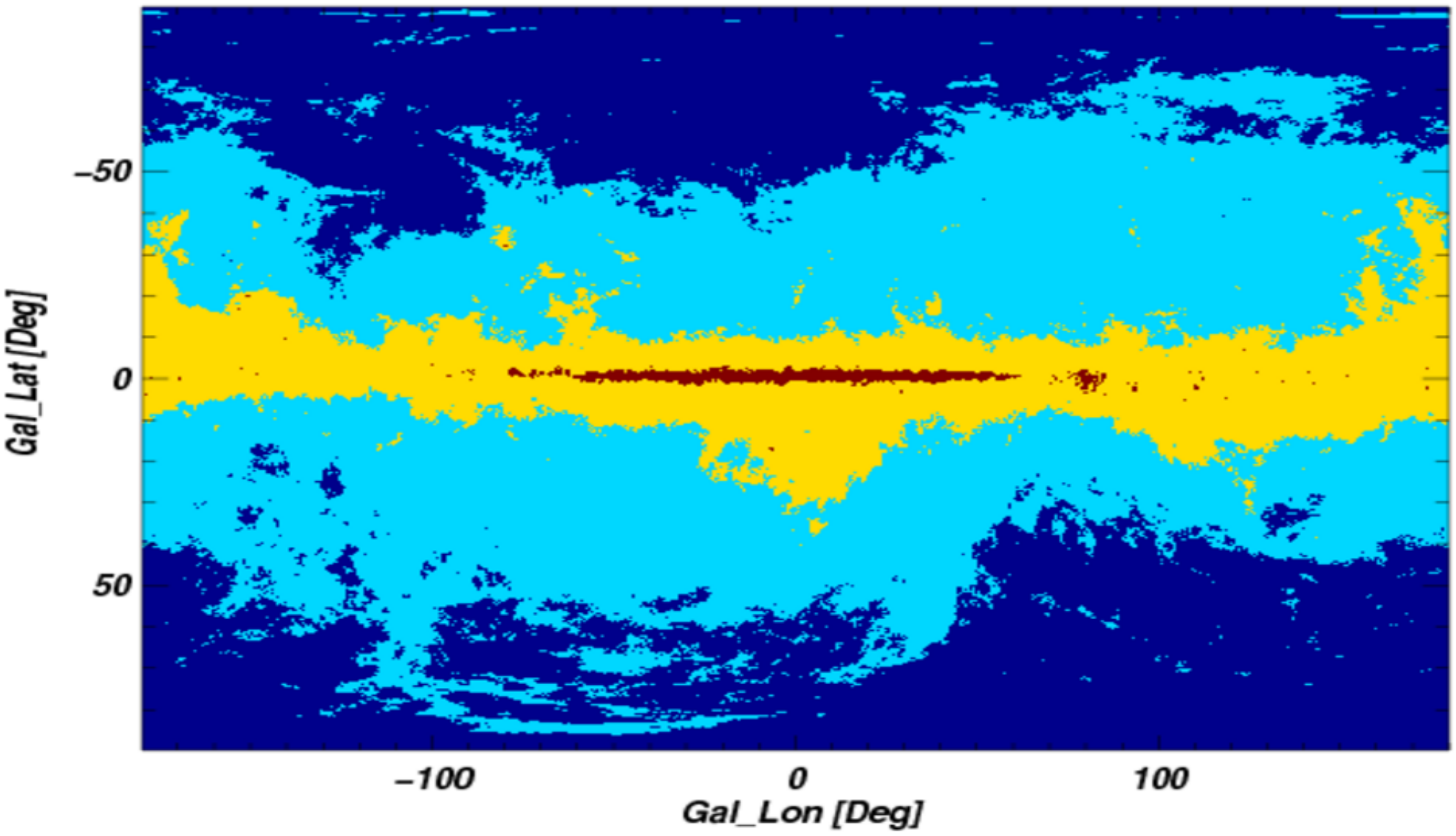

Fig. 1.8. Illustration of relative sensitivity of high-frequency dust polarimetry of EPIC vs. Planck. Planck will make accurate ($\sigma_P < 0.3\%$), 5′-resolution polarization measurements of the Galactic plane at 350 GHz, illustrated in red on the full-sky map above. With a 40 K telescope, EPIC at 850 GHz can significantly improve the sky coverage with accurate polarimetry (yellow), when smoothed to the resolution of Planck. With a 4 K telescope, no smoothing is necessary, and more than half the sky can be mapped accurately in polarization with 1′ resolution (cyan).

**Table 1.2.** Sensitivity of EPIC and Planck to Polarization of Galactic dust.

| Mission | Band | Angular Resolution arcmin FWHM | $\sigma(Q)$ kJy/sr/beam | polarization depth $A_V$ |
|---|---|---|---|---|
| Planck | 350 GHz | 5 | 24 | 4 |
| EPIC | 500 GHz | 2 | 0.9 | 0.06 |
| EPIC | 850 GHz | 1 | 0.7 | 0.01 |

The right column shows the minimum column density of dust, measured in visual wavelength optical depth $A_V$, for which a mission makes accurate polarization measurements.

Assuming a sufficiently cold telescope, pushing to the highest possible frequency afforded by EPIC brings enormous gains in sensitivity to dust due to the steeply rising spectrum. We propose an upper band at 850 GHz, which will offer 5 times better angular resolution and probe 100 times deeper, per beam, than the 350 GHz polarization survey of Planck (Fig. 1.8 and Table 1.2).

The 1' resolution of EPIC at 850 GHz is sufficient to take an important first look at the magnetic field structure in the neutral ISM of nearby spiral galaxies (Fig. 1.9). For the Andromeda galaxy, 200 pc resolution can be achieved, enough to resolve spiral arms and, e.g., compare field structure on opposite sides of density waves. Nearby face-on and inclined galaxies offer a favorable perspective on large-scale magnetic structure that is difficult to deduce for our own Galaxy.

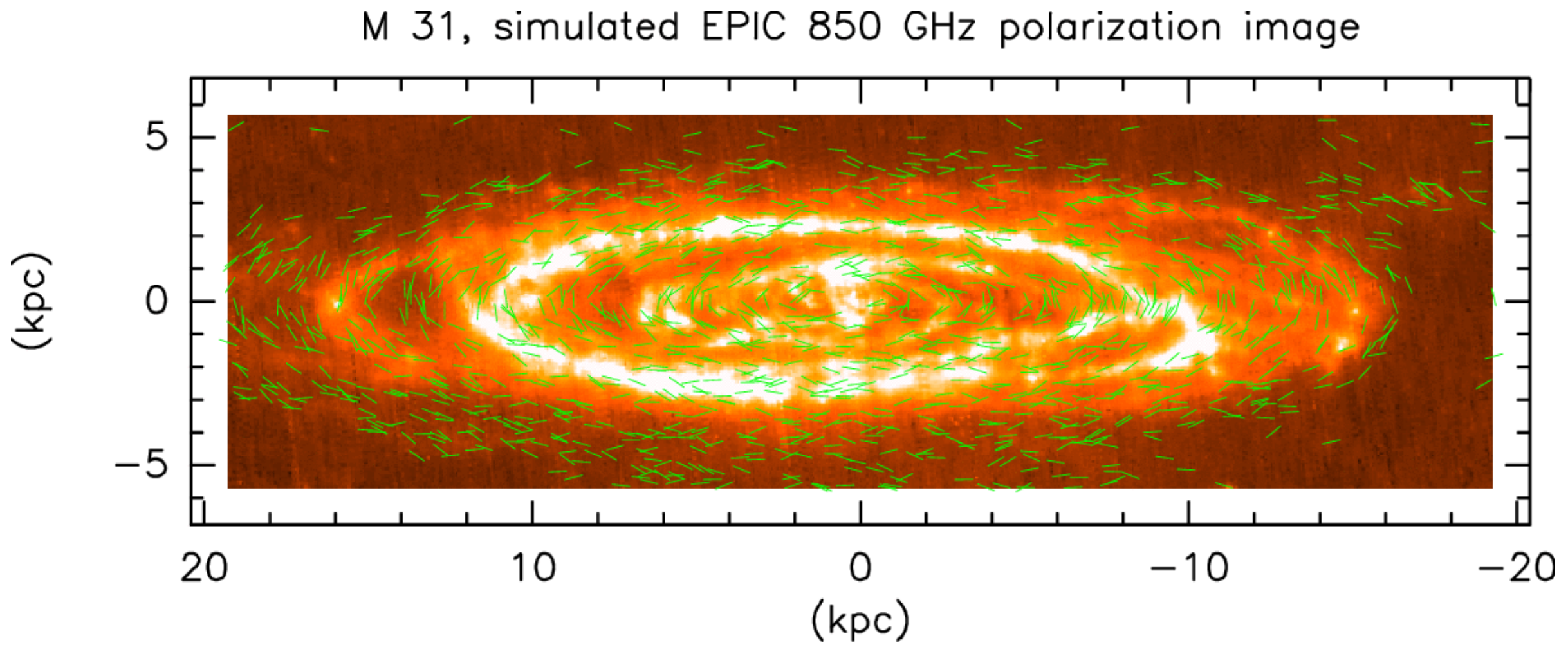

Fig. 1.9. Simulated dust polarization map of the Andromeda Galaxy with EPIC. The color scale shows Spitzer 160 μm emission [80], which has about the same resolution as EPIC at 850 GHz. Randomized green magnetic field vectors are shown where EPIC has enough sensitivity to detect polarization with accuracy $\sigma_P < 0.3\%$; for clarity, only one vector per 5 resolution elements is shown. No instrument in the field or in development has the sensitivity and resolution to map dust emission polarization for other spiral galaxies, until EPIC.

### 1.4 Angular Resolution and Sensitivity Requirements

The cosmic shear, scalar polarization and interstellar magnetic field science themes are especially dependent on the choice of angular resolution. While these themes are important, and robustly predicted by standard cosmology, they are outside the main science goal, namely to probe the IGW B-mode signal to at least r = 0.01. Though a deep search for IGW B-mode polarization, at least until confusion with cosmic shear B-mode polarization becomes problematic [81-82], does not require high angular resolution, a search for IGWs below 0.01 and for important secondary science goals centered around cosmic shear, non-Gaussianity, and scalar fluctuations, we have considered a higher resolution mission option than the *Low-Cost* mission that was considered in the previous EPIC concept study [18].

The new intermediate design of EPIC is optimized to carry out 1) a deep search for IGW B-mode polarization after lensing subtraction, and 2) the full secondary science themes described in section 1.3. We calculate the sensitivity and resolution required for a clear detection of sum of the neutrino masses down to 0.05 eV with lensing B-modes making use of a lensing potential power spectrum extraction using higher-order statistics from temperature anisotropy. For the lensing reconstruction, the signal-to-noise ratio for the lensing power spectrum detection is a function of both sensitivity and resolution. Most of the lensing detection is realized for a sensitivity $w_p^{-1/2}$ = 1-2 μK-arcmin and a resolution of ~6 arcmin, which allows a cosmic variance limited detection of lensing B-modes out to $\ell \sim 1800$ (Fig. 1.10).

**Table 1.3** Mapping NASA Objectives to EPIC Instrument Requirements

<table>
<tr><th>NASA Research Objective*</th><th>NASA Targeted Outcome*</th><th>Measurement Criteria</th><th>Instrument Criteria</th></tr>
<tr><td rowspan="3">What are the origin, evolution, and fate of the universe?</td><td>Test the Inflation hypothesis of the Big Bang</td><td>Measure inflationary B-mode power spectrum to astrophysical limits for $2 < \ell < 200$ at $r = 0.01$ after foreground removal</td><td>• All-sky coverage<br>• $w_p^{-1/2} < 6$ μK-arcmin<br>• 30 – 300 GHz<br>• 1° resolution<br>• Control systematic errors below $r = 0.01$</td></tr>
<tr><td>Precisely determine the cosmological parameters governing the evolution of the universe</td><td>Measure EE to cosmic variance into the Silk damping tail</td><td>• 10' resolution</td></tr>
<tr><td>Improve our knowledge of dark energy, the mysterious cosmic energy that will determine the fate of the universe</td><td rowspan="3">Measure lensing BB to cosmic limits to probe dark energy equation of state and dark matter</td><td rowspan="3">• $w_p^{-1/2} < 3$ μK-arcmin<br>• 6' resolution</td></tr>
<tr><td rowspan="4">How do planets, stars, galaxies and cosmic structures come into being?</td><td>Investigate the seeds of cosmic structure in the cosmic microwave background</td></tr>
<tr><td>Measure the distribution of dark matter in the universe</td></tr>
<tr><td>Determine the mechanism(s) by which most of the matter of the universe became reionized</td><td>Measure EE to cosmic variance to distinguish reionization histories</td><td>• Primary mission parameters above</td></tr>
<tr><td>Study the birth of stellar and planetary systems</td><td>Map Galactic magnetic fields via dust polarization</td><td>• 500 and 850 GHz bands</td></tr>
</table>

*Taken from NASA 2007 Science Plan

Primary Objective

Secondary Objective

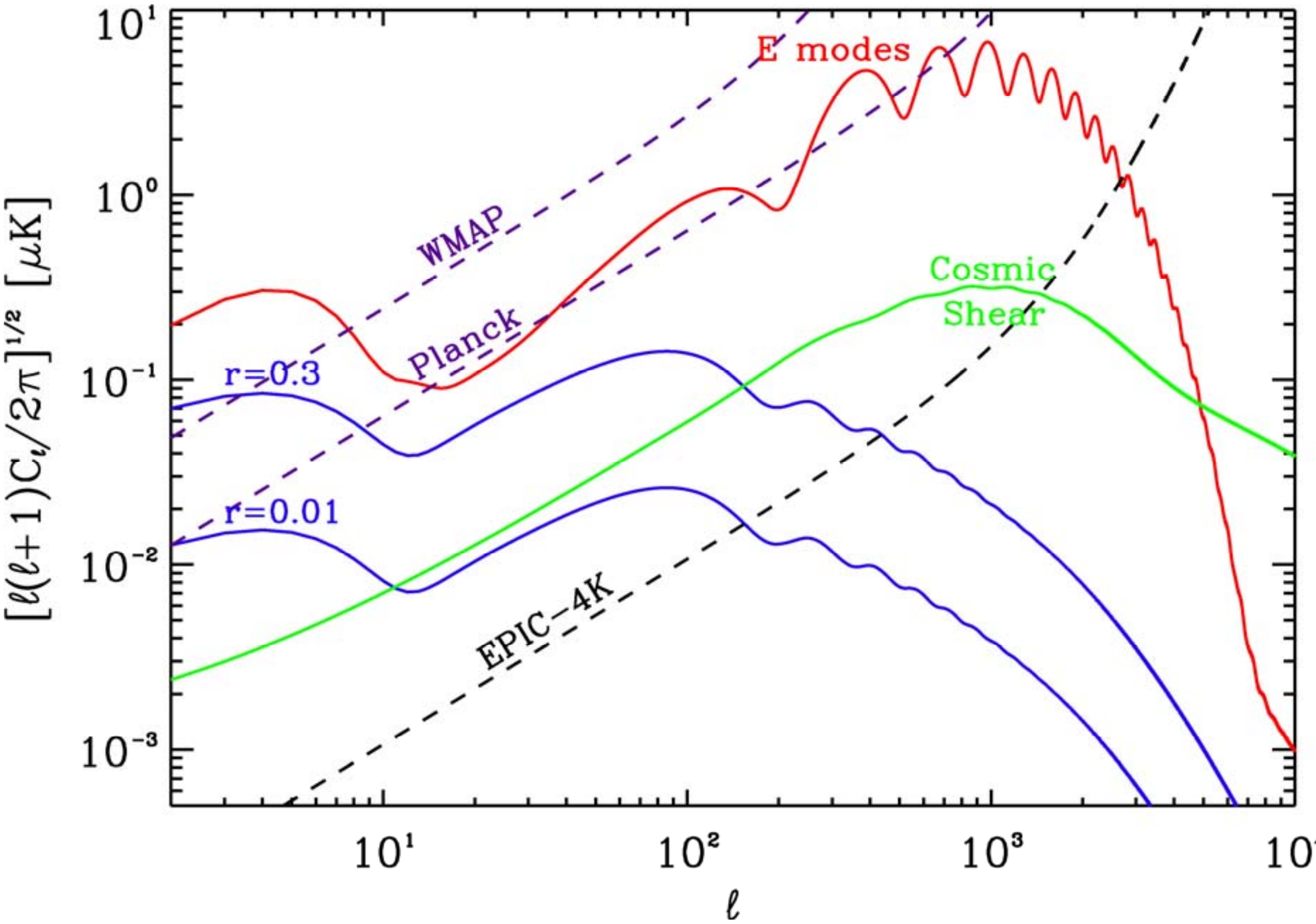

Fig. 1.10. The noise $C_\ell$s of EPIC-IM, WMAP and Planck, with polarization spectra the same as Fig.1.1. A comparison of these noise power spectra and the signals, such as primordial B-mode spectra shown in blue for r = 0.01 and r = 0.3 reveals the angular scale, or the multipole moment, where cosmic variance of primordial signals dominate the measurement. As shown, the detection of low-multipole reionization bump is dominated by cosmic variance while for low r models, the recombination bump at degree angular scales is the transition between noise domination to cosmic variance domination.

## 2. Foreground Removal

Polarized Galactic emission will likely set the practical limit to detecting primordial B-mode polarization. We have designed EPIC to have the best possible prospect of distinguishing the large angular scale E- and B-mode signals from Galactic emission.

There are two characteristic signals due to primordial B-modes. The first signature is due to rescattering of the primordial B-modes after reionization, and yields a peak at $\ell \approx 8$. The second, truly primordial, signature is a peak in the power spectrum at $\ell \approx 100$. The first signal is thus present on the largest scales while the second is present on small scales – of order 2˚. There are thus two very different regimes for estimating the foregrounds that may contaminate these signals. On large scales, Galactic emission is expected to be bright, roughly comparable to the largest expected IGW signal, and thus must be deeply subtracted. On degree scales however, one can restrict to very clean patches of sky, where the foregrounds may even be as faint as the r = 0.01 IGW B-mode signal, and still obtain sufficient cosmic variance precision to provide a good measurement.

As detailed models of foreground polarization and techniques to minimize foregrounds are discussed in Dunkley et al. [1], we present a summary of key methods and results here, focusing on the EPIC design. A comparison of foreground levels with the CMB signal is presented in Fig. 2.1.

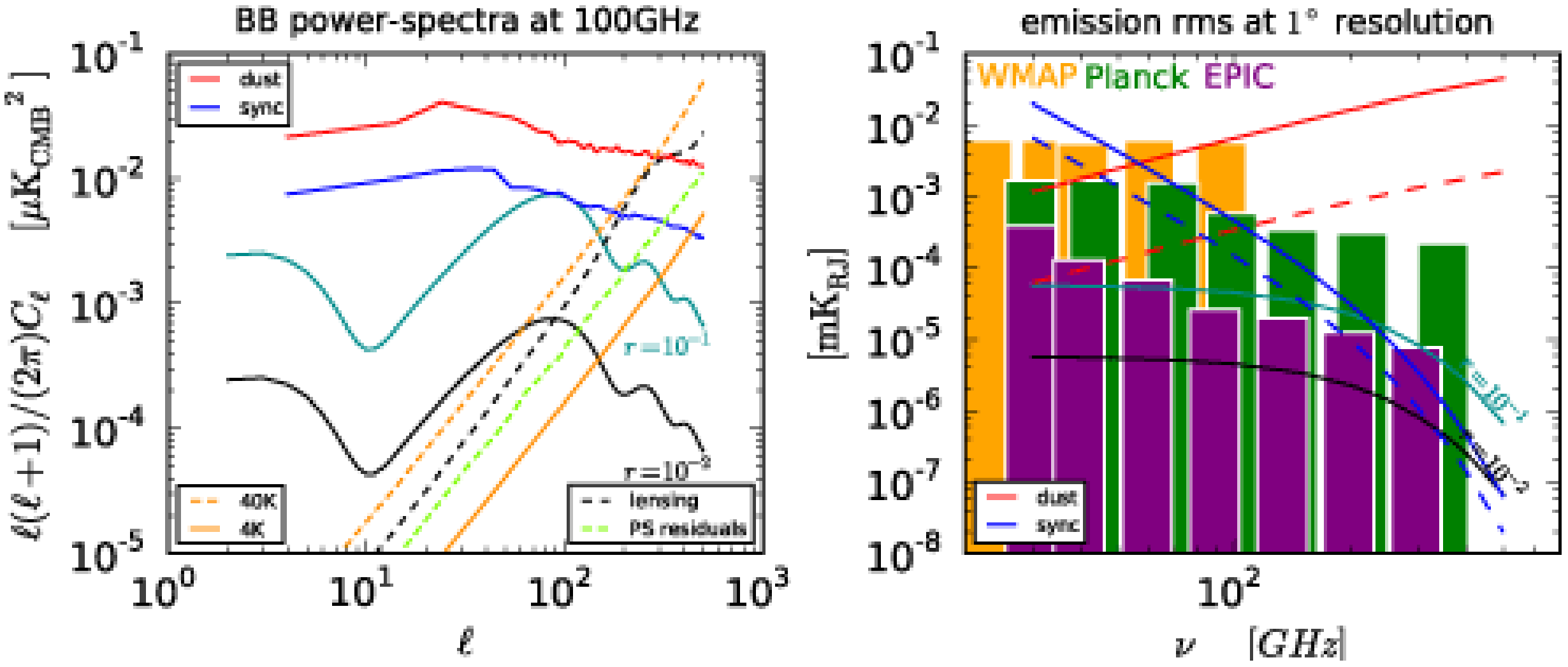

Fig. 2.1. Predicted emission levels based on the Planck Sky Model used for foreground calculations here. In the left panel, we show the B-mode polarization power spectra of dust (red) and synchrotron (blue) at 100 GHz, compared to CMB B-modes with r=0.1 (cyan) and r=0.01 (black solid) and lensing B-modes (orange). The power spectra of foregrounds are computed for the cleanest 55 % of the sky. The dust polarization fraction is assumed to be 12%. The power spectrum for the residual sources assumes all sources brighter than 500 mJy in temperature maps are masked. We also show the unbinned instrumental noise level from the two EPIC options, the 30 K and 4 K telescopes detailed in section 3. The right panel shows the typical frequency dependence of the foregrounds and CMB at pixels of one degree. The solid lines show the mean level using full sky, while dashed lines show the mean level of fluctuations in the cleanest 55 % region used for the power spectrum calculation in the left panel. Note that there is more variation in the dust than in the synchrotron, and that the minimum frequency changes depending on the region of sky. In bands we show the EPIC (30 K telescope option), WMAP, and Planck sensitivities, where the heights of the bands correspond to the per-pixel rms instrumental noise for pixels at one degree resolution. Figures adapted from [22].

## 2.1 Foregrounds Zoology

### *2.1.1 Synchrotron Emission*

Synchrotron radiation is emitted by electrons spiraling in galactic magnetic fields. The emission is an approximate power law in frequency with exponent (in Rayleigh-Jeans temperature) $\beta_s$ and polarization fraction $\Pi_s$, and both of these parameters may vary spatially. Geometrical projection suppresses polarization in a manner not necessarily correlated with $\beta_s$. WMAP provides the best measurement of polarized synchrotron over the full sky [8]. Investigating small 900 $deg^2$ fields in the multipole range $\ell = 30$ to 100, we observe a factor of 20 variation in synchrotron brightness. In the cleanest portions of the sky, we expect foreground levels as low as $\Delta T_{rms} \sim 60$ $nK_{RJ}$ at 100 GHz [1].

### *2.1.2 Thermal Dust Emission*

Dust grains emit blackbody radiation modified by a frequency-dependent emissivity, and becomes polarized because the grains preferentially align perpendicularly to magnetic fields. There is good evidence that the frequency dependence of the emissivity is not well described by a power law[5], and that the dust temperature is not described by a single temperature component.

Finkbeiner *et al.* [2] ('FDS') fit a two-component dust model to unpolarized IRAS, DIRBE, and FIRAS data. FDS interpret the components as silicate grains with $\langle T_1 \rangle = 9.4$ K and small (~10 nm) graphite grains with $\langle T_2 \rangle = 16.1$ K, although other materials cannot be ruled out. Although this model has some shortcomings, it does provide a useful starting point for modeling Galactic emission removal. Again for small fields of size 900 $deg^2$, we find significantly reduced dust emission levels in clean patches of the sky. A typical "clean" field has, according to the FDS maps, $\Delta T_{rms} \sim 10$ $nK_{RJ}$ at 100 GHz [1].

### *2.1.3 Dust "Exotic" Emission*

There are many claimed detections of spinning dust grains, which emit via rotational and vibrational modes and produce a spectral bump at tens of GHz with a non-negligible tail extending past 100 GHz [3]. The WMAP team currently uses the K-Ka template to remove synchrotron and spinning dust at the same time [4]. The expected spinning dust emission is only a few % polarized [5,8] as opposed to 50-75% for the synchrotron emission.

Another possible polarized microwave emission component may come from the magnetic dipole emission of thermally vibrating grains containing magnetic materials (e.g., iron) [7]. The polarized emission from these grains is predicted to reach 30 - 40 % at 100 GHz. Current models for this emission have large uncertainties such that current total emission measurements cannot rule out contributions from this mechanism. However, at the present time, correlations of the excess emission with H-alpha emission [22] and the 12 - 100 μm IRAS data [23, 24] clearly favor electric dipole emission from spinning dust over magnetic dipole emission from vibrating dust.

### *2.1.4 Sub-dominant Foregrounds*

Extragalactic radio and infrared compact sources are sufficiently diluted on the angular scales of interest that only the brightest sources need be removed. Tucci *et al.* [9] estimate that one can use the Planck compact source catalog to remove the brightest radio sources (> 200 mJy at 100 GHz) and leave a polarized contamination level of less than 10 nK at $\ell = 8$. Infrared point sources are expected to be largely unpolarized.

[5] For discussion of power law variations, see [2,8]

Free-free emission is intrinsically unpolarized, though the edges of HII regions may appear polarized via the same effect that gives rise to E-mode polarization of the CMB [10]. The effect would be small compared to other galactic emission.

## 2.2 Foreground Removal Strategies

We have investigated several scenarios for foreground removal, based on both pixel space and Fourier space techniques. The pixel-based technique is primarily insensitive to the amplitude of the foregrounds, but becomes less effective if the spectral indicies have significant spatial variation, because more free parameters are required for removal. The spectral technique only assumes that the CMB spectrum is precisely known, and removes components that do not match this spectral template. The spectral technique is insensitive to variations in spectral index, but degrades if the foregrounds are larger in amplitude.

### *2.2.1 Pixel-Based Foreground Removal*

The pixel-based foreground separation technique [11] uses a Bayseian formalism for parametric models of the individual foreground components using a MCMC algorithm to find the best-fitting parameters and errors for each pixel on the sky. We use a semi-realistic model for the sky and fit for only the 2 dominant foregrounds expected in polarization (synchrotron and thermal dust). We model the spectra as simple power-laws as a function of frequency, fitting for both the amplitude and spectral index simultaneously, along with the CMB amplitude for the Stokes parameters I, Q and U. For this investigation, we fit for a given foreground model, and evaluate errors after 1000 realizations of CMB and noise.

The sky model consists of CMB and 4 foreground components (synchrotron, free-free, thermal dust and spinning dust emissions). The amplitudes and spectra were chosen based on our current best knowledge of foreground emissions from recent works [12, 8]. The foregrounds are known to vary considerably from pixel-to-pixel on the sky and the details of each foreground component are still not well characterized, particularly in polarization [8]. One example is the assumption of that the synchrotron spectral index is constant with frequency. It is known to steepen with frequency due to spectral-ageing of the CR electrons [2]. However, most of the steepening is expected to occur at lower frequencies than those considered for EPIC (< 30 GHz). The polarization fractions are typical values expected at high Galactic latitudes and position angles for each component (i.e. distribution of Stokes Q and U) were given a random distribution in each realization. Little is known about the "anomalous dust" component, which emits strongly at frequencies < 60 GHz. For this study, we chose to use a typical spinning dust model [3] and assumed a relatively low polarization fraction as expected from spinning dust grains [5].

We then evaluated the removal of foregrounds assuming the band frequency coverage between 30 and 300 GHz. We estimate the residual uncertainty in our measurement of the CMB emission in each pixel after foreground signals have been removed using the multiple bands. We follow schematically the technique [11] of first fitting the nonlinear model parameters (power-law indices and dust temperature) on large pixels, then smoothing the nonlinear parameter fields spatially and fixing them when fitting for the amplitude components on smaller pixels. The technique is particularly sensible given that Galactic emission anisotropy power is primarily on large scales.

For brevity, we quote resulting uncertainty in the tensor-to-scalar ratio for r = 0.01 IGW signal in both the recombination and the reionization peaks for pixel-based methods and for other methods in Table 2.1.

**Table 2.1.** The forecast 1σ uncertainty in r for r=0.01 signal for a variety of methods

| **Method** | **Ave. Polarized Dust Fraction (%)** | **Description** | **Reionization Peak (ℓ < 15)** | **All Info (ℓ < 150)** |
|---|---|---|---|---|
| *EPIC-IM 30 K Option* | | | | |
| Pixel-based parameteric | 1 | Spectral indices fixed | $1.5x10^{-3}$ | $4.9x10^{-4}$ |
| Pixel-based parameteric | 1 | Power-law indices fitted | $2.5x10^{-3}$ | $6.2x10^{-4}$ |
| ILC | 5 | See text | $1.5x10^{-3}$ | $5.2x10^{-4}$ |
| SMICA | 4 | See text | $1.1x10^{-3(a)}$ | $3.2x10^{-4}$ |
| *EPIC-IM 4 K Option* | | | | |
| ILC | 5 | See text | $7.8x10^{-4(b)}$ | $1.9x10^{-4}$ |

Notes: Estimates based on the pixel-based parameteric fitting converts the residual foreground + instrumental noise in a pixel to an error on the tensor-to-scalar ratio through a Fisher matrix calculation. (a) For ℓ < 20; see Ref. [21]. (b) with lensing included as a noise; if lensing removed, $1.2x10^{-4}$.

*2.2.2 Fourier-Space Removal (ILC)*

In addition to pixel-based methods, foregrounds can also be removed in the harmonic space especially along the same manner that Tegmark *et al.* [14] used to produce a foreground-cleaned WMAP map (TOH map). We computed EPIC efficiency to remove foregrounds (we limited ourselves to the 2 dominant emissions: dust and synchrotron polarization), which combines optimally the $a_{lm}$ coefficients of the different frequencies to reduce the overall power spectrum while preserving the CMB signal:

$$a_{lm} = \sum_i w_l^i a_{lm}^i \qquad C_l = w_l^i C_l^{ij} w_l^j \qquad \sum_i w_l^i = 1$$

The CMB part of the correlation matrix $C^{ij}_\ell$ is determined with CAMB with standard cosmological parameters. The instrumental noise part of this matrix follows the NET and angular resolution values for the 30 K telescope option. The dust and synchrotron correlation is obtained through simulated maps of these emissions. We simulated these foreground maps as observed by EPIC between 30 and 300 GHz using data from WMAP at 23 GHz [4,8]. Assuming this channel is dominated by synchrotron emission (we reduced ℓ > 40 power to remove some noise), we extrapolated this map at higher frequencies using the software provided by the WOMBAT project to obtain our synchrotron maps. The WOMBAT project uses the spectral index β obtained from combining the Rhodes/HartRAO 2326 MHz survey [15], the Stockert 21cm radio continuum survey at 1420 MHz [16-17], and the all-sky 408 MHz survey [18].

In order to simulate the dust polarization, we assumed that the synchrotron signal is a good tracer of the galactic magnetic field and that the dust grains align very efficiently with this magnetic field. We used the synchrotron polarization angle to describe the dust polarization angle, consistent with the model presented by WMAP team [8]. For the intensity, we crudely assumed a constant overall polarization fraction of 5 % relative to the total dust intensity at a given frequency. Using this fraction, we used the model 8 interpolation [2] of the dust maps [19] to simulate the polarized dust emission over EPIC's frequency range [25]. Results for both 30K and 4K configurations of EPIC-IM are summarized in Table 2.1 and the estimated foreground residual for EPIC-IM 4K configuration is shown in Fig 2.2.

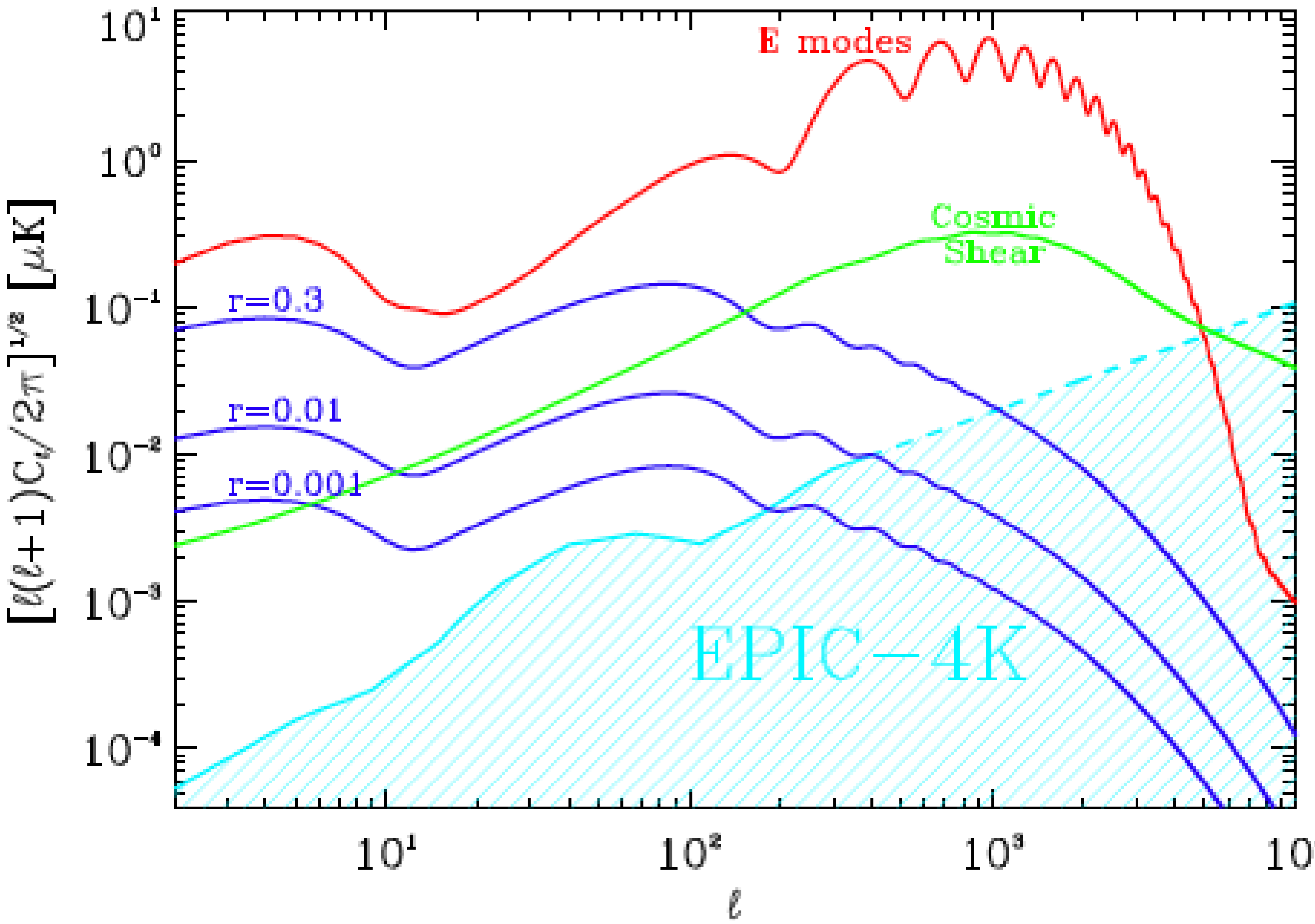

Fig 2.2 Estimated foreground residuals (shaded region) compared to primordial spectra of scalar polarization (red), CMB lensing (green), and B-modes with r=0.3, r=0.01, and r=0.001 (blue). The plot is for the EPIC-IM 4K configuration. The residuals are estimated using the Foruier-based cleaning techniques as described in Section 2.2.2.

### *2.2.3 Fourier-Space Removal (SMICA)*

An alternative approach to component separation in Fourier space comes from the SMICA framework [20]. An extension of this method, applied to B-mode spectra, can be used to estimate performances of current and future CMB experiments in determining the tensor-to-scalar ratio r [21].

The advantage of this method consists in the fact that it is able to recover the B-mode power spectrum in the presence of *unknown* foregrounds. The method minimizes the distance between the observed data covariance matrix and the sum of the covariance of the CMB and a pre-defined number of otherwise non-specified foregrounds. As for the CMB, the method assumes the blackbody frequency dependence and the knowledge of the spectrum's shape. The spectrum's amplitude is determined by the procedure, which automatically returns an estimate for the tensor-to-scalar ratio r.

In order to test performances, the method has been applied to simulated maps of the sky with different CMB B-mode levels and foregrounds as computed in the Planck Sky Model (PSM). An example of the diffuse foreground B-mode maps used in the simulations is reported in Fig. 2.3.

The main results are summarized in Table 2.1 where the effect of foregrounds and noise level for each experimental configuration is reported, together with the relative weight of small and large -scales in the total error budget for the tensor-to-scalar ratio. When only diffuse foregrounds are considered in the analysis, EPIC-30K provides measurements of r =0.01 (r=0.001) with a precision better than 27 (5) sigma. This result is not varied significantly by the inclusion of the lensing signal for the r=0.001 case. For tensor-to-scalar ratio values at or below 0.01, the uncertainty in the detection of r is dominated by foregrounds and the foreground residuals degrade the sensitivity by a factor of 3(2) for r=0.001 (r=0.01). However, increased instrumental performance, in terms of wide frequency coverage and spatial resolution are key elements for foreground subtraction, and EPIC is designed to maximize the sensitivity to foregrounds that are expected to dominate both at low and high frequencies. In the case of EPIC-IM 4K configuration, the errors on r determination for an input r value of 0.01 is 0.0002, 50 sigma, when no lensing is considered.

The constraining power moves from small scales to larger scale when r decreases down to the detection limit of the experiment, but no information on the CMB is derived when $\ell > 150$. Higher multipoles are still giving constraints on the foreground parameters, effectively improving the component separation also on large scales. It should be kept in mind that these results are obtained fixing all parameters bur r. Even in the absence of foregrounds, when all parameters are allowed to vary, the precision on r degrades by a factor 3 for r = 0.01 for both the EPIC-IM 30 K and 4 K configurations, with the main uncertainty coming from the possible variation of the tensor spectral index $n_t$.

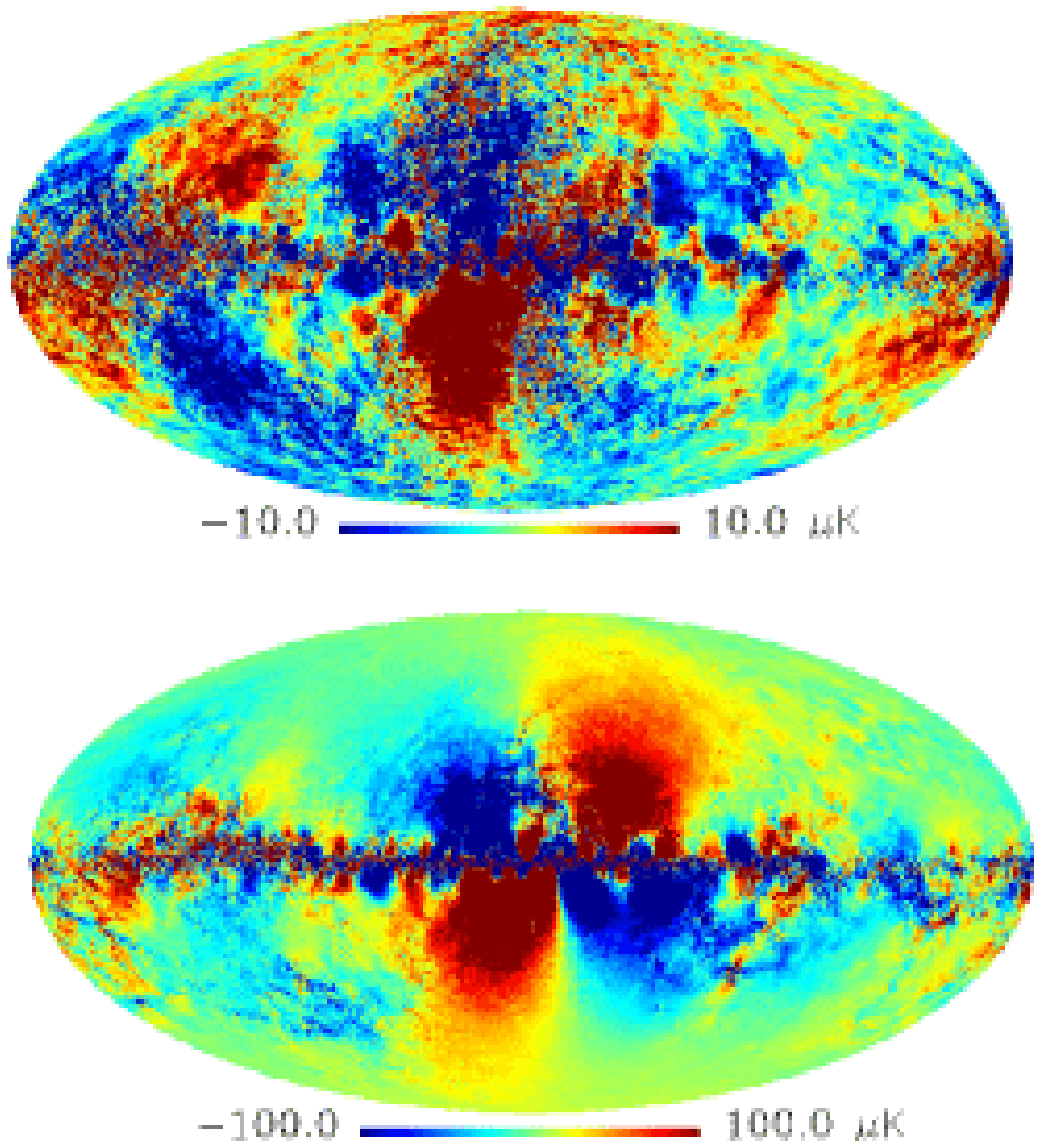

**Fig. 2.3.** B-modes of the Galactic foreground maps (synchrotron and dust) as simulated using V16.4 of the PSM. Top: synchrotron-dominated emission at 30 GHz. Bottom: dust-dominated emission at 340 GHz. From Betoule et al. [21].

While the PSM foreground model summarizes our current knowledge on the subject, when preparing for future missions it is worth also considering "worst case scenarios" within the currently allowed range of variability of foregrounds. Allowing for a running spectral index in the synchrotron component or multiplying the dust polarization level by a factor of two and increasing the dust small scale power from a nominal index of -2.5 to -1.9 do not imply a performance degradation for EPIC. The low noise level allows to adequately recover the foreground parameters and hence the tensor-to-scalar ratio r.

In addition to the diffuse foregrounds, unresolved point sources may cause problems in recovering the spectrum on small scales. With quite conservative assumption about the level of point source detection, it is found that unresolved point sources do not present a major problem for recovering small values of r (~0.001), as long as their statistical properties are known. For EPIC, point sources do not bias the results and increase the error bars by 3 %. Missing to model the point sources, however, may induce a strong bias and increase the error by 20 %.

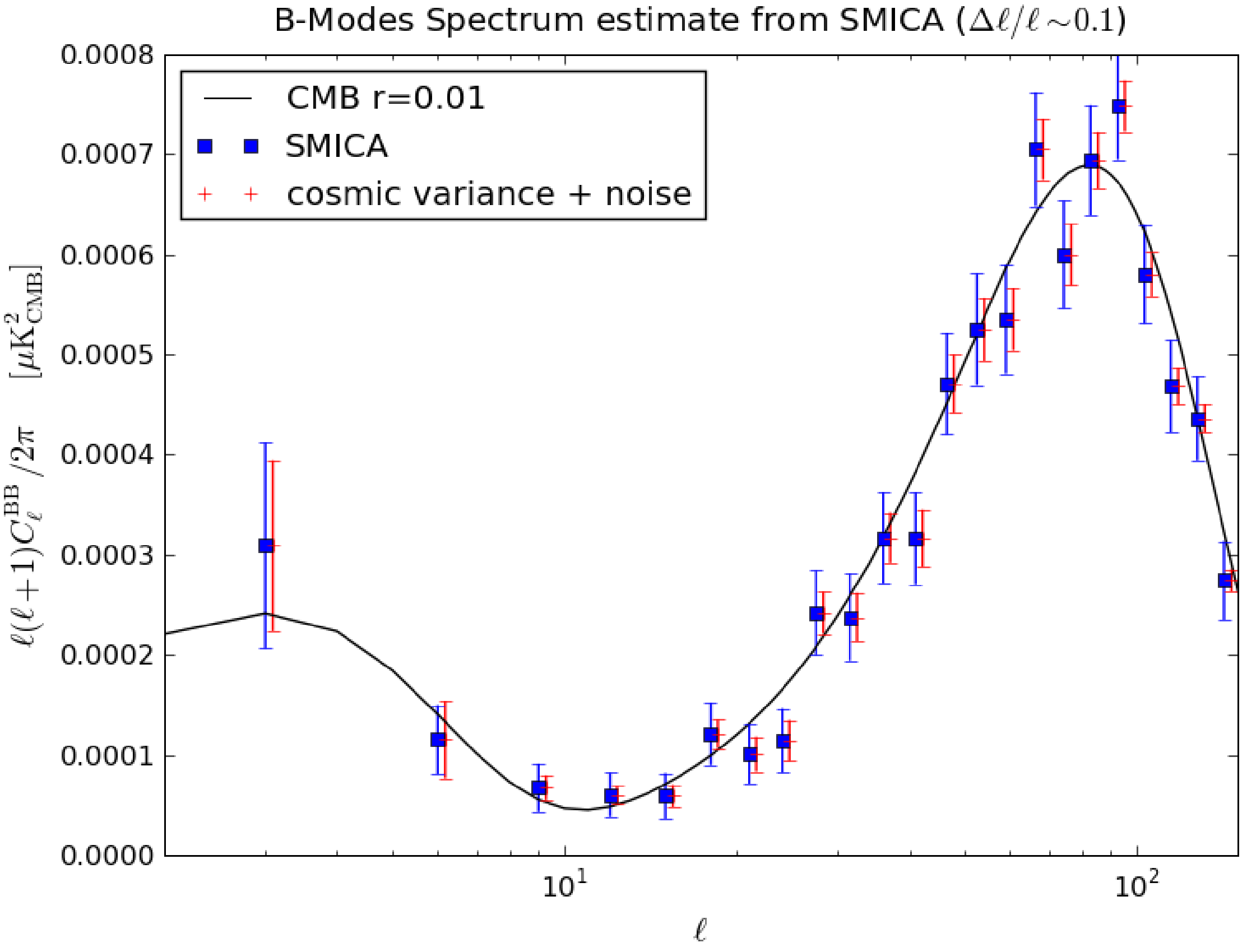

Fig. 2.4. BB power spectrum reconstruction for EPIC-IM 4K configuration assuming r = 0.01 and using the SMICA framework. Simulations include cosmic variance and raw instrument noise only (green) or also foregrounds (blue). The fraction of sky used is 89%. The power spectrum is binned at $\Delta\ell/\ell \sim 0.1$.

The lensing of the CMB signal may also reduce the ability to recover the small-scale primordial spectrum and therefore r. For a large-area coverage mission as EPIC however, the constraining power on small r values (~0.001) mainly comes from the large scale power spectrum ($\ell < 20$). As a result, lensing only marginally impacts the r determination, slightly degrading the information coming from $\ell > 20$ and impacting the overall precision of the measurement by about 15 %. This would not however be the case if the area coverage were to be drastically reduced to about 1 % of the sky. In such a case, an appropriate de-lensing reconstruction technique should be applied.

In the case foreground subtraction is efficient, the SMICA framework also allows us to perform CMB power spectrum reconstruction. An example of such case is reported in Fig. 2.4, where we show the reconstructed BB power spectrum both in presence and absence of foregrounds for the r = 0.01 case in the EPIC-IM 4 K configuration. The reconstruction is performed up to $\ell = 300$ as no lensing signal is included in the analysis. The spectrum is very well reconstructed and both bumps are clearly detected. At $\ell = 100$, the total noise including foregrounds is approximately double that of cosmic variance. Foreground removal dominates the errors at large $\ell$.

# 3. Mission Overview

The EPIC - Intermediate Mission (IM) configuration was developed as a follow-on to the initial EPIC mission study [1]. EPIC-IM has sufficient angular resolution, sensitivity, frequency coverage, and control of systematic errors to make a precise measurement of the inflationary B-mode spectrum to astrophysical limits, likely set by the removal of Galactic foregrounds, and will extract all of the cosmological information available in the E-mode and lensing B-mode polarization signals (see Figs. 3.1 and 3.2).

## 3.1 The Role of Space in CMB Measurements

The scientific objectives of EPIC, listed in Table 1.3, require highly sensitive and precise measurement of CMB polarization on all angular scales from the quadrupole to the Silk damping tail. These demanding observations require a space-borne instrument.

**Table 3.1.** Measurements Requiring a Space-borne Experiment

| Measurement Criteria | Attribute | Why Space is Needed |
|---|---|---|
| Measure inflationary B-mode power spectrum to astrophysical limits for $2 < \ell < 200$ at $r = 0.01$ after foreground removal | All sky coverage | High-fidelity measurements of low spatial multipoles |
| | Frequency coverage | Full access to the electromagnetic spectrum without degradation from Earth's atmosphere |
| | Systematic error control | Superior control, stability, redundancy and monitoring of systematic errors |
| Measure EE and lensing BB to cosmic limits | Sensitivity | Large improvements due to low backgrounds, large system throughput, long integration time |

Primary Objective

Secondary Objective

**All-sky Coverage:** While ground-based experiments can push to a sensitivity of $r \sim 0.01$ in limited regions of sky in the coming decade, the role of a future space mission is to carry out a precise measurement of the inflationary B-mode spatial power spectrum from $2 < \ell < 200$. Measuring the power spectrum on large angular scales requires an experiment that maps the entire sky with high fidelity, controlling systematic errors on large spatial scales. The only credible platform for such an all-sky measurement is from space.

**Frequency Coverage:** Subtracting polarized Galactic emission will clearly be necessary to uncover the B-mode spatial power spectrum. At low multipoles, Galactic emission will have to be modeled and subtracted to better than 10 % to reach $r = 0.01$. At higher multipoles, regions with low Galactic emission are known to exist which require significantly less subtraction to reach this goal. A multi-band frequency approach spanning 30 – 300 GHz is needed [2] to monitor and remove Galactic foregrounds. This subtraction is aided by the fact that the CMB electromagnetic spectrum is known to extremely high precision, and any component which is not CMB can be classified as a contaminant. Atmospheric absorption limits the full exploitation of this band from the ground, and even prevents observing in selected bands (60 GHz, 120 GHz) from balloon altitudes. Coverage of the full range of bands at the required sensitivity is only possible from space.

**Systematic Error Control:** Instrumental systematics must be measured and controlled to a new level of precision, particularly those effects that can convert the relatively bright CMB temperature and E-mode polarization signals into B-mode signals. The environment of an observatory at L2 allows for exquisite thermal stability. We employ a scanning pattern, only possible in space, which completely rotates the instrument with respect to the sky, while allowing continuous calibration on the CMB dipole. Space-borne measurements produce well-characterized and uniform data sets, with lasting legacy value, as witnessed by the watershed advances in cosmology provided by COBE and WMAP. A space-borne platform offers quantitatively superior control, stability, and assessment of systematic errors in redundant and uniform observations compared with any sub-orbital platform.

**Sensitivity:** Achieving the science goal of measuring B-mode polarization to r = 0.01 requires at least 10x greater sensitivity over the upcoming Planck satellite experiment. The EPIC-IM concept combines a low-background environment, unprecedented optical throughput, and a 4-year integration time to realize a large gain in sensitivity over Planck or any planned sub-orbital experiment [3]. Only space offers the combination of high instantaneous sensitivity and long integration times needed to reach the required sensitivity over the full sky.

### 3.2 The EPIC-IM Mission Concept

EPIC-IM (see Fig. 3.3) is based on an off-axis crossed-Dragone telescope (also called a compact range antenna) that provides a very large unabberated field of view with low polarization artifacts. The design provides a flat and telecentric focal plane, so that all the focal plane feeds look straight up into the telescope over the field of view. A 1.4 m cooled absorbing stop sets the effective projected aperture, and the primary and secondary mirrors are oversized to minimize spillover. This design avoids refracting optics, and is thus ideal for a single large multi-color focal plane. Far-sidelobe response is controlled by a combination of under-illuminating the mirrors, and using absorbing baffles at the primary, secondary, and aperture stop. These baffles must be appropriately cold and temperature stable so as to not introduce appreciable photon noise and signal drifts.

A large 100 mK focal plane provides 11,000 focal plane detectors operating in 9 frequency bands with unprecedented system sensitivity (see Tables 3.2 and 3.3), and a large improvement over Planck (see Table 3.4). The detectors are assumed to operate in matched pairs in a single band, with one detector sensing vertical polarization and one sensing horizontal polarization. The orientation of these pairs is varied over the focal plane to provide uniform instantaneous sampling of Q and U. More elaborate focal plane schemes (multi-color detectors, focal plane modulators, simultaneous Q and U polarization per pixel) are not assumed, but could offer systems advantages if successfully developed. The detectors are mounted in hexagonal tiles, packed with the highest frequencies located at the center of the focal plane where aberrations are lowest. The detector field of view must be collimated at the focal plane, either by antennas or feeds, as there is no possibility for beam control using a baffle or stop with the crossed Dragone telescope.

The telescope is cooled by a combination of passive cooling and a mechanical 4 K cryocooler, and the focal plane is cooled to 100 mK by either and adiabatic demagnetization refrigerator or a closed-cycle dilution refrigerator. The passive cooling system uses a 3-stage V-groove radiator that cools to a minimum temperature of ~30 K. The instrument is mounted to the spacecraft on a low-conductivity bipod, with each V-groove radiator mounted to the bipod in sequence. The optics are enclosed in a light-weight optics shield, and the shield is actively cooled to 18 K by an expansion stage of the 4 K cryocooler. The sub-K cooler operates from 4 K to provide continuous cooling at 100 mK, with an intermediate stage at ~1 K to reduce parasitic conductive and optical loads.

A large deployed 4-stage sunshield keeps radiation from the sun, earth and moon from reaching the instrument or optics. The sunshield is actuated on hinged booms and tensioned with cables. The booms are in a vertical configuration at launch to fit within the payload fairing, and deploy early in the flight while the instrument is at approximately room temperature. Each stage of the sunshield uses a double kapton membrane layer to protect against micro-meteorite penetrations. While each stage of the sunshield is progressively cooler, the sunshield itself does not effectively extract heat from the instrument due to the poor conductivity of the kapton membranes. Radiation from the inner sunshield accounts for only a small fraction (~6 %) of the total power on the 18 K stage of the cryocooler. The main cooling function of the sunshield is simply to allow thermal radiation from the V-groove to escape to space after multiple reflections on the sunshield membranes.

Power is provided by rigidly mounted solar panels attached to the bottom of the spacecraft bus. A deployed gimbaled Ka-band antenna is used for data downlink. The technical specifications of the instrument and spacecraft are summarized in Tables 3.5, 3.6, and 3.7.

EPIC-IM conducts science observations from an earth-sun L2 halo orbit. The orbit and cruise trajectory are described in detail in section 5.8 of the original study [1]. EPIC-IM uses a spinning and precessing scan strategy (see Fig. 3.5 and section 5.2.4 of [1]), rotating about the spin axis at ~0.5 rpm while precessing about the sun-line every hour. This strategy provides coverage patterns on the sky that are ideal for a polarization measurement, while keeping the solar illumination on the sunshield and solar panels constant. The scan strategy produces annual maps with a high degree of rotation of the instrument view angle on any given region of sky. It also produces daily maps covering more than half the sky with excellent angular coverage over most of this region. These maps can be used to track a variety of systematics effects on multiple timescales, and to form multiple difference maps for monitoring systematic errors. In order to perform telemetry communications during observations, the downlink antenna is gimbaled to maintain a constant line of contact to the earth.

**Technology Readiness (Section 4):** EPIC-IM is predicated on the continued rapid development of focal plane array technologies, and to a lesser extent on the development of improved cooling systems and optics. We present the heritage of developed technologies from Spitzer, Planck, and JWST, and a community plan [3] to rapidly develop emerging technologies in conjunction with a sub-orbital instrumentation program. EPIC-IM benefits from having several competing detector and cooling technologies to choose from during mission formulation.

**Systematic Error Mitigation (Section 5):** Controlling systematic errors is integral to our designs from the beginning. Detailed requirements for systematic error control are presented, as well as mitigations we have developed to monitor and control errors.

**Crossed Dragone Telescope (Section 6):** EPIC-IM is based on a crossed Dragone off-axis telescope, a telecentric design without a Lyot stop or refracting elements, that provides a very large field of view with minimal aberrations and polarization artifacts. We present a detailed analysis of main beam shapes and a physical optics calculation of the off-axis performance.

**Focal Plane Design (Section 7):** Large arrays of sensitive detectors lie at the heart of EPIC. New focal plane technology is emerging to provide the required sensitivities and formats, based on either SQUID-multiplexed transition-edge superconducting (TES) bolometers, or RF-multiplexed microwave kinetic inductance detectors (MKIDs). Antenna-coupled and feedhorn-coupled detectors are being developed which can accomplish the necessary beam collimation in compact focal plane architectures.

**Cooling (Section 8):** EPIC-IM uses a combination of passive cooling to ~30 K, a mechanical cryocooler to 4 K, and a continuous sub-K cooler operating from a 4 K base temperature to 100 mK. We present a detailed analysis of the passive cooling design, a 4 K cryocooler based on the JWST/MIRI cooler, and either a continuous adiabatic demagnetization refrigerator or a closed-cycle dilution refrigerator which is an outgrowth of the open-cycle dilution refrigerator flying on Planck.

**Deployed Sunshade (Section 9):** The EPIC scanning/precessing scan strategy requires a deployed multilayer sunshield to defeat solar input power to the instrument. We describe a design based on aluminized kapton sheets deployed on simple hinged booms.

**Spacecraft (Section 10):** The spacecraft requirements for EPIC, detailed in this section, can be met with existing technologies, and are close to several catalog commercial spacecraft buses available from multiple suppliers.

**Instrument Cost and Schedule (Section 11):** We have developed a grass-roots cost and schedule for the EPIC-IM instrument based on institutional experience of similar builds, developed by team members based on their involvement in Spitzer, Planck, and Herschel.

**Cost (Section 12):** We developed cost estimates for 4 mission options, taking the instrument design, and grassroots cost and schedule as inputs to a JPL team-x cost assessment.

Multiple technologies exist for the focal plane detectors, focal plane coupling optics, readouts, and cooling systems. At this time, no selection is being made for these technologies and the instrument architecture is scoped to allow flexibility. However for purposes of costing in sections 11 and 12 we assume for sake of definitiveness antenna-coupled TES bolometers read out with time-domain multiplexed SQUID amplifiers, and cooled with an adiabatic demagnetization refrigerator. The technology selection will occur at an appropriate time dictated by the project schedule and technology readiness.

### 3.3 Comparison with Previous Study

EPIC-IM was conceived to complement the previous mission study of a small low-cost configuration named EPIC-LC, and a large comprehensive-science mission named EPIC-CS, as shown in Fig. 3.6. EPIC-LC was developed as a low-cost and high-readiness mission targeting only the inflationary B-mode power spectrum. Since measuring the inflationary B-mode signal requires only modest ~1º resolution, EPIC-LC is based on six 30 cm refracting telescopes. EPIC-LC is designed to deliver the primary science goal of the IGW search on large angular scales at minimum cost and risk. This low-cost option uses largely proven technologies, including existing focal plane detectors and a liquid helium cryostat. EPIC-LC may be formulated in response to detection of inflationary polarization from a sub-orbital or ground-based platform, since it will very capably measure the entire B-mode power spectrum.

EPIC-CS was conceived to measure all the available cosmological information in the CMB E-mode and lensing B-mode signals. The angular resolution is sufficient to allow measurement and removal of the lensing B-mode signal for the deepest possible search for inflationary B-modes. EPIC-CS uses a 3 m Gregorian-Dragone telescope, based on a design developed for the Polarbear experiment.

This new study contrasts with the previous work to find an optimal ratio of scientific return to mission cost. This optimization hinges on having sufficient angular resolution to measure the E-mode and lensing B-mode signals to cosmic limits, and these criteria drive the aperture size to be ~1.5 m, in order to provide the ~5 arcminute resolution necessary to measure E-mode and lensing B-mode power spectra to cosmic limits into the Silk damping tail, as described in Table 1.4.1.

In contrast to the low-resolution EPIC-LC concept, EPIC-IM will thus completely map out the E-mode and lensing B-mode spectra, extracting all of the cosmological information available in CMB polarization. EPIC-IM includes new polarized bands at 500 and 850 GHz for Galactic science, providing unprecedented sensitivity and resolution in an all-sky measurement. The large available field of view gives EPIC-IM a marked system sensitivity advantage over the previous concepts, especially if the telescope is cooled to 4 K. As shown in Fig. 3.1, EPIC-IM with a 1.4 m aperture provides superior sensitivity to EPIC-CS with a 3 m aperture for $\ell < 5000$, and equivalent sensitivity for $5000 < \ell < 10000$, the sensitivity in the focal plane more than compensating for EPIC-IM's smaller aperture.

By a judicious choice of a 4 K cooler, EPIC-IM is more mass efficient than EPIC-LC, which uses a high-TRL liquid-helium cryostat. Thus EPIC-IM is only 25 % heavier than EPIC-LC but has 4.7x higher angular resolution and 2.8x higher sensitivity. EPIC-IM is 2.6x lighter than EPIC-CS but provides equivalent or better sensitivity for $\ell < 10000$.

### 3.4 Descope Option: 30 K Telescope

In developing this study we identified a descoped mission concept, called the "30 K Telescope Option". As shown in Fig. 3.7, compared to the baseline "4 K Telescope Option", the 30 K option does not actively cool the telescope, using only passive cooling to reach an optics temperature of ~25 K. This results in eliminating one V-groove radiator stage and one deployed sunshield stage. Because the 4 K cryocooler only cools the focal plane and not the telescope, the heat loads on the cooler are somewhat smaller. However the major difference is in the design of the focal plane. The baseline 4 K telescope option assumes a dense focal plane packing, with 2 $f\lambda$ spacing between feeds (or antennas). This packing necessarily results in higher edge spillover onto the cold 4 K absorbing primary, secondary, and aperture stops. With a higher telescope temperature, this spillover is no longer acceptable due to increased photon noise from the absorbing stops. Therefore the 30 K telescope option assumes a larger 3.25 $f\lambda$ spacing, with lower spillover but also a lower density of detectors. In addition to 2.6x lower density, the focal plane is also reduced in overall size. The scientific and cost tradeoffs are summarized in Fig. 3.1 and sections 11 and 12. While the cost difference between the two options is modest (only 7 %), it is worth noting these savings are dominated by the reduction in the focal plane size – fewer detectors, fewer readout channels, lower data rate, and reduced telecom and data storage requirements. In technical implementation the 30 K option provides somewhat lower risk with a smaller focal plane, though control of thermal radiation onto the focal plane is a greater challenge. The choice of cooling the telescope itself has comparatively smaller cost impact, largely derived from reduced structural mass and 4 K cryocooler mass and power. Therefore much of the cost savings could be realized by keeping the telescope at 4 K, and simply descoping the focal plane.

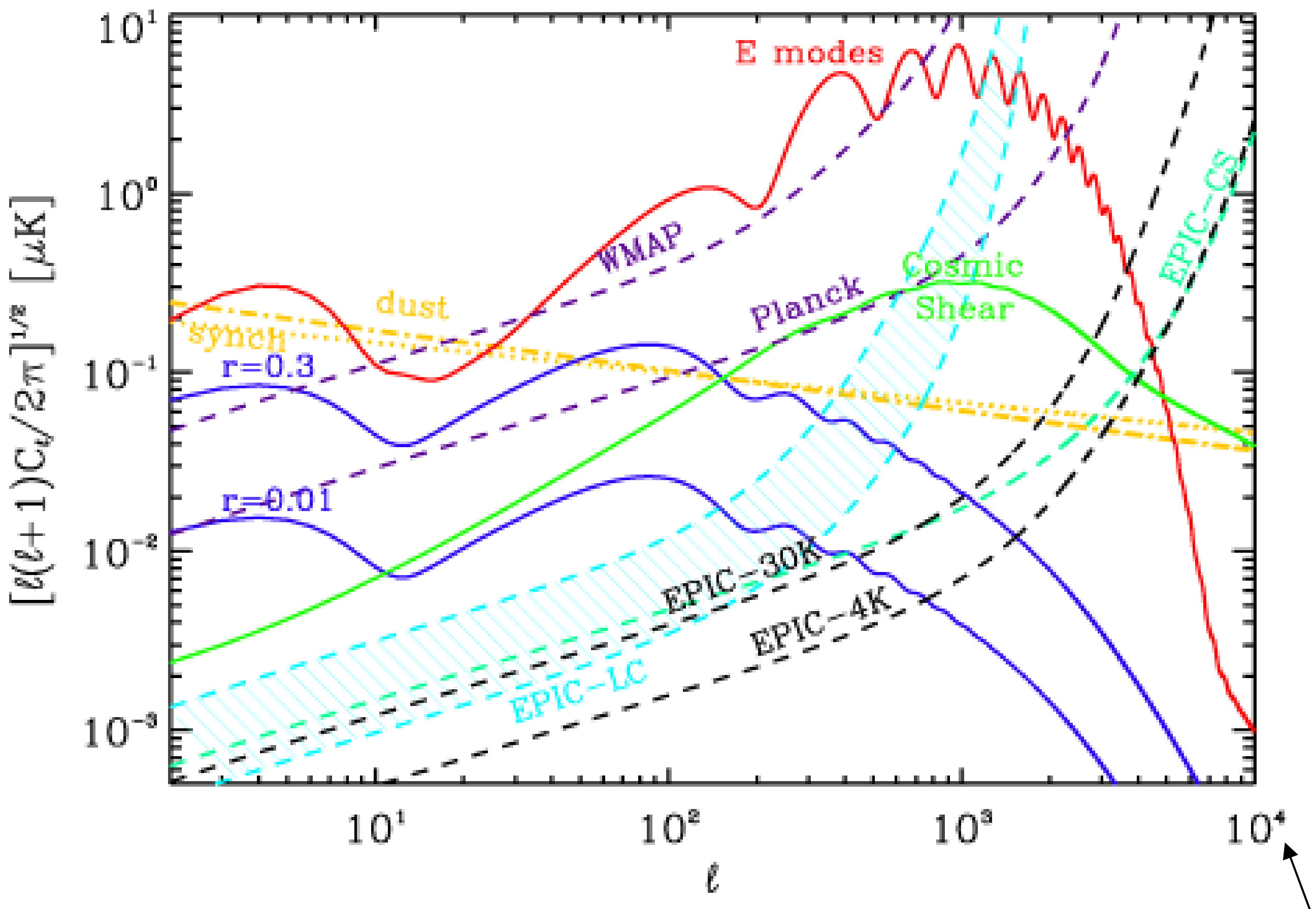

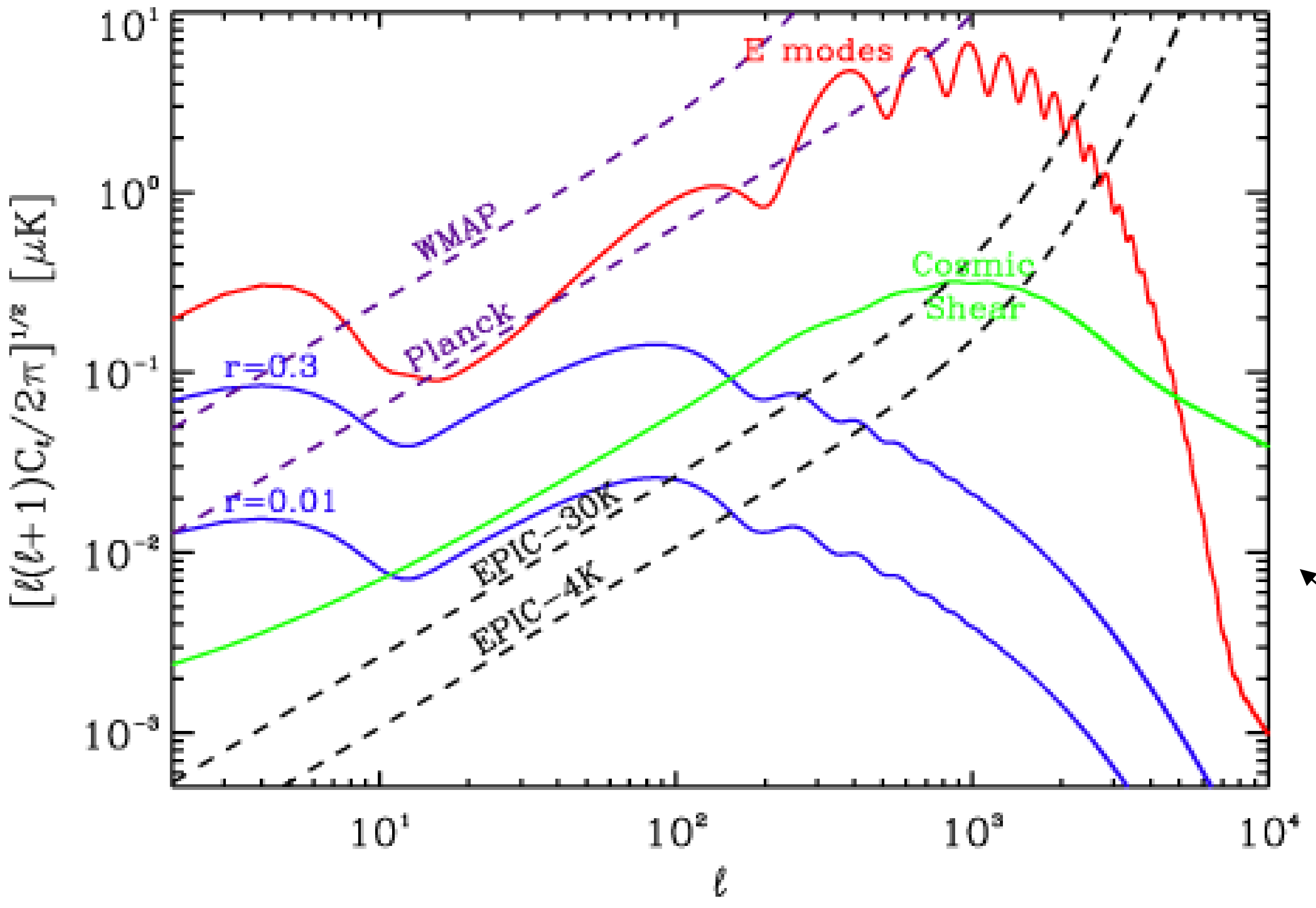

**EPIC-IM Science Objectives**

- Measure inflationary B-mode power spectrum to astrophysical limits for $2 < \ell < 200$ at $r = 0.01$
- Measure EE to cosmic variance into the Silk damping tail
- Measure lensing BB to cosmic limits to probe dark energy equation of state and dark matter
- Measure EE to cosmic variance to distinguish reionization histories
- Map Galactic magnetic fields via dust polarization

**Table 3.2 EPIC-IM Bands and Sensitivities**

| Freq [GHz] | $\theta_{FWHM}$ [′] | 4 K Telescope Option | | | | | 30 K Telescope Option | | | | |
|---|---|---|---|---|---|---|---|---|---|---|---|
| | | $N_{bol}$ [a] [#] | NET [μK√s] bolo [b] | NET [μK√s] band [c] | $w_p^{-1/2}$ [μK-′] [d] | $\delta T_{pix}$ [e] [nK] | $N_{bol}$ [a] [#] | NET [μK√s] bolo [b] | NET [μK√s] band [c] | $w_p^{-1/2}$ [μK-′] [d] | $\delta T_{pix}$ [e] [nK] |
| 30 | 28 | 84 | 84 | 9.2 | 14 | 83 | 24 | 83 | 17 | 26 | 150 |
| 45 | 19 | 364 | 71 | 3.7 | 5.7 | 34 | 84 | 70 | 8 | 12 | 69 |
| 70 | 12 | 1332 | 60 | 1.6 | 2.5 | 15 | 208 | 60 | 4.1 | 6.4 | 37 |
| 100 | 8.4 | 2196 | 54 | 1.1 | 1.8 | 10 | 444 | 55 | 2.6 | 4.0 | 24 |
| 150 | 5.6 | 3048 | 52 | 0.9 | 1.4 | 8 | 516 | 57 | 2.5 | 3.8 | 23 |
| 220 | 3.8 | 1296 | 59 | 1.6 | 2.5 | 15 | 408 | 77 | 3.8 | 5.8 | 34 |
| 340 | 2.5 | 744 | 100 | 3.7 | 5.6 | 33 | 120 | 220 | 20 | 30 | 180 |
| 500 | 1.7 | 1092 | 350 | 10 | 16 (140) [f] | 8 [g] | 108 | 1500 | 170 | 260 (2000) [f] | 140 [g] |
| 850 | 1.0 | 938 | 15000 | 280 | 740 (70) [f] | 7 [g] | 110 | 250k | 24k | 40k (3000) [f] | 340 [g] |
| **Total** [h] | | **11094** | | **0.6** | **0.9** | **5.4** | **2022** | | **1.5** | **2.3** | **13** |

[a] Two bolometers per focal plane pixel
[b] Sensitivity for a single bolometer to CMB temperature
[c] Sensitivity combining all bolometers in a band
[d] $[8\pi\, NET_{bolo}^2/(T_{mis}\, N_{bol})]^{1/2}(10800/\pi)$
[e] Sensitivity $\delta T_{CMB}$ in a 2° x 2° pixel
[f] Point source sensitivity in μJy (1σ) per beam without confusion
[g] Surface brightness sensitivity in Jy/sr in a 2° x 2° pixel (1σ)
[h] Combining all bands together

**Table 3.3** Sensitivity Model Input Assumptions

| | | | | | |
|---|---|---|---|---|---|
| Focal plane temperature | $T_o$ | 100 mK | Optical efficiency | $\eta_{opt}$ | 40 % |
| Blocker temperature | $T_{blkr}$ | 4 K | Fractional bandwidth | $\Delta\nu/\nu$ | 30 % |
| Optics temperature | $T_{opt}$ | 4 K / 30 K* | Noise margin† | | 1.414 |
| Mirror emissivity at 1 mm | ε | 1 % | Mission lifetime | $T_{life}$ | 4 years |
| Coupling to 4 K / 30 K stop | | 10 % / 0.5 %* | Heat capacity | $C_0$ | 0.15 pJ/K |
| Coupling to 4 K baffle | | 5 % | $\alpha = dln(R)/dln(T)$ | | 100 |
| Bolometer pitch | d/fλ | 2 / 3.25* | TES safety factor‡ | $P_{sat}/Q$ | 5 |

*Parameter for 4 K option / 30 K option †The total calculated sensitivity is multiplied by a safety factor of √2
‡The factors of safety are 20 for 500 GHz and 200 for 850 GHz (4 K) and 20 for 500 & 850 GHz (30 K)

Fig. 3.1 (The sensitivity of EPIC, WMAP and Planck to CMB polarization anisotropy. E-mode polarization from scalar perturbations is shown in red; B-mode polarization from tensor perturbations is shown in blue for r = 0.3 and r = 0.01; and B-mode polarization produced by lensing of the E-mode polarization is shown in green. Expected B-mode foreground power spectra for polarized dust (orange dash-dotted) and synchrotron (orange dotted) at 70 GHz. The band-combined instrumental sensitivity of EPIC is given assuming the 4 K and 30 K telescope options each with 4 years of observations. WMAP assumes an 8-year mission life; Planck assumes 1.2 years at goal sensitivities for HFI. Also shown are sensitivities of the previous low-resolution EPIC-LC and high resolution EPIC-CS shown in Fig. 3.1, described in [1]. EPIC-LC with a LHe cryostat is assumed to have a 2-year mission life at L2 and is calculated for high-TRL NTD Ge bolometers and larger-format TES bolometer arrays. All instrument sensitivities are shown band-combined in $\Delta\ell/\ell = 0.3$ logarithmic bins.

Fig. 3.2 The noise $C_\ell$s of EPIC-IM, WMAP and Planck, with polarization spectra the same as above. A comparison of these noise power spectra and the signals, such as primordial B-mode spectra shown in blue for r = 0.01 and r = 0.3 reveals the angular scale, or the multipole moment, where cosmic variance of primordial signals dominate the measurement. As shown, the detection of low-multipole reionization bump is dominated by cosmic variance while for low r models, the recombination bump at degree angular scales is the transition between noise domination to cosmic variance domination.

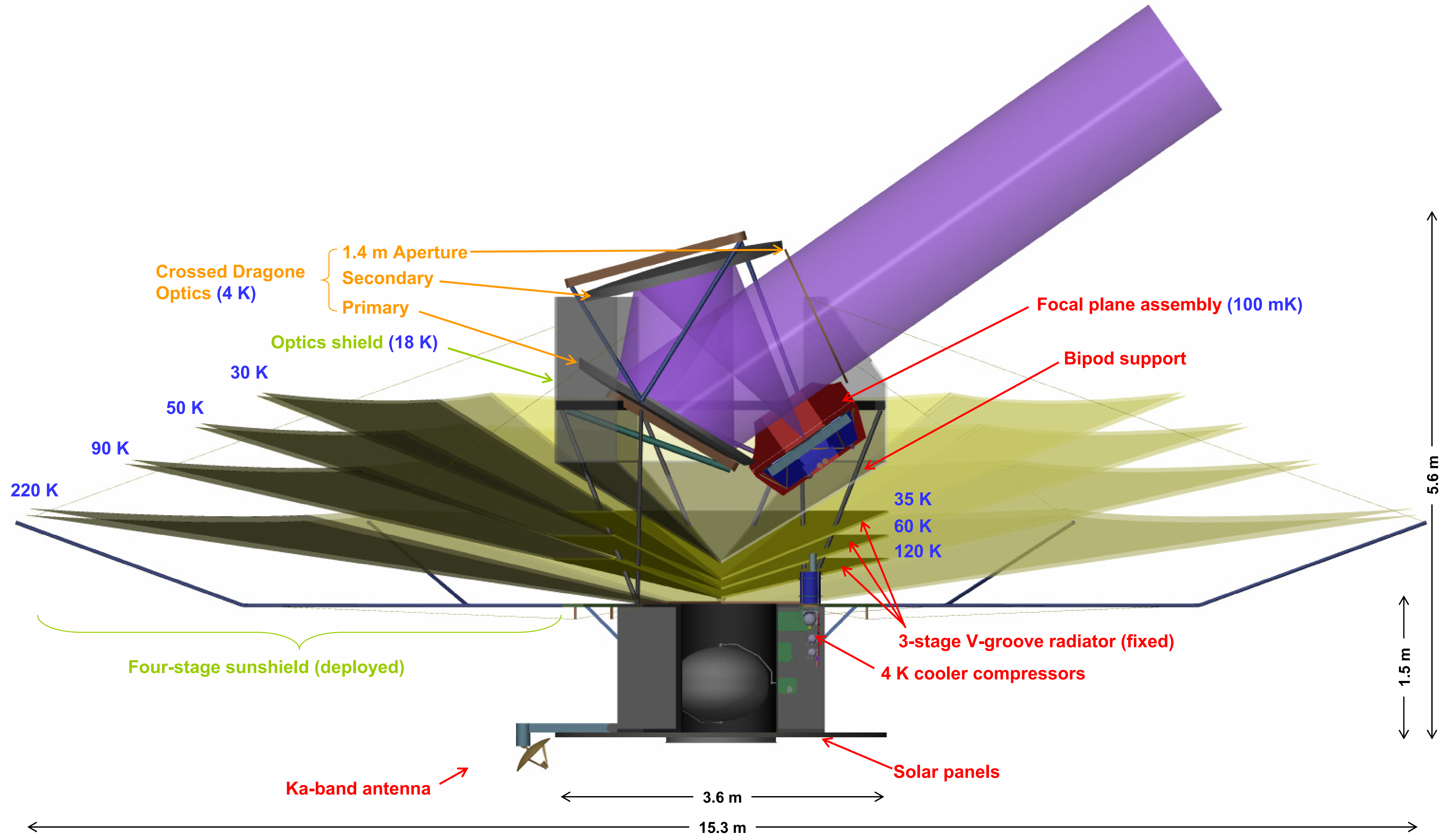

1.4 m Aperture
Crossed Dragone Optics (4 K)
Secondary
Primary
Optics shield (18 K)
Focal plane assembly (100 mK)
Bipod support
30 K
50 K
90 K
220 K
35 K
60 K
120 K
5.6 m
1.5 m
3-stage V-groove radiator (fixed)
4 K cooler compressors
Four-stage sunshield (deployed)
Solar panels
Ka-band antenna
3.6 m
15.3 m

Fig. 3.4 EPIC INTERMEDIATE MISSION: DEPLOYED AND LAUNCH CONFIGURATIONS

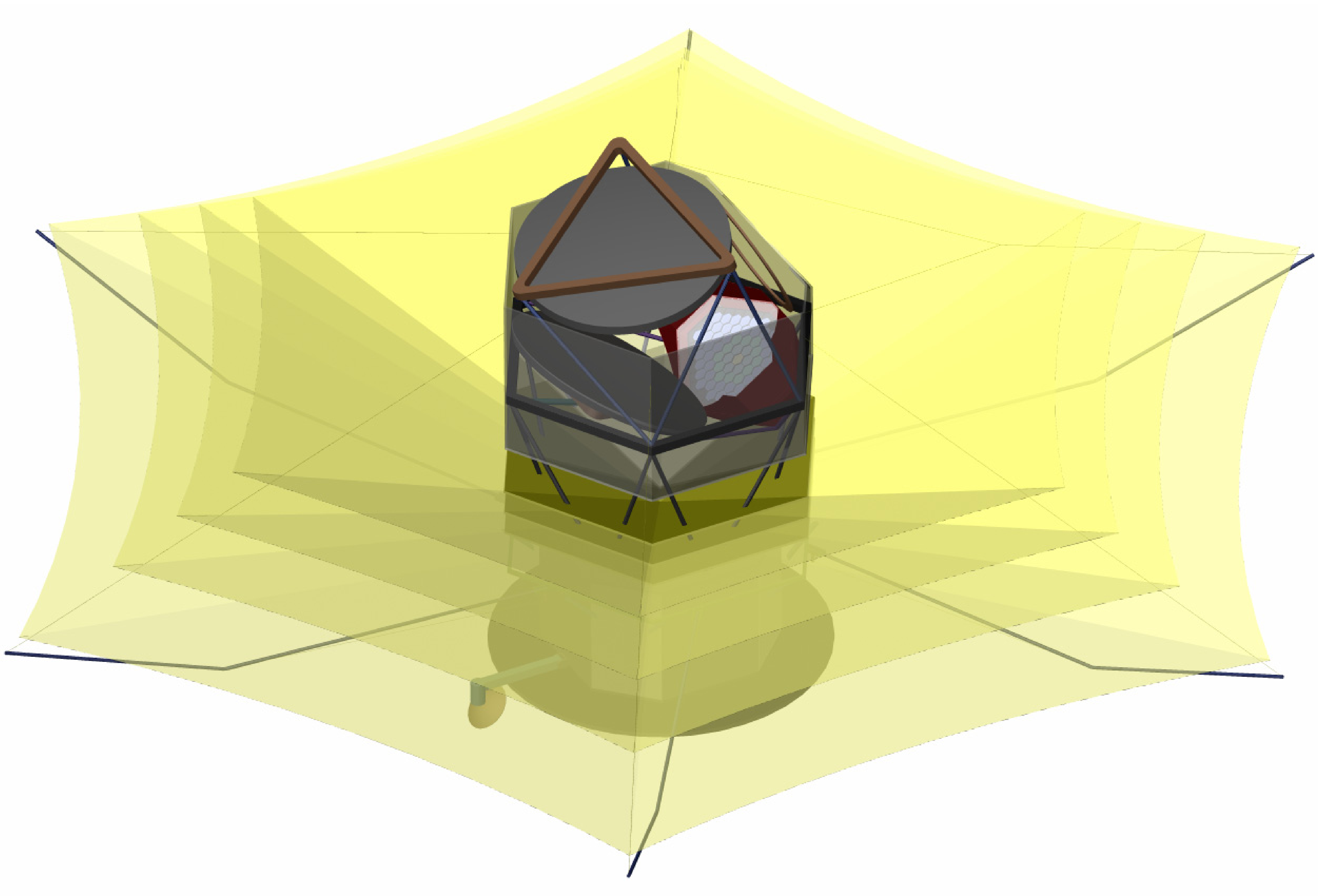

**Deployed Configuration**

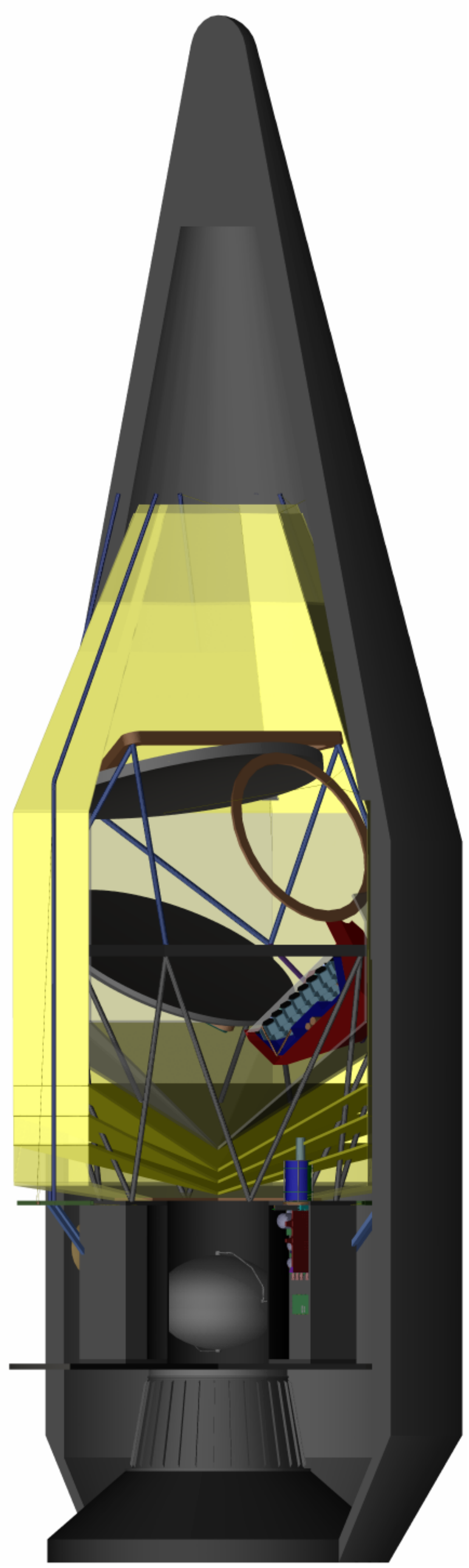

**Launch Configuration**
Atlas V 401

Fig. 3.5 EPIC INTERMEDIATE MISSION: SCAN STRATEGY

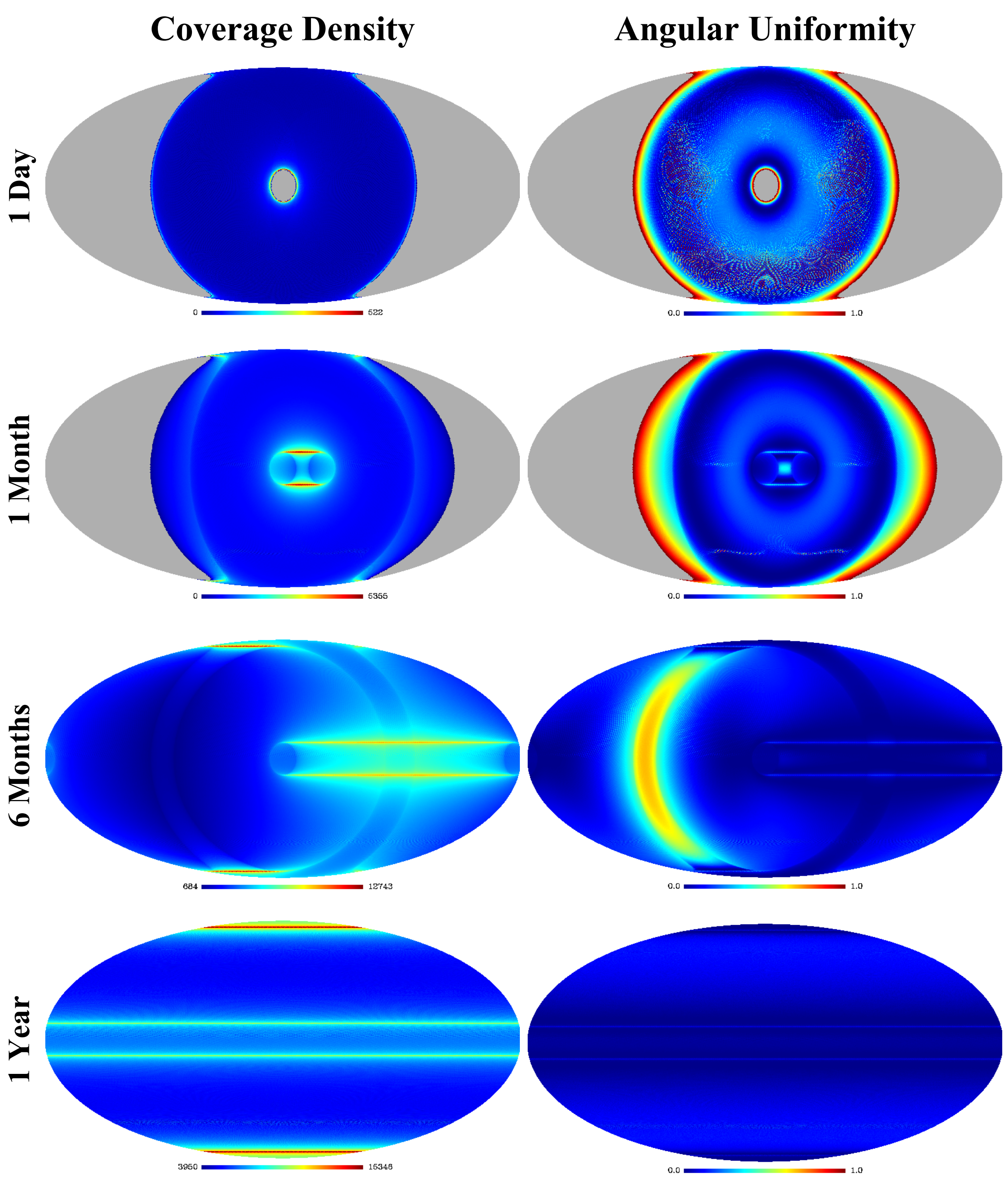

Coverage maps for a single pixel built up over the course of observations. The left column shows the coverage density in number of hits per sky pixel for a single detector. The right column shows the uniformity of the crossing angle on the sky, $<\sin 2\alpha>^2 + <\cos 2\alpha>^2$. For this metric of angular coverage, ideal uniform rotation gives 0 while poor angular coverage with not rotation gives 1. Coverage maps are shown from top to bottom for 1 day, 1 month, 6 months, and 1 year. After one year of observations, EPIC realizes highly uniform density and angular coverage, ideal for a polarization experiment, and a significant improvement over the scan patterns used by WMAP and Planck.

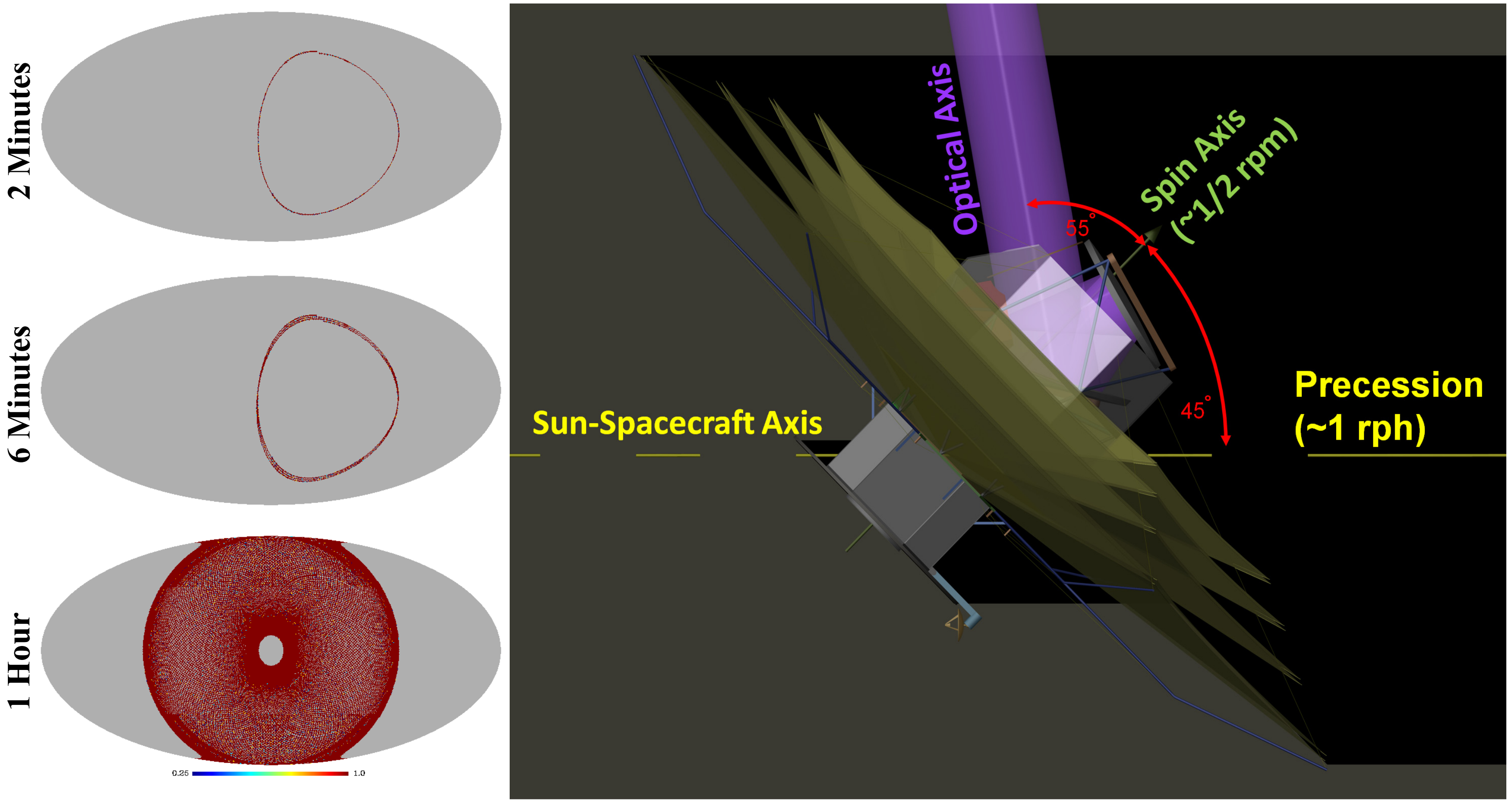

EPIC uses a scanning/precessing scan strategy to map the sky with uniform coverage, complete modulation of crossing angles, and high redundancy. The spacecraft spins about its axis of symmetry every 2 minutes. The optical axis of the telescope is canted 55˚ off the spin axis. The spin axis is tipped 45˚ from the sun-spacecraft axis. The spin axis precesses about the sun-spacecraft axis approximately every hour, but always maintaining a 45˚ angle. The deployed sun shield always keeps the sun and moon off the optics. The inner shields only view cold space for efficient passive cooling. The solar input power into the system is thus constant in order to maintain extreme thermal stability. The scan strategy requires despinning the scan strategy with a gimballed downlink antenna to Earth. At left, scan patterns are shown for a single detector in (top) one spin period of 2 minutes; (middle) 3 spin periods of 6 minutes; and (bottom) one complete precession period of1 hour.

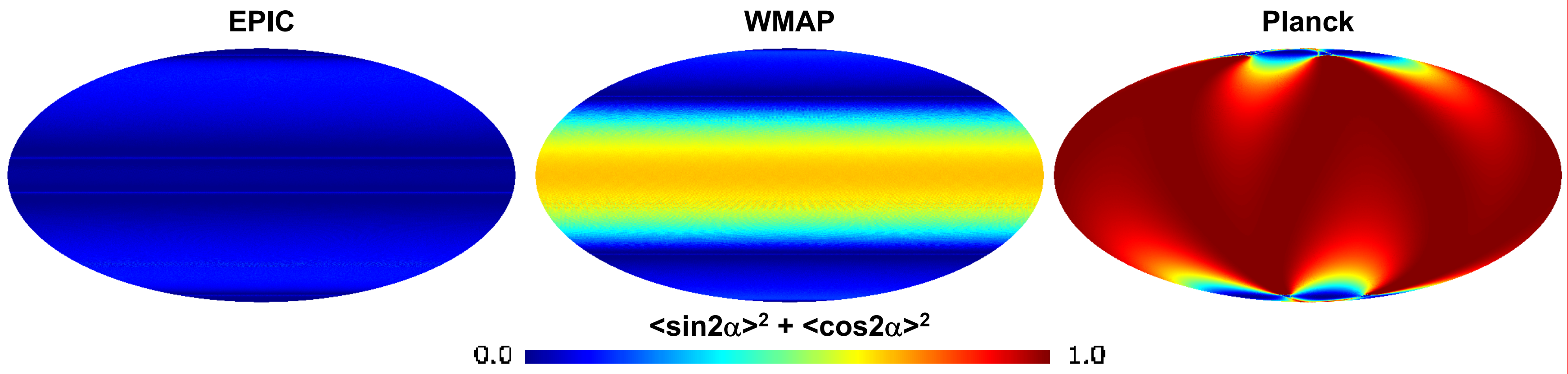

Angular coverage uniformity over the sky for the EPIC, WMAP, and Planck scan strategies after 1 year of observations. The quantity plotted, $<\sin 2\alpha>^2 + <\cos 2\alpha>^2$ is a measure of the variation in the crossing angle $\alpha$ over a given pixel on the sky. For polarimetery, this quantity is ideally zero, indicating uniform rotation about the telescope boresight on every patch of sky, removing some types of polarization signatures associated with the instrument. Planck, which scans in great circles crossing approximately at the ecliptic poles, has poor coverage across the ecliptic plane. WMAP uses a scanning and precessing strategy with improved angular uniformity. EPIC has nearly ideal angular coverage.

Fig. 3.6 COMPARISON OF EPIC-IM WITH PREVIOUS EPIC-LC / EPIC-CS STUDY

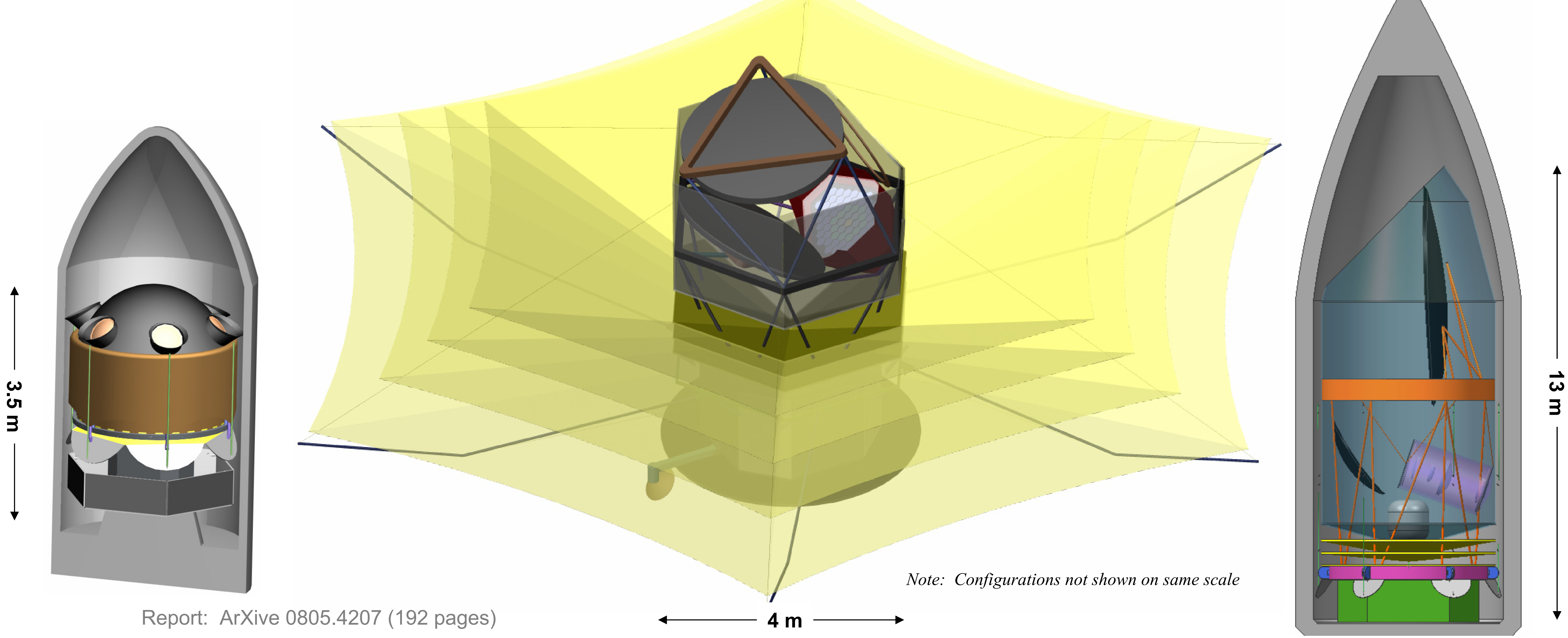

| *EPIC-* | *Low Cost* | *Intermediate Mission 4 K Option* | *Comprehensive Science* |
|---|---|---|---|
| **Science** | Inflationary B-mode polarization only | Inflationary B-modes, E-modes to cosmic variance, gravitational lensing to cosmic limits, neutrino mass, dark energy, Galactic astronomy | Inflationary B-modes, E-modes to cosmic variance, gravitational lensing, neutrino mass, dark energy, Galactic astronomy |
| **Speed** | 500 Plancks | 3600 Plancks | 250 Plancks |
| **Detectors** | 2400 | 11,000 (TES bolometer or MKID) | 1500 |
| **Aperture** | Six 30 cm refractors | 1.4 m Crossed Dragone telescope | 3 m Gregorian Dragone |
| **Bands** | 30 – 300 GHz | 30 – 300 GHz + 500 & 850 GHz | 30 – 300 GHz |
| **Cooling** | LHe cryostat + ADR | 4 K Cryo-cooler + ADR | TBD |
| **Mass** | 1320 kg CBE | 1670 kg CBE | 3500 kg CBE |
| **Cost** | $660M (FY07) | $920M (FY09) | No cost assessed |

Fig. 3.7 EPIC INTERMEDIATE MISSION: 4 K TELESCOPE OPTION VS 30 K TELESCOPE OPTION

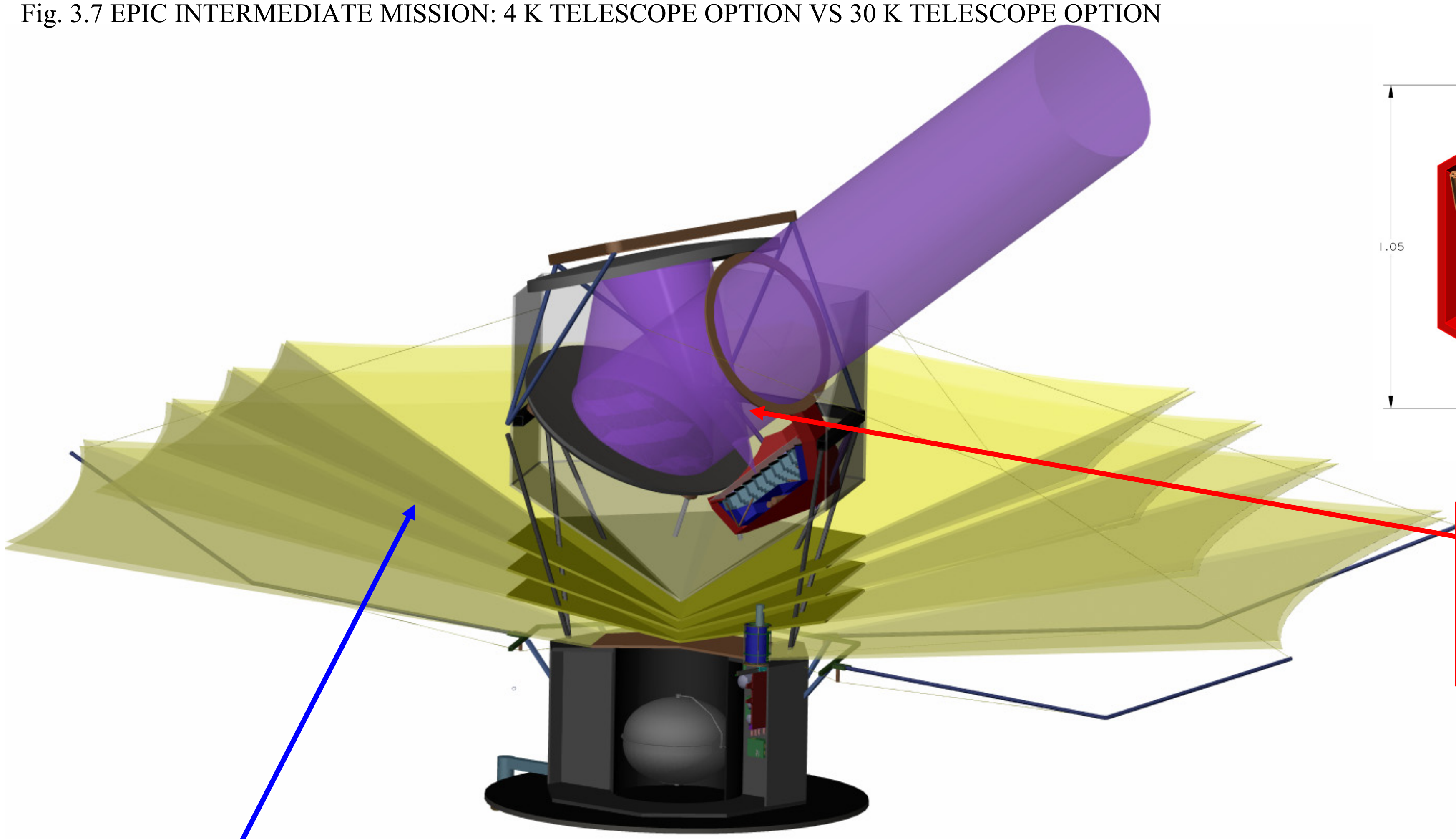

‘4 K Telescope’ Option

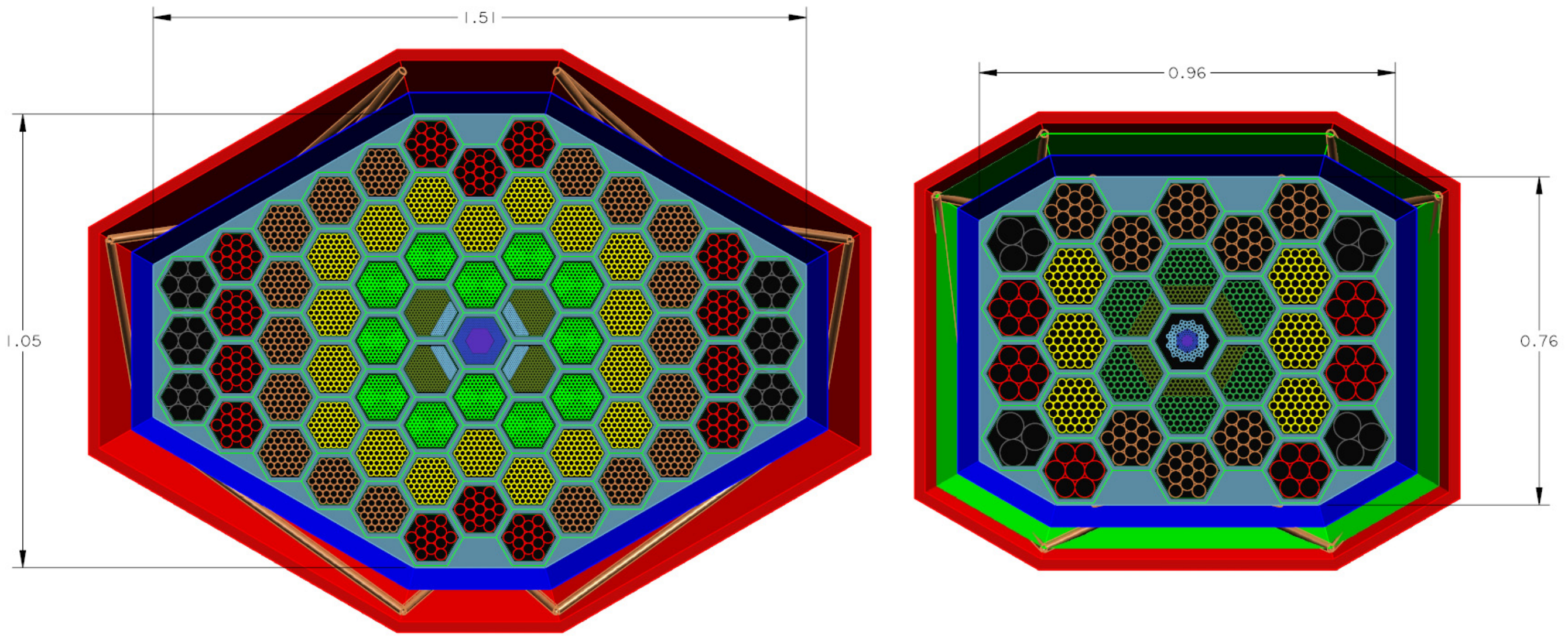

‘4 K’ Focal Plane

‘30 K’ Focal Plane

**Focal Plane**

| 4 K Option | 30 K Option |
|---|---|
| 11094 detectors @ 2-fλ packing | 2022 detectors @ 3.25-fλ packing |
| Higher pixel density; higher spillover | Lower pixel density; lower spillover |
| Radiation shields at 4 K and 1 K | Radiation shields at 18 K, 4 K and 1 K |

**Passive Radiator**

| 4 K Option | 30 K Option |
|---|---|
| 3 V-grooves and 4 deployed sunshields | 2 V-grooves and 3 deployed sunshields |
| Optical shield actively cooled to 18 K | Optical shield passively cooled to 35 K |
| Telescope actively cooled to 4 K | Telescope passively cooled to 25 K |

**4 K Cryocooler**

| 4 K Option | 30 K Option |
|---|---|
| 21 mW @ 4.4 K (CBE load) | 11 mW @ 4.4 K (CBE load) |
| 67 mW at 18 K (CBE load) | 8 mW at 18 K (CBE load) |
| 270 W bus power (2x margin on loads) | 165 W bus power (2x margin on loads) |

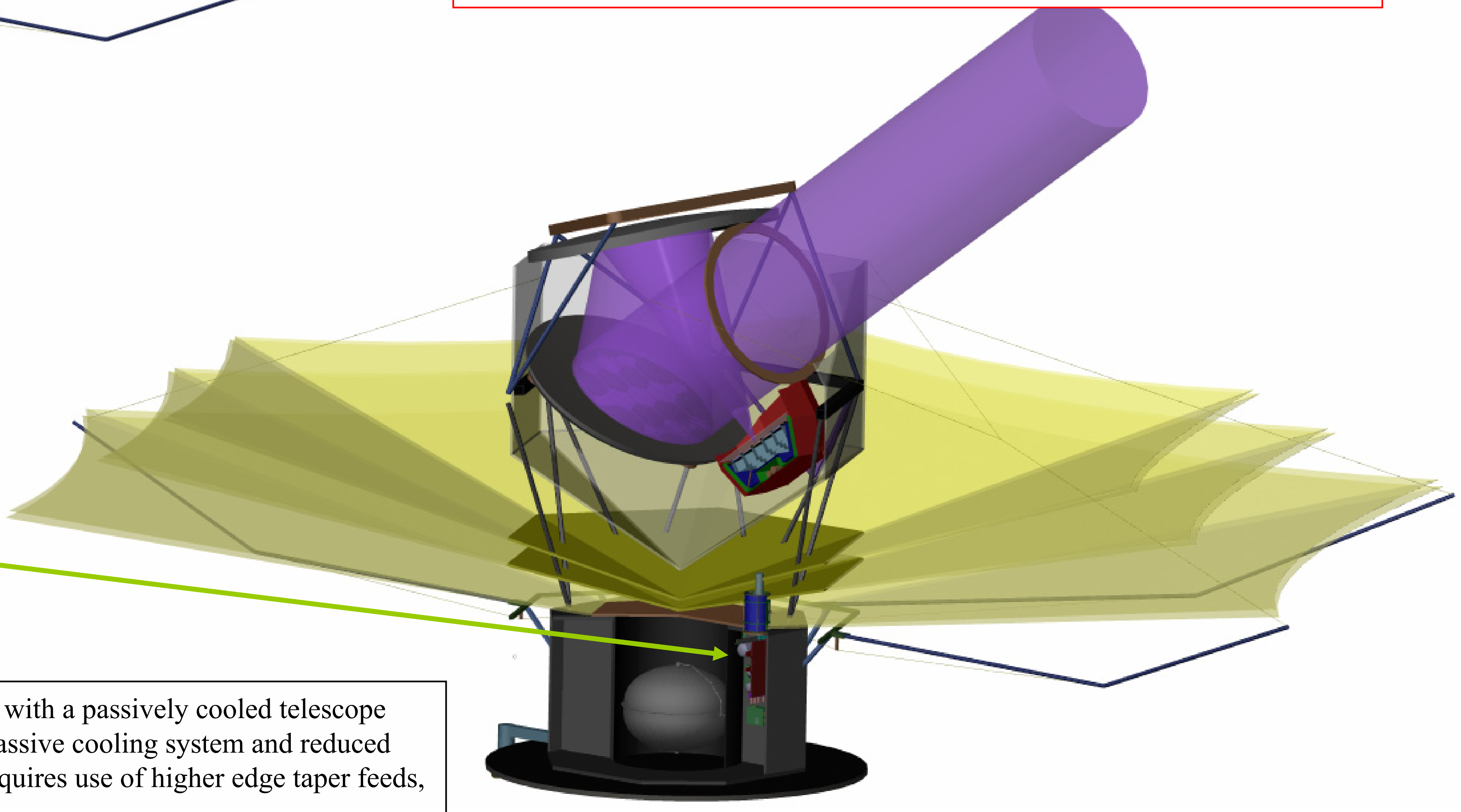

‘30 K Telescope’ Option

We study the cost and resource implications of a simpler cryogenic implementation with a passively cooled telescope with a physically smaller focal plane. The ’30 K telescope’ option uses a simpler passive cooling system and reduced heat loads on the 4 K cryocooler. Thermal emission from the telescope surround requires use of higher edge taper feeds, resulting in a lower density packing of feedhorns, and reduced system sensitivity.

## EPIC INTERMEDIATE MISSION: TECHNICAL SPECIFICATIONS

**Table 3.4 Comparison of EPIC and Planck Sensitivity $w_p^{-1/2}$**

| Freq [GHz] | EPIC 4 K Option | EPIC 30 K Option | Planck Goal[1] |
|---|---|---|---|
| 30 | 14 | 26 | 350 |
| 45 | 5.7 | 12 | 350 |
| 70 | 2.5 | 6.4 | 380 |
| 100 | 1.8 | 4.0 | 100 |
| 150 | 1.4 | 3.8 | 80 |
| 220 | 2.5 | 5.8 | 130 |
| 340 | 5.6 | 30 | 400 |
| 500 | 17 | 260 | |
| 850 | 400 | 40,000 | |
| **Total[2]** | **0.9** | **2.3** | **54** |

[1]Planck combined sensitivities in polarization for 1.2 year mission lifetime. Planck bands are shifted slightly to match the closest EPIC band.

[2]Total $w_p^{-1/2}$ is combined $w_p^{-1/2}$ from all bands in μK-arcmin

**Table 3.5 EPIC-IM 4 K Mission Summary**

| | | | |
|---|---|---|---|
| **Optics** | 1.4 m wide-field crossed Dragone | **Total Delta-V** | 170 m/s |
| **Orbit** | Sun-earth L2 halo | **Payload Power** | 440 W (CBE) |
| **Mission Life** | 1 year (required); 4 years (design goal) | **Spacecraft Power** | 533 W (CBE) |
| **Launch Vehicle** | Atlas V 401 | **Total Power** | 1392 W (w/ 43 % cont.) |
| **Detectors** | TES bolometer or MKID | **Payload Mass** | 813 kg (CBE) |
| **Bands** | 30, 45, 70, 100, 150, 220, 340, 500 & 850 GHz | **Spacecraft Mass** | 584 kg (CBE) |
| **Sensitivity** | 0.9 μK arcmin; 3600 Planck missions | **Total Mass** | 2294 kg (w/ 43 % cont.) |
| **# Detectors** | 11094 | **Vehicle Margin** | 1287 kg (36 %) |
| **Data Rate** | 7.7 Mbps | **Cost** | $920M FY09 |

**Table 3.6 On-Orbit Power Summary (During Telecom)**

| **Sub-Assembly** | **4 K Telescope Power [W]** | **30 K Telescope Power [W]** |
|---|---|---|
| Detector Readout Electronics | 150 | 75 |
| ADR Control Electronics and Housekeeping | 29 | 29 |
| 4 K Cryocooler Electronics | 61 | 48 |
| 4 K Cryocooler Compressors | 209 | 150 |
| **Payload Subtotal (CBE)** | **449** | **302** |
| **Payload Subtotal with 43 % Contingency** | **643** | **432** |
| Attitude Control System | 102 | 102 |
| C&DH | 136 | 35 |
| Power | 128 | 90 |
| Propulsion (dry) | 16 | 16 |
| Structures and mechanisms | 0 | 0 |
| Telecom + Antenna | 40 | 71 |
| Thermal | 46 | 37 |
| **Spacecraft Subtotal (CBE)** | **468** | **351** |
| **Spacecraft Subtotal with 43 % Contingency** | **669** | **502** |
| **Total Power** | **1311** | **934** |
| **GaAs triple junction for 45º at 80 ºC EOL** | **1430** | **1206** |
| **Margin** | **119 (8 %)** | **272 (23 %)** |

**Table 3.7 Mass Summary**

| Group | **Sub-Assembly** | **4 K Telescope Mass [kg]** | **30 K Telescope Mass [kg]** |
|---|---|---|---|
| **Focal Planes** | 100 mK assembly | 26.1 | 11.1 |
| | 1 K and 4 K filter assemblies | 23.2 | 9.8 |
| | 1 K and 4 K radiation shields | 24.0 | 10.2 |
| | Struts | 20.0 | 8.5 |
| | Cabling | 12.1 | 5.5 |
| | **Total Focal Plane Assembly** | **105.4** | **45.1** |
| **ADR** | Sub-K cooler | 16.7 | 16.7 |
| | Thermal straps | 3.0 | 3.0 |
| | Leads and cables | 1.6 | 1.6 |
| | **Total Focal Plane Assembly** | **21.3** | **21.3** |
| **4 K Cooler** | Compressors and cooler assembly | 49.2 | 49.2 |
| | Thermal straps | 10.0 | 10.0 |
| | Cabling | 5.0 | 5.0 |
| | **Total 4 K Cryocooler** | **64.2** | **64.2** |
| **Telescope** | Mirrors | 85.0 | 85.0 |
| | Aperture stop | 5.0 | 5.0 |
| | Absorbing primary and secondary stops | 15.0 | 15.0 |
| | Telescope struts | 28.5 | 28.5 |
| | Housekeeping harness | 2.0 | 2.0 |
| | **Total Telescope Assembly** | **135.5** | **135.5** |
| **Sun-shield** | Deployed sunshield | 105.0 | 101.0 |
| | Optics shield | 28.0 | 28.0 |
| | **Total Sunshield** | **133.0** | **129.0** |
| **Structure & Passive Cooling** | Fixed V-groove radiators | 123.0 | 82.0 |
| | Telescope support ring | 81.6 | 64.5 |
| | Bipod supports | 79.1 | 63.3 |
| | **Total Structure and Passive Cooling** | **283.7** | **209.8** |
| **Warm Electronics** | ADR control electronics | 20.0 | 20.0 |
| | 4 K Cryocooler electronics | 30.2 | 17.8 |
| | Detector readout electronics | 20.0 | 20.0 |
| | **Total Warm Electronics** | **70.2** | **57.8** |
| **Payload Subtotal (CBE)** | | **813.3** | **662.7** |
| **Payload Subtotal with 43 % Contingency** | | **1163.0** | **947.7** |
| | Attitude Control System | 72.1 | 72.1 |
| | C&DH | 36.0 | 20.4 |
| | Power | 47.3 | 44.0 |
| | Propulsion (dry) | 41.3 | 41.3 |
| | Structures and mechanisms | 196.0 | 170.6 |
| | Launch adapter | 16.7 | 16.3 |
| | Cabling | 47.7 | 43.6 |
| | Telecom + Antenna | 22.6 | 20.1 |
| | Thermal | 74.4 | 71.5 |
| **Spacecraft Subtotal (CBE)** | | **554.1** | **499.9** |
| **Contingency (43 %)** | | **238.3** | **215.0** |
| | Propellant [ΔV = 170 m/s] | 295.1 | 295.1 |
| **Spacecraft Subtotal** | | **1087.5** | **1010.0** |
| **Total Launch Mass** | | **2250.5** | **1957.7** |
| **Atlas V 401 Maximum Payload Mass to L2 (C3 = -0.45)** | | **3580** | |
| **Launch Vehicle Margin** | | **1329.5 (37 %)** | **1622.3 (45 %)** |

## 4. Technology Readiness

EPIC-IM is designed around high TRL systems. In cases where the necessary technology has not yet reached design goals, we identify several concrete options for further development and define a path to reach TRL-6. This development benefits enormously from implementing these technologies in sub-orbital and ground-based polarization experiments. Our projections suggest that in ~5 years the entire mission will be at TRL 6, the threshold necessary for entering Phase A.

Planck (see Fig. 4.1) provides a basis for the technology and methodology identified for the EPIC mission. EPIC subsystems such as the 4 K cooler, sub-K cooler, thermal blocking filters, and optics have direct technology heritage from Planck. The signal detection scheme used in EPIC is based on the Planck HFI system of pair-differenced scan-modulated polarized bolometers. The 1.5 m telescope aperture of Planck is essentially the same as EPIC-IM at 1.4 m, as is the highest operating frequency band (850 GHz). EPIC-IM combines passive cooling to ~30 K, a 4 K mechanical cryocooler, and a highly stabilized 100 mK continuous cooler, and all of these elements are found in Planck.

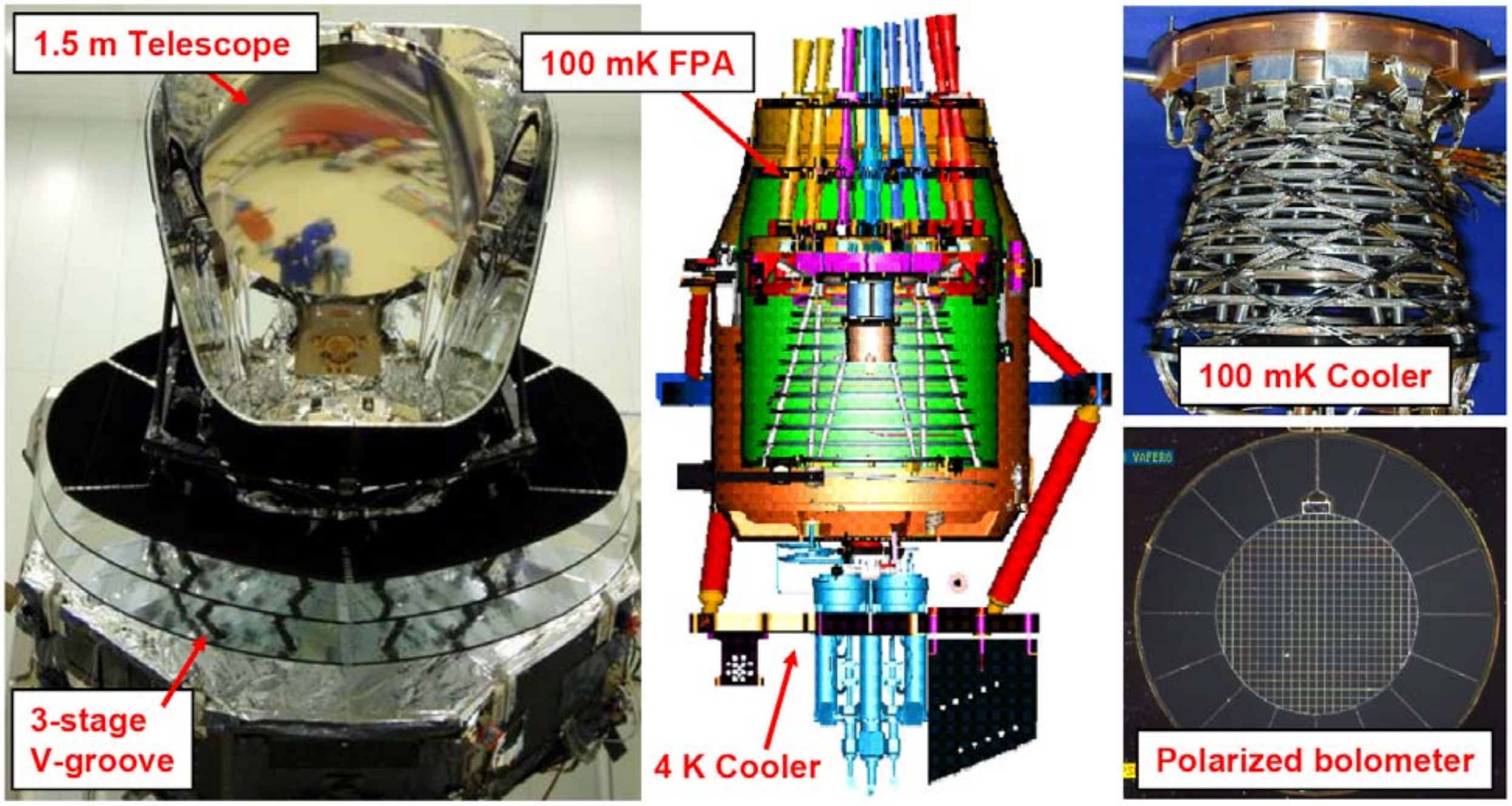

Fig. 4.1. Key technologies pioneered by Planck that provide heritage for EPIC-IM. (Left) Planck uses a 1.5 m CFRP telescope operating at $\lambda > 350$ µm, cooled by a 3-stage passive radiator system to ~35 K. (Center) The High Frequency Instrument has a large ~2 kg 100 mK focal plane cooled by a combination of a 4 K mechanical cooler and a continuous open-cycle dilution refrigerator (upper right). The focal plane is temperature stabilized using active and passive regulation consistent with the EPIC-IM requirements. Planck HFI relies on pair-differenced scan-modulated polarization-sensitive bolometer; EPIC-IM uses the same signal detection strategy but with large focal plane arrays.

The main technology developments over Planck required by EPIC-IM are first the large-format focal plane arrays, which are the basis of EPIC's improved sensitivity. The telescope provides a larger field of view with more rigorous control of far-sidelobes. Finally EPIC-IM uses a scan strategy optimized for polarization measurements, and this requires the multi-stage deployed sunshield.

Fig.4.2 EPIC-IM Technology Readiness and Development Plan

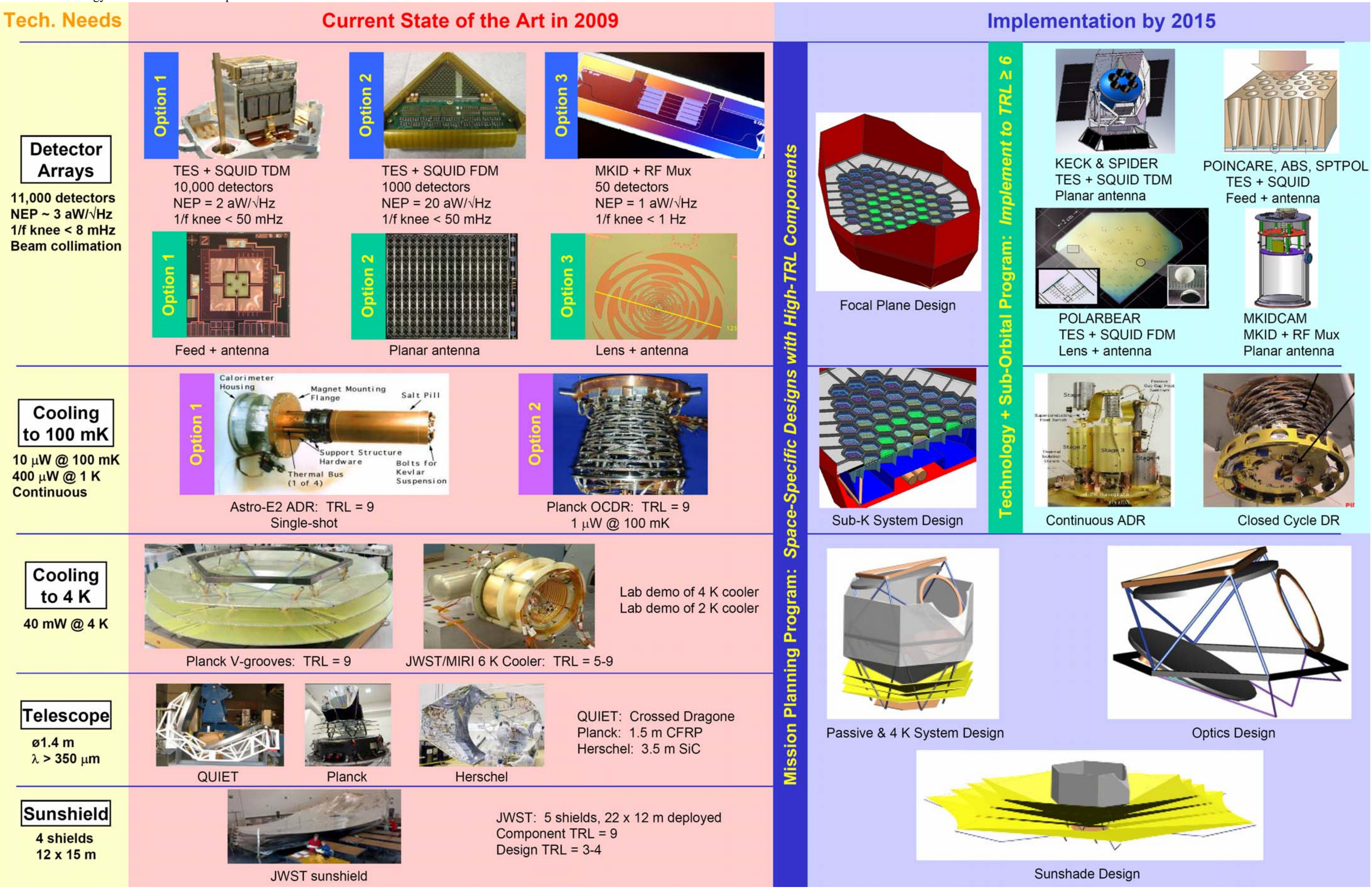

It is worth noting that in some important respects the EPIC-IM design is simpler than Planck. Planck was developed as a merger of two instruments, the Low Frequency Instrument (LFI) originally proposed as the COBRAS experiment, and the High Frequency Instrument (HFI), originally proposed as the SAMBA experiment. This merger placed demanding requirements on the cooling system, a combination of high power dissipation from the LFI and minimum temperature required by the HFI. Furthermore the interaction of the instruments, without precedent in any sub-orbital experiment, placed difficult requirements on systems engineering. EPIC-IM is based on a single detector technology optimized for polarization, which provides a simpler systems optimization. Planck (like WMAP) was conceived as a temperature anisotropy experiment, with polarization capability enhanced only after much of the design, including the scan strategy, was frozen. EPIC is designed for polarization measurements from the beginning.

In Sections 4.1 – 4.4 we discuss the technology readiness of several of the EPIC-IM subsystems. We concentrate on critical instrument specific subsystems. For the spacecraft, we received three responses to a request for information that was issued to industry. They all indicated that EPIC's requirements on standard satellite-bus services, such as pointing and telemetry are relatively standard. Therefore these systems are not discussed in this section. In Section 4.5 we describe the path from the current round of experiment to a technology level that is appropriate for a start of a mission in the middle of the decade.

### 4.1 Focal Plane Detector and Readout Technology

The focal plane includes the detectors and elements that couple radiation from free space. The detector and focal plane technology is also associated with specific readout technology, necessary for large-format arrays. Therefore we include both topics in this section. The sensitivity of bolometric detectors has progressed dramatically over the last 70 years. The gain in mapping speed from bolometer development, a combination of improved sensitivity and array format, as shown in Fig. 4.3 and Fig. 4.4, has been comparable to the familiar Moore's law for computer speed and memory, doubling every 2 years.

Simultaneous with the improvement in sensitivity, there has also been a dramatic increase in the number of detectors that are being implemented within any given instrument, see Fig. 4.4. This increase is the result of using standard silicon micro-fabrication techniques to mass produce detectors. EPIC reaps the benefits of both of these advances by using thousands of state-of-the-art, high sensitivity bolometric detectors.

Two detector technologies are candidates for EPIC-IM, transition edge sensors (TES) and microwave kinetic induction detectors (MKIDs). SQUIDs and HEMTs are used as front-end amplifiers to readout each of these detector technologies, respectively. In all cases the readouts from the thousands of detectors are multiplexed. SQUID multiplexing in the time (TDM) and in the frequency (FDM) domain are equally meritorious candidates at this point. Chapter 7 discusses these differing technologies in more detail.

Currently three distinct technologies are being considered as candidates to couple the radiation from space into the detectors. They are listed in Table 4.1 along with other focal plane technologies, their current state of development and expected TRL in ~2015.

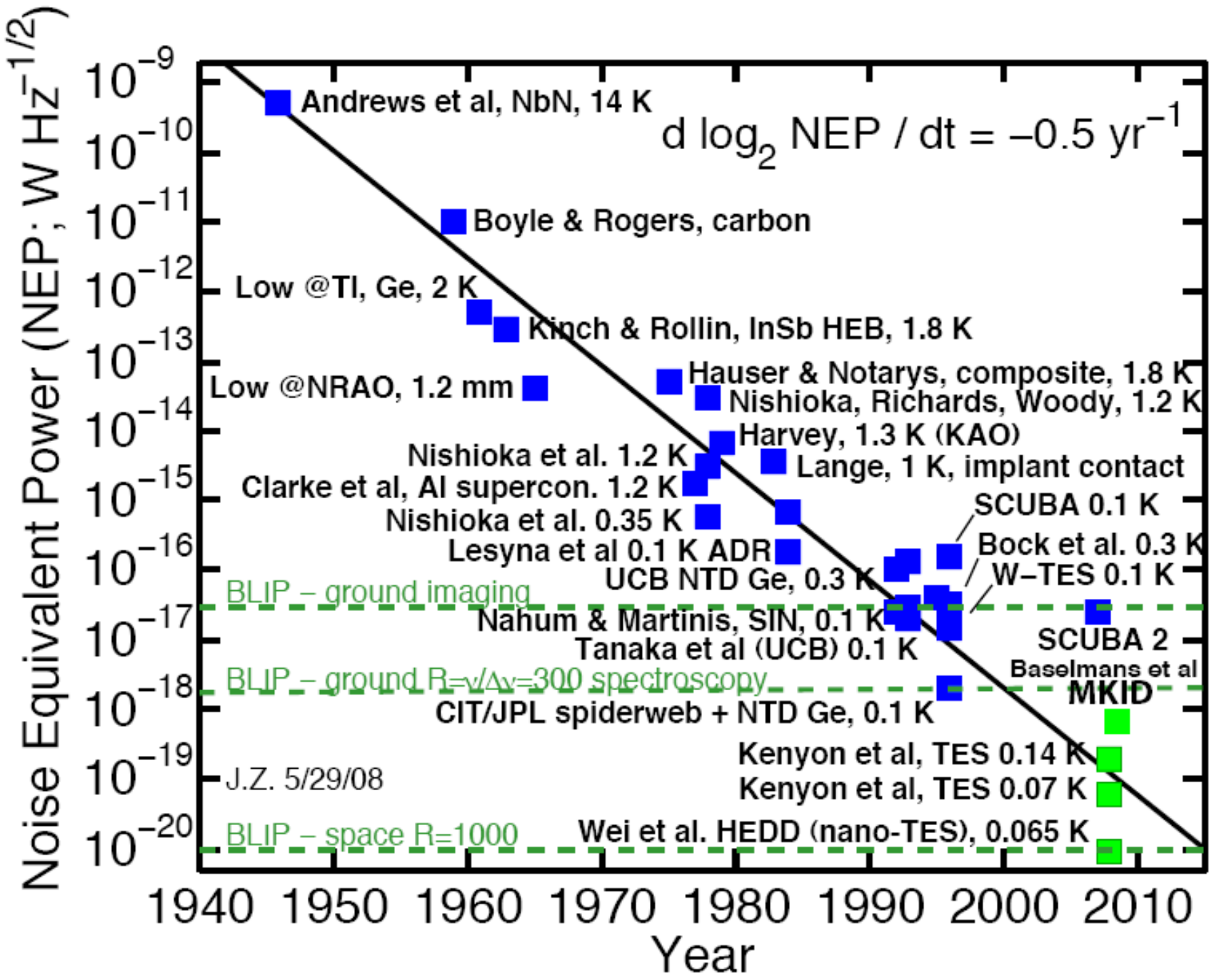

Fig. 4.3. Reduction in Noise Equivalent Power (NEP) over time for bolometric detectors. A reduction in NEP is an appropriate measure for the increase in sensitivity for detectors that are not background limited. Green symbols indicate laboratory devices which have not yet been fielded in instruments. The time required to make an observation scales as $NEP^{-2}$, doubling every year since the 1940s. Horizontal dash lines show several milestone capabilities that this progression in sensitivity provides (the background limit for CMB observations from space is $NEP \approx 3 \times 10^{-18}$ W/√Hz. Once individual detectors become background limited, further gains rely entirely on larger arrays (Figure provided by J. Zmuidzinas).

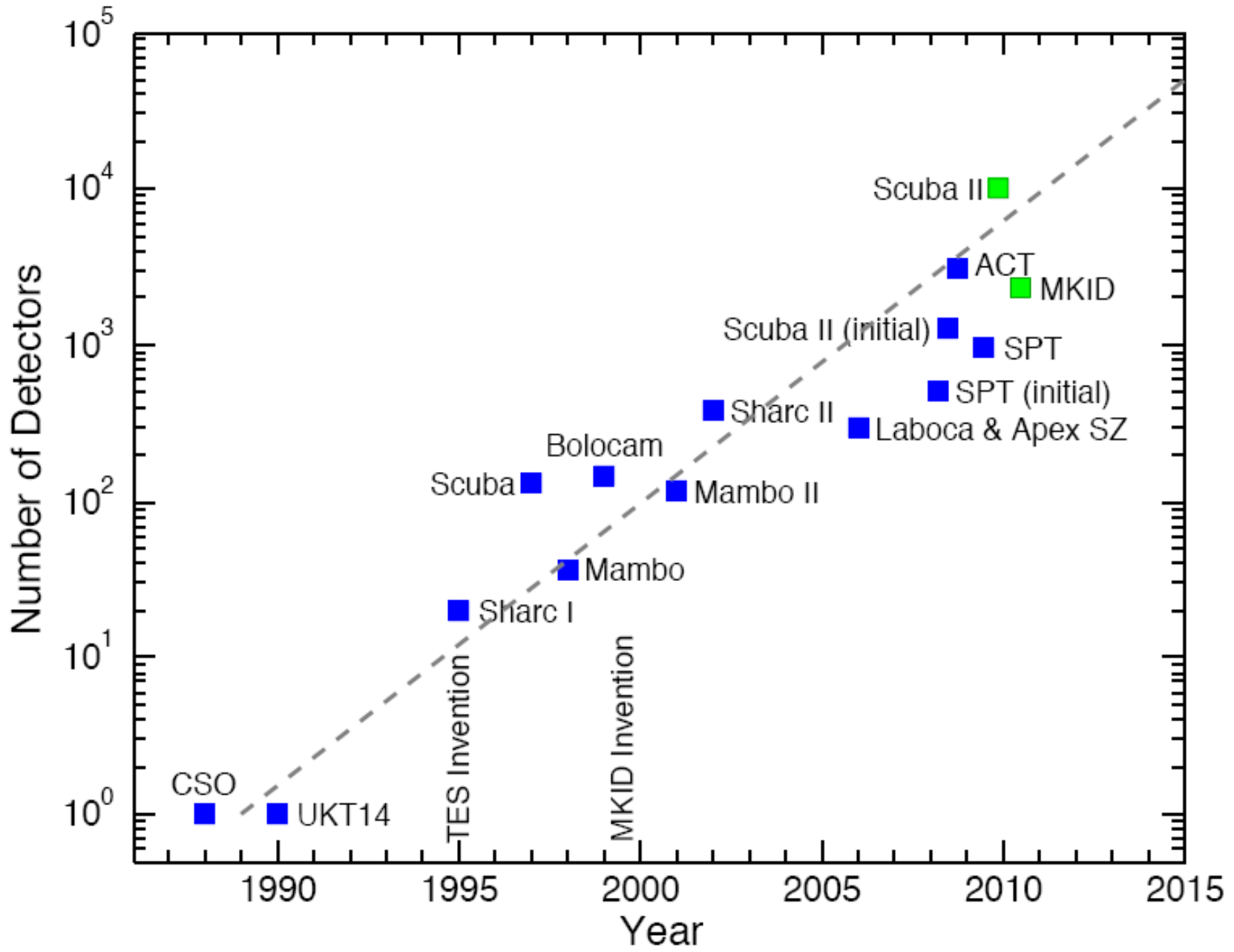

Fig. 4.4. There has been a dramatic increase in the number of detectors employed by instruments since the invention of the TES and MKID bolometric detectors, which lend themselves to mass production using standard silicon micro-fabrication techniques. Blue points indicate arrays operating in instruments, green indicate planned instruments which have not yet realized operation. (Figure provided by J. Zmuidzinas).

**Table 4.1.** Technology readiness of focal plane technologies

| Subsystem | Option | TRL[1] | Heritage[1] |
|---|---|---|---|
| Detectors | TES | 5 (6) | APEX, ACT, SCUBA-II, SPT (EBEX, PIPER SPIDER) |
| | MKID | 4 (5) | (MKIDCam) |
| Readouts | SQUIDS – FDM | 5 (6) | APEX, SPT, (EBEX) |
| | SQUIDS – TDM | 5 (6) | ACT, SCUBA-II, (SPIDER, PIPER) |
| | MKID+ HEMTs | 3 (5) | (MKIDCam) |
| Radiation Coupling | Phased Arrays of Antennas | 4 (6) | (Keck Array, SPIDER) |
| | Lens-Coupled Antennas | 4 (5) | (Polarbear) |
| | Horn-Coupled Antennas | 3 | (ABS, Poincare, SPTPOL) |

[1]For each subsystem and its appropriate option we list the current TRL as achieved by currently operating experiments, as shown in the Heritage column. In parentheses we give the expected TRL by the middle of the next decade after implementation by currently funded (but not yet deployed) experiments, as shown by the experiments in parentheses under Heritage. Note that we judge balloon experiments to be somewhat more representative of space-borne observations than ground-based experiments in assessing TRLs.

There are already funded programs in place that will mature many of the focal plane technologies to TRL 6 by the middle of the next decade. Ground based instruments such as APEX, ACT, and SPT, are already using specific implementations of TES arrays to observe astrophysical objects. The Keck Array and other instruments that will come on line will use somewhat different implementations. The MKIDcam will use 576 detectors and will be coupled to the CSO telescope in Mauna Kea. Elements of the TES technology are poised to advance to TRL 6 with the launch of the EBEX, SPIDER and PIPER balloon-borne payloads within about two to three years. To reach TRL-6 by the middle of the next decade an MKID camera will need to be developed for a balloon borne payload at more representative loadings.

## 4.2 Cooling

The bolometric detector system for EPIC requires cooling to sub-Kelvin temperature. These are achieved by passive cooling of the instrument shell to temperatures around 35 K, using an active cooler to achieve a temperature of about 4 K in a second stage, and a sub-Kelvin cooler that will maintain the focal plane at a temperature of about 0.1 K. The cooling system is described in more detail in Chapter 8.

### *4.2.1 Passive Cooling*

Radiative cooling has been successfully applied to solve cooling needs of many previous space missions. Several commercial software packages are available to model the temperature of a spacecraft and its components under user defined orbital conditions. Passive radiative cooling should therefore be considered as TRL 9.

The V-Groove radiator design of EPIC follows the footsteps of the Planck mission. The design consists of multiple stages of radiators, each having a V-shape cross section. Extensive analysis of the passive cooling of EPIC is presented in Section 8. We show that with standard techniques EPIC achieves its design goals. The design of this subsystem is based on known parameters of materials and on modeling that have been applied successfully in space, most recently with Planck and Spitzer Space Telescope.

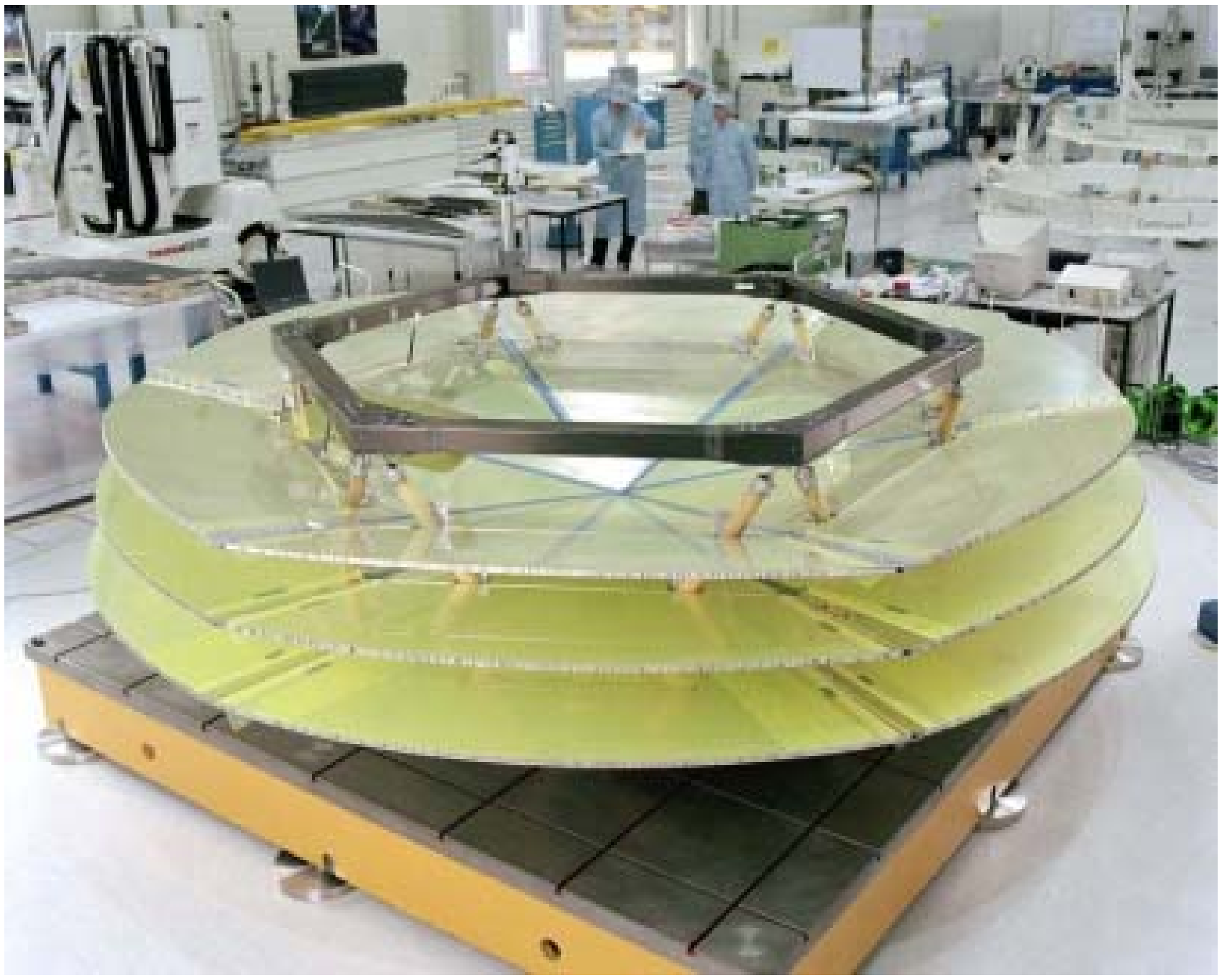

Fig. 4.5. A picture of Planck's 3-stage V-groove radiator at an early stage of assembly .

### *4.2.2 4 K Mechanical Cooler*

The EPIC cryocooler requirements are very similar to the requirements of the MIRI instrument cooler being developed by NGST for NASA's JWST. As a result the cooler is currently relatively mature (technology currently deemed TRL 6 by NASA for JWST) and includes major subsystems that are TRL 9 with flight heritage. The MIRI cooler program is scheduled to be competed well prior to EPIC's need. As a consequence of the hardware similarity to the MIRI cryocooler, the EPIC cryocooler cost and schedule risk will be considerably reduced by the opportunity to derive heavily from the flight hardware development on the leading MIRI program and the availability from the MIRI program of flight drawings, flight manufacturing process documents and flight product assurance documents.

Fig. 4.6 is a block diagram of the proposed EPIC cooler and its differences from the MIRI cooler, containing photos of key components indicating their maturity. The chief hardware differences between the two designs is the addition of a second compressor stage to the JT cooler, removal of the JWST configuration specific CTA components required for the JWST mission and changes to thermal and mechanical interfaces. The second compressor stage is required to attain a higher pressure ratio ~10:1 vs ~3:1 for MIRI single stage. The higher pressure ratio provides a lower gas inlet pressure allowing 4.4 K operation while maintaining the outlet pressure that establishes the mass flow rate for the required level of cooling. With these changes low risk modifications to areas such as the recuperator tubing diameter and JT restriction L/D will be required to achieve the 4.4 K cooling without impacting overall cooler system efficiency relative to the MIRI cooler. With these few changes and modification of the cooler operating conditions the EPIC cooling requirements are readily accommodated.

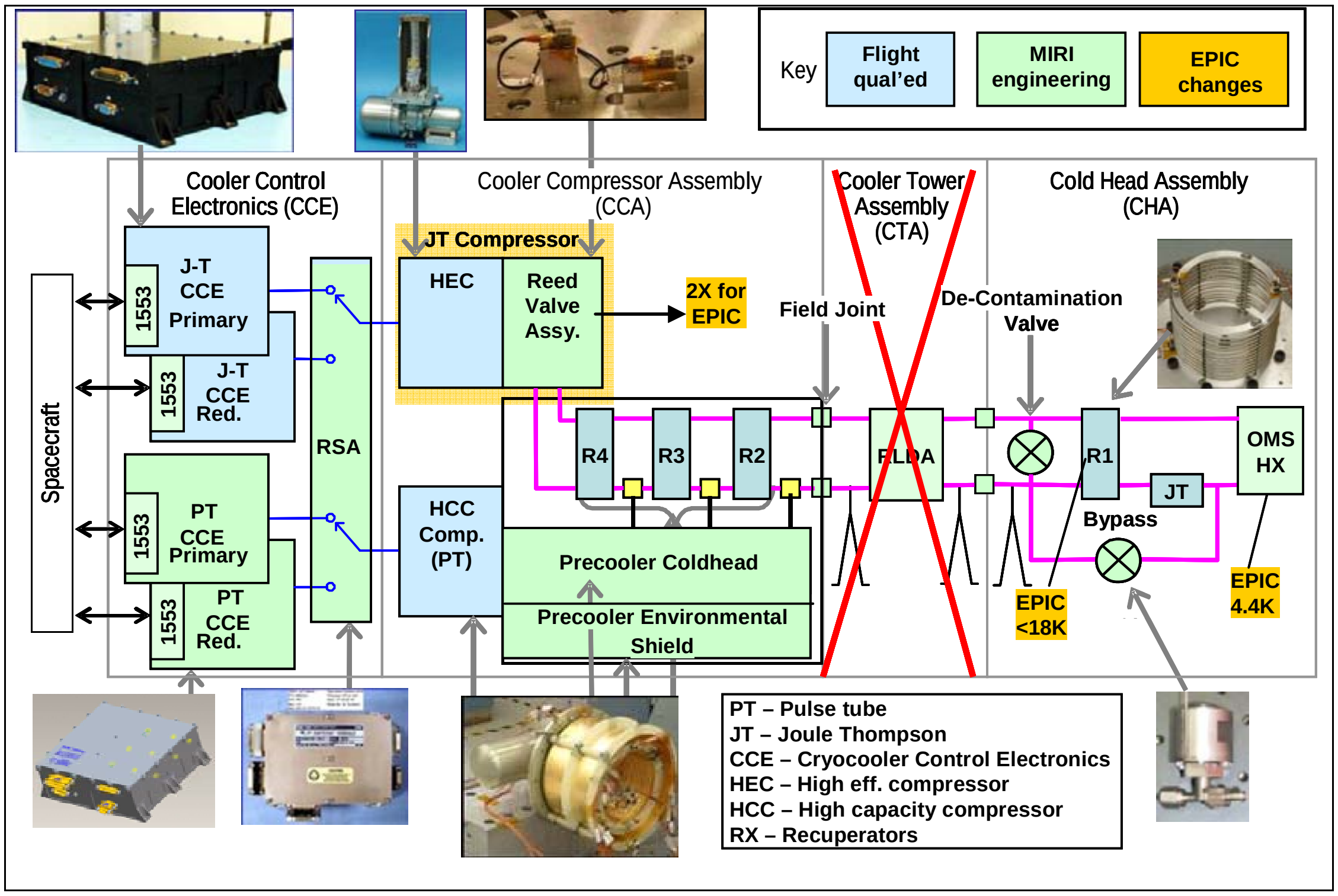

Fig. 4.6. EPIC Cooler Block Diagram Differences from MIRI Cooler

**Table 4.2.** Hardware maturity of the EPIC 4 K cooler[1]

| Sub-System | Current TRL | TRL in 2015 | Comments |
|---|---|---|---|
| **Electronics** | | | |
| CCE (for JT) | 8 | 9 | EM unit in life test at AFRL. 17 Flight units delivered on HTP, NEWT and ABI programs through 2008. |
| Software | 8 | 9 | Flight software delivered on HTP/ABI project. |
| CCE (for Pre-Cooler) | 5 | 9 | Increased power output version of TRL 8 design |
| Accelerometer Preamp | 9 | 9 | 11 flight units in orbit |
| **Pulse Tube Pre-Cooler** | | | |
| Compressor | 6 | 9 | EM acceptance tested and in life test at AFRL |
| PT Coldhead | 6 | 9 | EM passed launch vibe on MIRI program |
| **JT Cooler** | | | |
| Compressor | 9 | 9 | 26 flight units built, 2 in orbit, 1 more to be launched in 2009 |
| Valves | 6 | 9 | Qualified on MIRI program, Based on RAL Planck cooler valves (TRL 8) |
| Recuperators | 6 | 9 | Qualified on MIRI program |
| 4.4 Kelvin Cold Head | 6 | 9 | Similar 6K cold head qualified on MIRI program |
| **Bypass Valve** | 5 | 9 | Purchased component |

[1]Based on the cooler for the MIRI instrument on board JWST; many of its components have TRL 9

Table 4.2 provides the current TRL level of the heritage hardware envisioned for EPIC and the expected TRL level in five years. All the hardware is planned to be TRL 9 in five years because this time period extends past the planned launch date in 2013 for JWST.

*4.2.3 Sub-Kelvin Cooler*

EPIC-IM requires ~10 μW continuous cooling to 100 mK with high temperature stability, coupled to a large multi-stage focal plane structure. This development benefits from the many single shot coolers[6] used on past experiments to achieve sub-Kelvin temperatures (50-300 mK) for far-IR and mm-wave detectors. Examples of such sub-orbital and orbital experiments include BICEP, ACBAR, Python, Boomerang, MAXIMA, ZSPEC and IRTS, XRS, and Herschel, respectively.

One technological path towards a continuous sub-Kelvin cooler is the open cycle dilution refrigerator [1] (OCDR) used on Planck. The OCDR will cool the Planck High Frequency Instrument to 100 mK in space continuously for 2-3 years. Thus the OCDR is at TRL 8 and will soon be at TRL 9. The OCDR requires tanks of high pressure $^3$He and $^4$He gas that are expendable, similar to the stored cryogen in cryostats. To achieve the higher heat lift required for EPIC, a promising alternative is a closed cycle version of the Planck OCDR (CCDR) recently demonstrated by Puget et al. [2]. This system has achieved a base temperature of 39 mK and several μW of lift at 100 mK. The cooler requires a mechanical pump, similar to those already in use for the MIRI and SPICA coolers, to circulate $^3$He and a $^4$He phase separator, which dissipates a few mW at $T \leq 2$ K. A space qualified 2 K cooler with sufficient heat lift for this intercept is already available. This stage can also be engineered into the CCDR system using the circulating $^3$He for use with an intercept temperature of 4-6 K. The CCDR is currently at TRL 4. Advancing to higher TRL requires operation with flight model pumps, system engineering and design of a CCDR for specific flight applications, and operation of the CCDR at an environment similar to the one in space.

A second option for the 100 mK cooler is a continuously cycling adiabatic demagnetization refrigerator (cADR) [3,4,5]. The cADR works by coordinated cycling of several single shot ADRs to produce a fixed temperature at the detector and heat intercept stages. A prototype cADR has been demonstrated in the laboratory so is at TRL 4. Many of the technologies for the cADR have been adapted from the TRL 9 XRS single shot ADR. Three technologies needed for a space-qualified cADR require development. These are: (1) reliable heat switches with on/off switching ratios larger than 1000 and operated at temperatures as low as 50 mK, (2) magnetically shielded, high field (4-6 Tesla) magnets with quench protection, and (3) low thermal impact high temperature superconducting cable harnesses to carry the 1-15 A of current needed to drive the magnets and possibly the heat switches. In addition to individual component development, system level testing of the cADR with prototype instruments is important. This technology poses risk of EMI from the magnet drive and the presence of cycling magnetic fields nearby the sub-Kelvin instrument that is likely to contain superconducting electronics. Demonstration of a cADR sub-Kelvin with an instrument in the next 5 years would advance this technology to TRL 6.

The size of proposed kilo-pixel bolometric detector systems operated at 100 mK is unprecedented even for ground-based systems. EPIC-IM has a focal planes distributed over a one meter diameter cryogenic space. The size and mass will require careful sub-Kelvin structural and thermal engineering. This engineering effort should demonstrate the required base temperatures at the detector focal plane with a light weighted mounting structure and a ~10 nK temperature control over a focal plane. It is worth noting that Planck has already achieved the necessary temperature stability, albeit with a smaller and less massive focal plane.

[6] A single shot cooler needs periodic recycling.

## 4.3 Optics

The optical system of EPIC-IM consists of the telescope and band defining filters. The telescope comprises of two reflectors and an aperture stop, all mounted on struts and actively cooled to 4 K. The coupling of the electro-magnetic radiation into the detectors is discussed in Chapter 7.

### *4.3.1 Telescope*

From an optical point of view the aperture stop and reflectors have a TRL 9 because reflecting systems with dual mirrors have flown successfully in the past, most recently at these wavelengths with WMAP and now with Planck and Herschel. The Planck telescope, with a projected aperture of 1.5 m and a minimum wavelength of 350 μm, was fabricated from carbon-fiber reinforced plastic composite. The Herschel telescope, with a 3.5 m aperture and a minimum wavelength of 60 μm, was fabricated from silicon carbide, significantly exceeding the requirements for EPIC. Both of these telescopes were tested at cryogenic temperatures (40 – 80 K), but not quite to the 4 – 30 K range planned for EPIC-IM. A detailed trade and materials study would occur early in the project to assess the relative merits of CFRP and SiC, including mass, performance, cost and risk. Although the specific crossed-Dragone design has not yet been tested in space, its implementation does not impose new challenges.

Actively cooling of the mirrors and the entrance aperture is not expected to pose a substantial challenge. The 4 K cooler, which is described in more detail in Section 8.2, is already sized to provide cooling for the telescope and has 100 % margin on thermal loads. It is a scaled and slightly modified version of the cooler for the MIRI instrument on board JWST.

### *4.3.2 Filters*

The large throughput of EPIC leads to a focal plane of about 1 m in diameter that is tiled with hexagonally shaped detector wafers. The band defining filters are located on the sky side of each of the detector wafers and therefore their maximum size is determined by the size of these wafers. Filters that are currently implemented on sub-orbital experiments have sizes of up to 30 cm and technology exists to increase this size to 45 cm. These sizes already exceed the planned 15 cm hexagonal cells planned for the focal plane. Therefore filtering technology has TRL 6.

### *4.3.3 Optical Design, Simulations and Software*

There are a number of software packages capable of carrying detailed optical analysis of the reflector-only optical system of EPIC-IM. These include ZEMAX and CodeV, for ray optics, and DADRA and GRASP for physical optics. (To our knowledge neither DADRA nor GRASP can include lenses). These packages have become industry standards and their results have been validated in numerous applications most recently with WMAP and Planck. The most significant challenges may be in the detailed design for far-sidelobe suppression, and whether absorbing baffles at 4 K could provide an overall systems benefit.

### 4.4 Deployed Sunshade

There is an extensive list of space missions that have successfully deployed solar panels, mesh antennas, and membrane-like structures on-orbit. Thus, most or all components of the sunshade have TRL 9. Although the EPIC system is built around known and well-tested components, it is in conceptual form and thus has a 'design TRL' of 3-4. By 2015 JWST will have demonstrated deployment of a large, multi-layer sunshade on-orbit, albeit in a different configuration. This sunshade is more than twice the size and has a more complex deployment scheme than the proposed EPIC sunshade.

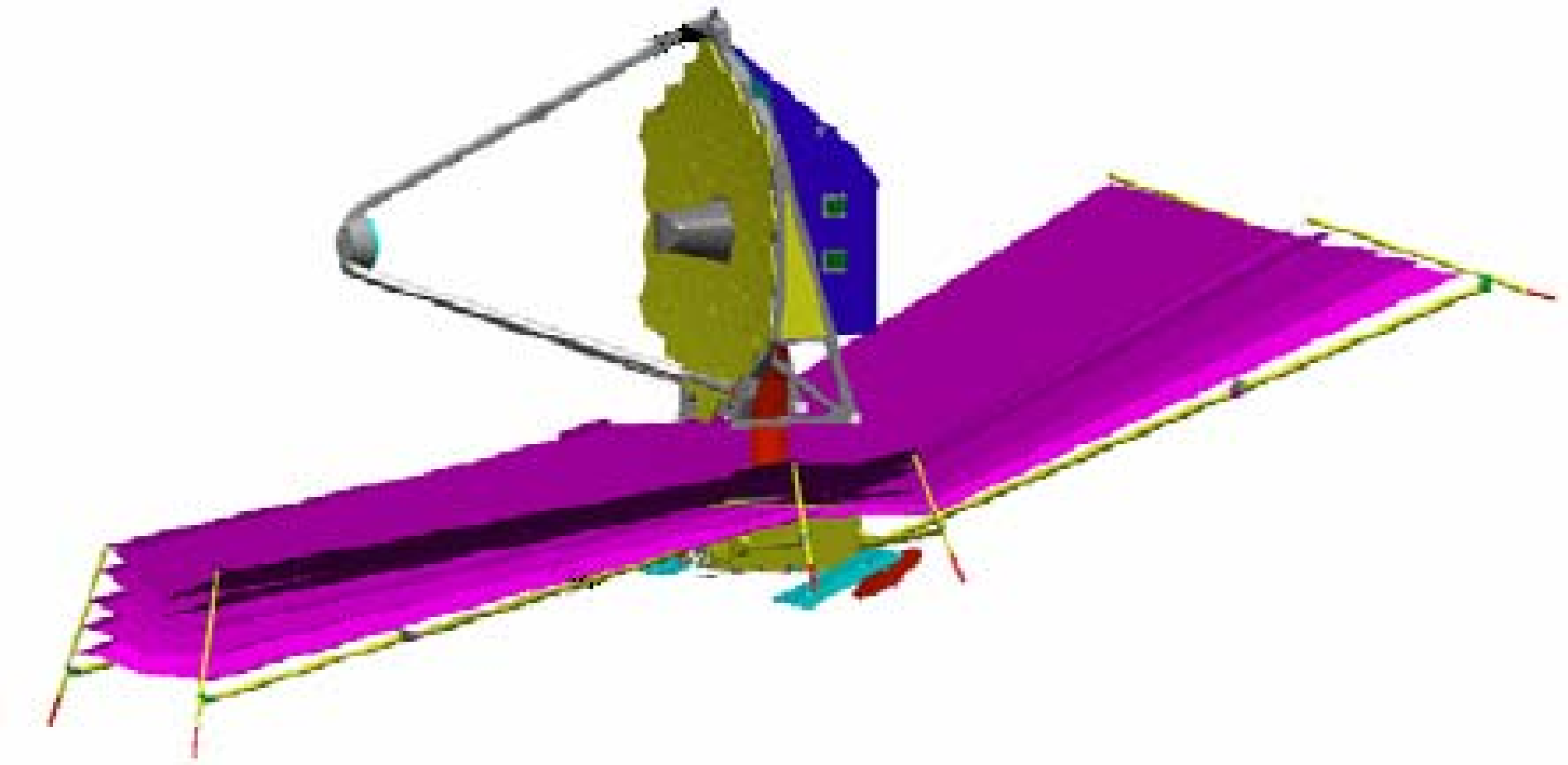

Fig. 4.7. JSWT Telescope with Deployable Sunshade

The EPIC sunshade deployment can be driven by one of a wide variety of electric motor actuators that have been proven on orbit multiple times on many different spacecraft. The deployable hinges are based on Aquarius and MER heritage designs, and although many deployable, locking hinges have already been used in space, the EPIC versions will require qualification at some level in order to be deemed flight-worthy. The composite struts that support the sunshade are made from material that has been flown extensively. Likewise, the EPIC optics 'tent' is also TRL 9, as it does not deploy and contains structural aluminum tubes, metallic joints, and MLI blankets that have flown on countless missions.

We have investigated an alternative option for a deployable sunshade system for EPIC. This alternative is based on the successful wrapped-rib style deployable antennas. Wrapped-rip designs are typically used for very long struts that will not fit into a rocket fairing using a simple folding approach. For a sunshade of the diameter proposed for EPIC, a wrapped-rib design would add complexity and be comprised of mechanisms with less heritage.

### 4.5 The Technology Path to Phase A for EPIC-IM

The CMBPol Strategic Mission Concept Study team has recommended that NASA pursue an active program of sub-orbital experiments and technology development over the next decade [6]. Fig. 4.8 shows the recommended funding profile of two programs to develop and implement technologies for CMBPOL, a technology development component which is tighly coupled to ground-based and sub-orbital experiments, and a mission planning component looking forward to design issues that are unique to a space mission. The funding profile holds the expense level over the decade roughly commensurate with the level now. The flow of technologies from their current state through these programs is illustrated in Fig. 4.2.

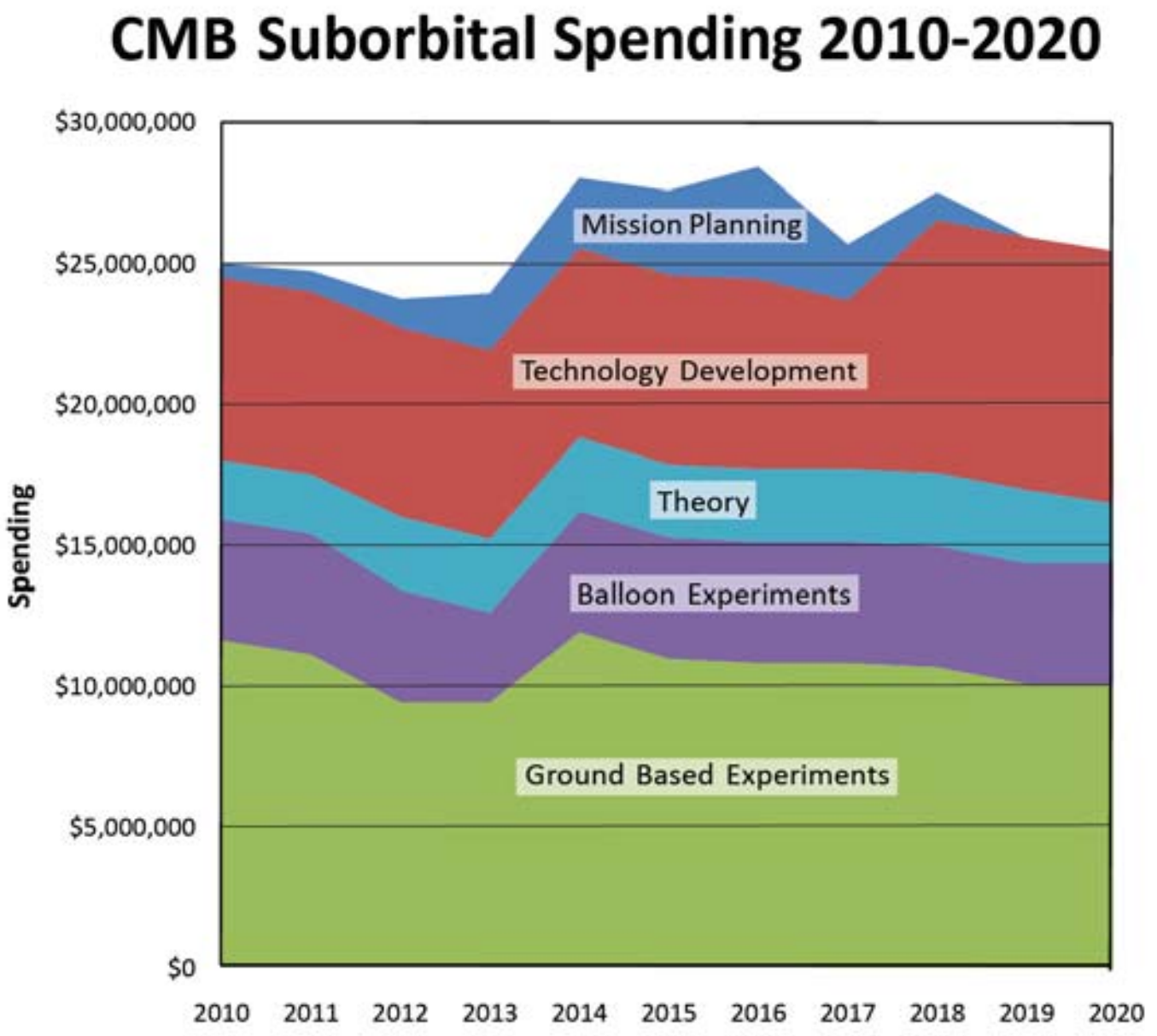

Fig. 4.7. Funding profile recommended for the next decade by the CMBPol Strategic Mission Concept Study team

The development program, funded by NASA and DOE, will feed component technologies into a vigorous ground-based observations program, funded by NSF, and a sub-orbital balloon program, funded by NASA. The observing program will rapidly implement focal plane detectors, readouts, and sub-K coolers in working science instruments. This experience is invaluable for not only carrying forward CMB polarization science, but for real-world experience with the methodology of CMB measurements, including systems design and techniques for suppressing systematic errors. This approach has a proven heritage from the sub-orbital pathfinders that preceded COBE, WMAP, and most recently Planck. The results of these programs will guide a final selection of technologies for the satellite mission.

In parallel, the Mission Concept Study team has also recommended the establishment of a funding wedge for Mission Planning, funded by NASA and possibly under the auspices of a project office, prior to a phase-A start. The Mission Planning component will advance the systems designs required for a space mission by understanding the impact of design choices such as scan strategy and frequency band allocation, and studying the issues across all technologies that are unique to the space environment. Mission Planning supports an assessment of the systems issues with different components by carrying out an analysis of different options. This design work guides the development of those systems with high TRL components but need adaptation to the EPIC-IM architecture, such as the sun-shield, passive and 4 K cooling systems, and the telescope design and structure. This program is invaluable for spotting systems risks early on in the development, and transitioning technologies developed in the laboratory and sub-orbital experiments to the unique requirements of a space mission.

## 5. Systematic Errors

Polarimetric fidelity is integral to EPIC's design. Detecting the nano-Kelvin CMB signals imprinted by the inflationary gravitational wave background requires advances, not only in detectors and optics, but also in systematic error mitigation. Fortunately, all of these systematics have been confronted, and in most cases mitigated, by many first-generation CMB polarimeters. Some of these effects are avoidable by experiment design, and some can be corrected for at the post-data-acquisition phase as part of the data analysis. We have modeled the impact of optical and thermal systematic effects on the EPIC mission. Many of these, such as thermal and electrical gain drifts, 1/f noise, far-sidelobes, and pointing errors, are already familiar from experiments designed for CMB temperature anisotropy. For polarimetry, a new class of error arises from the polarimetric fidelity of the optical system, which produces false B-mode polarization signals from much brighter temperature anistropy and E-mode polarization.

Throughout this report we have defined our *requirement* on control of systematic errors such that the impact of the effect is below the science target of r = 0.01 after correction. Specifically, as Table 5.1 indicates, we *require* that the residual systematics level be less than one-third of the cosmic variance of the binned signal power expected for r = 0.01 for $\ell \leq 100$ where the modes are binned in bands $\Delta\ell/\ell = 0.3$. Our requirement is that we have sufficient knowledge of systematic errors so that they can be controlled, post-correction, allowing EPIC to detect an r = 0.01 gravitational-wave B-mode polarization power spectrum. The requirement also allows an essentially systematics-free detection of the B-mode lensing signal [1]. Our more-ambitious *design goal* is to suppress the raw amplitude of systematic effects below EPIC's binned instrumental noise for $\ell \leq 100$ so that the effect is negligible without correction. The goal and requirement levels are illustrated in Fig. 5.2. These levels are intended to apply to the residuals of all systematic errors. However, estimating the combined effect of all systematics requires a full simulation, so for simplicity we instead investigate the levels applied to individual effects later in this section.

**Table 5.1** Systematic Error Requirement and Goal

| Measurement Criteria | Requirement | Design Goal |
|---|---|---|
| Measure inflationary B-mode power spectrum to astrophysical limits for $2 < \ell < 200$ at r = 0.01 after foreground removal | Suppress systematic errors below cosmic variance level for $\ell \leq 100$ at r = 0.01 after correction | Suppress raw systematic effects below instrument noise for $\ell \leq 100$ |

Note: The systematic criteria are specified for $\ell \leq 100$, due to the rapidly rising B-mode lensing signal which prevents a cosmic-varience limited measurement at $\ell > 100$.

### 5.1 Description of Systematic Effects

Systematic errors in the measurement of polarization can be induced by imperfection in the optical beams, temperature drifts of the optics and detectors, scan synchronous signals from various sources including far-sidelobe response to local sources such as the sun, earth, moon and Galactic plane, 1/f noise in the detectors and readouts, and calibration errors. We pay particular attention to polarization and shape imperfections of the main telescope beams, since these are effects particular to polarimetry. Throughout this report we assume that the polarization is measured by the difference of matched detector pairs, where each bolometer is sensitive to linear vertical (V) or horizontal (H) polarization. Differencing the signals from the matched pair reduces common-mode signals from unpolarized radiation, as well as thermal drifts, pick-up, and stray magnetic fields. Furthermore we assume the signals are modulated by scanning the spacecraft at a relatively low spin rate ~0.5 rpm. We first provide a short taxonomy of the numerous sources of systematic error listed in Table 5.2.

***Main Beam Effects:*** The optical system can produce a variety of effects associated with polarization and shape deviations in the main beams. Instrumental polarization effects leak CMB temperature anisotropy (T, $\nabla$T, $\nabla^2$T, etc) into polarization (characterized by both E- and B-modes), while cross-polarization effects leak CMB E-mode polarization into B-mode polarization. Of the two effects, instrumental polarization is generally more important since temperature anisotropy is always brighter than E-mode polarization anisotropy. Nevertheless because B-mode polarization is so much fainter than E-mode polarization, cross-polarization effects must still be treated carefully. The requirements on beam effects are translated to the end-to-end optical chain, including the telescope design.

***Scan-Synchronous Effects:*** We define any effect that does not average down from the scan strategy over the duration of the mission as a scan synchronous effect. Scan-synchronous signals typically arise from a geometry that is external to the spacecraft or optics. Far sidelobe response to the sun, earth, moon and Galactic plane will produce a scan fixed pattern. Thus the optical system needs to have a very high degree of off-axis rejection to these sources of emission. Solar heating can also give a scan-synchronous signal. Although EPIC always holds the sun at a fixed angle so that the average solar power is constant, shadows on the sun side of the spacecraft can produce a scan-synchronous signal by inducing a slight temperature variation associated with the ecliptic poles, which are observed in-phase with the shadows. Pickup from magnetic fields at L2 can also produce a scan synchronous signal.

***Thermal Drifts:*** Temperature drifts in the optics can produce time-varying optical signals on the detectors due to variations in thermal emission. To first order, this largely unpolarized signal is removed by the common-mode rejection of the detector pair difference, but since the common mode rejection is not perfect, the temperature of the emitting optic must be sufficiently stable, either through passive design or active control. Temperature fluctuations of the 100 mK focal plane also produce false bolometer signals which mimic optical power but are due to variations in the thermal power flowing through the detector's isolating supports. These fluctuations are removed by differencing detectors - to the extent that pairs of detectors are matched. We assume that thermal drifts must be controlled on the time scale of a spin period of the spacecraft. This assumption is conservative since drifts are less serious for smaller angular scales.

***Other:*** In addition to the effects listed above, a wide variety of systematics can potentially result in spurious B-mode polarization signals. 1/f noise in the detectors and readouts can cause stripes in the map resulting in a loss of sensitivity to particular $\ell$-modes. This effect is at least partially mitigated by having a highly cross-linked scan strategy; reducing the effect of stripes in the cross-scan direction. The focal plane can either be designed with sufficient intrinsic stability, as in the case of Planck, or the polarization signal can be actively modulated. Mismatched passbands between detector pairs will leak intensity from unpolarized foregrounds into polarization. The relative gain of detector pairs are calibrated on the CMB dipole, so passband differences will cause differential gain to foregrounds. Table 5.2 summarizes the most challenging systematics, and how they imprint signals into the data stream.

**Table 5.2.** Summary of Systematic Effects and Mitigations

| Systematic Error | Description | Potential Effect | Mitigation |
|---|---|---|---|
| *Main Beam Effects – Instrumental Polarization* | | | |
| $\Delta$ Beam Size ($\Delta\mu$) | $FWHM_V \neq FWHM_H$ | $\nabla^2 T \rightarrow B$ | Telescope design[b]<br>In-flight beam measurements[a]<br>Orbit-modulated dipole[a]<br>Scan crossings[a] |
| $\Delta$ Gain ($\Delta g$) | Mismatched gains | $T \rightarrow B$ | |
| $\Delta$ Beam Offset ($\Delta\rho/\sigma$) | Pointing V ≠ Pointing H | $\nabla T \rightarrow B$ | |
| $\Delta$ Ellipticity ($\Delta e$) | $E_V \neq e_H$ | $\nabla^2 T \rightarrow B$ | |
| Satellite Pointing | Q and U beams offset | $\nabla E \rightarrow B$ | Dual analyzers[b]<br>Pointing specification[a] |
| *Main Beam Effects – Cross Polarization* | | | |
| $\Delta$ Rotation | V & H not orthogonal | $E \rightarrow B$ | In-flight measurements on polarized sources[a] |
| Pixel Rotation ($\varepsilon$) | V ⊥ H but rotated w.r.t. beam's major axis | | |
| *Scan Synchronous Signals* | | | |
| Far Sidelobes | Diffraction, scattering | Pickup from sun, earth, moon and Galactic plane | Optical baffling[c]<br>In-flight measurements on moon, 6-month jackknives[a] |
| Thermal Variations | Solar power variations | Temperature variation in optics, detectors | Passive thermal design[a] |
| Magnetic Pickup | Susceptibility in readouts and detectors | Residual signal from ambient B field | Focal plane shielding[c] |
| *Thermal Stability* | | | |
| Optics Temperature | Varying optical power from thermal emission | Residual signals from temperature variations | Dual analyzers[b]<br>Temperature control[a] |
| Focal Plane Temperature | Thermal signal induced in detectors | | |
| *Other* | | | |
| 1/f Noise | Detector and readout drift | Striping in map | Stable detectors and readouts[b] |
| Passband Mismatch | Variation in filters | Differential response to foregrounds | Measure to the required level[b] |
| $\Delta$ Speed of Response | Different time response between bolometers | $\nabla T \rightarrow B$ | Measure to the required level[b], Scan crossings[a] |

[a]Proven in space or to be demonstrated by Planck
[b]Sub-orbital demonstration planned
[c]Sub-orbital demonstration planned, but significant adaptation required for space
Notation: Differential gain $\Delta g \equiv (g_1-g_2)/g$
Differential beam size $\Delta\mu \equiv (\sigma_1-\sigma_2)/\sigma$ where $\sigma = (\sigma_1+\sigma_2)/2$
Differential beam offset $\Delta\rho/\sigma \equiv (\theta_1-\theta_2)/\sigma$
Differential ellipticity $\Delta e = (e_1 - e_2) / 2$ where $e = (\sigma_x-\sigma_y) / (\sigma_x+\sigma_y)$
Pixel rotation $\varepsilon$ in arcmin

Spacecraft pointing errors depend on the technique used for polarization analysis. With a single analyzer, these errors will couple temperature gradients into polarization, similar to differential beam offset. With a dual analyzer, there is no temperature leakage and these errors only convert E polarization into B polarization.

## 5.2 Main-Beam Systematic Effects

Polarimeters require more stringent control over optical systematics effects than temperature anisotropy measurements. Main beam systematics can be classified according to instrumental polarization (IP) and cross-polarization (XP) effects and their behavior under rotation of the beams around their symmetry axes. This is the approach first taken by Hu, Hedman, & Zaldarriaga (2003) [2] who simulated systematic effects for coherent (RF amplifying) polarimeters. Unlike Hu *et al*., the approach of Shimon et al. (2008) [3] assesses the impact of systematic effects on the Stokes parameters directly, and is more relevant to bolometric polarimeters such as EPIC. We performed calculations using two separate techniques; one assessing the systematics effects in the map domain and the other in the Fourier domain. Results from both approaches are in good agreement.

### *5.2.1 Instrumental Polarization Effects*

We parameterize optical systematic effects by their distortion of two nominally Gaussian beam patterns associated with each of the two linearly polarized antenna planes that correspond to two matched bolometers. The antenna planes are referred to as vertical 'V' and horizontal 'H' and thus each bolometer is sensitive to either a V or an H polarization. Each antenna pattern is given by $G(\theta) = \exp(-\theta^2/2\sigma^2)$, where $\theta$ is the boresight angle and $\sigma$ is the beamwidth.

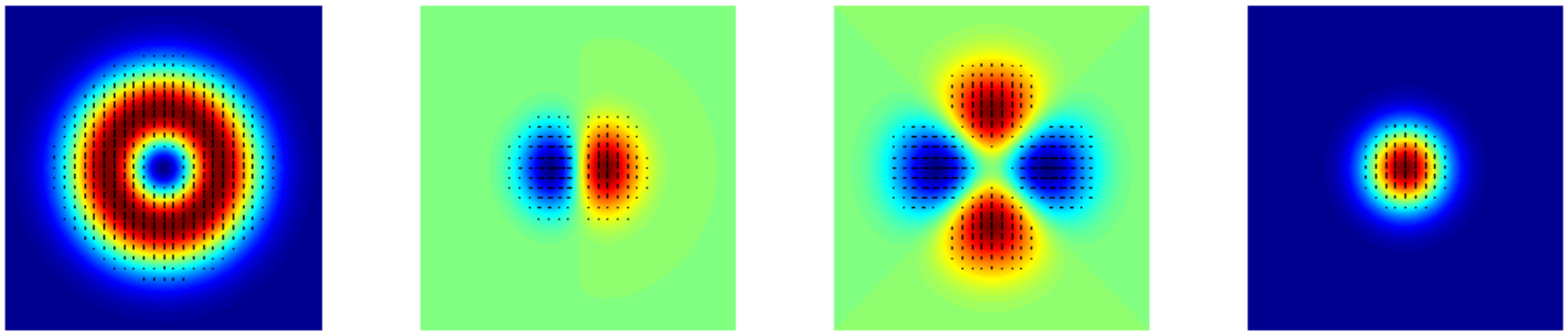

Fig. 5.1. Systematic effects in real space. From left to right: differential FWHM (monopole effect), differential beam offset (dipole IP effect), differential ellipticity (quadrupole effect) and differential gain (monopole effect).

Main beam effects are shown graphically in Fig. 5.1. 'Differential Gain' occurs when the two detectors have unequal optical transmission or gain. Differencing the bolometer signals associated with each antenna leads to an apparently polarized signal. 'Differential Beam Width' occurs when the two beams are circularly-symmetric Gaussians, but have different beam widths $\sigma_V \neq \sigma_H$. If each antenna in the pair produces an elliptically shaped beam then 'Differential Ellipticity' corresponds to the effect arising from the difference in ellipticities. The effect of 'Differential Beam Offset' is caused when the direction of the centroid of the two beam patterns on the sky is not identical, and couples gradients in the CMB temperature anisotropy into polarization. Beam systematics induced by differential ellipticity and beamwidth depend on the second gradient of the underlying temperature anisotropy on scales comparable or smaller than the beamwidth. As a result, higher-resolution experiments will have lower systematics on scales relevant to the inflationary B-mode peak, and so general level of control required for EPIC-IM with a 1.4 m aperture is relaxed compared to EPIC-LC with a 30 cm aperture. For EPIC-IM these systematics peak on scales of the lensing-induced B-mode (few arcminute scales).

Rotating the optics view angle on a patch of sky helps to separate polarization systematics from the instrument from polarization on the sky. True sky polarization has a quadrupolar symmetry, so that the signal in a pair of detectors varies as sin 2θ, where θ is the instrument view angle on the sky. Both differential gain and differential beam width possess monopole symmetry, i.e. the signal does not change by rotating the instrument. Differential beam offset has dipolar symmetry and varies as sin θ. Both of monopolar and dipolar effects can be entirely removed with an ideal scan strategy. Differential ellipticity however has a quadrupolar symmetry and is not separable from true polarization with instrument rotation.

*5.2.2 Cross-polarization Effects*

Main beam cross polarization effects can be caused by several sources. 'Differential Rotation' causes the two polarizations measured in a detector pair to be non-orthogonal. Since this effect is second-order (it converts polarization), it is ignored by Hu *et al*. [2], but we include it. Differential rotation can be caused by misalignment between detector pairs. 'Pixel Rotation' corresponds to the axes of a pixel staying orthogonal but rotating on the sky. This effect may be caused by rotation of the optics with respect to the focal plane, or uncertainties in the satellite rotation angle. Both effects also enter simply due to uncertainty in measuring the polarization axes on the sky. 'Optical Cross-Polarization' enters into all optical systems due to distortion, which changes the magnification over the field of view but also rotates angles on the sky.

A challenge with cross-polarization is that the effects have the same quadrupolar rotational symmetry as that of the true B-mode signal. Thus cross-polarization effects cannot be distinguished through rotation of the instrument. Spurious second-order cross-polarization effects can also arise, producing cross-correlations between E and B (as well as TB cross-correlations) and their respective power spectra, effectively introducing "forbidden spectra" such as EB. These effects can interfere with cosmological measurements of non-standard cosmology which exhibit intrinsic TB and EB correlations (Appendix B).

*5.2.3 Main Beam Requirements*

We flow down the high level systematics goal and requirement levels in Table 5.1 to main beam requirements. We first assume for simplicity that each effect must be suppressed individually below these levels rather than the combination. Then we apply the required level only to the bands with useful CMB sensitivity at 70, 100, 150 and 220 GHz. We scale the goal level so that the effect is sub-dominant to the sensitivity level in each band. Note that the requirement and goal are based on the primary inflationary B-mode science requirement and thus cover $\ell < 100$ and do not extend to higher multipoles required for secondary science such as lensing. Finally because beam effects rise rapidly with $\ell$, we take the goal and requirement at $\ell = 100$. This generally keeps beam systematics below the required and goal levels, although is not exact and some effects do violate the goals at lower $\ell$. While approximate, these levels allow us to assess the level of control and knowledge needed to mitigate main beam effects.

In Fig. 5.2 we illustrate the results of our multipole-based calculations. In addition to the beam effects described above, we simulated the effect of satellite pointing errors after reconstruction. Note that all of these effects are calculated for a single focal plane pixel at 100 GHz in Figs. 5.2 and 5.3. To the extent that parameters vary over the focal plane, these effects will partially average down to give a smaller residual signal. Beam effects have various dependencies on the beam width σ. In power spectrum units ($\mu K^2$), differential gain and differential rotation are independent of beam size, but differential beam width and differential ellipticity scale as $\sigma^4$. Differential pointing scales in a complicated manner, but for our uniform scan strategy we found it scales as $\rho^2$ (where ρ is the pointing offset). Consequently, for a given requirement on the level of systematics induced by differential beamwidth and ellipticity at a given ℓ (e.g. ℓ ~ 100), the allowed differential beamwidth and ellipticity increase in higher resolution experiments compared to lower resolution ones.

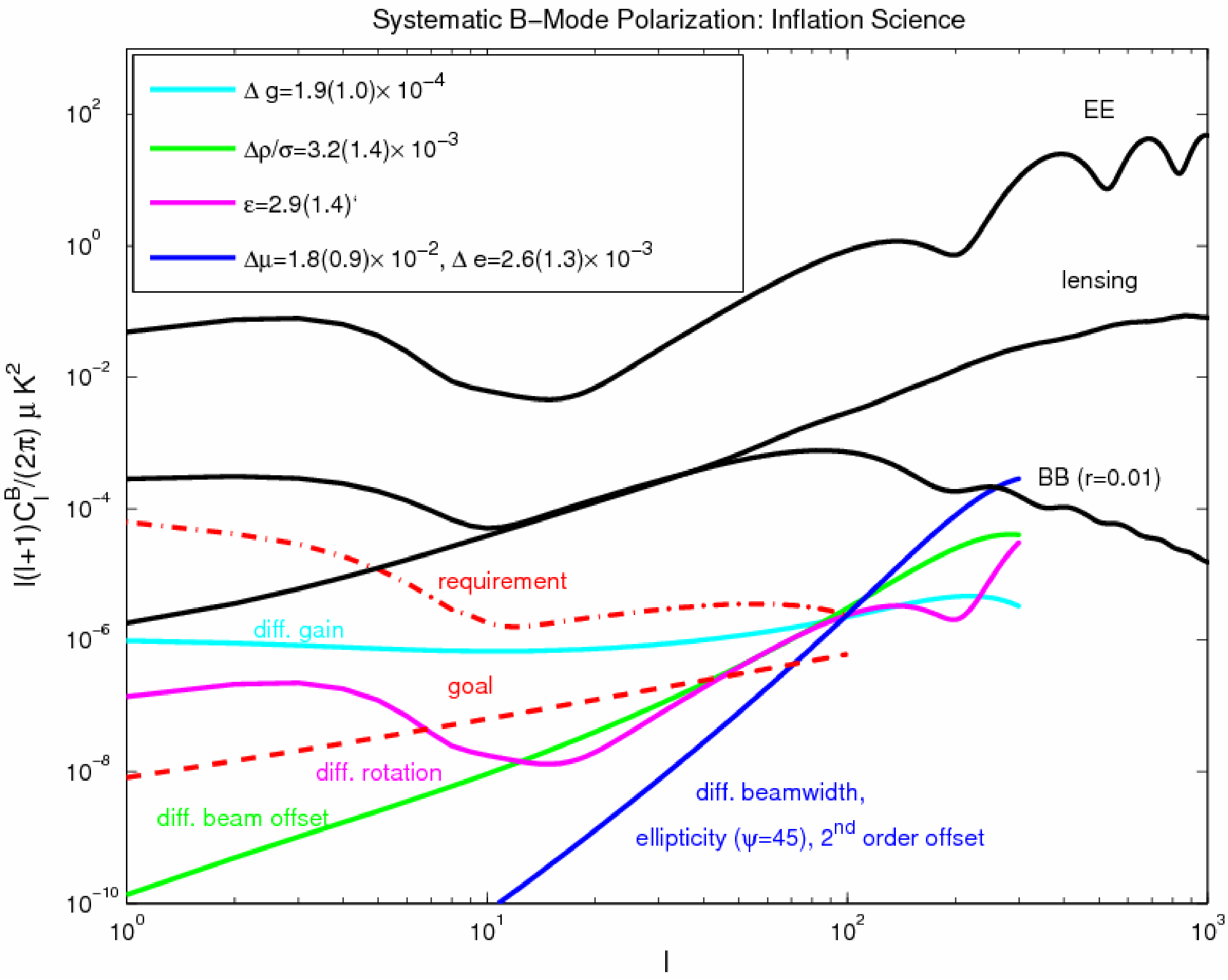

Fig. 5.2. Spurious B-mode instrumental polarization power spectra (T, ∇T, $\nabla^2$T → B) for EPIC-IM 4 K at 100 GHz. The amplitudes of the effects are all chosen to be equivalent at ℓ = 100 and coincide with the requirement curve on beam systematics at this ℓ, somewhat higher than the goal curve. Note that the solid blue curve corresponds to three separate effects which have the same power spectrum. Differential ellipticity is shown for ψ = 45°, which only produces B-mode polarization. Differential beam offset is shown for two cases, one case for the EPIC scan strategy, and the other case is for an idealized scan pattern covering all scan angles uniformly over the entire sky. These spectra indicate the level of the raw effect, or the level of knowledge required for removal. The threshold requirements (goals) for those beam parameters are summarized for each effect in the plot.

The tolerances we derive coincide at ℓ = 100, set by the systematics requirement and goal. Table 5.4 summarizes the beam parameters tolerances required for meeting this criterion in the seven CMB frequency bands. We note that the requirement on differential beam offset is very stringent. We think the data in the present scan pattern can be manipulated to further reduce the effects of differential beam offset by taking advantage of the wide range of scan angles available on any pixel on the sky, eliminating the remaining small non-uniformity in angular coverage from the scan pattern.

### *5.2.4 Error Tolerancing and Science Impact*

The inflationary B-mode signal peaks on horizon scales ($\ell \approx 100$) and decays very quickly. The energy scale of inflation is proportional to the amplitude of this primary B-mode peak, which in turn is nearly independent of all other cosmological parameters. An additional peak (at $\ell \approx 10$) in the B-mode power spectrum due to inflationary gravitational waves which re-entered the horizon at reionization ($z \approx 6$). Since most of the main beam effects couple to CMB temperature gradients they typically increase with $\ell$ and, if constrained to be low at $\ell \approx 100$, are generally negligible at $\ell \approx 10$. A refined, Fisher-matrix-based, analysis for EPIC-IM (4 K) which propagates the beam systematics into bias on r is presented in Appendix A.

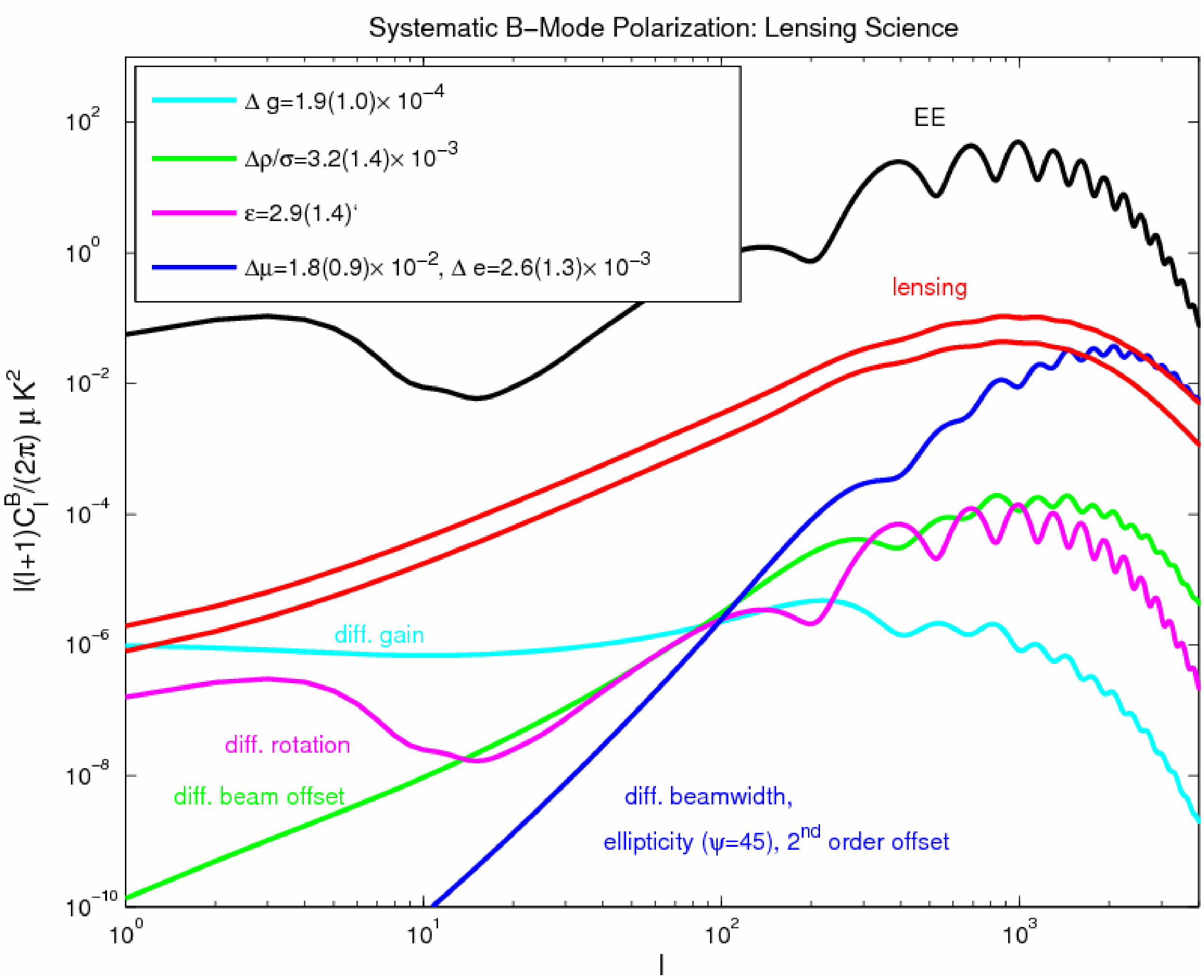

Fig. 5.3. The impact of spurious B-mode instrumental polarization power spectra (T leakage to B-mode) on CMB lensing science for the same beam systematics parameters as in Fig. 5.2. Red curves that peak at $\ell \sim 1000$ correspond to the lensing-induced B-mode signal for two different cosmological models 1) with massless neutrinos (top red curve) and 2) with total neutrino mass of 2 eV (bottom red curve). In both cases the beam systematics (optimized for and calibrated by the inflationary signal) are at least an order of magnitude smaller than the lensing signal in all multipoles of interest.

CMB lensing by the intervening large scale structure (the distribution of which depends on several cosmological parameters, most notably neutrino masses and dark energy equation-of-state) distorts the CMB. Lensing is best *detected* with the measurement of B-mode correlations on arcminute scales, which are non-existent on these scales in the absence of lensing. In contrast to the inflationary-induced B-mode polarization whose amplitude depends on the, yet unknown, energy scale of inflation, the B-mode generated by lensing is guaranteed and lies within the standard cosmological model framework. Fig. 5.3 extends the plot of Fig. 5.2 beyond $\ell = 300$. It is clear from this plot that the systematics do not rise fast enough from $\ell = 100$ to $\ell = 1000$ (the peak of the lensing B-mode signal) to compromise the lensing science - irrespective of neutrino mass. The continuous red curves bracket the minimum and maximum lensing signal for massless neutrinos and 2eV total neutrino mass. This demonstrates that beam systematics optimized for inflation science (assuming r = 0.01) give negligble errors to the lensing B-mode signal.

## 5.3 Systematic Error Mitigations

EPIC-IM is designed to provide high polarization fidelity for detecting B-mode polarization. Many of the effects described have been controlled in previous experiments, and we intend to follow these practises in the development of the experiment. Furthermore we have designed EPIC to allow measurement of many associated errors in orbit, providing a degree of assurance that systematic errors that are not negligible in amplitude can still be controlled and removed. Thus our approach combines heritage gained from experience in previous and on-going CMB measurements, with techniques that can only be realized in space.

### *5.3.1 Systematic Error Mitigation Strategy*

We have designed multiple levels of systematic error suppression into EPIC-IM, exploiting the natural advantages provided by a differential polarimeter. As shown in Table 5.2, main beam effects are first reduced by the crossed-Dragone telescope, which has very low abberations and main beam polarization effects. The scan pattern provides a high degree of variation in the view angle on the sky, ideal for mitigating some of these effects, particularly differential beam offset. The relative gains of detector pairs will be measured in-flight on the annual dipole, the unpolarized signal due to the earth's orbit around the sun, using the daily dipole as a transfer standard which is measured every 2 minutes. The Planck estimate that the orbit-modulated dipole can provide an absolute calibration accurate to 0.4 % [4]; its accuracy for relative calibration has not been simulated. Furthermore the beams can be measured very accurately in flight on bright unpolarized point sources. The latter two techniques are being used by both WMAP and Planck.

Cross-polar effects can be corrected *post-facto* by measurement of known polarized sources. We note in particular that diffuse polarized emission in the Galactic plane is relatively bright and will be precisely characterized by sub-orbital and ground-based measurements well in advance of EPIC. These sources have a non-CMB spectrum and place requirements on knowledge of the pass-bands.

Far sidelobes are reduced by the optical design, which benefits from the use of cold absorbing baffles located at the aperture stop, and surrounding the primary and secondary mirrors. The off-axis response will be monitored in flight by measurements of the moon during the cruise to L2. Differencing 6-month maps can be used to estimate the level of asymmetric far-sidelobe pickup from the Galaxy [5]. The EPIC-IM sun-shield prevents the sun, earth or moon from illuminating the optics. The solar input power into the instrument is stable, and any scan synchronous signals from the sun are rapidly attenuated through the passive cooling chain. Attenuation of magnetic fields is a new requirement arising from the susceptibility of TES and MKID sensors. EPIC will be guided by sub-orbital experiments spinning in Earth's field (50 μT) which is $10^4$ times larger than the very low field enviroment at L2 (5 nT). These experiments use a combination of high-mu and superconducting shielding, as well as focal plane technologies (gradiometric SQUIDs, bias switching) to reduce susceptibility. The primary hurdle for EPIC is to obtain less-stringent overall shielding, principally shielding of locally generated fields, without adding significant instrument mass.

EPIC analyzes polarization using the difference signal between matched detector pairs, an approach developed for Planck with heritage from BOOMERANG, QUAD, and BICEP. This differencing reduces susceptibility to unpolarized optical emission, particularly varying emission associated with the cold telescope and absorbing baffles. Furthermore, differencing also reduces induced signals from in variations in the focal plane temperature. The temperature of the 100 mK focal plane, as well as the 1 K intercept and 4 K optics, will be stabilized using precision thermometry and thermal control in conjunction with passive thermal filtering. Planck employs a scheme of active and passive control to realize stability close to the specifications required by EPIC. Finally dual analyzers relax the requirement on spacecraft pointing. With a single analyzer, spacecraft pointing errors convert temperature gradients to polarization. With dual analyzers, pointing errors only convert E-mode polarization to B-mode polarization, a much smaller effect.

EPIC mitigates striping due to 1/f noise in the focal plane due to its high intrinsic stability detectors and readout electronics. While the level of stability required for EPIC-IM has already been demonstrated with NTD Ge detectors in ground-based tests of the Planck High Frequency Instrument, for TES detectors the level of 1/f noise suppression has been demonstrated at system level. Passband mis-matches between detectors complicate extracation of foregrounds, and propagate through calibration measurements based on non-CMB sources. These differences can be accounted and corrected given sufficient knowledge of the spectral response, and require careful measurement during ground-based instrument characterization. Finally differential speed of response between detector pairs converts temperature to polarization signal, in an analagous fashion to differential beam offset. This effect can be measured by scanning over unpolarized sources at varying scan speeds during the cuise phase, and is reduced by rotating the scan angle over a region of sky by the scan strategy.

### *5.3.2 Scanning Strategy*

The scan strategy is a central consideration for removing systematics. Rotating the polarization angle on the sky allows us to separate systematics associated with a preferred direction in the focal plane, and allows us to mitigate many of the polarization artifacts associated with main beam mismatches. Furthermore, scan redundancy provides an important check on many systematic effects, by allowing us to compare maps on identical regions of sky over multiple time scales.

EPIC's scan strategy consists of spinning the payload about the boresight axis, and precessing the boresight axis about the anti-solar direction, as described in Fig. 3.5. This pattern provides a highly uniform angular coverage over the entire sky. We obtain fully sampled independent maps of more than half the sky after several precession cycles (< 24 hours), and complete maps of the entire sky in six months. This redundancy allows for the application of multiple statistical tests. For example, by making maps in fixed scan angles, and by comparing maps before and after rotation, we can assess the amplitude of main beam polarization effects before they are removed by polarization-angle rotation. We can construct difference (jackknife) maps on multiple time scales (hours, days, weeks, months, years) to accurately assess instrument noise. The absolute and relative gain of each detector is measured from the CMB's dipole on the same region of the sky on the time scale of several hours. Over the course of the mission we can produce maps in fixed spin angle or fixed precession angle to assess spin synchronous signals. Finally the high-redundancy of the scan pattern mitigates against data interruptions, loss of pixels, and loss of arrays.

We note that the requirement on differential beam offset is very stringent. As evident from Fig. 5.2, this effect is greatly reduced by an ideal scanning pattern, which covers all scan angles uniformly on each piece of sky. We think it is possible that the data in the present scan pattern can be manipulated to significantly reduce the effects of differential beam offset by taking advantage of the wide range of scan angles available on any pixel on the sky to remove any remaining non-uniformity. For example, we note that angular dependence of the measured signal at each pixel is a superposition of the lowest few multipoles of the polarization angle, alpha, viz:

$$I(p) = T(p) + R \cos\alpha + S \sin\alpha + Q \cos 2\alpha + U \sin 2\alpha + \text{higher order multipoles},$$

at a pixel "p". The second and third terms in this expression arise from the dipole, or first order differential pointing contribution to the instrumental polarization. "R" and "S" play a similar role to the true polarization terms Q and U. It is clear that the $R \cos\alpha + S \sin\alpha$ cannot represent true cosmological polarization, which is quadrupolar in nature, being modulated twice for each single physical rotation in α. The R and S terms result from convolution from CMB temperature with the beam. Since the polarization angle α is recorded for each pixel, one can remove all data taken at pixel p with α values that *fail* to measure at both the polarization angle α *and* α + 180. Discarding all measurements which don't have their mirror-counterpart does not mitigate higher order spurious modes such as the quadrupole or octopole, but removes the most pernicious (in practice) main-beam systematic - the 'dipole'. A more refined strategy might be to weight scans mathematically to recover ideal scanning. This is possible because the scan strategy has good coverage of α and α + 180 degrees, for a large number of α values. Of course removing some data comes with a noise penalty. In practice, however, this is a negligible, percent-level effect.

### 5.4 From Science Requirements to Instrument Specifications

We have developed requirements for the precision to which each systematic effect must be suppressed or measured in EPIC-IM to meet the high-level requirements listed in Table 5.1. Requirements are flowed down to specific systematics effects by analyzing each effect and holding it to the required level. Beam effects specific to the angular resolution chosen for EPIC-IM are calculated for each band in Table 5.4. For thermal fluctuations, we calculate instantaneous requirements by requiring that the leakage temperature noise is less than 10 % of the detector NEP when all noise sources (detector, photon, and systematic) are added in quadrature, assuming 99 % matching in gain. For focal plane temperature fluctuations we assume 95 % matching between pairs of TES bolometers. These requirements are already within the level of demonstrated thermal stability by Planck [6] using active and passive control.

For scan-synchronous effects, we take the suppression to be simply flat at 3 $nK_{rms}$ (required) and 1 $nK_{rms}$ (goal), which is the approximate level required without accounting for the spatial signature of the particular effect. Note again that scan-synchronous effects are conservative – effects associated with the instrument alone average down over the course of entire EPIC observing campaign as the satellite maps out a large range of spin and precession angle over the scan strategy. Common mode temperature fluctuations, which may escape detection in individual pairs, will tend to quickly average down. Since different detectors view different parts of the sky, effects that are common to the entire focal plane also benefit from this additional averaging, particularly on small scales.

**Table 5.3** Systematic Error Goals and Requirements for EPIC-IM (4K)

| **Systematic Error** | **Description** | **Suppression to Meet Goal** | **Knowledge to Meet Requirement** |
|---|---|---|---|
| Main Beam Effects – Instrumental Polarization | | | |
| Δ Beam Size (Δμ) | $FWHM_E \neq FWHM_H$ | See Table 5.4 | See Table 5.4 |
| Δ Gain (Δg) | Mismatched gains | | |
| Δ Beam Offset (Δρ/σ) | Pointing E ≠ Pointing H | | |
| Δ Ellipticity (Δe) | $e_E \neq e_H$<br>$\Delta e = (e_1-e_2)/2$ | | |
| Satellite Pointing | Q and U beams offset | < 12" | < 38" |
| Main Beam Effects – Cross Polarization | | | |
| Δ Rotation | E & H not orthogonal | $\theta_1-\theta_2 <$ 8.7' | $\theta_1-\theta_2 <$ 13.4' |
| Pixel Rotation (r) | E ⊥ H but rotated w.r.t. beam's major axis | See Table 5.4 | See Table 5.4 |
| Scan Synchronous Signals[1] | | | |
| Far Sidelobes | Diffraction, scattering | < 1 $nK_{CMB}$ | < 3 $nK_{CMB}$ |
| Thermal Variations | Solar power variations | | |
| Magnetic Pickup | Susceptibility in readouts and detectors | | |
| Thermal Stability[2,3] | | | |
| 4 K Optics[4,5] | Varying optical power from thermal emission | 175 μK/√Hz; 0.8 μK s/s | 525 μK/√Hz; 2.4 μK s/s |
| 0.1 K Focal Plane[6] | Thermal signal induced in detectors | 130 nK/√Hz; 0.5 nK s/s | 400 nK/√Hz; 1.5 nK s/s |
| Other | | | |
| 1/f Noise[7] | Detector and readout drift | 0.008 Hz (0.5 rpm) | 0.1 Hz (0.5 rpm) |
| Passband Mismatch[8] | Variation in filters | $\Delta\nu_c/\nu_c < 1 \times 10^{-4}$ | $\Delta\nu_c/\nu_c < 1 \times 10^{-3}$ |
| Time Constant Mismatch[9] | Differential speed of response | $< 1 \times 10^{-4}$ | $< 1 \times 10^{-3}$ |

[1]Scan-synchronous signals assumed to have a flat power spectrum and match the average goal/requirement for ℓ < 100. More detailed estimates depend on the shape of the systematic error.
[2]Assumes 1 % matching to unpolarized optical power, calculated to give 1(3) $nK_{CMB}$(rms).
[3]Assumes TES bolometers with 5 % matching to focal plane drifts, calculated to give 1(3) $nK_{CMB}$(rms).
[4]Planck achieves < 30 μK/Hz at 4 K regulated on Sterling-cycle cooler stage.
[5]Planck achieves < 5 μK/√Hz at 1.6 K regulated on open-cycle dilution refrigerator J-T stage.
[6]Planck achieves < 40 nK/√Hz at 0.1 K regulated on focal plane with open-cycle dilution refrigerator.
[7]SPIDER: MacTavish et al. (2008) [7].
[8]BICEP achieves $\Delta\nu_c/\nu_c < 1 \times 10^{-3}$ Takahashi et al. (2008) [8].
[9]BICEP achieves $< 3 \times 10^{-3}$ Takahashi et al. (2008) [8].

We assessed spacecraft pointing errors by adding a constant shift in spacecraft coordinates into a map-based simulation. The effect is similar to that of differential beam offset, except that the B-mode signal comes from $\nabla$E instead of $\nabla$T. The resulting goal specification of 12" (1$\sigma$ post reconstruction) is conservative because actual pointing errors will be time variable, and will average down over EPIC's highly redundant observations. Note that the spacecraft is specified to deliver 36" (3$\sigma$ post reconstruction).

### *5.4.1 Main Beam Effects*

To tolerance the required optical fidelity for detection of the IGB signal we propagated simulated main beam systematic effects, like those shown in Fig. 5.2, combined with our scan strategy. Two independent simulations pipelines were developed to appraise the impact of systematic effects associated with deviations of the main-beam from ideal, first in map space and later Fourier space, as described in detail in the previous EPIC study [9], and the two methods agree nearly exactly.

**Table 5.4** Summary of Main Beam Requirements and Goals

| $\nu$ [GHz] | $\theta_{FWHM}$ [arcmin] | $\delta T$ ($\ell$=100) [$nK_{CMB}$][b] | | $\Delta g$ [$10^{-4}$] | | $\Delta\mu$ [$10^{-3}$] | | $\Delta\rho/\sigma$ [$10^{-3}$][c] | | $\Delta e$ [$10^{-3}$][d] | | $\varepsilon$ ['] | |
|---|---|---|---|---|---|---|---|---|---|---|---|---|---|
| 30 | 28 | | 4.2 | | 5.2 | | 4 | | 2.3 | | 0.6 | | 7.8 |
| 45 | 19 | | 2.0 | | 2.5 | | 5 | | 1.6 | | 0.7 | | 3.8 |
| 70 | 12 | 1.6 | 1.1 | 1.9 | 1.4 | 9 | 7 | 2.3 | 1.3 | 1.3 | 0.9 | 2.9 | 2.1 |
| 100 | 8.4 | 1.6 | 0.8 | 1.9 | 1.0 | 18 | 9 | 3.2 | 1.4 | 2.6 | 1.3 | 2.9 | 1.4 |
| 150 | 5.6 | 1.6 | 0.7 | 1.9 | 0.8 | 40 | 18 | 4.9 | 2.1 | 5.8 | 2.5 | 2.9 | 1.3 |
| 220 | 3.8 | 1.6 | 0.9 | 1.9 | 1.1 | 90 | 50 | 7.3 | 4.2 | 13 | 7.2 | 2.9 | 1.6 |
| 340 | 2.5 | | 2.3 | | 2.9 | | 300 | | 14 | | 43 | | 4.3 |

Requirements on: Differential gain $\Delta g \equiv (g_1\text{-}g_2)/g$
Differential beam size $\Delta\mu \equiv (\sigma_1\text{-}\sigma_2)/\sigma$ where $\sigma = (\sigma_1+\sigma_2)/2$
Differential beam offset $\Delta\rho/\sigma \equiv (\theta_1\text{-}\theta_2)/\sigma$
Differential ellipticity $\Delta e = (e_1 - e_2) / 2$ where $e = (\sigma_x\text{-}\sigma_y) / (\sigma_x+\sigma_y)$
Pixel rotation $\varepsilon$ in arcmin

[a]Requirement (blue) and goal (red) levels are referred to band-averaged beams.
[b]Required and goal level $[\ell(\ell+1)\, C_\ell/2\pi]^{1/2}$ at $\ell$ = 100 for EPIC-IM 4 K for a 4-year mission.
[c]Differential beam offset assumes the raw scan pattern. Scan symmetrization relaxes this requirement by approximately a factor of 100.
[d]Differential ellipticity calculated for the worst-case $\psi = 45^o$. EPIC-IM is ~100x less prone to the more typical optical effect at $\psi = 0^o$ which to first order converts T $\rightarrow$ E.

## 5.5 In-Flight Measurements

While the hardware mitigation methods are powerful, and greatly reduce the raw systematic impact, many systematic effects can be removed entirely if sufficiently characterized. We describe several techniques that serve to reduce residual systematic effects post observation.

**Table 5.5** Systematic Error Measurements in Flight

| Systematic Effect | In-Flight Checks |
|---|---|
| Main Beam Effects | Measure beams on polarized and unpolarized sources<br>Combine data in fixed view angles |
| Instrument Noise Model | Construct difference maps |
| Spin Synchronous Signals | Combine data in fixed spin and precession angles |
| Relative Pair Gains | Orbit-modulated CMB dipole using dipole as a transfer standard |
| Instrument Gain Model | |

### *5.5.1 In-Flight Main Beam Measurements*

Main beam systematics ultimately result from uncertainties in the beam parameters, and these effects can be removed through accurate measurements. EPIC will naturally measure unpolarized sources (planets, asteroids) and polarized sources during scientific observations. How well can the beam ellipticity and other parameters be measured? This is a function of detector noise and the density of calibration point-sources on the sky. The details of calculating beam errors from a 2-D map of a point-like source is described in Smith et al. (2008) [10]. An example of measured beam differences is shown in Fig. 5.4.

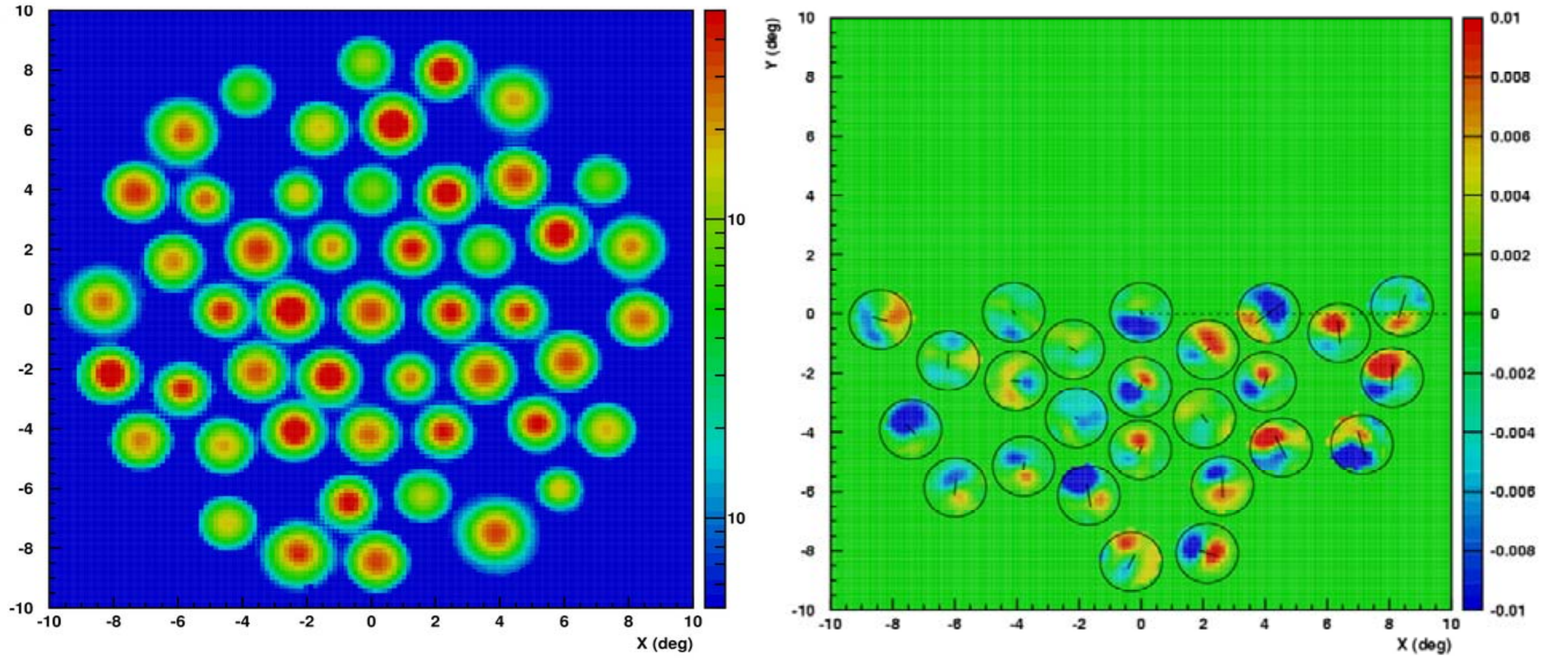

Fig. 5.4. Precision main beam maps from the BICEP experiment, measured with an unpolarized source. The raw unpolarized response of detectors in the focal plane are shown on the left, and the difference beams between detector pairs are shown on the right. Note the scale on the right-hand figure is intentionally expanded to illustrate the residuals in the difference beams. The beams are measured with sufficient accuracy to allow removal of false B-mode signals down to a level of $r \leq 0.01$.

For the case of EPIC-IM (4 K), we estimate the residual uncertainty on main intrument polarization beam effects based on measurments of Jupiter, which is observed daily over a wide range of view angles for approximately 6 months every year. These measurements generally satisfy the goals and requirements, except in the extreme case of the 30 GHz band which has relatively poor CMB sensitivity. Note that this estimation is conservative, since EPIC will be able to carry out similar measurements on other planets and other bright point sources. Furthermore any artifacts from this estimation are unlikely to be confused with true CMB polarization since the residuals will have a very different spectrum from the CMB.

**Table 5.6** In-Flight Measurements of Band-Averaged Beams on Jupiter

| $\nu$ [GHz] | $\theta_{FWHM}$ [arcmin] | $\Delta$ [$10^{-3}$] | $\Delta\mu$ [$10^{-3}$] | | $\Delta\rho/\sigma$ [$10^{-3}$] | | $\Delta e$ [$10^{-3}$] | |
|---|---|---|---|---|---|---|---|---|
| 30 | 28 | 0.9 | | 4 | | 2.3 | | 0.6 |
| 45 | 19 | 0.5 | | 5 | | 1.6 | | 0.7 |
| 70 | 12 | 0.3 | 9 | 7 | 2.3 | 1.3 | 1.3 | 0.9 |
| 100 | 8.4 | 0.2 | 18 | 9 | 3.2 | 1.4 | 2.6 | 1.3 |
| 150 | 5.6 | 0.2 | 40 | 18 | 4.9 | 2.1 | 5.8 | 2.5 |
| 220 | 3.8 | 0.2 | 90 | 50 | 7.3 | 4.2 | 13 | 7.2 |
| 340 | 2.5 | 0.5 | | 300 | | 14 | | 43 |

Note: Following Table 5.4, requirements are shown in blue text; goals are shown in red text on differential beam size ($\Delta\mu$), beam offset ($\Delta\rho/\sigma$), and ellipticity ($\Delta e$).
Jupiter assumed to be 170 K and 50" in apparent diameter.
Cells are shaded red where measurement $\Delta$ does not achieve the required accuracy.

### *5.5.2 Other Measurements*

CMB measurements require an accurate model for the instrument noise. In the case of EPIC this model will be built up over the course of scanning the sky. The model can be checked by fomulating difference maps from the highly redundant scan strategy, testing the validity of the noise model on different time scales.

The instrument gains will be most accurately measured in flight. Following WMAP, we will use the annual dipole modulation produced by the earth's orbital motion around the sun as a calibration tool, and use the dipole on any given day as a transfer standard. The orbit-modulated dipole is unpolarized, and may be used to calibrate out the relative gain between channel pairs to high precision, following an ongoing study by the Planck team to use the dipole to measure relative gains.

Mismatched detector time responses will leak temperature signals into polarization. While the time response can be measured on the ground, the definitive measurements will occur in flight. During the cruise phase we will scan over bright point sources at various scan speeds to measure detector time responses and separate these effects from beam shapes.

In addition to main beam measurements on compact sources, the scan strategy allows us to construct maps in view angles that intentionally exacerbate beam effects. Given our precise knowledge of CMB temperature, these maps can be used to measure the polarization signal produced by beam artifacts, and then estimate and remove these effects in optimally combined maps [11].

# 6. Crossed-Dragone Telescope

For the EPIC-Intermediate Mission, we have chosen the study a dual reflector antenna system coupled directly to the focal plane without reimaging optics. The particular mirror configuration chosen is the Crossed Dragone, also known as the Side-Fed, or Compact Range Antenna. The motivation for choosing this configuration is discussed in [1]. To summarize, the Crossed Dragone provides a very large FOV and is compatible with a flat, telecentric focal plane without the need for refractive reimaging optics. This simple feature avoids many of the issues with refracting elements such as AR coatings, index uncertainty and dispersion, birefringence and surface roughness.

The specific design considerations for EPIC IM are:

1. Compact size compatible with the launch shroud
2. Boresight angle to match 55° angle in scan strategy
3. Large FOV
4. Cold aperture stop
5. Telecentric focal plane
6. Beam scale polarization distortions
7. Low far sidelobe level to minimize Galactic contamination
8. Weight
9. Cooling

The mirrors are fabricated by either lightweight carbon fiber or silicon carbide with aluminized coatings. Cooling considerations dictate that the focal plane be placed close to the spacecraft bus, and that the optical system be surrounded by a optics box to reduce thermal radiation from warmer stages. With this design choice, much of the optical design can be evaluated using geometric optics, with the polarized beams and sidelobes requiring more involved simulation.

By forcing the focal plane to be rigorously telecentric, we have moved the aperture stop from the primary to a location in front of the primary, far enough from the beam so that a cold physical aperture can be used to intercept sidelobes from the focal plane pixels. The geometric design and aberration performance is discussed in section 6.1.

Given that the EPIC-IM design also does not include any refracting elements, the polarization performance of the mirrors can be accurately simulated using physical optics. The Dragone condition for a dual reflector pair produces a central feed with very low cross-polarization. The condition at the same time produces the very large FOV with a side effect of low polarization errors. It is likely that the polarization performance of the optical system will be limited by the focal plane coupling technology. The results of the Physical Optics analysis of the polarized beams appear in Section 6.2, as well as a comparison of the systematics requirements for main beam effects.

While the current EPIC-IM optical design was chosen to maximize throughput, several measures were undertaken to minimize and analyze the far sidelobes produced by diffraction from edges. The baffling plan incorporates an absorbing aperture stop, absorbing rings around the mirrors, and a reflective optics box. In addition to the baffling, the focal plane detectors under-illuminate the mirrors, with an edge taper that depends on the focal plane packing. The far-sidelobes are simulated using both intensive Physical Optics calculations and using the Geometric Theory of Diffraction in section 6.3. Finally, the Galactic contamination from the far-sidelobes is evaluated by convolving the simulated beams with a galactic foreground model as presented in section 6.4.

## 6.1 Geometric Design and Analysis

### *6.1.1 Optical Layout*

It is well known that dual reflector mirrors pairs can be completely specified by five parameters [2]. The exact mirror parameters in this study were chosen to produce an ultra-compact mirror pair, with 15 cm clearance between the beam and the focal plane for filters. The location of the aperture stop was also tuned to be close to the secondary without interfering with the focal plane. This particular realization of the Crossed Dragone is a "local-minimum" in parameter space, and it is possible that a better realization can be found with a more exhaustive search.

The overall scale of the telescope is constrained by the desire to place the focal plane near the spacecraft bus, and at the same time orient the bore sight at 55º from the spacecraft rotation axis. Setting the focal plane near the bus allows for more efficient cooling, and the bore sight requirement is set from the desired scan pattern in Fig. 3.5. Applying the orientation constraint and leaving space for mirror supports and the hexagonal optics box allows a final aperture at 1.4 m. The physical parameters for the EPIC-IM optical elements are listed in Tables 6.1 - 6.3, and a geometric raytrace produced by Zemax is shown in Fig. 6.1.

It is important to note that the focal plane and mirrors have an elliptical outline to maximize throughput. The width of the focal plane in the short dimension is limited by vignetting, while the width in the long dimension is limited by geometric aberrations relative to 30 GHz diffraction. The sizes of the mirrors in Fig. 6.1 were chosen such that the extreme fields were not vignetted by the mirror rims, leaving the cold aperture as the limiting stop.

**Table 6.1** Parameters used to specify the EPIC-IM Crossed Dragone

| Parameter | Value | Description |
|---|---|---|
| Effective focal length | 3.0 m | Sets the plate scale at the Gregorian focus |
| Primary focal length | 5.6 m | Focal length of the paraboloidal primary |
| Primary offset | 5.11 m | Displacement of the optical axis from the symmetry axis of the primary, in other words, the distance that the primary is off center |
| Secondary semi focal spacing | 3.55 m | Half the separation of the foci of the ellipsoidal secondary |
| Horn divergence | 13.13 deg | Half the opening angle at the focal plane, defining the aperture and the marginal rays |

**Table 6.2** Optical Parameters for the EPIC-IM Optical system

| Parameter | Value | Description |
|---|---|---|
| Effective Focal Length | 3.0 m | Focal length of mirror system, also used to calculate plate scale |
| Effective Aperture | 1.4 m | Defined by entrance aperture |
| F/# | 2.14 | |
| FOV X | 30.3 deg | Available FOV in long dimension at 30 GHz |
| FOV Y | 19.9 deg | Available FOV in short dimension at 30 GHz |

**Table 6.3** Derived physical parameters used to construct the EPIC-IM Crossed Dragone

| Parameter | Value | Description |
|---|---|---|
| Dragone Tilt | 61.821 deg | Angle between symmetry axis of primary and secondary |
| Feed Tilt | 28.179 deg | Angle between axis of secondary and axis of the feed (normal to focal plane) |
| Secondary Radius | 7.2214 m | Radius of Curvature of the Secondary, used in Zemax to specify surface |
| Conic Constant | -5.9704 | Conic Constant of the Secondary |
| Primary rim Y | 2.02 m | Projected Minor axis of primary aperture |
| Primary Rim X | 2.34 m | Projected Major axis of primary aperture |
| Secondary offset | 1.11 m | Offset of secondary rim from symmetry axis |
| Secondary Rim Y | 2.2 m | Projected Minor axis of secondary aperture |
| Secondary Rim X | 2.6 m | Projected Major axis of secondary aperture |
| Aperture Diameter | 1.4 m | Entrance aperture is circular |
| Aperture location | 3.10 m | Distance between the center of the entrance aperture to the vertex of the primary. In the modified Crossed Dragone, the aperture is not the primary. Once the focal plane is forced to be telecentric, the aperture moves away from the primary toward the far edge of the secondary. |
| Aperture tilt | 10 deg | Tilt of the entrance aperture. The crossed dragone is highly decentered, and the ideal entrance pupil is tilted. |

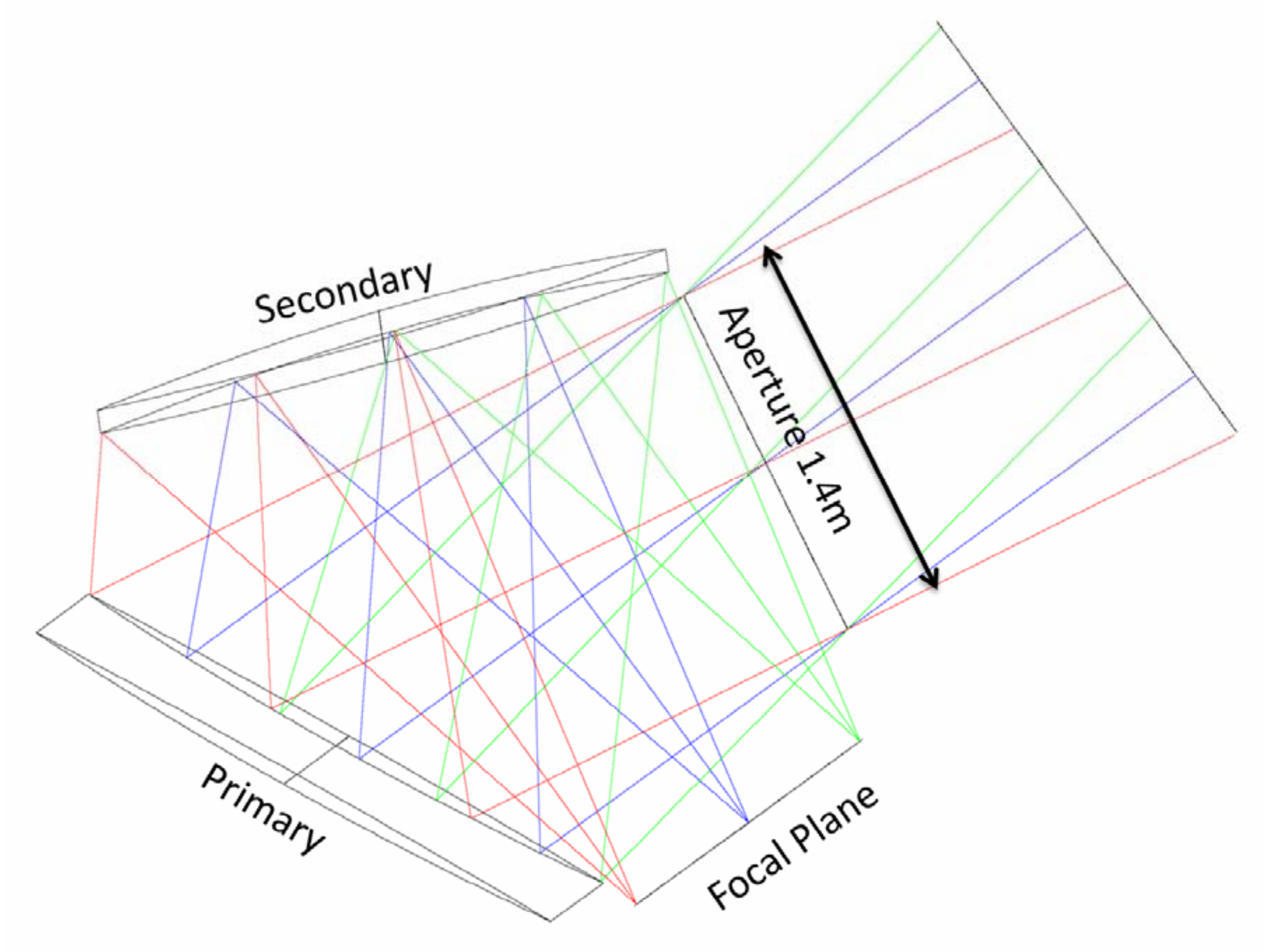

Fig. 6.1. Raytrace diagram of EPIC-IM. The four main components are shown, along with rays from the extreme fields. The marginal rays in this figure represent the -10dB tapered rays from each pixel for the 2 fλ case. For the 3.25 fλ case, the marginal rays are the -25 dB tapered rays. The focal plane and mirror projection here shows the short dimension. The focal plane and mirrors are longer into the plane of the figure.

*6.1.2 Aberration Performance*

The Crossed Dragone has very low geometric aberrations, leading to a very large field of view and throughput. The EPIC-IM cold mission concept takes advantage of this performance with a large focal plane, leading to high system sensitivity. There is enough aberration free area on the focal plane to accommodate 11,000 focal plane detectors in nine frequency bands. The central pixel in the focal plane is perfectly corrected, and aberrations grow slowly with radial feed position. The EPIC-IM focal plane takes advantage of this slow degradation by using the center of the focal plane for higher frequency feeds as shown in Fig. 6.2.

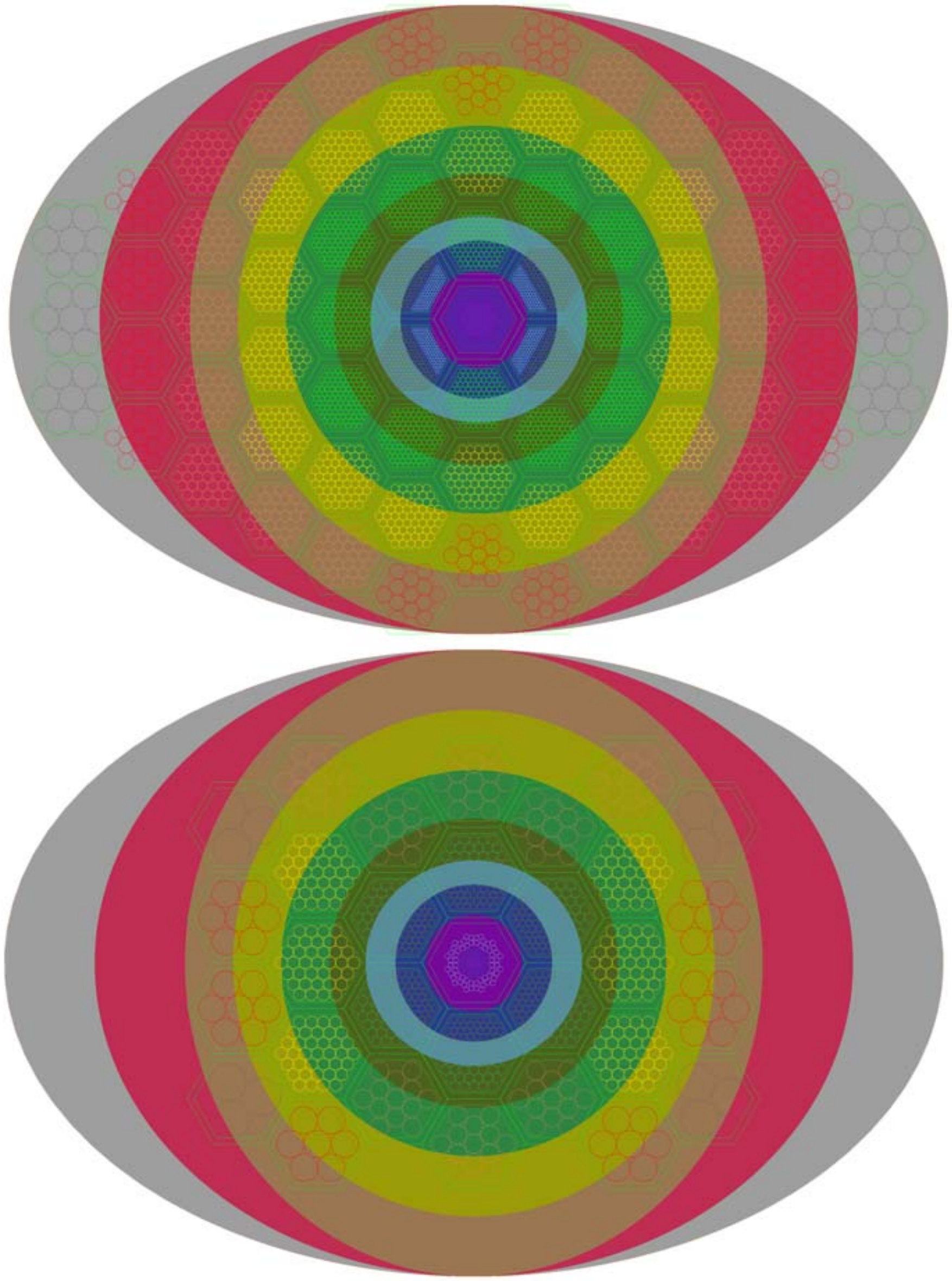

Fig. 6.2 Available focal plane area as a function of position for both the 4K option (top) and the smaller warm option (bottom). The colored regions denote the available area, and the outline of the wafer is overlaid. The extent of the shaded region is bounded by aberrations in the long dimension, defined by Strehl=0.8. The extant in the short dimension is set by vignetting. The colored bands (from inside to outside) are 850, 500, 350, 220, 150, 100, 70, 45 and 30 GHz.

## 6.2 Main Beam Effects

The effects of diffraction on mirror based optical systems can be accurately simulated with physical optics. Without reimaging lenses, the accuracy of physical optics modeling of the EPIC-IM beam performance is limited only by knowledge of the focal plane coupling technology. For the purpose of evaluating the telescope performance, we have modeled the focal plane elements as ideal circular Gaussian tapered feeds.

The EPIC-IM optics were simulated using GRASP-9. One 2 fλ feed at the appropriate wavelength was simulated at the center of each focal plane hex shown in Fig. 6.3. For the purpose of this simulation, it is assumed that a 2 fλ feed spacing corresponds to an aperture beam taper of -10dB. Both orthogonal linear polarizations were simulated to evaluate polarization artifacts. The mirrors were also assumed to have a finite conductivity of 2.5e7 S/m, appropriate for aluminum, allowing us to simulate instrumental polarization.

### *6.2.1 Geometrical Optics vs. Physical Optics*

Fig. 6.3 shows a comparison between the co-polarized output of the simulation and the geometric spot diagrams from raytracing. From the plot, it is clear that the PO simulation adequately reproduces the geometric aberrations, and furthermore that the frequency at each field locations was properly chosen to remain dominated by diffraction effects.

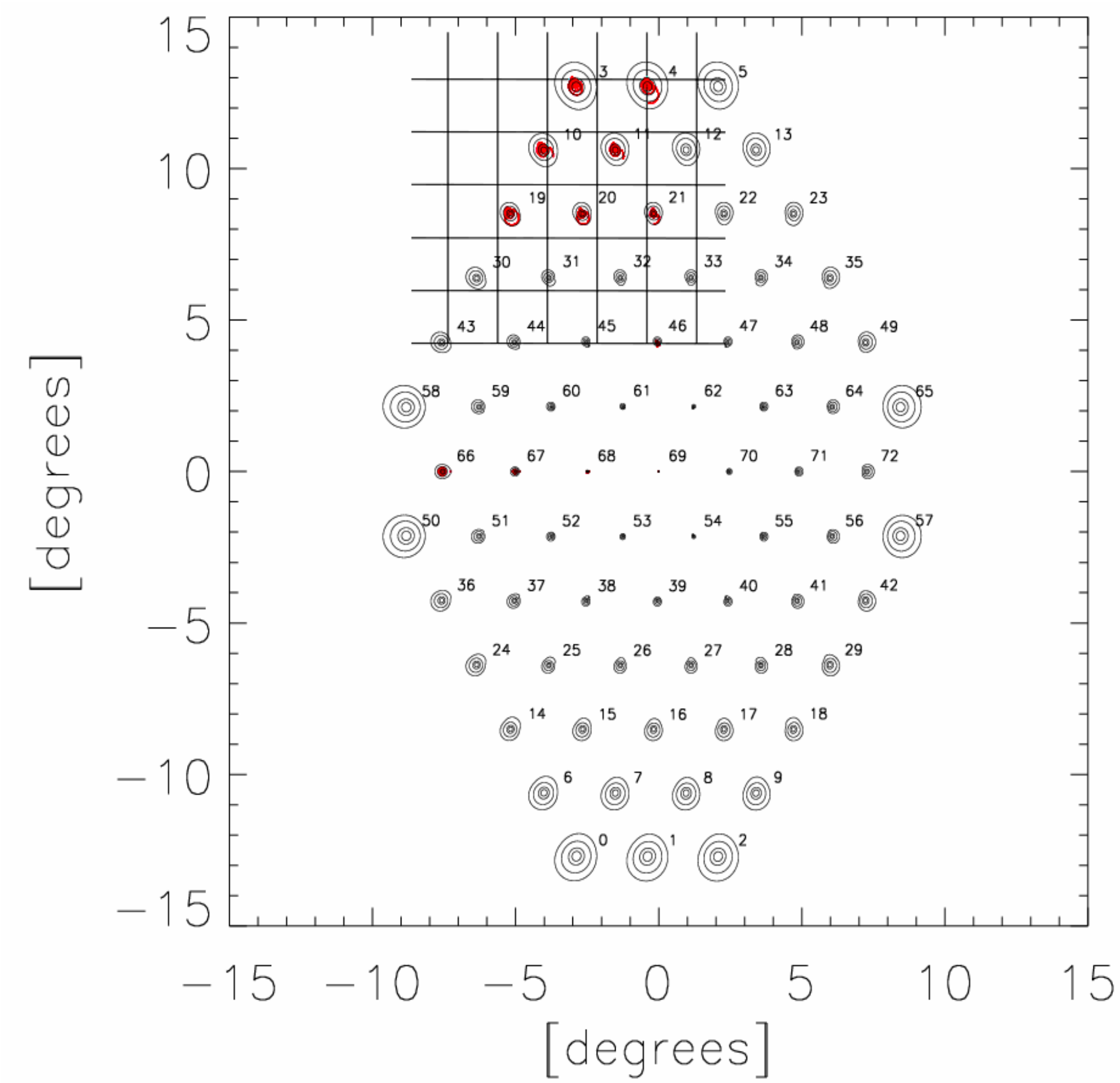

Fig. 6.3. Geometric Optics vs Physical optics simulation of the beams. The contour plot is the co-polarized result of the Physical optics simulations. Each beam shown here is from a feed centered at each wafer shown in figure 6.2. The lowest contour shown is -20dB. The spot diagram from ZEMAX for selected fields is overlaid in red.

*6.2.2 Comparison with Systematics Requirements*

The results of the Grasp-9 main beam simulations can be evaluated for polarization effects induced by imperfect optical performance. Within a feed, the two orthogonally polarized beam patterns are fit to two-dimensional Gaussian models, and the resulting fit parameters are used to calculate the main beam effects.

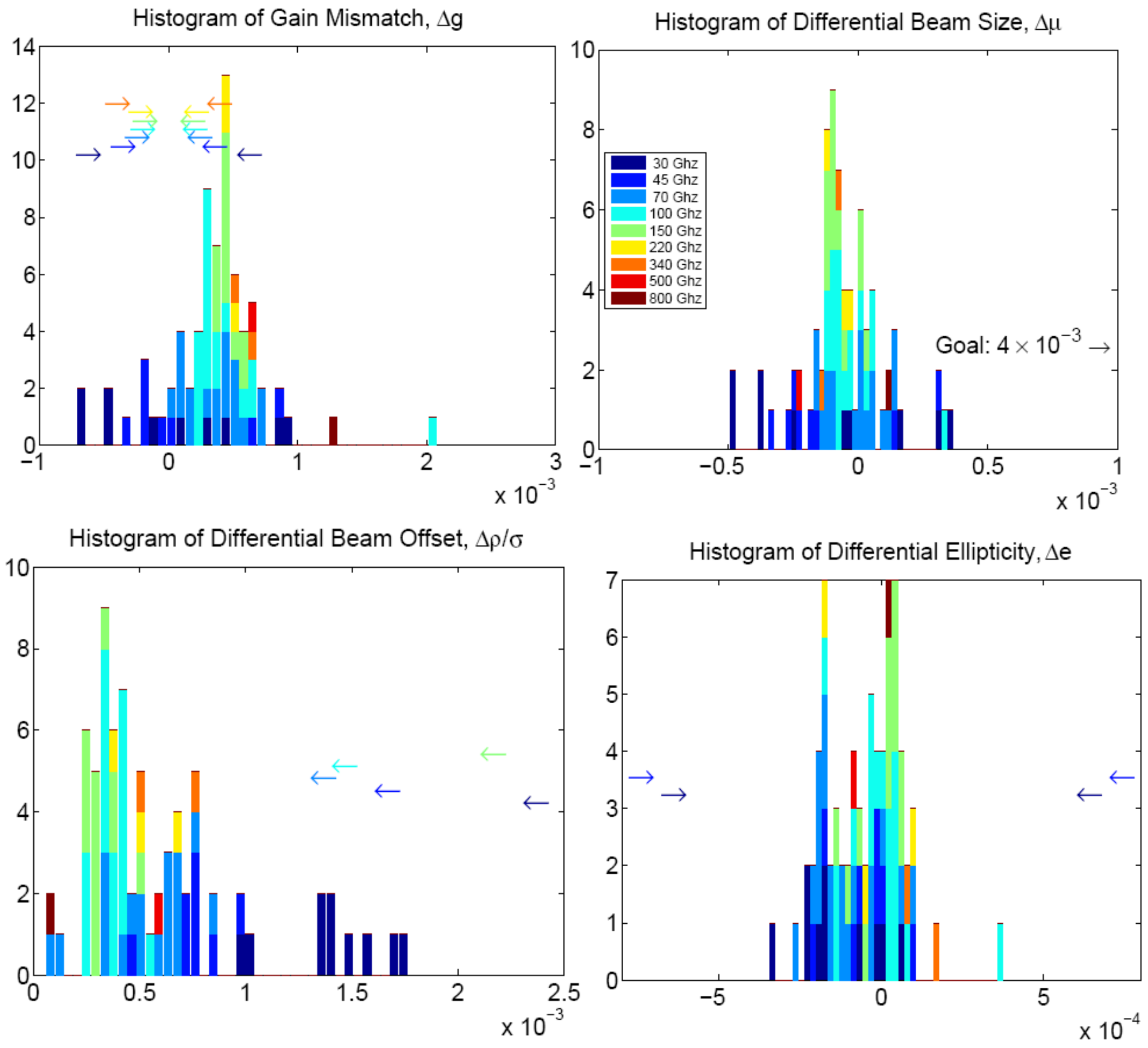

Fig. 6.4 Histograms of main beam effects. Refer to Section 5.4.1 for definitions of each effect. Histograms are color coded by frequency. Colored Arrows denote the frequency dependant goals from table 5.4. Goals for some frequencies are outside of plot area. Most simulated effects off the telescope are smaller than the goals, with the exception of the differential gain. The Gain mismatch is caused by the finite conductivity of the reflector material, and is degenerate with calibration between bolometer pairs, which will be removed by in-flight measurement.

Definitions for the main beam effects appear in section 5.2. The results of the physical optics simulations show that the telescope itself has an uncorrected raw performance that is within the systematic goals in Table 5.4, with the exception of the gain mismatch. The gain mismatch is caused by one linear polarization being preferentially absorbed due to oblique reflection of a slightly resistive reflector surface. This mismatch in gain is masked by any difference in gain or efficiency of the bolometers, and must be calibrated during in-flight measurements.

### 6.3 Far Sidelobes

A ray trace analysis of the telescope optics shows that the Crossed Dragone configuration should produce several different far side lobes. Example far side lobes can be seen in the cross-sectional view of the telescope in Fig. 6.5. While looking at the left panel in Fig. 6.5 and thinking in the reverse time sense, the far side lobes are correlated with any rays that originate from the "center feed" and terminate on the sky away from the main beam. Sky signals entering the telescope through these far side lobes can produce systematic errors in the observations that may be large compared to the sought-after gravity wave B-mode signal. The magnitude of any systematic error produced is related to the level of the associated far side lobes. It is not possible to determine these levels with ray plots. Therefore, to ascertain the overall performance of the telescope, we simulated a full-sky map of the telescope beam pattern using GRASP-9, which is a commercial physical optics software package.

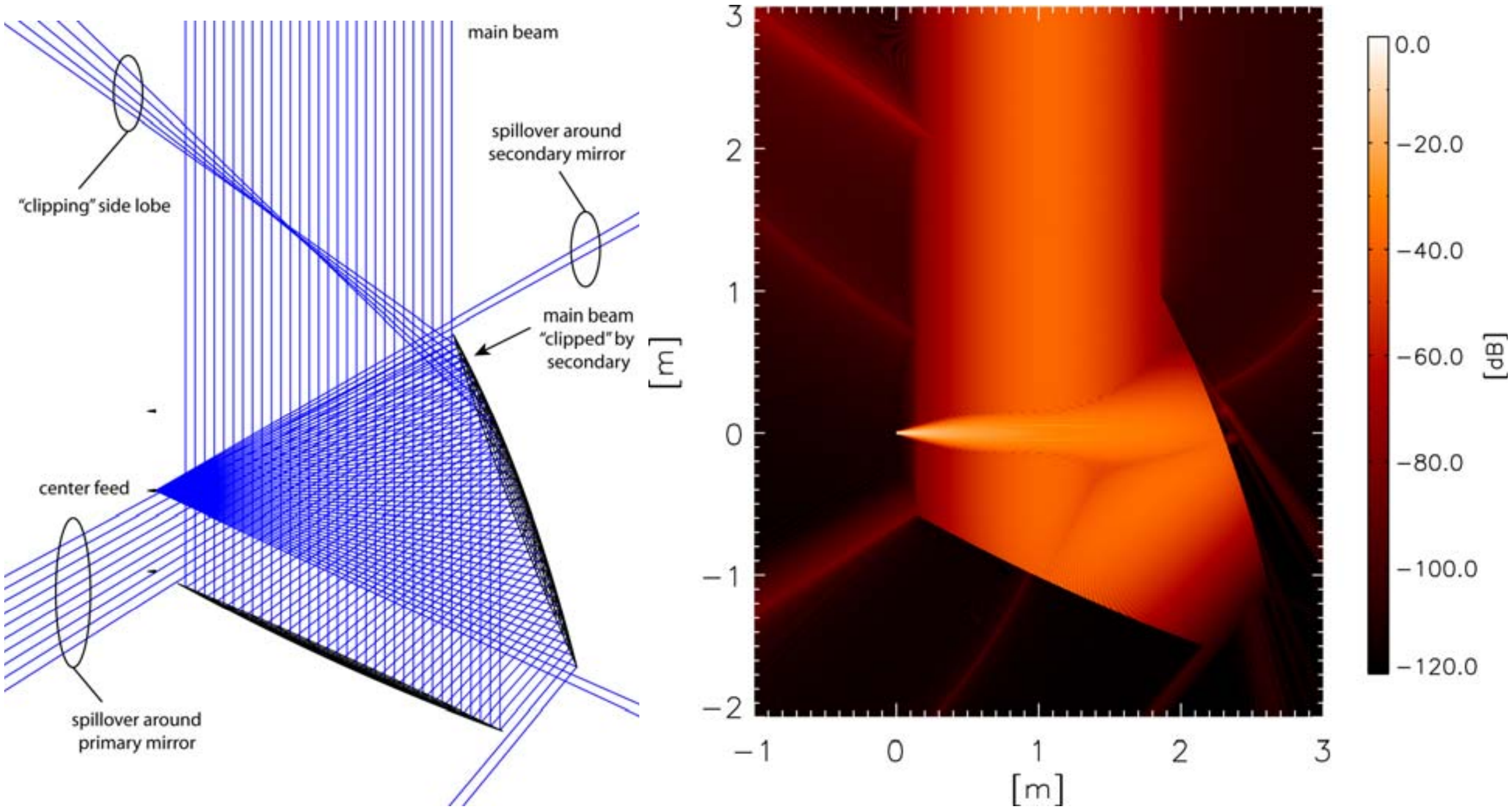

Fig. 6.5. A cross-sectional view of the telescope optics. A ray plot for the center feed is shown in the left panel. A small fraction of the rays propagating from the primary mirror to the sky in the reverse time sense are "clipped" by the secondary mirror. These clipped rays produce a far side lobe. A physical optics representation of this effect is shown in the right panel. Here, the magnitude of the Poynting vector is plotted, and the signal is normalized to the maximum signal in the plot. The clipping side lobe is clearly visible in the upper left corner of this panel.

#### *6.3.1 Physical Optics Analysis*

We computed the full-sky beam pattern of a hypothetical 150 GHz feed located at the center of the focal plane of the telescope. The feed beam we used was linearly polarized and had an ideal Gaussian angular distribution. For the majority of the simulations, the taper of the feed beam was set to -25 dB at a half angle of 13.13 deg. This feed taper corresponds to a horn with a 3.25 fλ aperture diameter. Both polarization orientations of the feed beam were studied. The resulting far-side-lobe maps are shown in Figs. 6.6 and 6.7 and equatorial cuts of these maps are shown in Fig. 6.8.

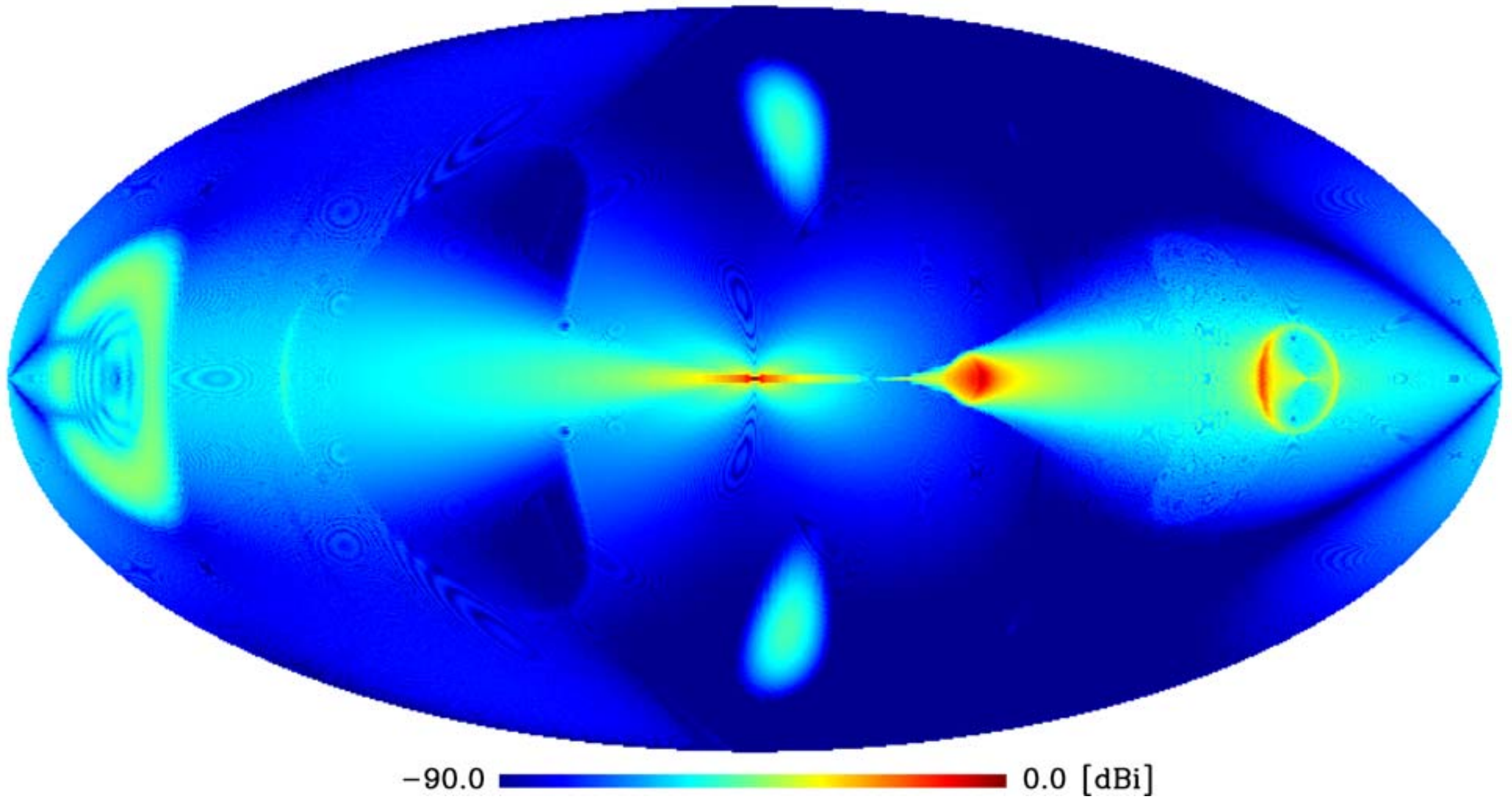

Fig. 6.6. A mollview projection of the all-sky telescope pattern assuming the telescope has no baffling. The main beam of the telescope is the bright dot in the center of the figure. The side lobes produce by spillover around the secondary and primary mirrors are the circular features at the left and right sides of the figure, respectively. The "clipping" side lobe appears just to the right of the main beam. The plot range was selected to highlight the far side lobes; the forward gain of the telescope is 65.1 dBi. The equatorial profile of this map is shown in Fig. 6.8.

We were particularly concerned with the "clipping" far side lobe highlighted in Fig. 6.5. Consequently, the physical optics simulations were designed to optimally study this effect. Simulations were done using two different methods. Both methods were computationally expensive with run times lasting up to approximately one month for one polarization of a single feed. The first method yielded the far-field beam pattern of the telescope assuming no telescope baffling. The second method *simulated* the effect of the telescope baffling by using the aperture integration method (see Section 6.3.2).

For both simulation methods, the feed was used as the radiation source. Surface currents driven by the fields emanating from the feed were computed on both the primary and secondary mirrors. These surface currents subsequently became radiation sources that ultimately radiated to the sky. To simulate the clipping effect, the currents computed on the secondary mirror were driven by two sources: the feed in the focal plane and radiation emanating from currents on the primary mirror. The number of points used to simulate the surface currents on the mirrors was set so that the far field beam patterns would be accurate to at least -120 dB. The map computed using the first method is shown in Fig. 6.6. This map was computed using an x-polarized feed. The x axis in the focal plane is parallel to the symmetry plane of the telescope.

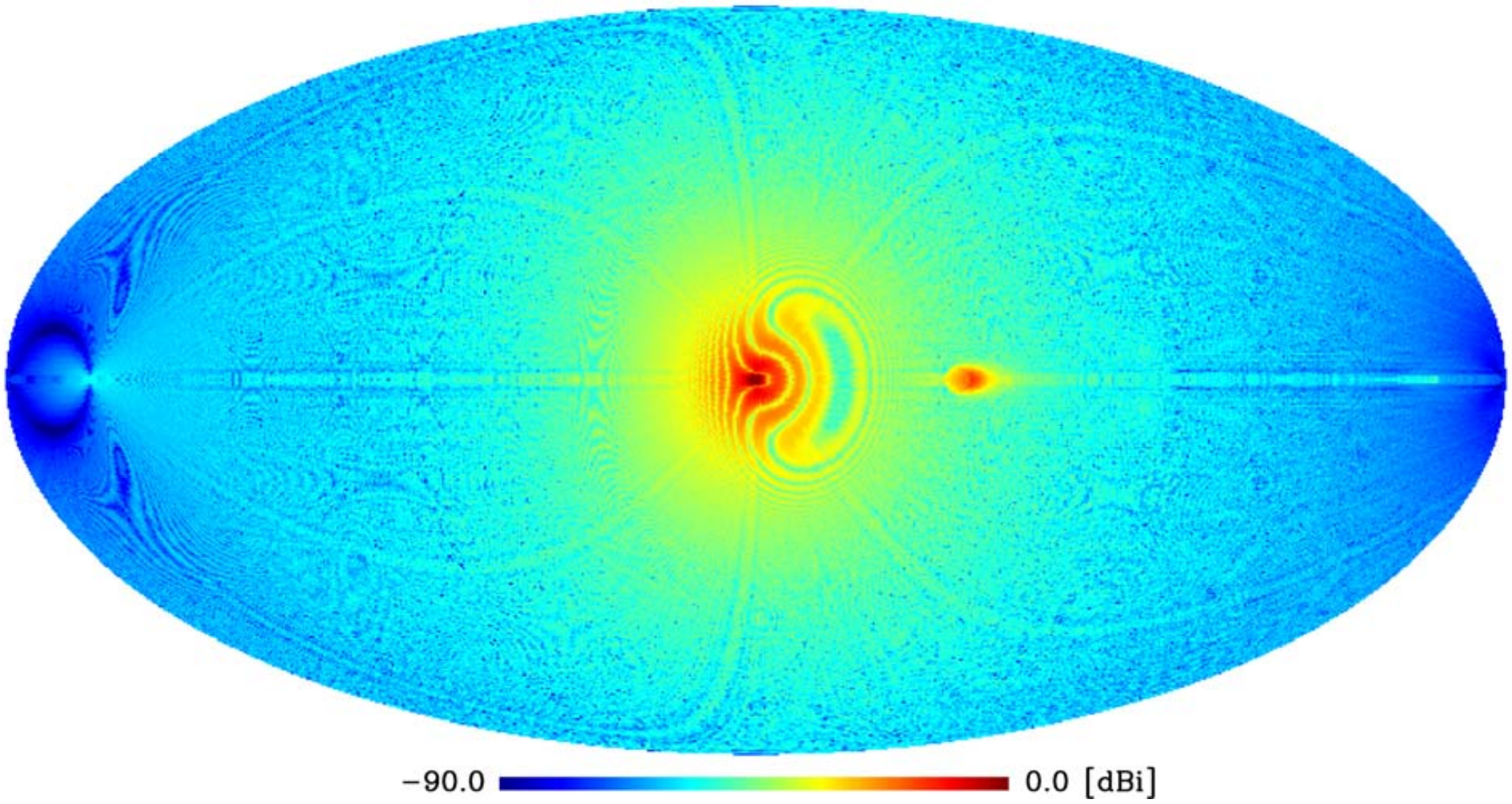

Fig. 6.7. A mollview projection of the all-sky far side lobe pattern of the telescope with simulated baffling. For this computation, the effect of the telescope baffling was simulated using the aperture integration method. The equatorial profile of this map is shown in Fig. 6.8.

### *6.3.2 Aperture Integration Method*

For the second method (the aperture integration method), the objective was to accurately calculate the stray light suppression of an absorbing baffle, and subsequently optimize the baffling design for maximum suppression. Suppose J and M -- the electric and magnetic current sources -- radiate in presence of some scatterer inside a closed surface S. The resultant field is denoted by $\boldsymbol{E_o}$, $\boldsymbol{H_o}$. The Equivalence Theorem states that in order to solve for the true fields outside, one can replace the field on the closed surface S, by a set of equivalent sources $\boldsymbol{J_{eq}}$, $\boldsymbol{M_{eq}}$. These equivalent sources can be chosen to make the field inside zero while giving rise to the exact field of ($\boldsymbol{E_o}$,$\boldsymbol{H_o}$) outside S.

The aperture field integration method produces an excellent approximation to the Equivalence Theorem analysis, and was used for this calculation. Here, the aperture field was computed using physical optics and the physical theory of diffraction by including the following field contributions: radiation from the main reflector, radiation from the subreflector, direct radiation from the source feed, and secondary radiation from the subreflector, which results from the main reflector re-illuminating the subreflector (i.e the clipping side lobe). The aperture field was only calculated in the opening of the telescope baffle, and the field everywhere else was set to zero, since the baffle was modeled as a perfect absorber. The aperture field, consisting of both and E and H fields, was then radiated to the sky to produce the telescope beam pattern in Fig. 6.7.

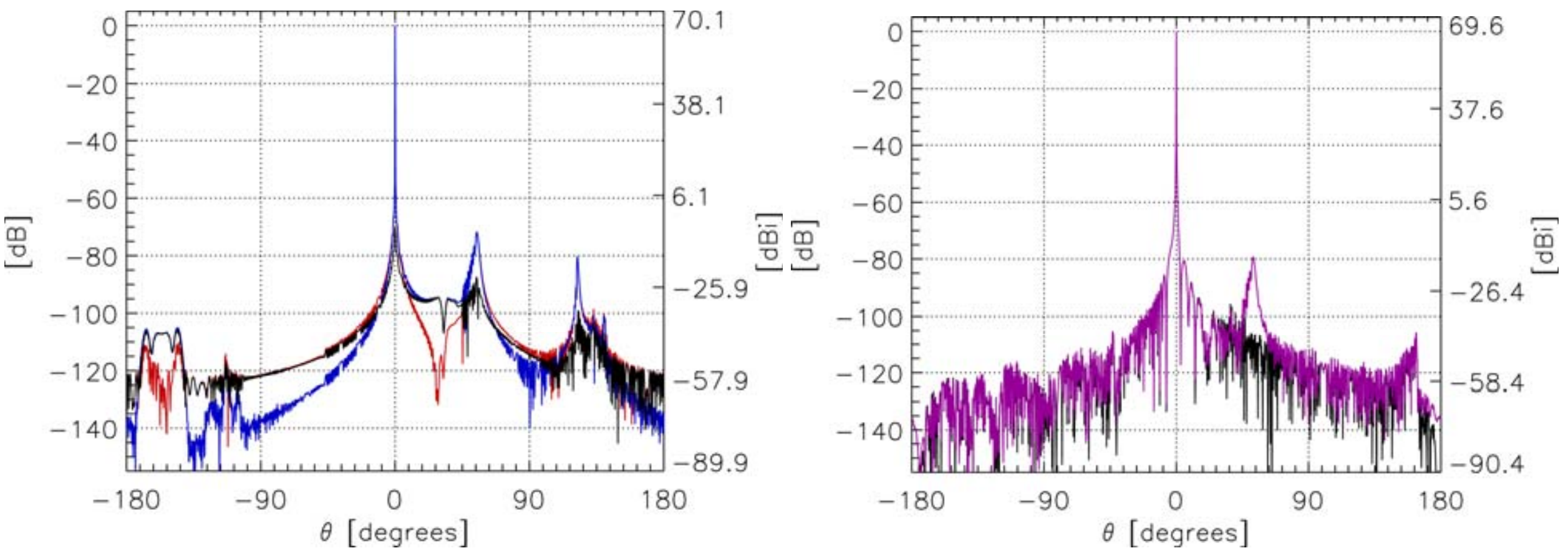

Fig. 6.8. Telescope beam pattern profiles. The red curve plotted in the left panel is an equatorial cut of the map in Fig. 6.6, which is the telescope beam pattern with no telescope baffling for an x-polarized feed. The simulated pattern for a y-polarized feed is plotted as the blue curve, while the difference between the two is the black curve. An equatorial cut of the map in Fig. 6.7 is plotted as the purple curve in the right panel, which is the telescope beam pattern computed with the aperture integration method. Notice the far side lobes that appear near θ = -150 deg and θ = 120 deg in the left panel do not appear in the right panel.

### *6.3.3 GTD Analysis*

The EPIC-IM cold version has a pixel spacing of 2 fλ, implying roughly 15 dB higher edge taper than the 3.25 fλ case. This adds significant computation time to far side lobe simulation. The geometric theory of diffraction (GTD) is a fast alternative to using physical optics to calculate the far side lobes, but it has the disadvantage that it potentially produces spurious artifacts.

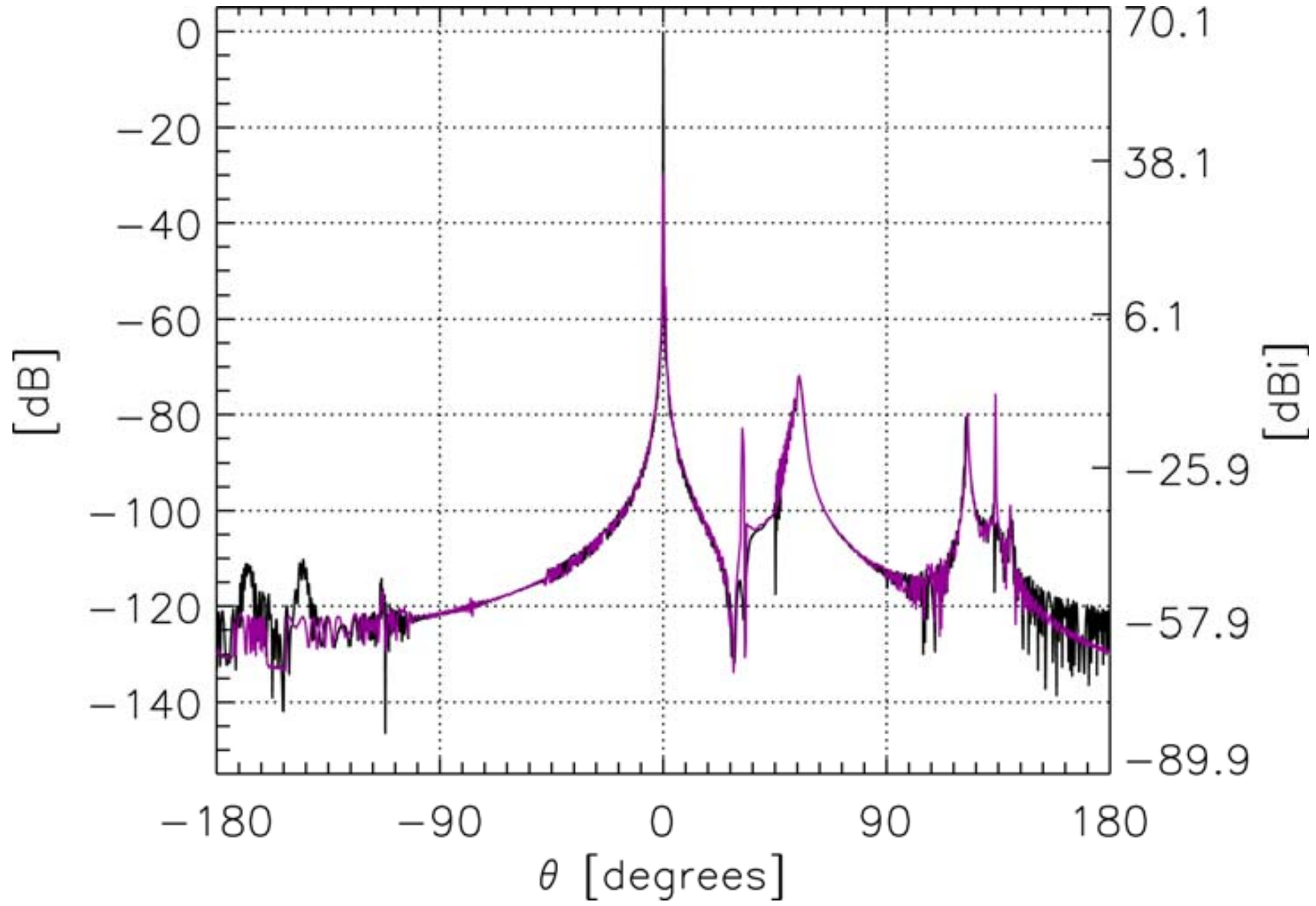

Fig. 6.9 GTD vs PO+PTD for a 3.25 fλ beam at the center of the EPIC-IM focal plane. The black curve is from a PO+PTD simulation, and the purple curve is from GTD. While the GTD results show some artifacts, the "clipping" side lobe at 60 degrees appears in both simulations.

The EPIC-IM 3.25 fλ optics were analyzed by both GTD and PO+PTD with GRASP-9. GTD showed a reliable reproduction of the "clipping" side lobe, as shown in Fig. 6.9. This result provided enough confidence to analyze the 2 fλ case using the GTD. Fig. 6.10 is a plot of the GTD results for a 2 fλ feed centered on the focal plane. For the purpose of this simulation, it is assumed that a 2 fλ feed corresponds to a -10 dB edge taper at the aperture, although the aperture itself is not included in this simulation. Both x and y linearly polarized beams were simulated for the central feed. The plots clearly show the "clipping" side lobe. The results additionally show that both polarizations are roughly equally sensitive to this side lobe, meaning that the side lobe is un-polarized.

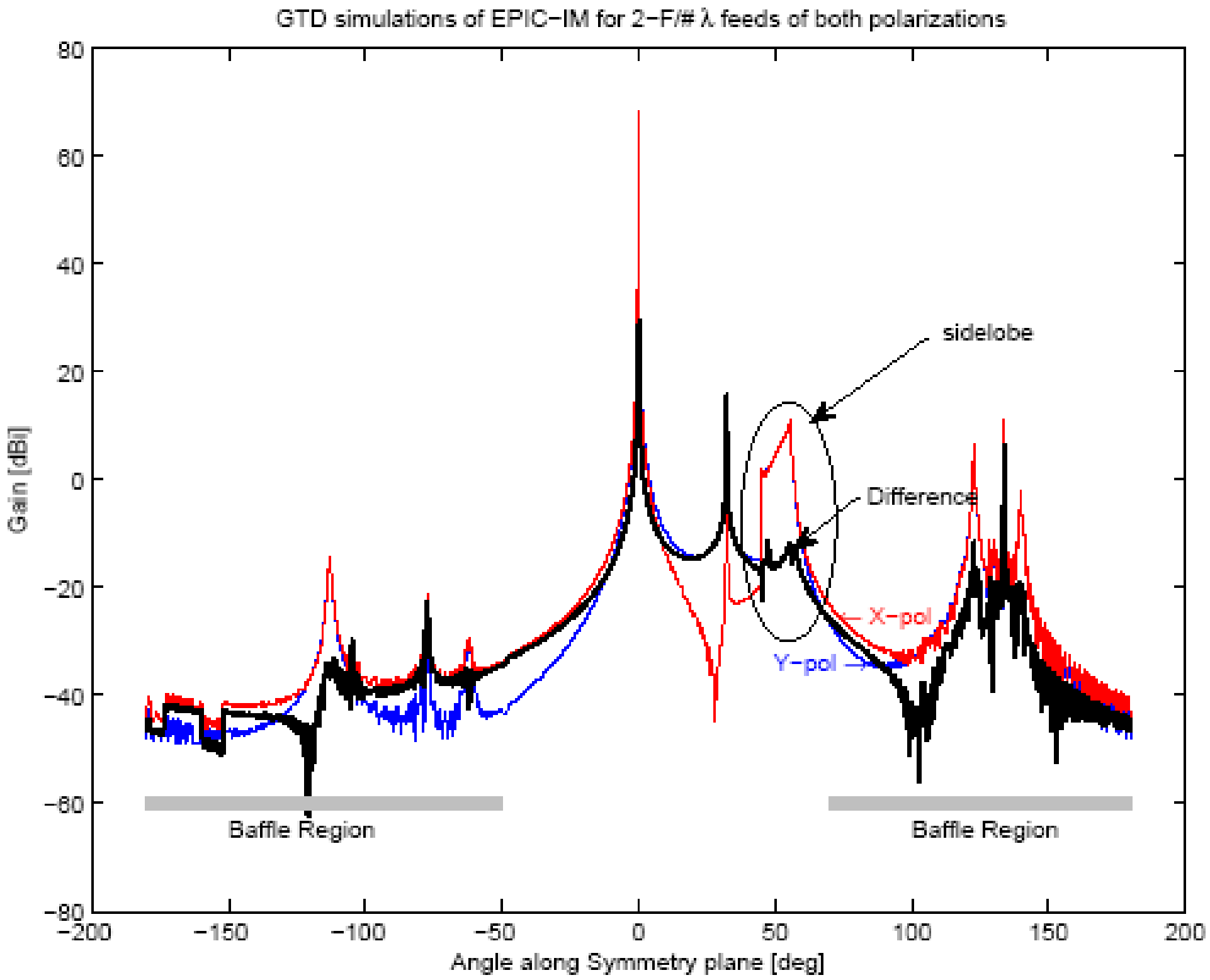

Fig. 6.10. GTD simulation of the EPIC-IM (cold), 2 fλ optics. The x-polarized results are shown in red, and the y polarized in blue. The difference between the two polarization is shown in heavy black. The ``clipping'' side lobe is circled. From the difference, it is clear that the clipping side lobe is highly unpolarized. The feature at +25 deg does not appear in the more rigorous physical optics simulations, and is therefore likely an artifact. All other features will be absorbed by baffling, roughly denoted by the gray bars.

The x and y-polarized far side lobe maps computed with method 1 were combined to produce Stokes parameter maps. The polarization map, which is defined here as $(QT^2 + UT^2 + VT^2)^{1/2}$, is plotted in Fig. 6.10. This map indicates the level to which unpolarized sky signals become polarized. For reference, the maximum signal in the map is approximately 0 dB, and the forward gain of the associated TT map is 54.1 dB. This information suggests the expected instrumental polarization is $10^{-3}$ % for the center feed.

To determine whether this far side lobe performance is sufficient, the Stokes parameter maps were convolved with a map of the Galactic signals that are anticipated to appear at 150 GHz. For this convolution, the main beam of the telescope was suppressed using a mask. Therefore, the convolved sky signal map shows how the Galactic signals are redistributed via the far side lobes of the telescope. This redistributed signal is the systematic error we want to characterize. After the convolution, the magnitude of the spurious signal in each pixel was compared to the performance requirement, which assumes a tensor-to-scalar ratio of 0.001.

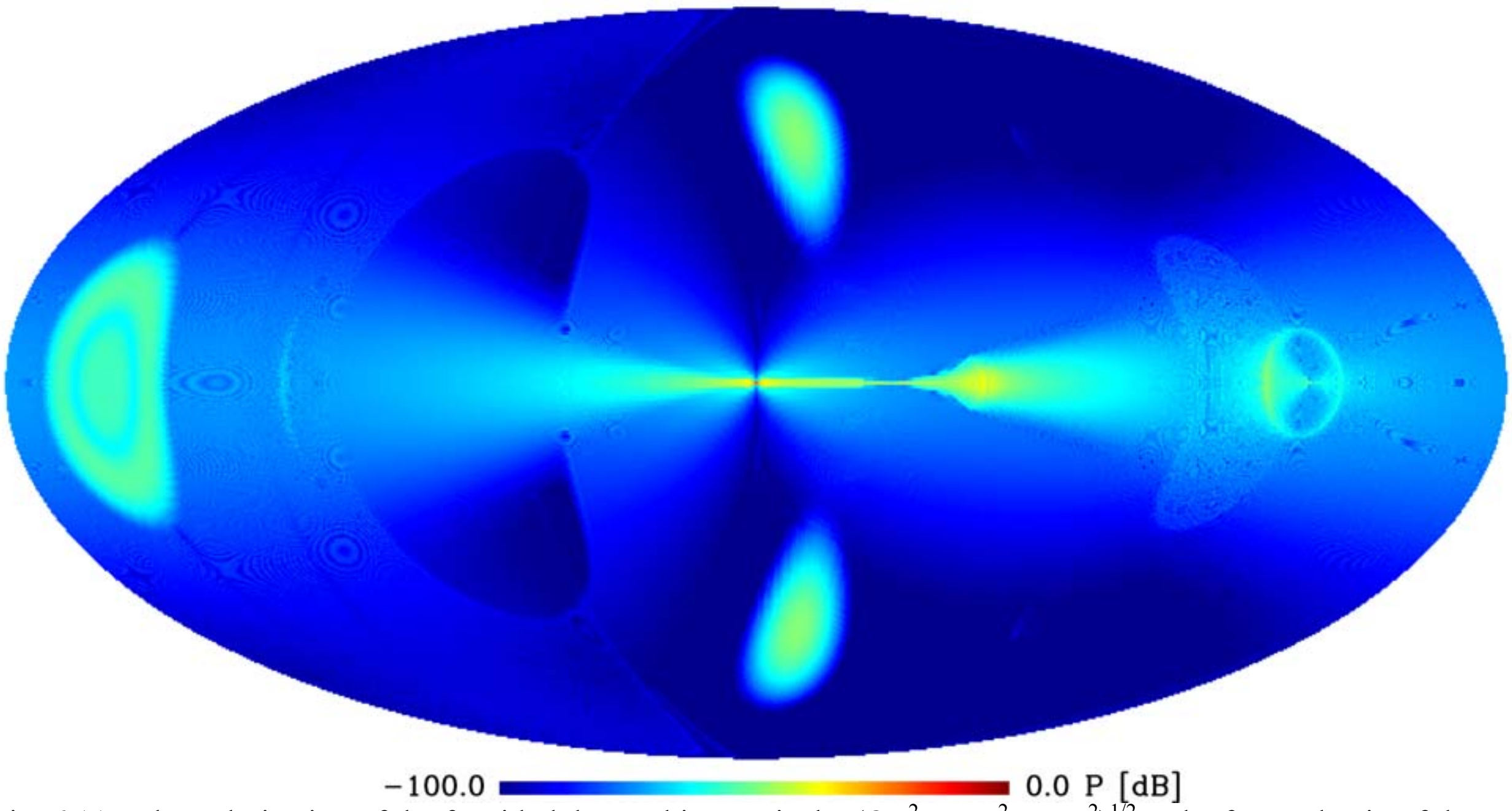

Fig. 6.11. The polarization of the far side lobes. This map is the $(QT^2 + UT^2 + VT^2)^{1/2}$. The forward gain of the TT beam is 54.1 dB.

### 6.4 Galactic Signal from Far-Sidelobes

In order to evaluate the effect of the signal from the far sidelobes, we convolve the QT beam maps with a 150 GHz sky model. The main beam is masked so only the contamination from the far sidelobes is calculated. Since the beam is asymmetric, it is necessary to rotate the beam with respect to the sky at each pointing to get the complete convolution.

In order to further quantify this result, we make two histograms: the number of pixels with a given intensity, and the integrated histogram giving the total number of pixels less than a given intensity. This is similar to the plots that are shown for site surveys. By inspection one can see that fully 90 % of the pixels are < 0.2 nK.

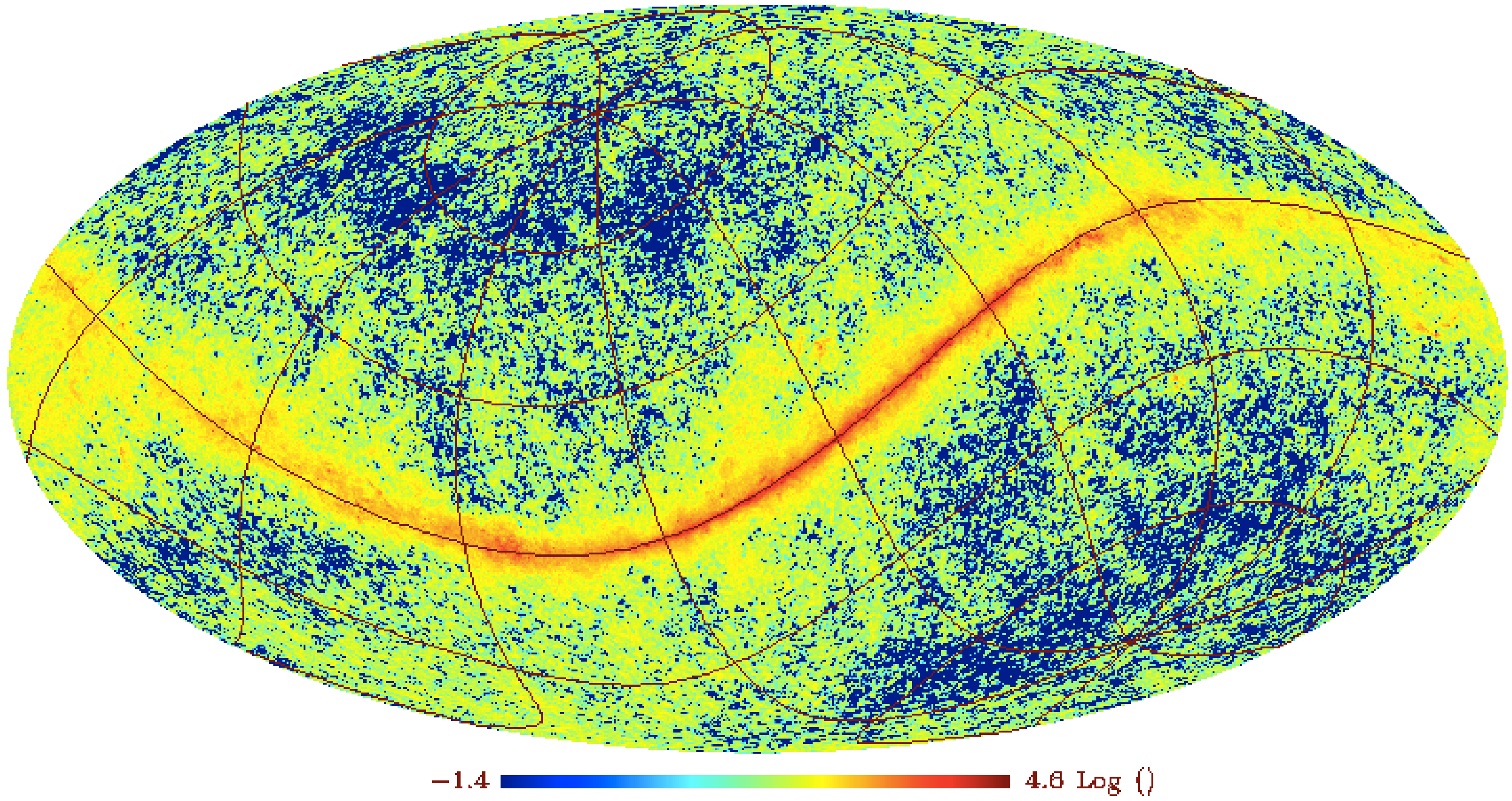

Fig. 6.12. The 150 GHz sky map used for convolution with the far sidelobe pattern. Units are in log(mK).

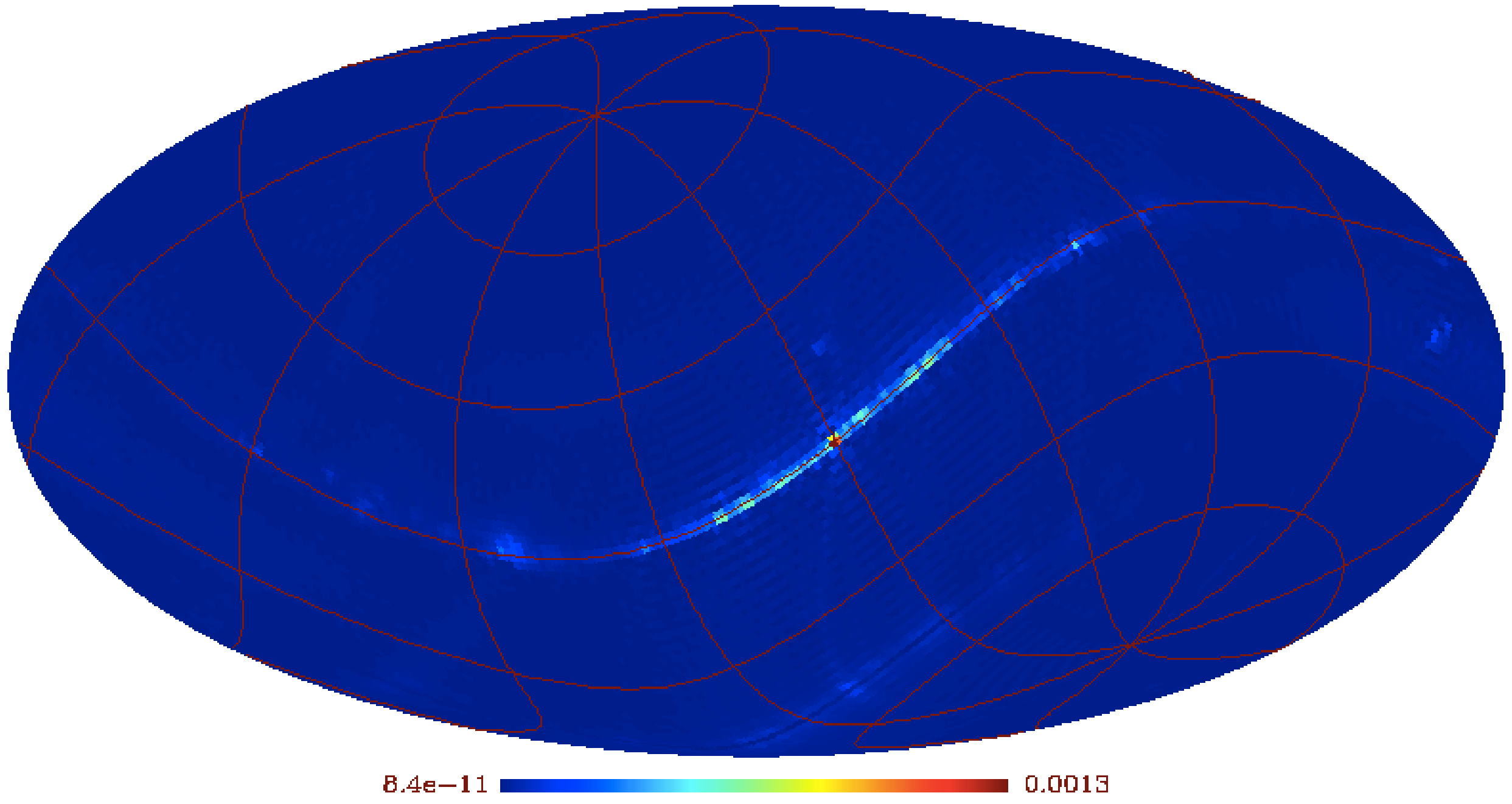

Fig. 6.13. The results of the convolution. Units are $\mu K_{CMB}$.

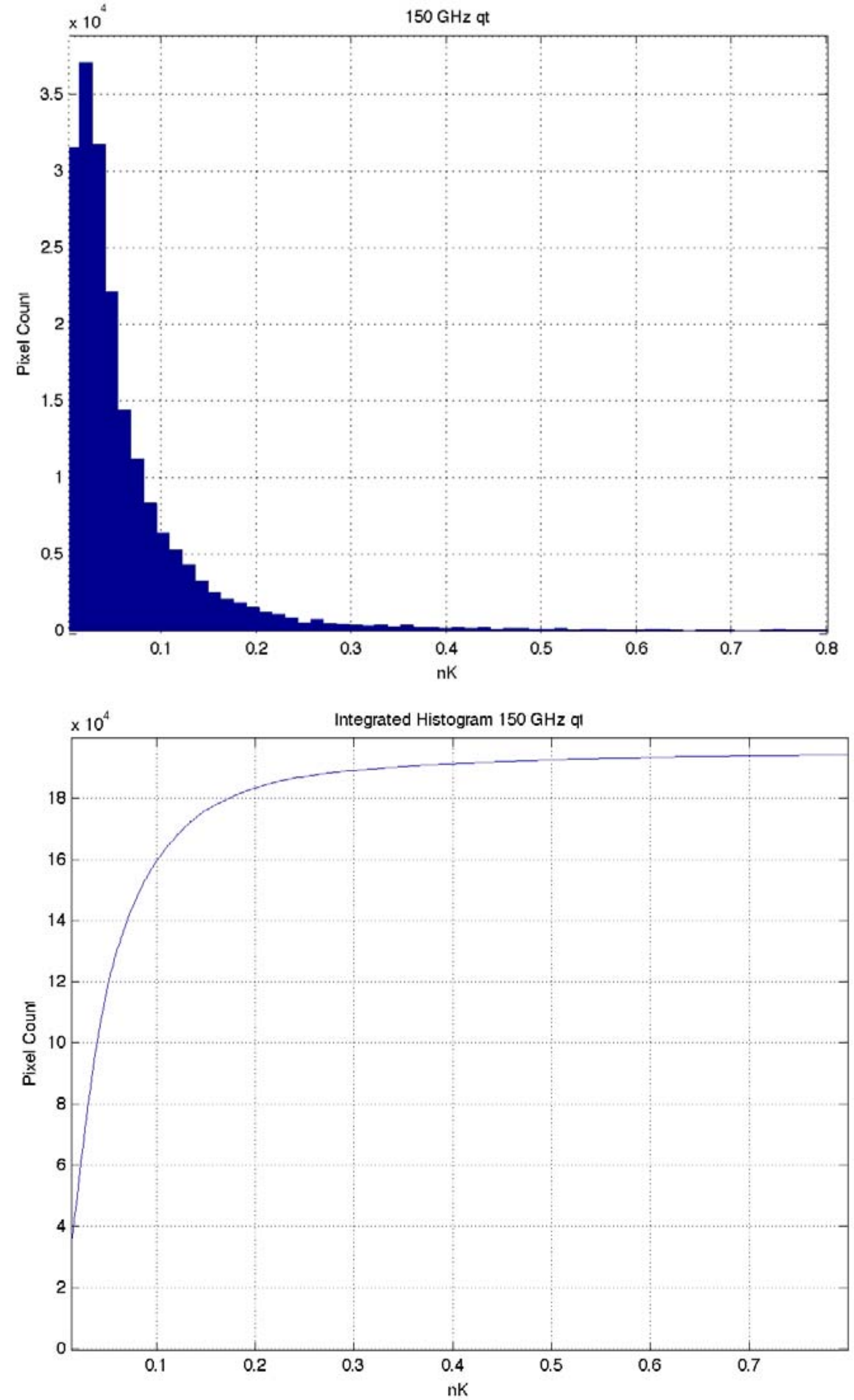

Fig 6.14. Histogram of signal values from Fig. 6.13. The top panel is a histogram of leakages of galactic foregrounds into the EPIC-IM sidelobes. The bottom panel shows the same data as a cumulative distribution.

## 6.5 Summary and Future Work

The level of sidelobe control shown in Fig. 6.14 gives ~0.1 nK rms polarized signals over the majority of the sky, well within the required level of 3 nK, and the goal level of 1 nK, which are quoted for flat power spectra after band combination. However, this analysis was calculated in a single CMB band at 150 GHz at the center of the focal plane assuming 3.25 fλ illumination.

These calculations must be extended to the 2 fλ case. The scaling shown in Fig. 6.10 would indicate the level of far-sidelobes should be roughly at the required level of performance for the case of 2fλ illumination. Furthermore the calculation needs to be extended to cover the field of view and bands in the focal plane. Finally we note that we have not analyzed the use of optimized baffles, such as tapered absorbers surrounding the primary and aperture stop, or baffles to reduce the clipping lobe. These may offer further improvement in far-sidelobe performance.

# 7. Focal Plane

EPIC's large sensitivity advantage over Planck and planned sub-orbital experiments comes entirely from improved capability in the focal plane, provided by large format detector arrays. Currently the CMB community is actively developing competing detector, optical coupling, and readout technologies, and are demonstrating these technologies in sub-orbital experiments. This parallel development provides robustness and will likely result in multiple technology choices for a space mission. Our study seeks to assess the designs and technical challenges intrinsic to a space-borne instrument that are independent of any specific technology implementation.

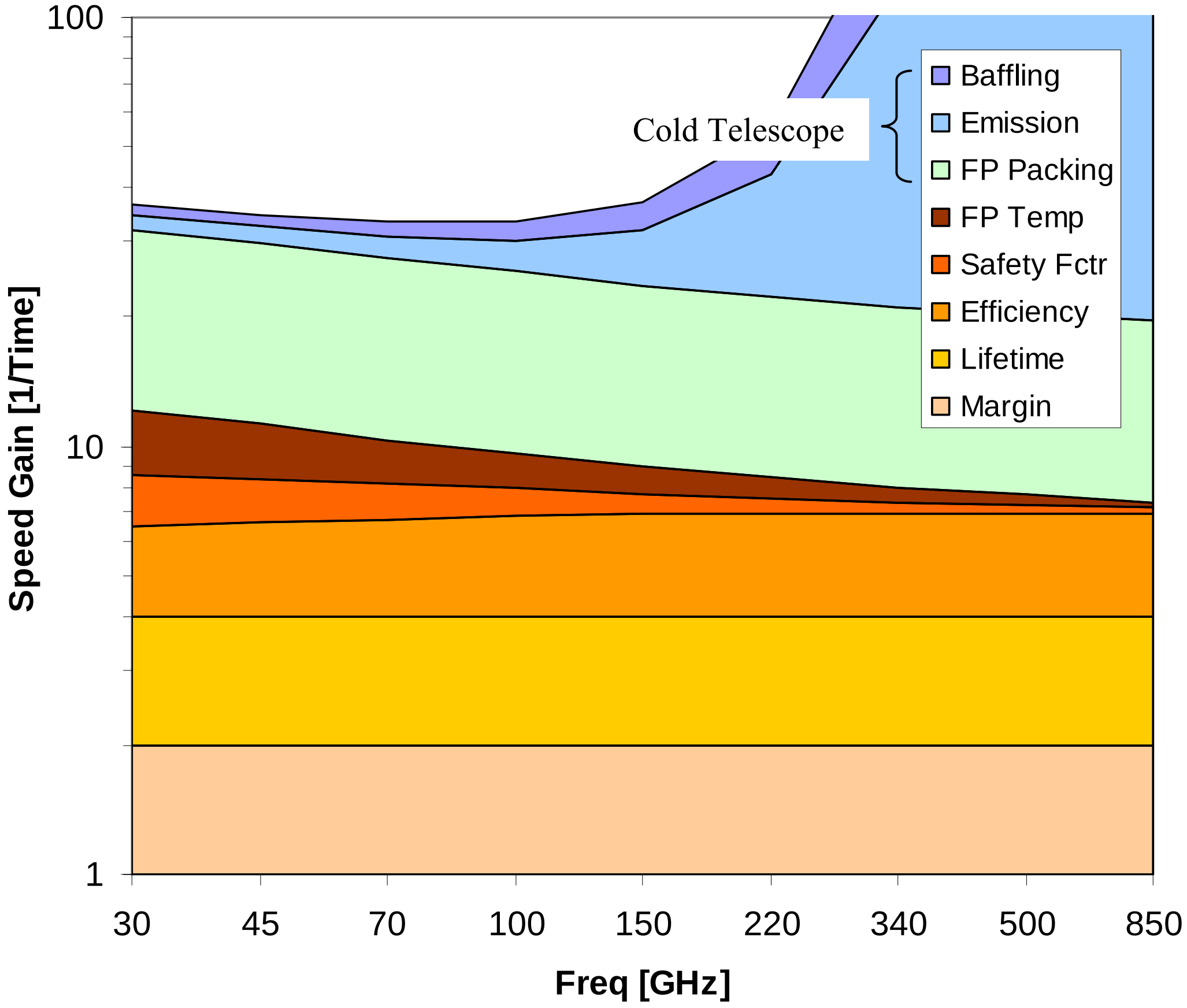

Fig. 7.1 Improvements in mapping speed that result from changing mission parameters from case 1 to case 2 for background-limited TES bolometers described in section 7.3. Note that changing the focal plane packing changes the spillover, and is thus linked with effective telescope emissivity. Increasing the packing density will not provide a net sensitivity gain due to increased photon noise, unless the telescope is cooled.

**Table 7.1.** Changes in Mission Parameters for Fig. 7.1

| **Option** | | **Case 1** | **Case 2** |
|---|---|---|---|
| Margin | | 1.4 | 1.0 |
| Lifetime | | 4 years | 8 years |
| Efficiency | | 40 % | 70 % |
| TES Safety Factor | | 5 | 3 |
| Focal Plane Temperature | | 100 mK | 50 mK |
| Cold Telescope | Focal Plane Packing | 3.25 fλ | 2.0 fλ |
| | Emission | 40 K, ε = 1 % | 4 K, ε = 10 % |
| | Baffling | | 4 K, ε = 1 % |

The EPIC-IM 4 K focal plane has been designed to take full advantage of the large system throughput of the crossed Dragone telescope and the low background available from space. The detectors for EPIC-IM are essentially background limited, offering modest improvements over Planck (see Table 7.7) in sensitivity per detector. Instead, the majority of EPIC-IM's sensitivity over Planck relies on larger-format arrays. We carried out a trade study of the improvements possible in mapping speed (which scales as $1/NET^2$) shown in Fig. 7.1 under the assumptions of the TES detectors in Table 7.5 operating within a fixed field of view. A significant total gain is possible, larger than a factor of 30, by making the individual trades listed in Table 7.1. However many of these trades involve either substantial resources, such as increased mission life, or reduced conservatism factors, such as noise margin or optical efficiency. Reducing the base temperature from 100 mK to 50 mK results in a relatively minor system gain, as does reducing the safety margin on TES detector power saturation. By far the largest system improvement comes from cooling the telescope from 40 K to 4 K. The system gain arises from two factors. A colder telescope allows for a higher density of detectors [1], since we can tolerate higher spillover off the aperture onto an absorbing surface assumed to be the same temperature as the telescope. The second factor results from reduced photon noise, with large improvements at higher frequencies where the emission from a 40 K telescope dominates over the CMB. This observation forms the basis of the two mission configurations: the baseline 4 K telescope with a large densely 2fλ-packed focal plane that provides maximum sensitivity, and the descoped 30 K telescope with a smaller and sparsely 3.25fλ-packed focal plane that provides somewhat reduced sensitivity.

## 7.1 Focal Plane Design

We have developed a modular focal plane design, illustrated in Fig. 7.1. The overall focal plane structure is designed for flexibility, so that the weighting of frequency bands, and the choice of polarization analysis direction over the focal plane, can both be adjusted while the mission is in development without affecting the overall design. The focal plane is composed of sub-array hexagonal 'tiles' fabricated from 150 mm wafers. Individual tiles are mounted in a hexagonal cell structure at 100 mK. We reserve an avoidance perimeter on each wafer to allow for customized optical filters for each tile, and to provide space for electrical connections along the edge of the sub-array wafer. There is ample room provided behind each wafer for wiring and cold multiplexing readout electronics.

**Table 7.2** Focal Plane Layouts

<table>
<tr><th rowspan="2">Freq [GHz]</th><th colspan="3">4 K Telescope Option</th><th colspan="3">30 K Telescope Option</th></tr>
<tr><th>Ntile [#]</th><th>Det/tile [#]</th><th>Detectors [#]</th><th>Ntile [#]</th><th>Det/tile [#]</th><th>Detectors [#]</th></tr>
<tr><td>30</td><td>6</td><td>14</td><td>84</td><td>4</td><td>6</td><td>24</td></tr>
<tr><td>45</td><td>14</td><td>26</td><td>364</td><td>6</td><td>14</td><td>84</td></tr>
<tr><td>70</td><td>18</td><td>74</td><td>1332</td><td>8</td><td>26</td><td>208</td></tr>
<tr><td>100</td><td>18</td><td>122</td><td>2196</td><td>6</td><td>74</td><td>444</td></tr>
<tr><td>150</td><td>12</td><td>254</td><td>3048</td><td rowspan="2">6</td><td>86</td><td>516</td></tr>
<tr><td>220</td><td rowspan="2">4</td><td>324</td><td>1296</td><td>68</td><td>408</td></tr>
<tr><td>340</td><td>186</td><td>744</td><td rowspan="3">1</td><td>120</td><td>120</td></tr>
<tr><td>500</td><td rowspan="2">1</td><td>1092</td><td>1092</td><td>108</td><td>108</td></tr>
<tr><td>850</td><td>938</td><td>938</td><td>110</td><td>110</td></tr>
<tr><td>Total</td><td>73</td><td></td><td>11094</td><td></td><td></td><td>2022</td></tr>
</table>

## Fig. 7.2 EPIC INTERMEDIATE MISSION:  FOCAL PLANE DESIGN

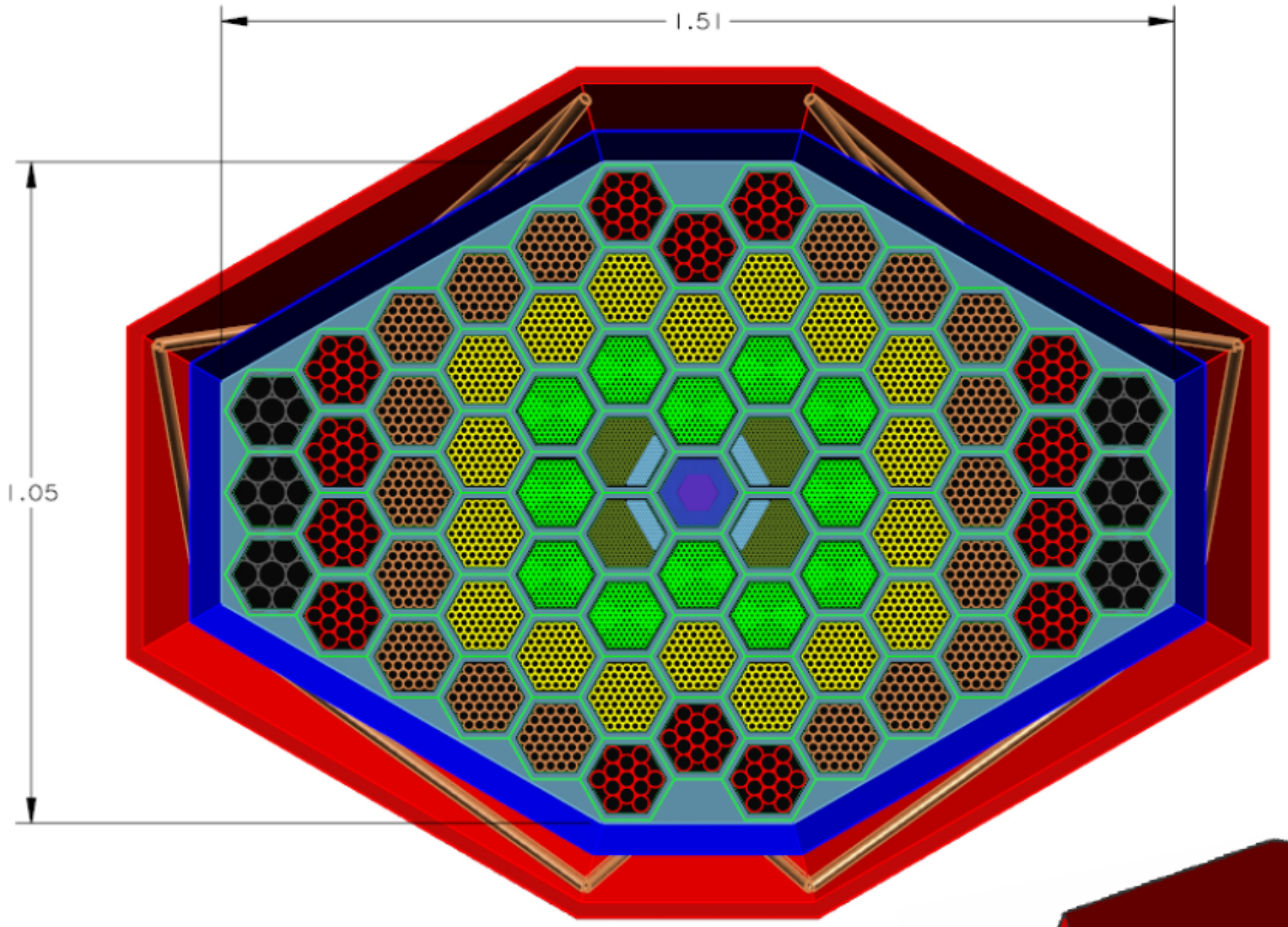

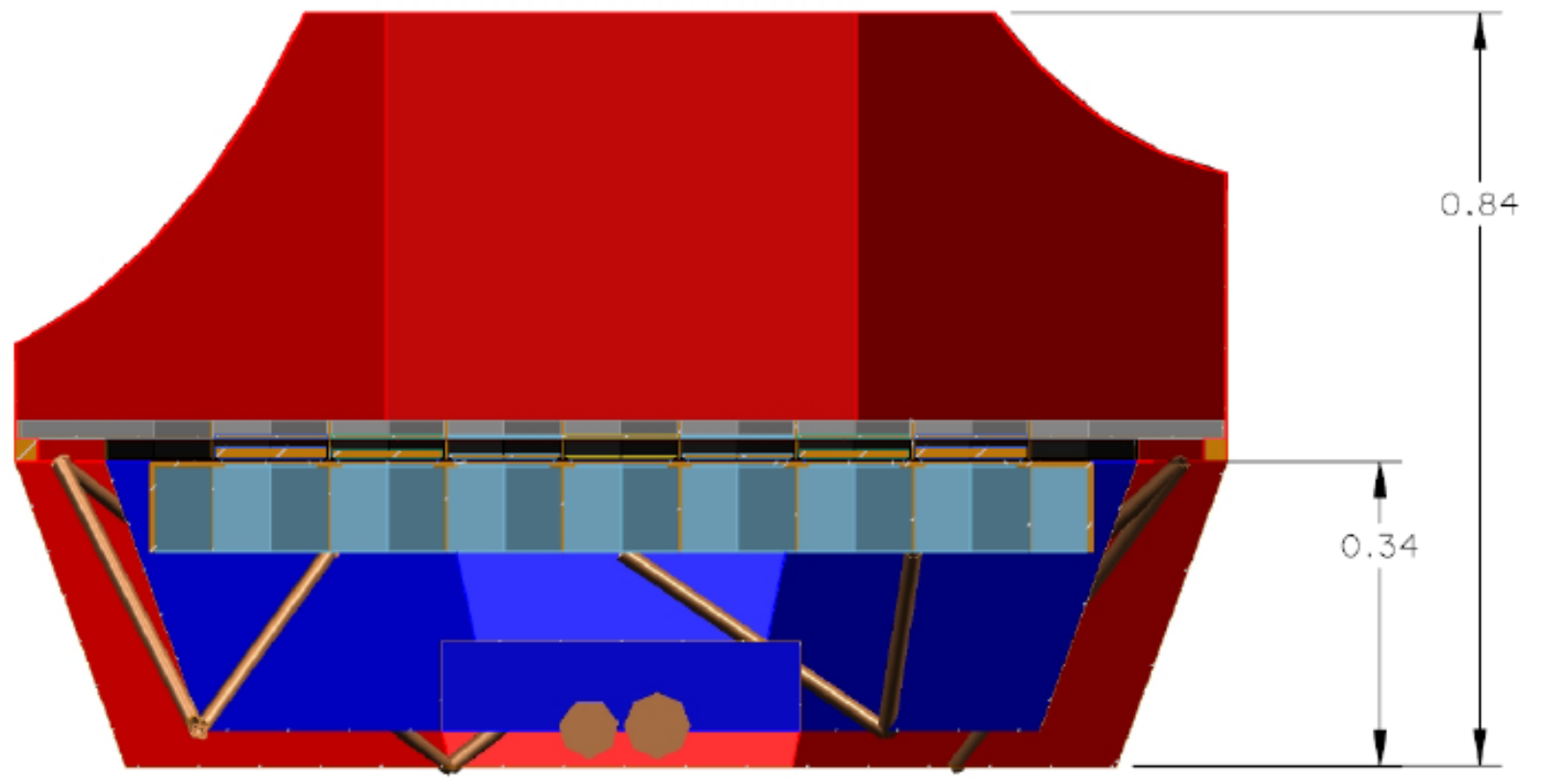

The EPIC-IM crossed Dragone telescope provides a very large focal plane, populated with detector sub-arrays operating (from the center to the edge) at 850, 500, 340, 220, 150, 100, 70, 45 and 30 GHz.  The detectors are packed at 2-f$\lambda$ spacing on individually multiplexed sub-arrays.

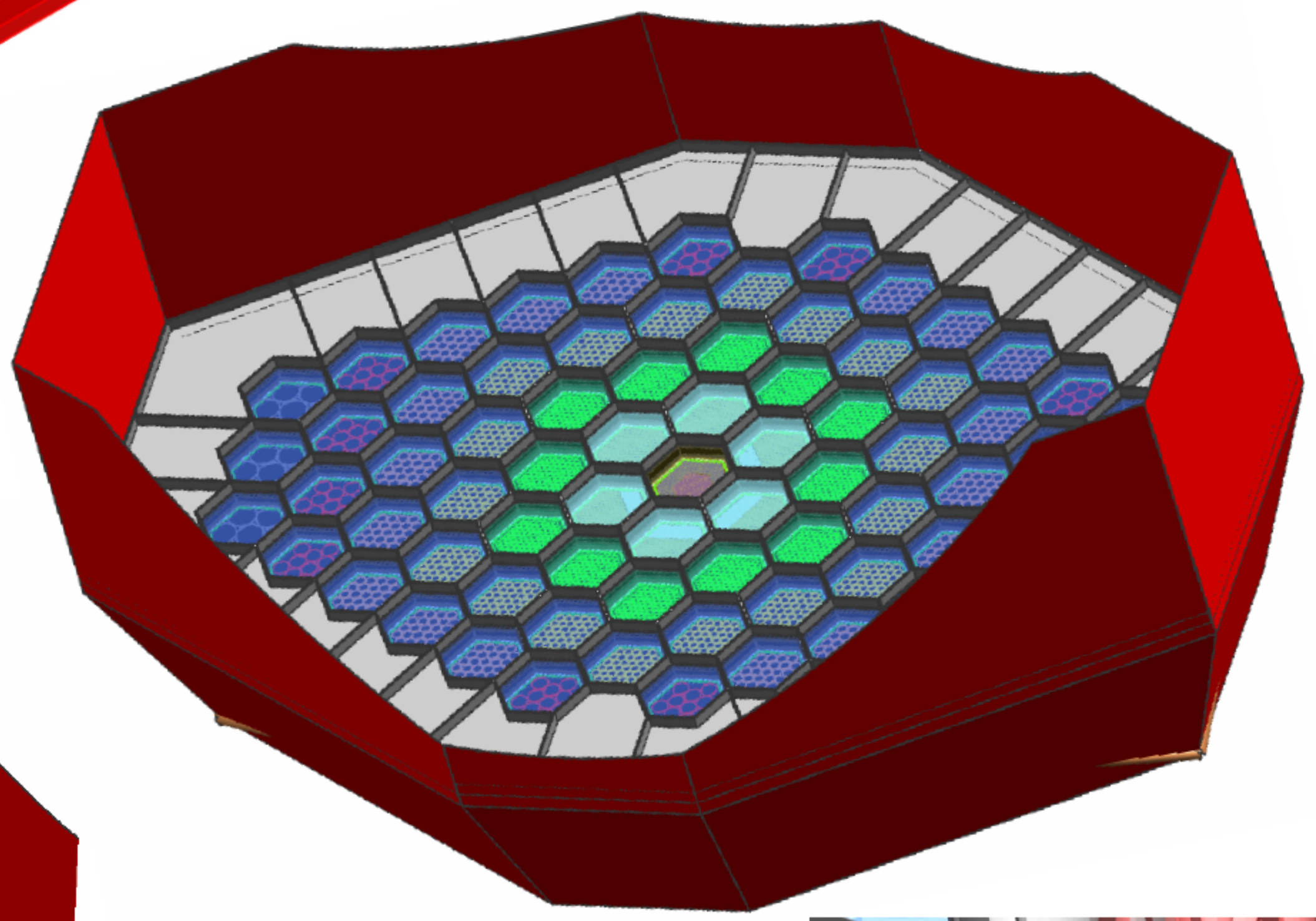

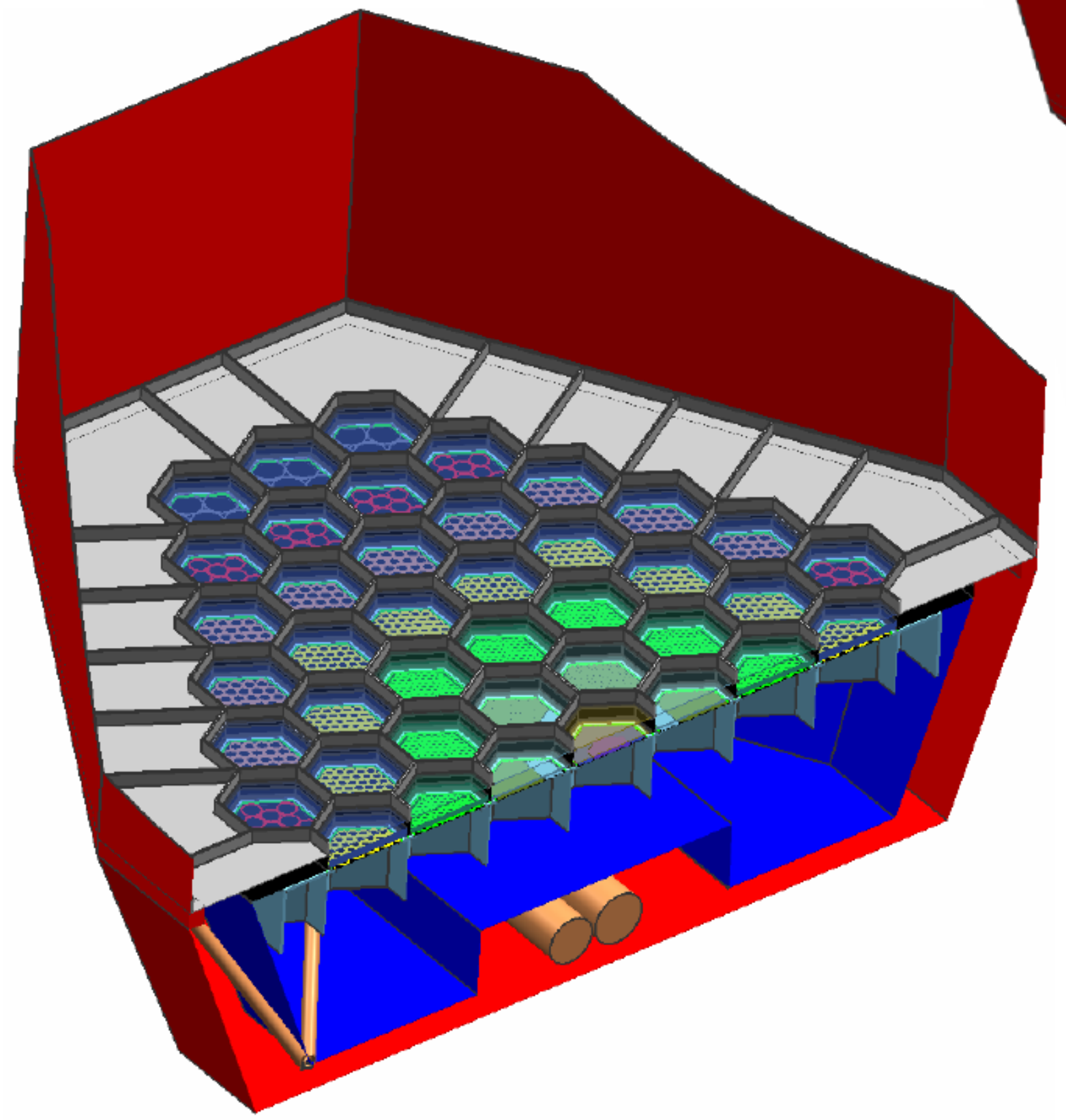

Cross sections showing modular assembly of focal plane sub-arrays, radiation shields at 4 K (red) and 1 K (blue), low-conductivity struts, and a continuous 100 mK ADR at bottom.  The sub-arrays are mounted in hexagonal cells at 100 mK (light blue). Modular blocking filters are located just above the 100 mK sub-arrays, in two stages at 4 K and 1 K.  The filters are sized to not vignette the detector field of view, as shown on the right.

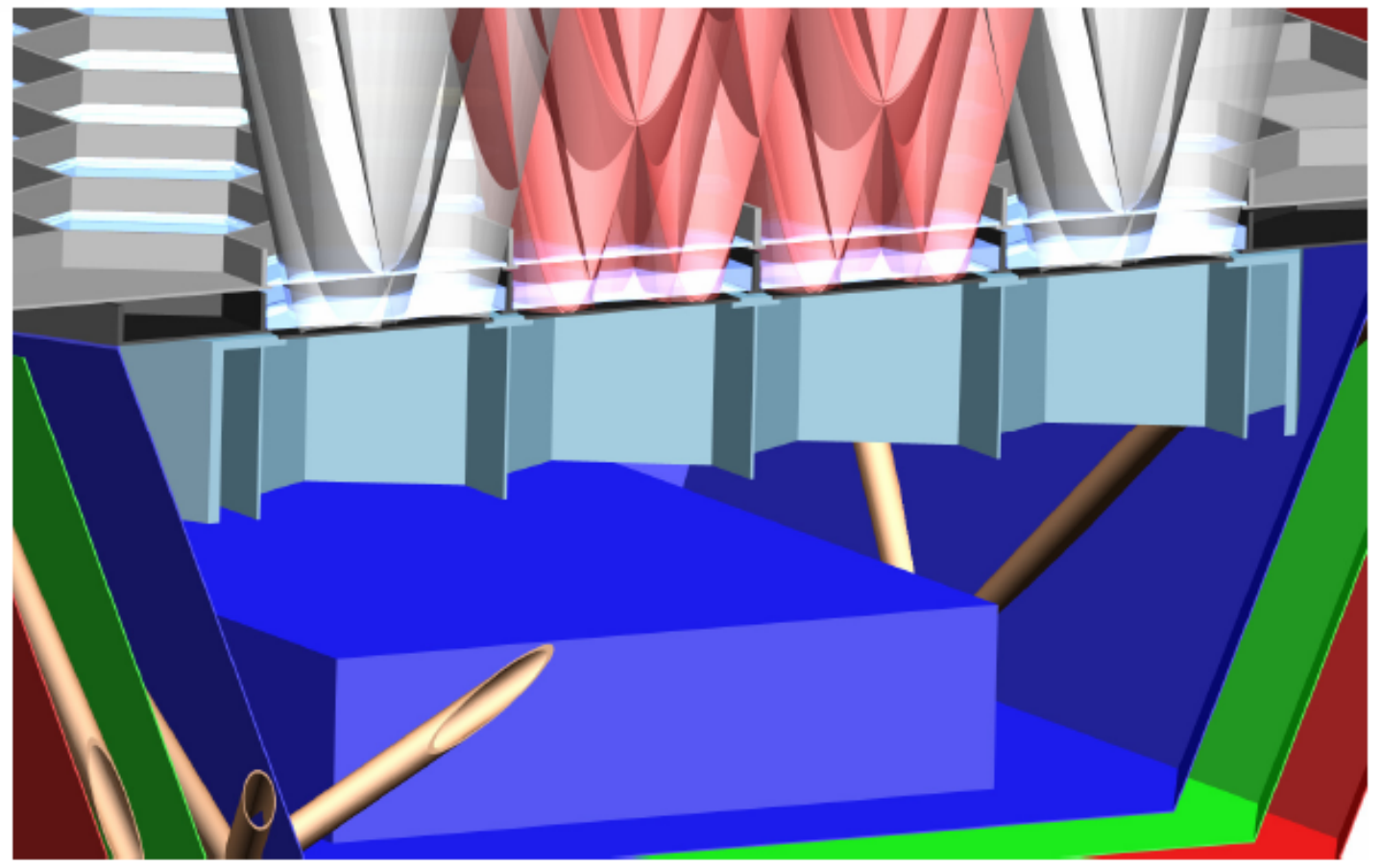

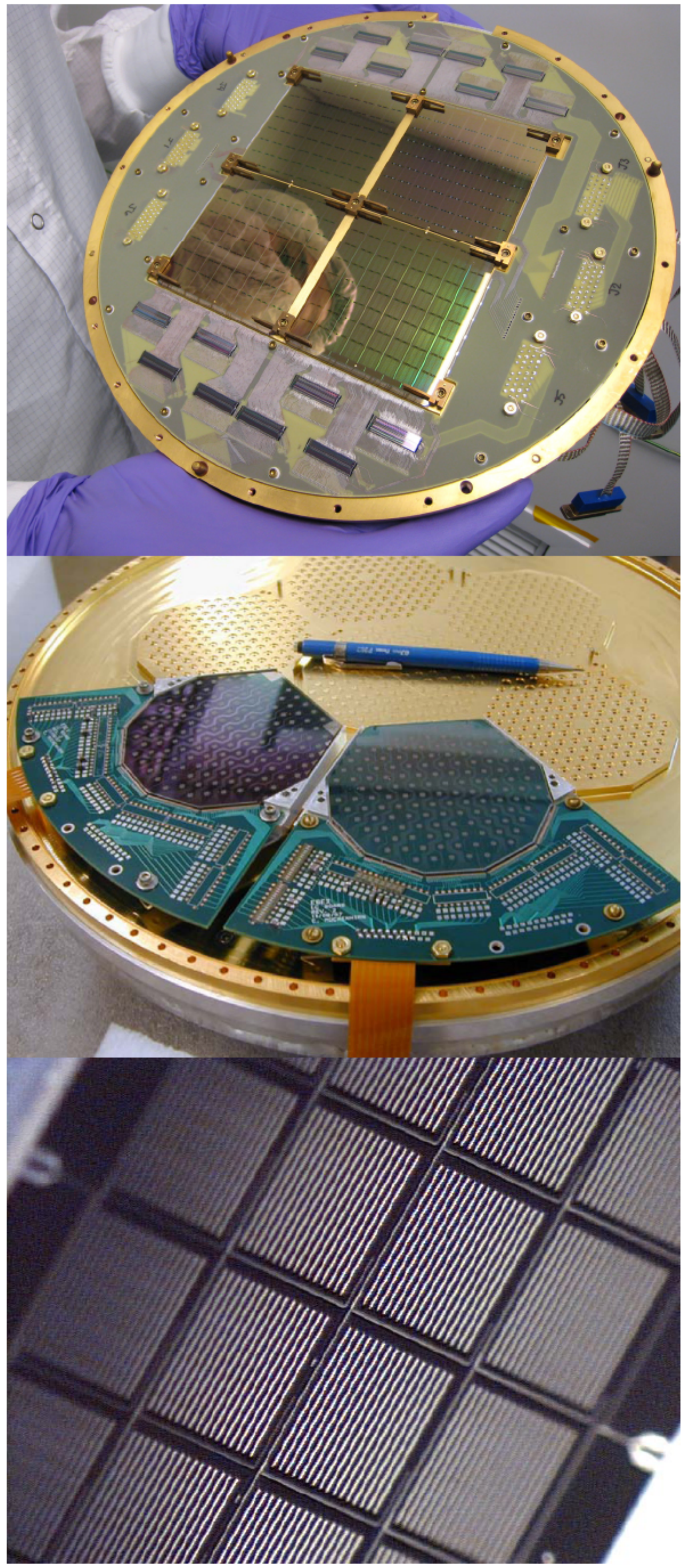

Examples of candidate focal plane technologies for EPIC-IM:  (top) TES bolometers readout out by TDM SQUIDs; (middle) TES bolometers readout by FDM SQUIDs; (bottom) planar antenna-coupled RF-multiplexed MKIDs.

We assume polarization is extracted by a detector pair in each pixel (a feed or antenna), so that differencing the signals gives an instantaneous measure of either Stokes Q or U depending on the orientation of the pixel. Multiple analysis orientations are then included over the focal plane to give equal instantaneous system sensitivity to Q and U. More elaborate polarization analysis is certainly possible, and can be considered within the framework of this focal plane architecture. With the very large throughput crossed-Dragone telescope, beam collimation must be intrinsic to the focal plane since there is no possibility for meaningful beam control with a Lyot stop or a baffle tube. Focal plane arrays with antennas and corrugated feedhorns are currently being developed that provide such on-array collimation. Although there are detailed tradeoffs to consider, our focal plane structure can accommodate all of the readout approaches described in section 7.2, and all of the optical coupling approaches described in section 7.6.

**Table 7.3** Assumptions in Calculating Radiative Loads

| | | | |
|---|---|---|---|
| Surround temperature | 18 K / 30 K | Optics f/# | 2.2 |
| Surround emissivity | 100 % | Filter absorption | 10 % |
| Telescope temperature | 4 K / 30 K | Filter reflection | 90 % |
| Telescope emissivity | 5 % | Focal plane absorption | 10 % |

**Table 7.4a** Estimated Radiative Heat Loads (4 K Telescope Option)

| **Freq [GHz]** | **Ntile [#]** | **$\nu_c$ (4 K) [GHz]** | **$\nu_c$ (1 K) [GHz]** | **P (4 K) [μW]** | **P (1 K) [μW]** | **P (0.1 K) [μW]** |
|---|---|---|---|---|---|---|
| 30 | 6 | 39 | 36 | 80 | 0 | 0 |
| 45 | 14 | 59 | 54 | 190 | 0.1 | 0 |
| 70 | 18 | 91 | 84 | 250 | 0.1 | 0.1 |
| 100 | 18 | 130 | 120 | 250 | 0.1 | 0.2 |
| 150 | 12 | 195 | 180 | 170 | 0.1 | 0.5 |
| 220 | 4 x 0.75 | 442 | 264 | 40 | 0.9 | 0.3 |
| 340 | 4 x 0.25 | 442 | 408 | 14 | 0 | 0.4 |
| 500 | 1 x 0.7 | 1105 | 600 | 8 | 1.5 | 0.6 |
| 850 | 1 x 0.3 | 1105 | 1020 | 3 | 0.1 | 0.8 |
| **Total** | **73** | | | **1000** | **3.0** | **2.8** |

**Table 7.4b** Estimated Radiative Heat Loads (30 K Telescope Option)

| **Freq [GHz]** | **Ntile [#]** | **$\nu_c$ (4 K) [GHz]** | **$\nu_c$ (1 K) [GHz]** | **P (4 K) [μW]** | **P (1 K) [μW]** | **P (0.1 K) [μW]** |
|---|---|---|---|---|---|---|
| 30 | 4 | 39 | 36 | 570 | 0 | 0 |
| 45 | 6 | 59 | 54 | 850 | 0 | 0 |
| 70 | 8 | 91 | 84 | 1140 | 0.1 | 0.1 |
| 100 | 6 | 130 | 120 | 850 | 0.1 | 0.2 |
| 150 | 6 x 0.75 | 286 | 180 | 640 | 1.1 | 0.4 |
| 220 | 6 x 0.25 | 286 | 264 | 210 | 0.1 | 0.4 |
| 340 | 1 x 0.35 | 1105 | 408 | 50 | 3.6 | 0.3 |
| 500 | 1 x 0.15 | 1105 | 600 | 20 | 1.3 | 0.4 |
| 850 | 1 x 0.05 | 1105 | 1020 | 7 | 0.1 | 0.5 |
| **Total** | **31** | | | **4300** | **6.4** | **2.2** |

The 100 mK focal plane assembly is housed inside a ~1 K shield, thermally connected to an intermediate stage of the sub-K cooler. The 1 K stage intercepts parasitic conducted thermal power from the supports, and attenuates radiation from warmer stages of the instrument. Finally a 4 K shield surrounds the 1 K shield. Low-pass filters are located at both the 1 K and 4 K stages, mounted in a hexagonal cell structure similar to that of the focal plane. These filters attenuate high frequency radiation, and depending on the technology could also provide band definition. As shown in Fig. 7.2, the filters and mounting structure are sized to avoid vignetting.

The filtering architecture is designed to reduce radiation loading on the large 100 mK focal plane structure. We have calculated the radiative power loading under the conservative assumption that the focal plane is surrounded by a 100 % emitting surface at the temperature of the optical shield (18 K for the 4 K telescope, 30 K for the 30 K telescope). The filters, assumed to be simple low-pass edge filters, reject this thermal power by reflection, and direct absorbed power to the 4 K and 1 K stages. The resulting loads (see Tables 7.3 and 7.4a&b) are acceptable for the sub-K cooler designs under consideration.

The focal plane structure is suspended on isolating Ti alloy mounts, and cooled by a sub-K cooler, either an ADR or closed-cycle dilution refrigerator. Due to the long thermal time constants associated with the structure, the sub-K cooler must provide continuous rather than 1-shot operation. The thermal requirements on the sub-K cooler, calculated in section 8.3, could be relaxed by using supporting the focal plane with thermally actuating launch locks during the launch when the instrument is at room temperature, and using a low-conductivity support system for flight operation at zero-g.

### 7.2 Focal Plane Technologies

A new generation of sensor and readout combinations is being fielded in sub-orbital and ground-based experiments and the first science publications are being produced. Transition-Edge Sensors (TES) have been deployed with two types of multiplexed SQUID-based readout, frequency-domain multiplexing (FDM) and time-domain multiplexing (TDM). Operational experiments using FDM are SPT, APEX-SZ, and EBEX. Experiments using TDM are ACT, GISMO, MUSTANG, and SABOCA. These experiments have cameras with 128-3000 pixels. Many experiments are in development using these technologies including the Sub-mm Common-User Bolometer Array-2 (SCUBA-2) sub-millimeter camera with 10,000 pixels.

The Microwave Kinetic Inductance Detector (MKID) is a recently developed sensor-readout combination that has the attractive feature that the sensor and multiplexer are combined onto a monolithic substrate. The MKIDs technology has been deployed with the 144-detector DEMOCAM receiver on CSO.

#### *7.2.1 TES Bolometer with SQUID multiplexing*

TES bolometers cooled to temperatures of 100 mK can have a sensitivity that is nearly limited by photon arrival statistics over much of the frequency range of interest. They have two properties that are essential for building large focal-plane arrays (i) They are simple to fabricate using optical photolithography, and (ii) their readout can be multiplexed so that a row of detectors can be readout using a single amplifier – this greatly reduces the complexity of the cryogenic wiring.

The TES is a superconducting film biased in the middle of its transition. It is voltage biased, and in this mode it has high stability and linearity due to negative feedback that occurs between the thermal and electrical "circuits" of the bolometer. The signal from a TES is measured using a Superconducting QUantum Interference Device (SQUID) ammeter, which can operate at cryogenic temperature. Our team has experience building TES detectors over the last decade, including Al/Ti, Cu/Mo and Au/Mo proximity sandwiches, and elemental Ti. Figures 7.3b and 7.3c show example noise data for multiplexed TES sensors.

# EPIC Multiplexed Readout Technologies

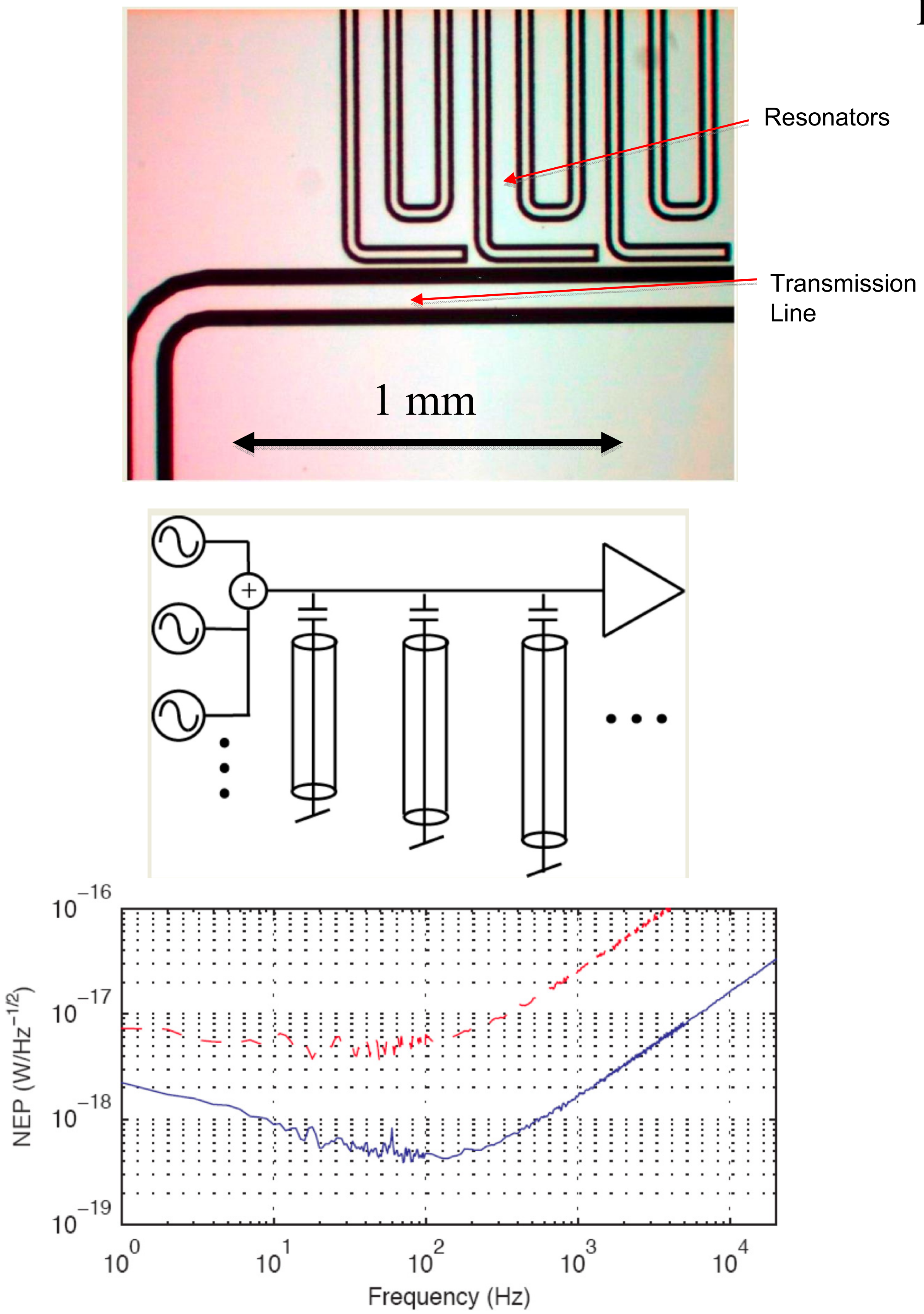

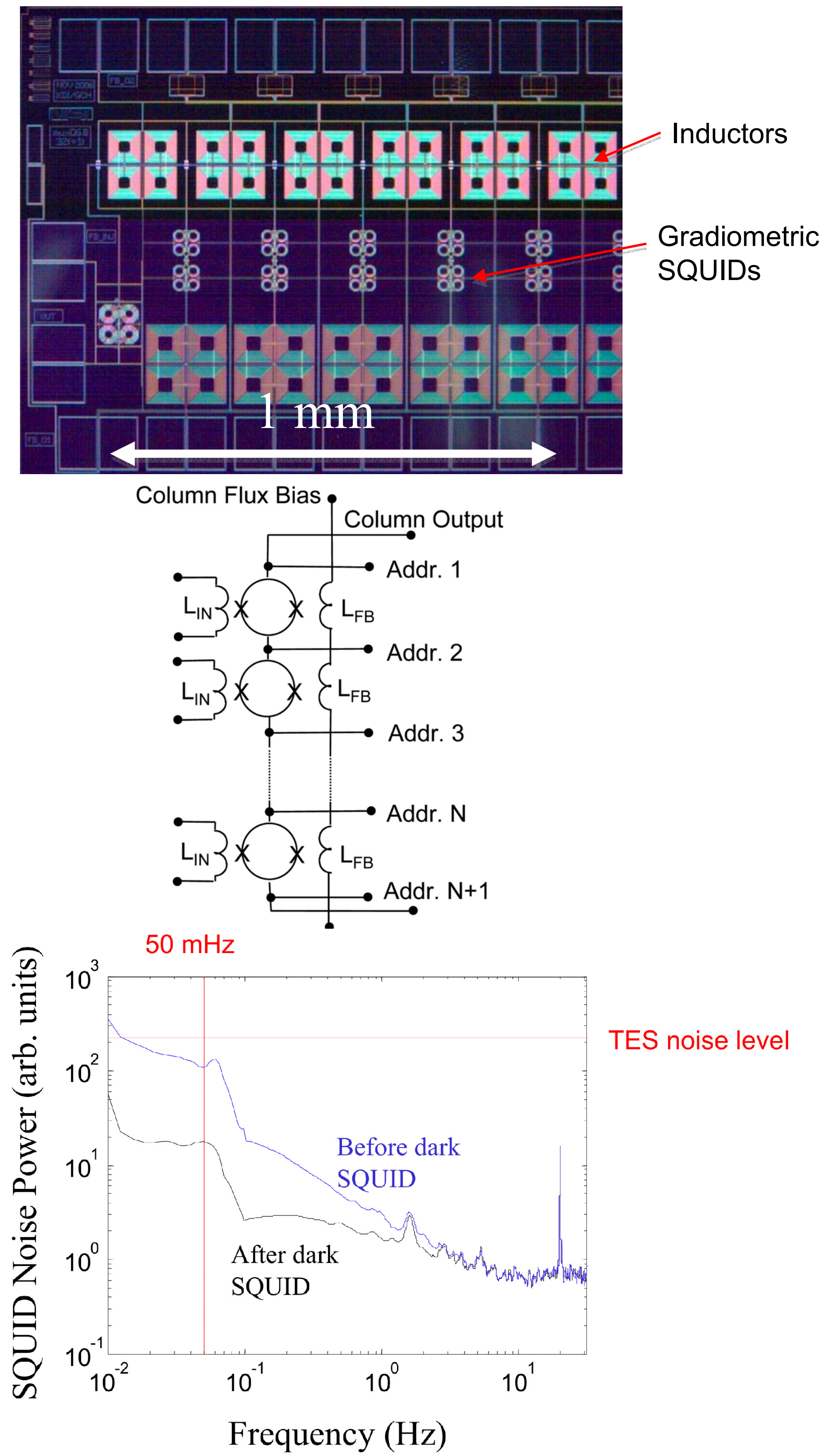

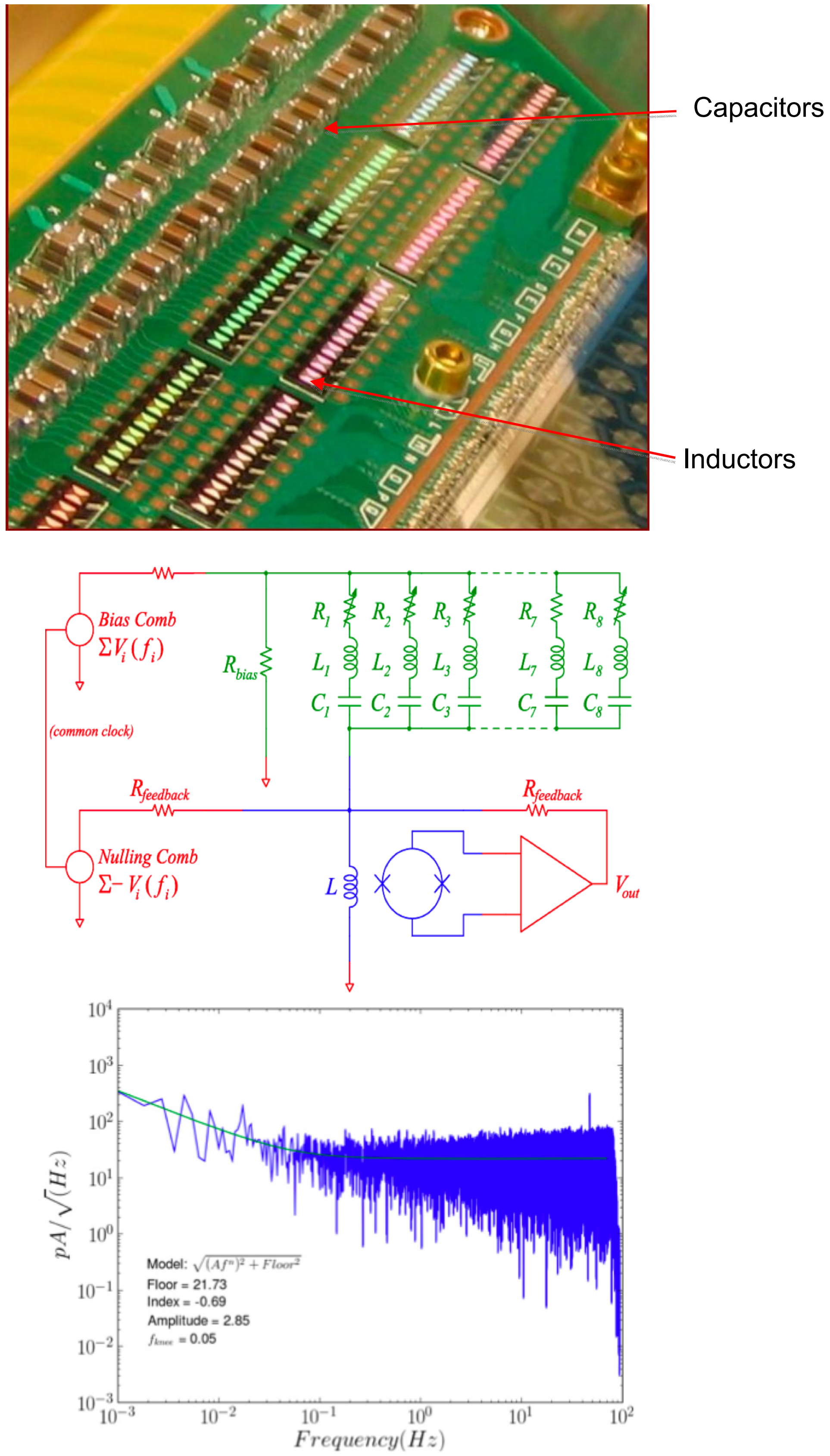

**Fig 7.3a Multiplexed Kinetic Inductance Detectors**

The MKIDs system features an integrated sensor and multiplexer element. Power deposited in a superconducting distributed resonator reduces it kinetic inductance and changes its frequency. A number of resonators separated by a constant-frequency intervals shunt a transmission line, and the impedance of the resonators are measured by detecting the change in transmission in amplitude or phase. (Top) Three resonators coupled to a transmission co-planar waveguide. (Middle) Schematic diagram of multiplexer. Biases are generated digitally and mixed up to GHz frequency. The signals are measured using a cold HEMT amplifier. (Bottom) Noise Equivalent Power for amplitude detection (red) and phase detection (blue). Figures and data courtesy CIT /JPL.

**Fig. 7.3b Time Domain Multiplexed TES**

The time-domain multiplexer uses SQUIDs as switches to sequentially sample a set of TES bolometers. Common address lines are shared between columns resulting in a low total wire count. (Top) Photograph of a 32-channel multiplexer chip. (Middle) Schematic of time-domain multiplexer. The TES bolometers are connected to the input inductors, and when the bolometer is read out when the associated SQUID is in the on condition (Bottom) Noise power for a time-domain multiplexer, which is lower than TES noise for f > 10 mHz. Figures and data courtesy NIST.

**Fig. 7.3c Frequency Domain Multiplexed TES**

The frequency-domain multiplexer uses a sinusoidal bias for each TES bolometer with a unique, identifying frequency for each bolometer. The signals for a set of bolometers are read by a single SQUID and the amplitude modulated signals are recovered at room temperature. (Top) Photograph of bolometer array with lithographed inductors and surface mount capacitors. (Middle) Schematic of multiplexer circuit. (Bottom) Noise plot for two differenced bolometers (as would be done in EPIC). The noise level corresponds to an NEP of ~ 3 x $10^{-17}$. All the Figures and data courtesy University of California, Berkeley.

There are several readout multiplexer technologies that are now mature, and they can be broadly divided into techniques that divide signals in either time or frequency domains. A time-domain readout multiplexer that uses SQUID switches to sequentially choose the detector that is read with the single output amplifier has been developed at NIST. The time-domain multiplexer can read 32 detectors with a single readout amplifier with no loss in bolometer noise performance or bandwidth.

Several groups are independently working on frequency-domain readout multiplexing. In this scheme, each detector is biased using a sine wave with a unique frequency, the bias signals are amplitude-modulated by the bolometers, and the sum of all the currents is measured using a single SQUID ammeter. The Berkeley frequency-domain multiplexer is operational in SPT, APEX-SZ, and EBEX and will be deployed in POLARBEAR.

### *7.2.2 Kinetic Inductance Detector with microwave multiplexing*

The Kinetic Inductance Detector (KID) operates on a different principle than a bolometer. In a KID, the RF power from the sky breaks Cooper pairs in a superconductor held well below its transition temperature changing its kinetic inductance. In the Microwave Kinetic Inductance Detector (MKID), a GHz resonator is built using the KID and changes in the kinetic inductance change the resonator frequency. Each of the multiplexed KIDs are built with a resonator at a unique frequency and therefore they can all be readout simultaneously using a GHz comb of excitations and HEMT amplifier.

The chief advantage of the MKIDS concept is the simplicity of the readout system. Up to 1000 detectors can be read using a single HEMT with negligible dissipation at the lowest focal-plane temperature. The sensitivity of MKIDs has steadily improved with time, but the sensitivity and low-frequency stability required for EPIC have not yet been demonstrated. NEP data are shown in Fig. 7.3a. An MKID focal plane would require a 100 mK cooler to achieve the needed sensitivity and therefore there is no advantage compared to a TES focal plane in this regard. A prototype MKIDs system called DemoCAM was deployed to CSO with a focal plane of 32 detectors. The first science camera will have 2304 detectors spread over four frequency bands.

## 7.3 TES Bolometer Sensitivity Calculation

We calculate detector sensitivities using a version [2] of the non-equilibrium noise model developed by Mather [3,4] adapted for transition-edge bolometers. We first estimate the optical power absorbed at the detector shown in Table 7.6 using the assumptions listed in Table 7.5. We note that the assumed 30 % bandwidth and 40 % optical efficiency are based on the achieved parameters for the Planck focal plane, and some future improvement in efficiency could be reasonably expected. We assume a fixed ratio of TES saturation power to optical power, $P_{sat}/Q = 5$, and optimize the transition temperature for maximum sensitivity. The noise equivalent power (NEP) is then calculated for a combination of photon and detector noise, and converted to a noise equivalent temperature (NET) in CMB temperature units. We neglect the noise contribution of the readout, which can generally be made negligible for both time-domain and frequency-domain SQUID multiplexing. Finally we multiply the total calculated sensitivities by a margin factor of $\sqrt{2}$. The TES bolometer sensitivities form the basis for the mission sensitivities listed in Table 3.2. However MKIDs offer the prospect for similar sensitivities, and it is premature to select either technology at this point.

**Table 7.5** Sensitivity Model Input Assumptions

| | | | | | |
|---|---|---|---|---|---|
| Focal plane temperature | $T_o$ | 100 mK | Optical efficiency | $\eta_{opt}$ | 40 % |
| Blocker temperature | $T_{blkr}$ | 4 K | Fractional bandwidth | $\Delta\nu/\nu$ | 30 % |
| Optics temperature | $T_{opt}$ | 4 K / 30 K* | Noise margin† | | 1.414 |
| Mirror emissivity at 1 mm | $\varepsilon$ | 1 % | Mission lifetime | $T_{life}$ | 4 years |
| Coupling to 4 K / 30 K stop | | 10 % / 0.5 %* | Heat capacity | $C_0$ | 0.15 pJ/K |
| Coupling to 4 K baffle | | 5 % | $\alpha = d\ln(R)/d\ln(T)$ | | 100 |
| Bolometer pitch | $d/f\lambda$ | 2 / 3.25* | TES safety factor‡ | $P_{sat}/Q$ | 5 |

*Parameter for 4 K option / 30 K option
†The total calculated sensitivity is multiplied by a safety factor of √2
‡The factors of safety are 20 for 500 GHz and 200 for 850 GHz (4 K) and 20 for 500 & 850 GHz (30 K)

**Table 7.6** Optical Loading Summary

| Freq [GHz] | 4 K Telescope Option | | | | 30 K Telescope Option | | | |
|---|---|---|---|---|---|---|---|---|
| | CMB [fW] | Baffle [fW] | Mirrors [fW] | Stop [fW] | CMB [fW] | Baffle [fW] | Mirrors [fW] | Stop [fW] |
| 30 | 100 | 8 | 1 | 17 | 100 | 8 | 5 | 7 |
| 45 | 130 | 11 | 1 | 22 | 130 | 11 | 8 | 11 |
| 70 | 160 | 15 | 1 | 30 | 160 | 15 | 16 | 16 |
| 100 | 170 | 17 | 2 | 34 | 170 | 17 | 26 | 23 |
| 150 | 140 | 18 | 3 | 35 | 140 | 18 | 47 | 33 |
| 220 | 81 | 15 | 3 | 30 | 81 | 15 | 78 | 46 |
| 340 | 23 | 8 | 2 | 16 | 23 | 8 | 140 | 64 |
| 500 | 3 | 2 | 1 | 5 | 3 | 2 | 210 | 81 |
| 850 | 0.2* | 0.1 | 0.0 | 0.2 | 0.2* | 0 | 330 | 100 |

*Includes interstellar dust emission at N(HI) = 5e20 $cm^{-2}$

**Table 7.7** TES Bolometer Parameters

| Freq [GHz] | $G_0$ [pW/K] | $\tau$ [ms] | $NEP_{photon}$ [aW/√Hz] | $NEP_{bolo}$ [aW/√Hz] | $NEP_{tot}$ [aW/√Hz] | NET [$\mu K_{CMB}$ √s] | $NET_{Planck}$ [$\mu K_{CMB}$ √s] |
|---|---|---|---|---|---|---|---|
| *4 K Telescope Option* | | | | | | | |
| 30 | 3.4 | 1.3 | 2.6 | 3.1 | 5.7 | 84 | 239 |
| 45 | 4.5 | 1.0 | 3.5 | 3.6 | 6.9 | 71 | 292 |
| 70 | 5.6 | 0.8 | 4.6 | 3.9 | 8.6 | 60 | 414 |
| 100 | 5.9 | 0.8 | 5.5 | 4.1 | 9.7 | 54 | 102 |
| 150 | 5.2 | 0.9 | 6.3 | 3.8 | 10.4 | 52 | 83 |
| 220 | 3.5 | 1.3 | 6.1 | 3.1 | 9.7 | 59 | 134 |
| 340 | 1.3 | 3.5 | 4.7 | 1.9 | 7.1 | 100 | 404 |
| 500 | 1.2 | 3.3 | 2.7 | 1.8 | 4.6 | 350 | |
| 850 | 0.6 | 2.0 | 0.8 | 1.2 | 2.1 | 15000 | |
| *30 K Telescope Option* | | | | | | | |
| 30 | 3.4 | 1.3 | 2.6 | 3.1 | 5.7 | 83 | 239 |
| 45 | 4.5 | 1.0 | 3.5 | 3.6 | 6.9 | 70 | 292 |
| 70 | 5.6 | 0.8 | 4.6 | 4.0 | 8.6 | 60 | 414 |
| 100 | 6.2 | 0.7 | 5.7 | 4.2 | 10 | 55 | 102 |
| 150 | 6.3 | 0.7 | 6.9 | 4.2 | 11 | 57 | 83 |
| 220 | 5.9 | 0.8 | 8.1 | 4.2 | 13 | 77 | 134 |
| 340 | 6.2 | 0.7 | 10 | 4.2 | 16 | 220 | 404 |
| 500 | 32 | 0.3 | 14 | 10 | 24 | 1500 | |
| 850 | 47 | 0.2 | 22 | 11 | 35 | 2.5e5 | |

Notes: NETs are per detector
Sensitivity requirements for the Planck LFI are shown in red, goals for the HFI in blue, approximate to the measured performance in instrument-level testing. Actual flight sensitivities for Planck await launch.

The margined sensitivities shown in Table 7.7 are only about a factor 2 better per detector than the Planck HFI bolometers, which are already close to the background limit. This simply emphasizes that the individual detector sensitivities are already within state of the art, and that EPIC must use large-format arrays to gain sensitivity over Planck. It is worth noting that in the development of Planck HFI, the instrument held a similar √2 noise margin in the goal sensitivity estimates, and this margin was allocated to the bolometer, optics, and readout components. However, the final Planck HFI bolometer NET sensitivities measured at instrument level on the ground exceeded the goal sensitivities, so we think the margin factor we have chosen is appropriate and possibly even a bit conservative.

**Table 7.8** Focal Plane Temperature Susceptibility

| **Freq [GHz]** | **$dT_{CMB}/dT_0$ [K/K]**[1] | **$NET_0$ [μK/√Hz]** | | **$dT_{CMB}/dT_{Opt}$ [mK/K]**[1] | **$NET_{Opt}$ [mK/√Hz]** | |
|---|---|---|---|---|---|---|
| | | **Bolo**[2] | **Band**[3] | | **Bolo**[2] | **Band**[3] |
| ***4 K Telescope Option*** | | | | | | |
| 30 | 72 | 1.7 | 3.6 | 105 | 1.1 | 12.3 |
| 45 | 65 | 1.6 | 1.6 | 108 | 0.9 | 4.9 |
| 70 | 55 | 1.5 | 0.8 | 113 | 0.7 | 2.0 |
| 100 | 46 | 1.6 | 0.7 | 122 | 0.6 | 1.3 |
| 150 | 37 | 2.0 | 0.7 | 144 | 0.5 | 0.9 |
| 220 | 30 | 2.8 | 1.6 | 195 | 0.4 | 1.2 |
| 340 | 26 | 5.5 | 4.0 | 360 | 0.4 | 1.4 |
| 500 | 130 | 3.9 | 2.3 | 880 | 0.5 | 1.7 |
| 850 | 5600 | 3.7 | 2.4 | 6500 | 3.2 | 10.6 |
| Total | | | 0.43 | | | 0.53 |
| ***30 K Telescope Option*** | | | | | | |
| 30 | 72 | 1.7 | 6.9 | 8 | 14 | 280 |
| 45 | 65 | 1.6 | 3.4 | 9 | 10 | 110 |
| 70 | 55 | 1.5 | 2.1 | 11 | 7.5 | 52 |
| 100 | 49 | 1.6 | 1.5 | 14 | 5.6 | 27 |
| 150 | 45 | 1.8 | 1.6 | 21 | 3.8 | 17 |
| 220 | 51 | 2.1 | 2.1 | 42 | 2.6 | 13 |
| 340 | 120 | 2.5 | 4.6 | 170 | 1.8 | 16 |
| 500 | 3400 | 0.7 | 1.4 | 1500 | 1.7 | 16 |
| 850 | 48000 | 0.8 | 1.4 | 270000 | 1.3 | 13 |
| Total | | | 0.65 | | | 6.3 |

[1]Raw CMB signal susceptibility for a change in base or optics temperature on a single bolometer.
[2]Equivalent single-detector sensitivity to base temperature variations, i.e. a base NET at this level produces noise at the detector equivalent to the instrument noise in a single bolometer.
[3]Band-combined sensitivity to base temperature variations assuming 5 % matching to base temperature variations and 1 % matching to variations in optical power

Given the detector model, we can assess our susceptibility to temperature variations. Bolometers are thermal detectors, and sensitive to fluctuations in the base temperature. Furthermore variations in the optics temperature will modulate the optical load on the detectors. In Table 7.8 we list the focal plane susceptibility to these temperature variations, which are summarized for the aggregate focal plane in Table 5.3. These requirements are generally similar to the temperature stability requirements for Planck. This is because EPIC uses pair differencing for all of its polarization measurements, which helps to relax the stability requirement. Although Planck has less system sensitivity, the instrument uses non-differenced channels used to measure temperature anisotropy which drives the required temperature stability. Planck has already demonstrated a combination of active and passive temperature stabilization at this level of performance.

### 7.4 MKID Sensitivity Calculation

Sensitivity calculations for a MKID focal plane are shown in Table 7.9 assuming a focal-plane temperature of 100 mK. The signal from an MKID can be readout as both a modulated amplitude and a modulated phase. Compared to HEMT noise, the phase modulation signal is larger than the amplitude modulation signal, which is reflected in the lower phase modulation NEPs in Table 7.9. Measurements of current phase and amplitude mode NEPs are shown in Fig. 7.3a. The MKID NEP for phase readout is roughly equal to the photon-noise NEP at 300 GHz, but gently increases with decreasing frequency becoming a factor of 1.4 higher than the photon noise NEP at 30 GHz. The ratio of total NEP for MKIDS compared to TESes ranges from a factor of ~1 to ~1.5.

An MKIDs focal-plane requires a 100 mK temperature just as the TES focal-plane does, but for a different reason. The low operating temperature is driven by the need to keep the energy to break a Cooper pair sufficiently low such that statistical fluctuations in quasiparticle production do not dominate the NEP. The quasiparticle fluctuation noise term is shown in Table 7.9 under the $NEP_{recomb}$ column.

EPIC-IM requires low-frequency stability to ~10 mHz. To achieve this specification with MKIDs will require further development. It can be seen in Fig. 7.3a that the phase noise has a rising spectrum at low-frequency. Current MKID resonators show excess phase noise at low-frequency due to “two-level systems” (TLS) noise due to excitable energy levels in the insulating layer of the resonator. It may be possible to substantially reduce or eliminate the TLS noise.

**Table 7.9** Sensitivity estimates for 100 mK MKIDs focal plane

| ν [GHz] | $Q_{tot}$ [pW] | $n_{qp}$/ photon | $h_e$ | $NEP_{photon}$ | $NEP_{recomb}$ | $NEP_{HEMT_amp}$ | $NEP_{HEMT_phase}$ | $NEP_{tot_amp}$ | $NEP_{tot_phase}$ |
|---|---|---|---|---|---|---|---|---|---|
| 30 | 0.12 | 2 | 1 | 2.8 | 2.2 | 7.2 | 1.4 | 8 | 3.8 |
| 40 | 0.14 | 2 | 0.8 | 3.3 | 2.8 | 9.2 | 1.8 | 10.1 | 4.7 |
| 60 | 0.18 | 3.2 | 0.8 | 4.2 | 3.0 | 9.8 | 2.0 | 11.1 | 5.5 |
| 90 | 0.20 | 4.4 | 0.7 | 5.1 | 3.3 | 10.8 | 2.2 | 12.4 | 6.4 |
| 135 | 0.17 | 6.2 | 0.7 | 5.7 | 3.2 | 10.5 | 2.1 | 12.4 | 6.8 |
| 200 | 0.14 | 8.8 | 0.7 | 6.1 | 2.9 | 9.6 | 1.9 | 11.7 | 7.0 |
| 300 | 0.09 | 12.8 | 0.6 | 6.1 | 2.4 | 8.0 | 1.6 | 10.3 | 6.8 |

Noise equivalent power (NEP) numbers are given in aW/√Hz, *ν* is the central frequency of the band, $Q_{tot}$ is the total optical loading, and $n_{qp}$/photon is the number of quasiparticle pairs created per photon which determines the level of statistical noise. MKIDs can be read using both amplitude and phase and both NEPs are tabulated here.

### 7.5 Sensor and Readout Tradeoffs

Focal plane resources for the TES sensor and TDM readout requirements are shown in Table 7.10. The TES/TDM combination was chosen to model resource needs and costs since it places the largest resource requirements on the sub-K stage of the three technologies considered, as shown in Tables 7.10 and 7.11. For the 4 K option, the 100 mK power dissipation for the TES/TDM combination is 1.9 μW, whereas the dissipation for both the TES/FDM and MKIDS cases is 20 nW. The TES/TDM 100 mK dissipation figure assumes 5 nW per SQUID compared to a current value of 10 nW/SQUID. The factor of 2 reduction in assumed power is conservative given recent SQUID developments and a factor of 5-10 may be possible. The TES/FDM and MKIDs 100 mK dissipation figures are based on achieved performance.

The TES/TDM case also has the highest number of wires to the 100 mK stage, 5277. The TES/FDM case requires 720 wires. The wire counts for TDM and FDM both assume a conservative design with no shared wires between array tiles. The MKIDS case requires 11 low-thermal conductance coaxial cables, assuming that the coaxial connections can be daisy chained between detector tiles. We note the 100 mK cooler is designed for these loads, and that the higher dissipation and wire number for TDM is an important but not dominant component of the total heat load on the 100 mK stage. The heat leak through the supports and IR flux are both larger as shown in Table 8.3.1.

The MKIDs system will use HEMT amplifiers at 20 K that dissipate 100 mW in the 4 K telescope case. The TES/FDM total power dissipation is 150 mW at 20 K. The HEMT or FDM buffer amplifier power dissipation would dominate the loads on the 20 K stage provided by the cryocooler (see Table 8.2.1), and this stage is separated by a significant distance from the shielded 4 K focal plane enclosure.

**Table 7.10** Focal Plane Power and Wiring for 32:1 SQUID Time-Domain Multiplexing

| Freq [GHz] | 4 K Telescope Option | | | 30 K Telescope Option | | |
|---|---|---|---|---|---|---|
| | Nmux [#] | Pmux [nW] | Nwire [#] | Nmux [#] | Pmux [nW] | Nwire [#] |
| 30 | 6 | 30 | 246 | 4 | 20 | 164 |
| 45 | 14 | 70 | 574 | 6 | 30 | 246 |
| 70 | 54 | 270 | 990 | 8 | 40 | 328 |
| 100 | 72 | 360 | 1116 | 18 | 90 | 330 |
| 150 | 96 | 480 | 1080 | 18 | 90 | 330 |
| 220 | 44 | 220 | 444 | 18 | 90 | 330 |
| 340 | 24 | 120 | 304 | 4 | 20 | 62 |
| 500 | 30 | 150 | 244 | 4 | 20 | 62 |
| 850 | 35 | 175 | 279 | 4 | 20 | 62 |
| **Total** | **375** | **1875** | **5277** | **84** | **420** | **1914** |

Notes: Assumes 5 nW dissipation per 32:1 multiplexer chip
Assumes independent bias, feedback and address wiring per tile, so each tile can operate independently, with commoned row address return.

All sensor/readout systems dissipate a significant fraction of the total spacecraft power budget at 300 K. For each of the technologies, further development in power dissipation is required beyond current ground- and balloon-based systems. All the warm electronics dissipation values in Table 7.11 include projected reductions. For TES/TDM, we assume a proven multiplexing ratio of 32:1 and demonstrated readout electronics developed for SCUBA2. The 4 K case requires a ~10x reduction, and the 30 K case requires ~5x reduction, in electronics power dissipation per detector compared with existing readout systems (these systems have generally not been designed to minimize power consumption). An initial study at JPL indicated a factor of 5 reduction in power is possible using optimized space-qualified components, and that another factor of 2 reduction may be possible with development.

For TES/FDM, the assumed multiplex factor is 32 compared to 8 in current use. We have achieved a factor 10 in readout bandwidth using a SQUID amplifier with local feedback called a LInearized SQUID Array (LISA), which should be more than adequate for the specified factor 4 increase in multiplexing factor. It is very likely that even higher multiplexer factors will be achieved within five years. The LISA also has a low output impedance suitable to drive the long cables from 4 K to 300 K. The TES/FDM 300 K dissipation in Table 7.11 also assumes the use of an ASIC for modulation and demodulation with a factor of 5 power reduction compared to the FPGAs that are currently used. The MKIDs specification assumes a multiplex factor of 1000 and that multiple tiles can be readout with a common transmission line. This multiplex factor is roughly a factor of 10 higher than what has been achieved, but could be done with existing bandwidth in both the HEMT and 300 K electronics.

All detector and readout systems require shielding of the focal plane from magnetic fields at some level. Reducing B-field susceptibility is an important systems consideration for all 3 options, with different levels of susceptibility and different solutions in each case. The ambient field at L2 is quite low (~ 5 nT), so susceptibility to local fields may be more of a concern, for example, varying fields from an ADR cooler. Currently, experiments use large cryoperm magnetic shields for shielding of Earth's field (50 μT). Of the three options, the first-stage of TDM SQUID readouts may currently be the most susceptible component. However the use of gradiometric-wound SQUIDs that null signals from magnetic fields have been developed that reduce this magnetic field sensitivity by a factor of 100. Differencing matched detectors to measure polarization also reduces susceptibility to magnetic fields.

**Table 7.11** Multiplexer Power Dissipation for Three MUX Technologies

| Temp | 4 K Telescope Option | | | 30 K Telescope Option | | |
|---|---|---|---|---|---|---|
| | TDM | FDM | MKIDs | TDM | FDM | MKIDs |
| 300 K | 150 W | 264 W | 100 W | 75 W | 53 W | 20 W |
| 20 K | 0 | 0 | 100 mW | 0 | 0 | 20 mW |
| 4 K | 375 μW | 540 μW | 0 | 85 μW | 108 μW | 0 |
| 1 K | 0 | 0 | 0 | 0 | 0 | 0 |
| 100 mK | 1875 nW | 20 nW | 20 nW | 420 nW | 4 nW | 4 nW |

### 7.6 Focal Plane Feed Architecture

A differencing polarimeter measures polarized radiation by subtracting the signals from two detectors that are sensitive to orthogonal polarizations, without an active polarization modulator. Many bolometric CMB polarization experiments, such as BOOMERanG, BICEP, QUAD, and Planck HFI use differencing polarimeters. Future experiments using differencing polarimeters KECK and SPIDER will use scan modulation in conjunction with a waveplate stepped on a timescale of many hours, used to mitigate main beam effects, not for signal stability. The EPIC-IM optical design precludes the use of a waveplate for all practical purposes, so the only possibility for polarization modulation would be to use active switching in the focal plane. While these devices have been explored, significant development is still required. Therefore our baseline signal modulation uses bolometer differencing and scanning, which in turn requires a very stable focal plane.

Presently all antenna designs brings out vertical and horizontal polarizations into two detectors, and the difference is used to extract a single Stokes parameter. However, one can extract both Q and U in a single pixel by splitting half the power in each polarization into a 180° degree hybrid, followed by a pair of detectors. This scheme requires 4 detectors per pixel, with half the power in each detector. TES detectors can be designed to this lower background with

# EPIC Focal Plane Technologies

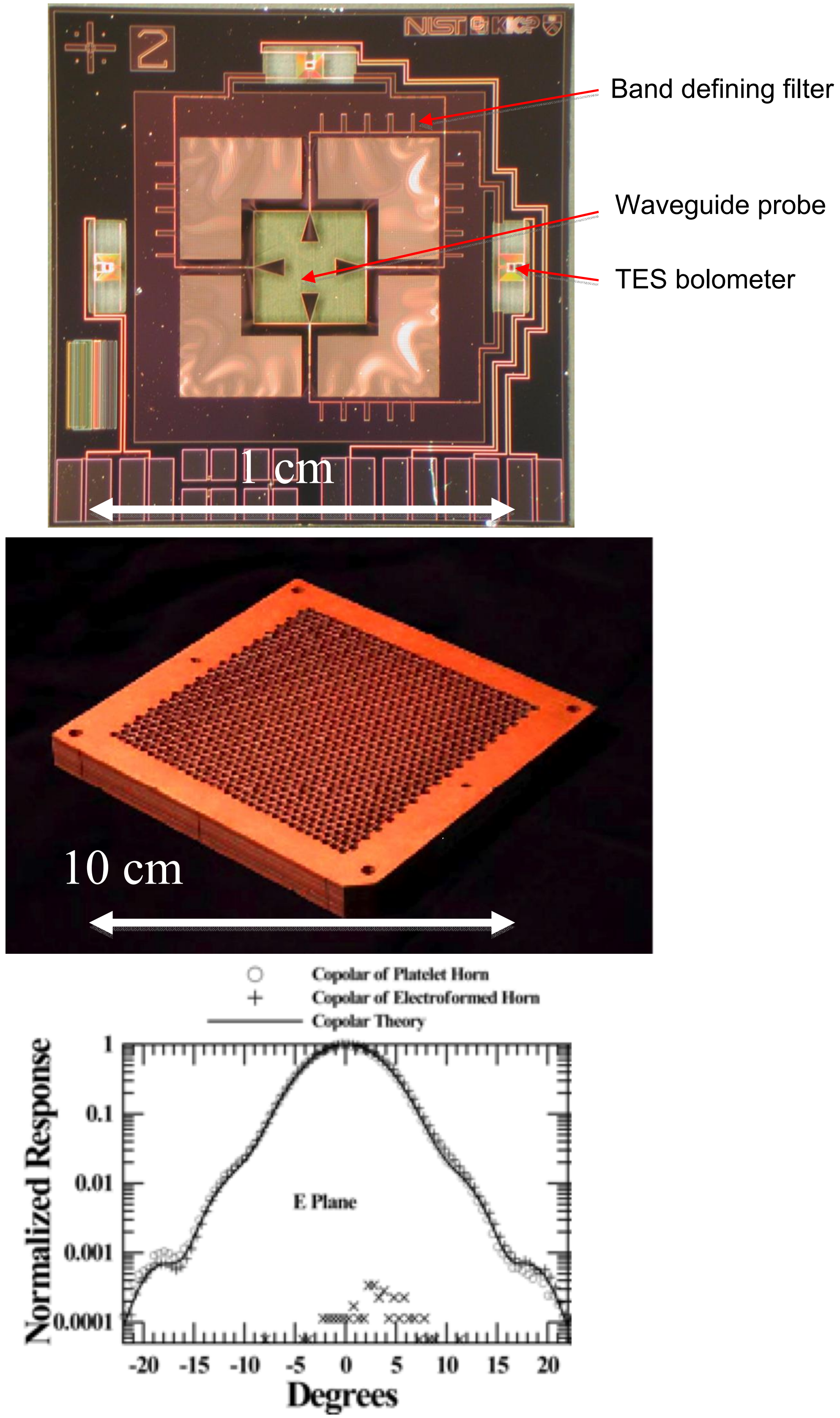

**Fig. 7.4a Scalar Horn Coupling**

Scalar horns have been used in many CMB experiments to date including WMAP and Planck. Traditional scalar horn-coupled instruments used hand-assembled single pixels, but new lithographic techniques are being developed for building large-scale arrays. (Top) Detector chip with waveguide probes to couple to waveguide. On-chip filters determine the frequency band, and transition-edge sensor bolometers detect power. Chip is built under a NIST/CU/KICP/Princeton collaboration. (Middle) Array of scalar horns built using stacked, lithographed "platelet" arrays. Array built by GSFC. (Bottom) Beam measurement of aluminum platelet array by U. Florida.

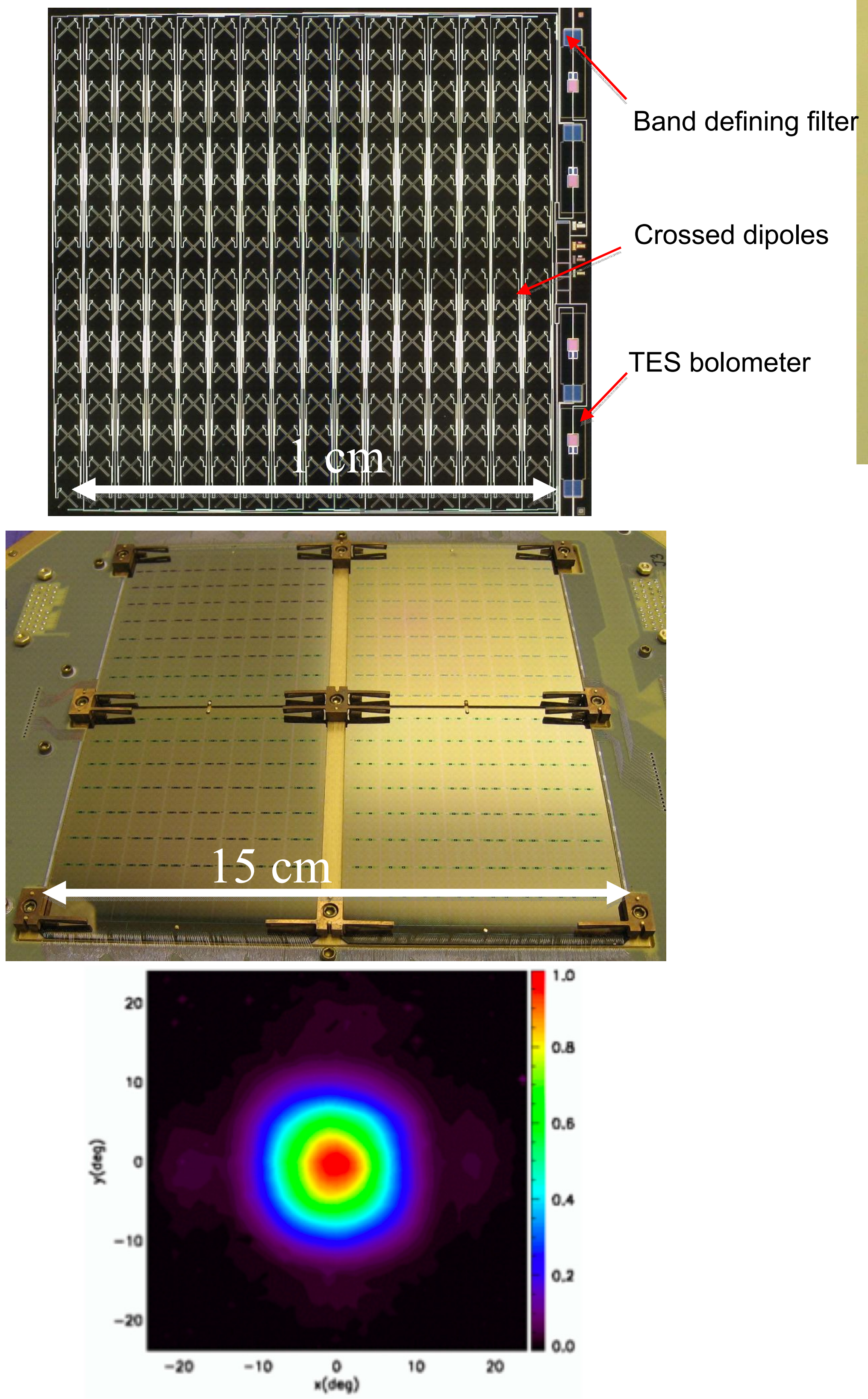

**Fig. 7.4b Phased-Array Antenna Coupling**

A focal-plane feed with a highly directional beam can be formed by a phased-array of wavelength-scale dipoles. This technique is common in radar and communication. (Top) Photograph of a single phased-array pixel. The "x" shapes are crossed dipoles that sense two polarization states. A transmission-line summing network connects all the dipoles. Band-defining filters and transition-edge sensor bolometers are located on the right of the chip. (Middle) Four 8 x 8 arrays of the same pixels shown at the top along with readout circuit board. (Bottom) Beam map made with a pixel such the one in the top photographs. Arrays and measurements by JPL/CIT.

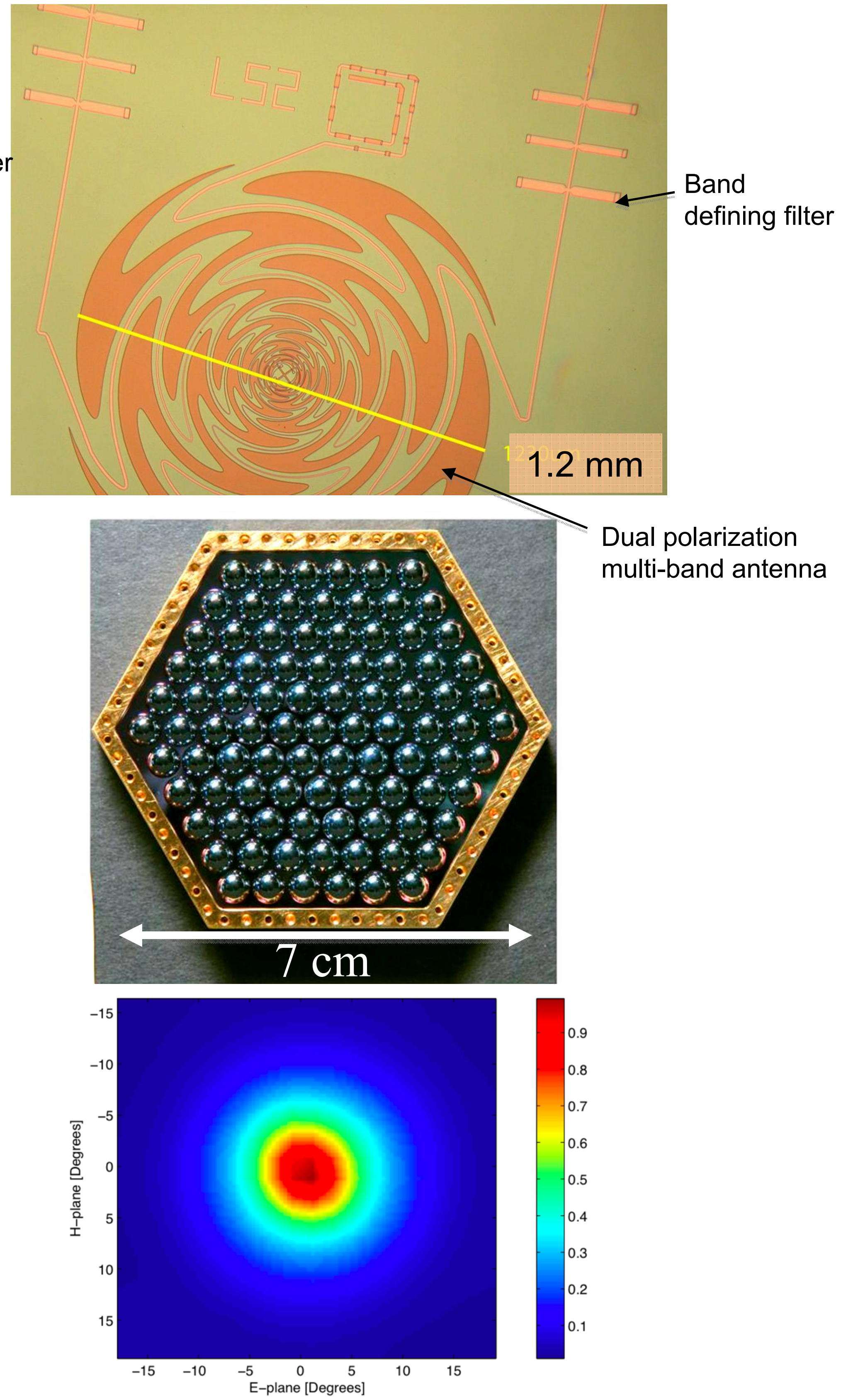

**Fig. 7.4c Antenna with Contacting Lens Coupling**

A dual-polarization multichroic pixel with 3:1 bandwidth can be built using a planar log-periodic antenna with a contacting lens to achieve high directivity. A focal-plane of multichroic pixels would reduce the size and weight of the focal plane, greatly reducing technical risk. (Top) Photograph of a dual-polarized log-periodic antenna with 80-240 GHz bandwidth and band-defining filters for a single band. Transition-edge bolometers (not shown) detect power. Channelizing RF filters will be used to detect 4-5 photometric bands from a single antenna. (Middle) Focal-plane array of hemispherical silicon lenses. (Bottom) Beam map of a 90 GHz pixel. Array and measurements by University of California, Berkeley.

negligible overall loss in sensitivity. Alternatively, the focal plane can be alternated between Q and U by using +/- 90˚ hybrids and two detectors per pixel. This arrangement gives better instantaneous Q/U coverage in a single scan. A detailed systematics simulation must be carried out to determine if these approaches bring worthwhile benefits.

We have studied three methods for optical coupling of the focal plane: (i) planar-probe-coupled scalar horns, (ii) phased-array planar antennas, and (iii) lens-coupled planar antennas. All three are viable, and further study will be required before a single technology can be chosen for EPIC. As described in the technology development plan section x, near term sub-orbital experiments using these technologies will help clarify the tradeoffs and give a basis for down selection for EPIC.

### *7.6.1 Scalar-horn Coupling*

Scalar-horn antennas have a strong heritage in CMB experiments. They have been used, for example, in COBE, WMAP, and Planck HFI. For CMB polarization experiments, scalar horns have advantages of highly symmetric beams, low cross-polarization, and low-sidelobes. Arrays of scalar horns can be coupled to a monolithic array of bolometers by use of planar, lithographed OMTs as are being developed by the GSFC and NIST groups. The scalar-horn is most advantageous for systems with minimal optics, e.g. those with no cold aperture stop, where the horn defines the optical performance rather than the rest of the optics. In EPIC-IM, the low sidelobe level relies on beam collimation in the focal plane, particularly in the 30 K telescope case. By using scalar horns, the cross-pol and ellipticity performance of the entire EPIC-IM system will be dominated by the contributions of the telescope. Therefore the high performance of the scalar horns is helpful for systematic error control.

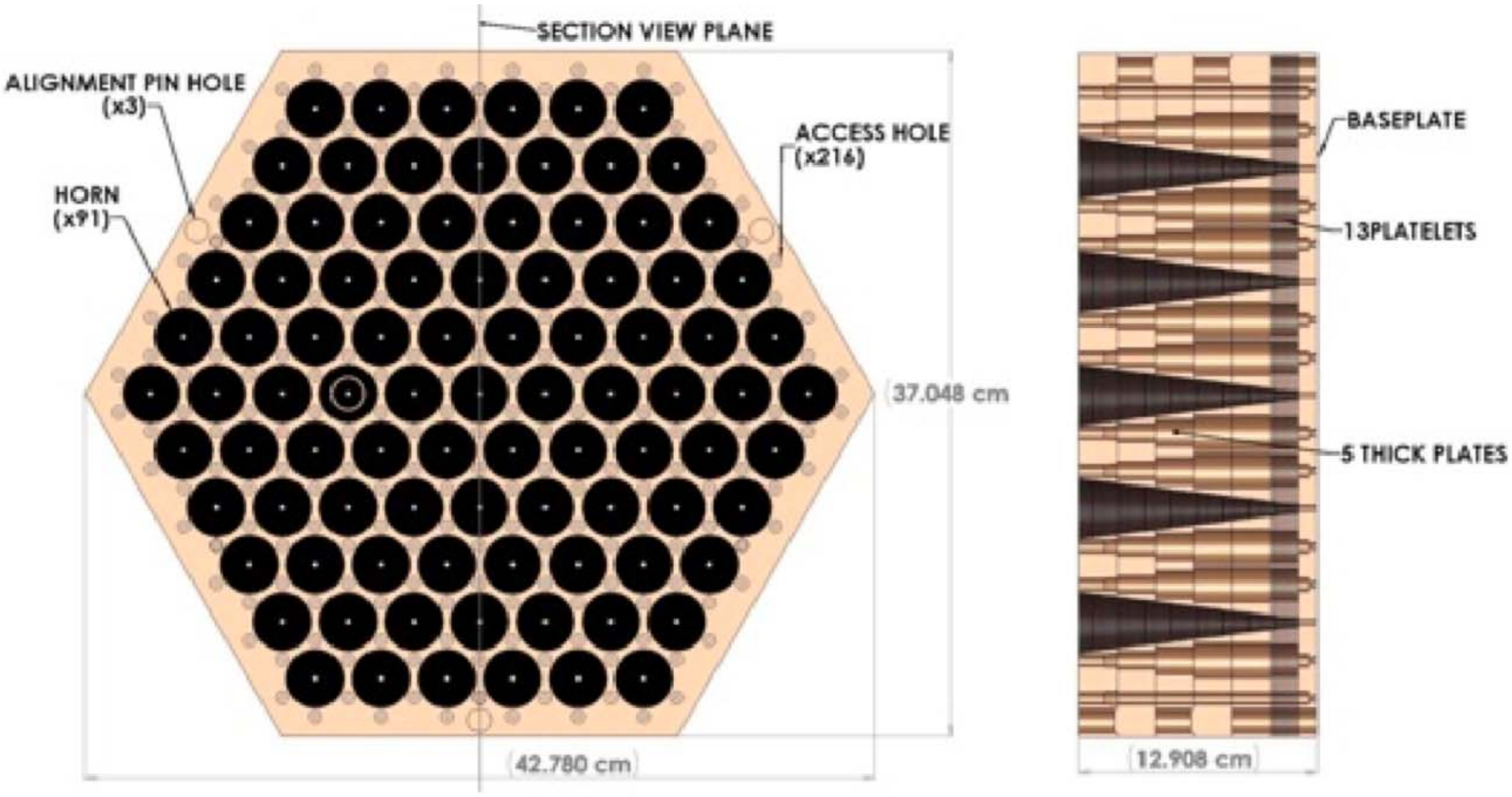

Figure 7.5. Front (Left) and side cross-section (Right) views of a platlet array of scalar horns. This design is a 91 pixel 90 GHz array built for the QUIET experiment. The array is made from 5 thick plates each with 18 corrugations per horn and fabricated from aluminum. The array weights 20 kg.

The main drawback of the platelet scalar horn array is the high mass. The 91 element 90 GHz horn array for QUIET shown below weighs 20 kg. This mass can be reduced by use of lighter materials and by removal of more material between the horns. The horn array would have to be supported at either 100 mK, or at 4 K by developing a temperature in waveguide, and therefore a considerable reduction in weight would be needed to be practical for EPIC-IM.

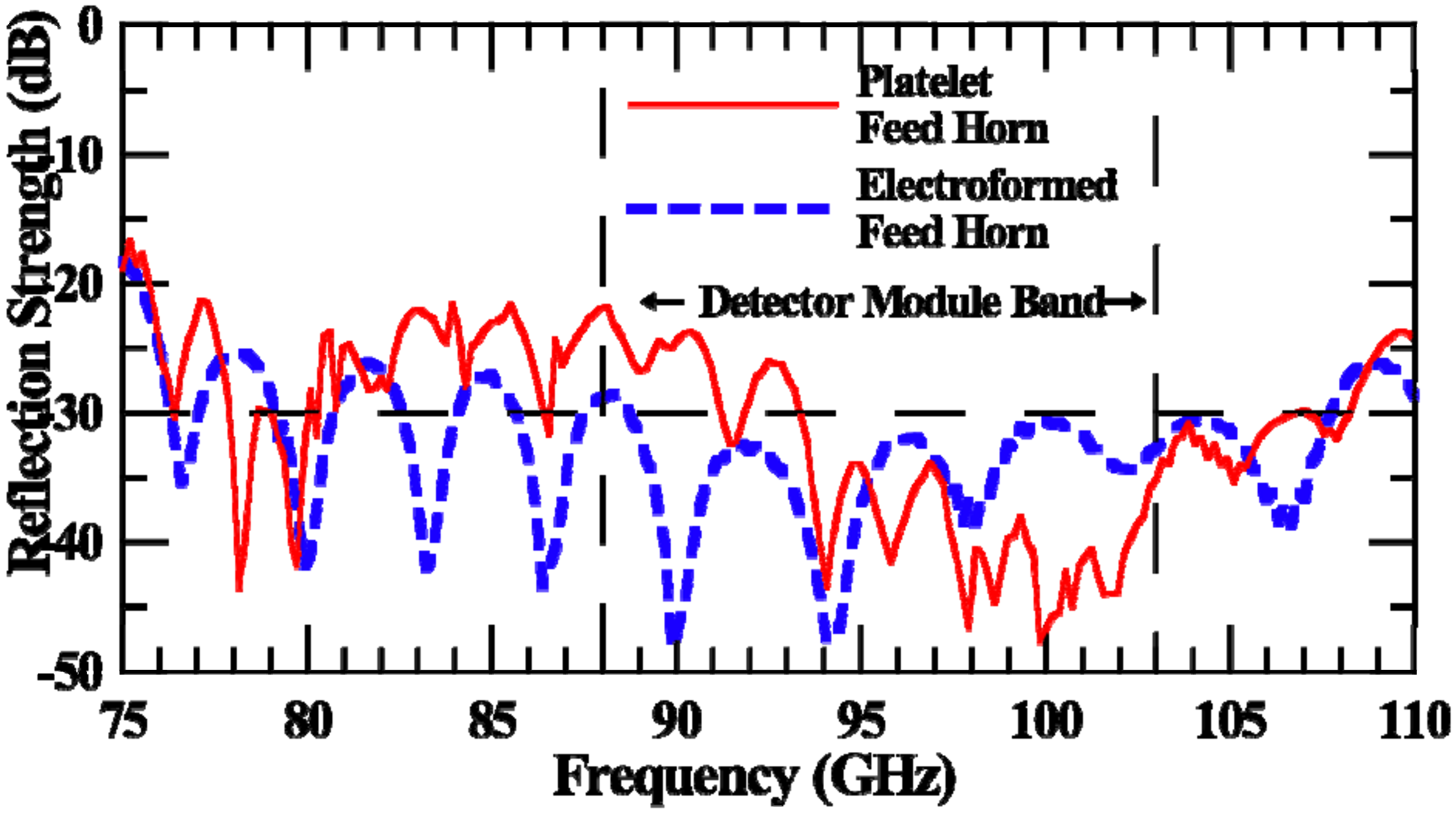

Fig. 7.6. Return loss of the platelet array shown in Fig. 7.5 compared to a traditional electroformed feed horn. The performance is excellent in the intended band.

### *7.6.2 Phased-Array Antenna Coupling*

The angular size of an antenna's beam becomes smaller due to diffraction as the effective area of the antenna grows. Most planar antennas have a size that is comparable to the wavelength of the radiation and a correspondingly broad antenna pattern. A phased-array antenna combines a large number of small antenna elements to form a larger antenna. Phased-array antennas are common at radio wavelengths for e.g. radar and communications.

The millimeter-wavelength monolithic in-phase array antenna has been developed recently by the JPL/CIT group. A photo of an array coupled to a circuit board with SQUID multiplexer readout chips is shown in Fig. 7.7. The antenna is made from a large number of slot dipoles, and the RF signals are added coherently by a network of microstrip transmission lines. After addition, the signals are bandpass filtered, and finally the signals are detected by TES bolometers.

Advantages of the phased-array antenna over other candidates include minimal focal-plane mass, efficient use of the focal-plane area, and a completely monolithic fabrication process which can be critical for making large arrays. The current development status is that single pixels with a complete antenna/filter/bolometer have been measured and show symmetric beam patterns closely matching theoretical predictions, low cross-polarization, a spectral bandpass with the expected width, and high optical efficiency. Figure 7.8 shows a measured antenna pattern and frequency band shapes. The engineering parameters for making a large array including TES uniformity and band placement have been studied in detail and appear to all be sufficiently controllable. An array has been developed and is now being integrated into a multiplexed focal plane (see Fig. 7.4b).

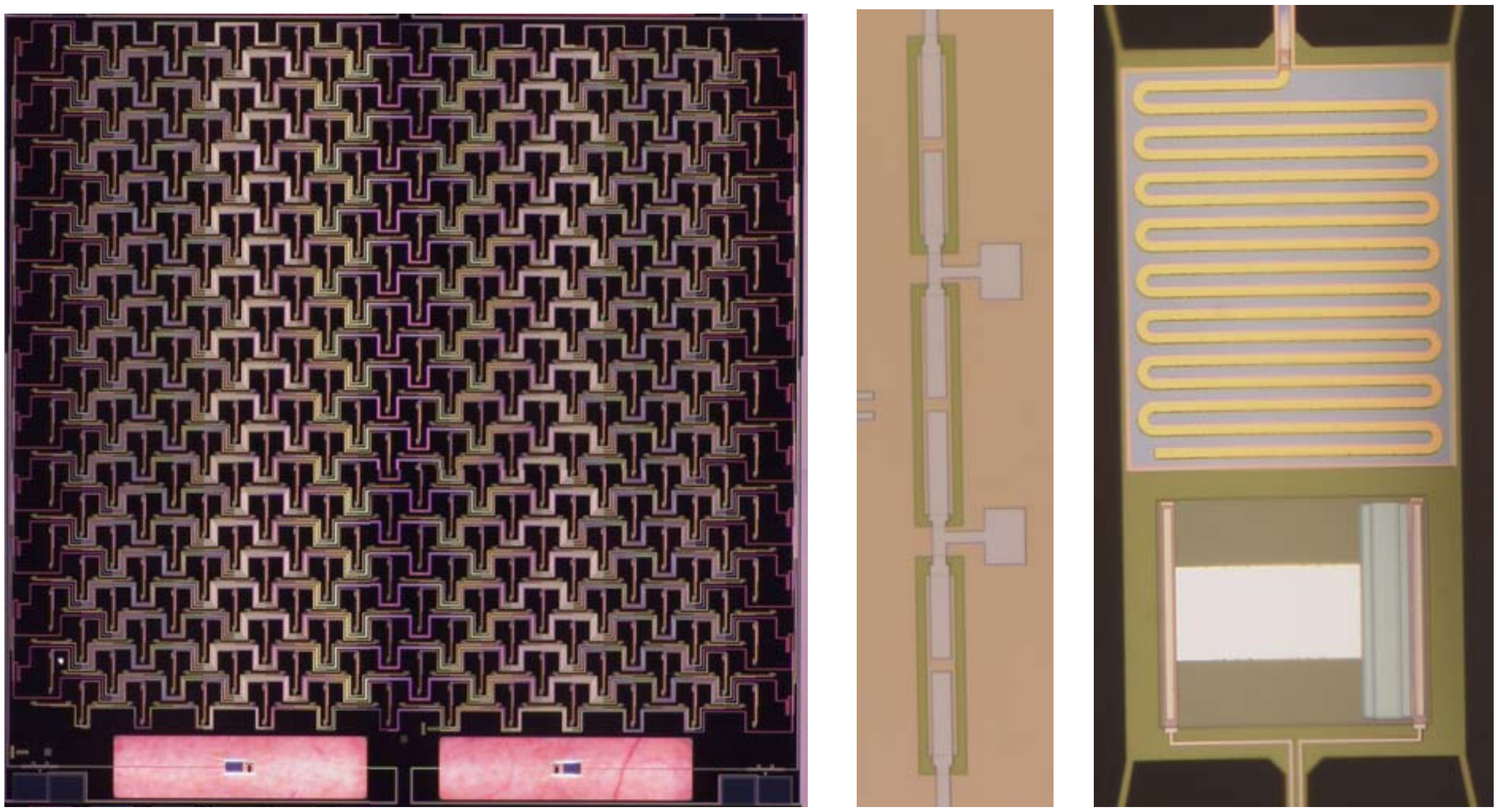

Figure 7.7. (left) Planar in-phase array antenna-coupled pixel at 145 GHz, an improved version of the design shown in Fig. 7.4b. The design uses lumped-element band-defining filter (middle). Radiation is dissipated in a lossy meander (right) thermally isolated on silicon nitride beams and read out with a TES bolometer. This layout illustrates the JPL planar antenna concept.

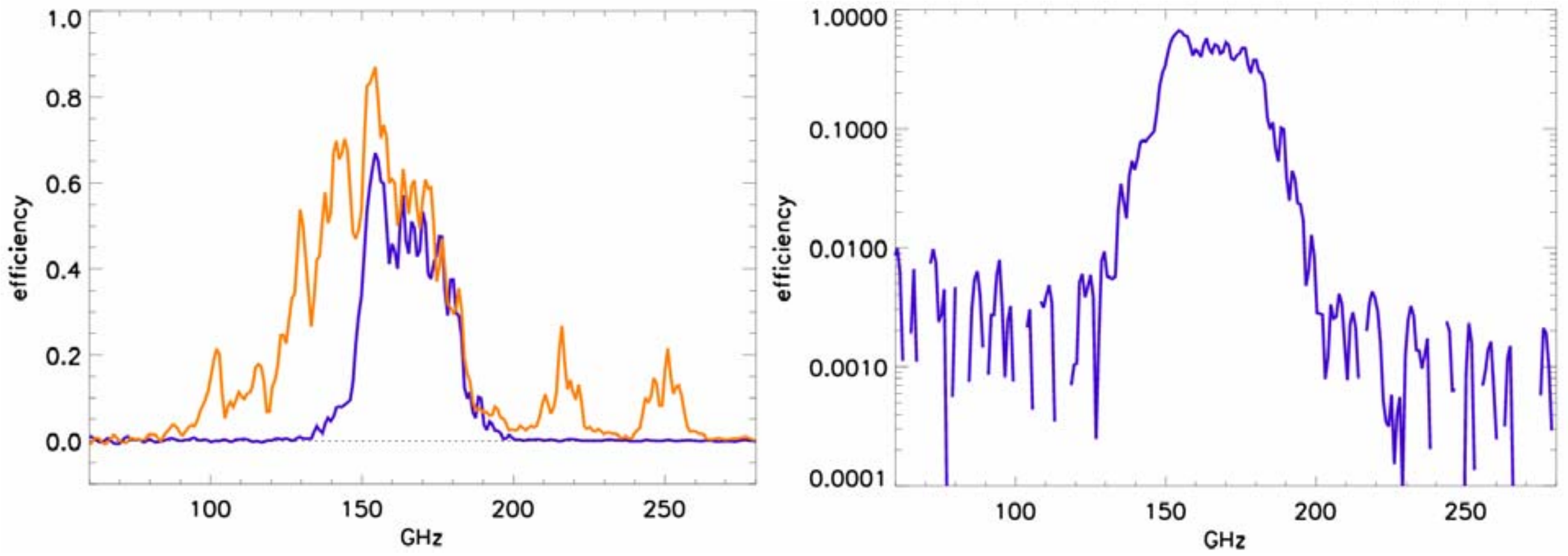

Fig. 7.8. Measured spectral response (left) of an antenna manufactured with (blue) and without (orange) the transmission-line filter shown in Fig. 7.7. The response is plotted in units of optical efficiency, determined by measuring the response to a cryogenic blackbody source. This device was tested without a backshort, which is expected to improve the efficiency by ~15 %. No leaks (right) are evident in the spectral response down to the measurement noise floor of ~1e-3.

*7.6.3 Lens-coupled planar antennas*

A small antenna that is comparable to a wavelength in size can be attached to a small contacting lens to give a suitable beam for coupling to a telescope. This approach has been well studied in the engineering and sub-mm mixer community. Much of the area under the lens is available for components such as filters, switches, and readout components. Although, the baseline design of EPIC-mid uses a large array of single-color pixels spread among multiple frequency bands, a long-term advantage of lens-coupled planar antennas is that they can be built to sense multiple frequency bands and two polarizations in a single pixel. The focal-plane area could be reduced by a factor up to 5, which greatly reduces the number of detector wafers, with reduced parastic loading from the supports and infrared radiation, easing the requirements on the sub-Kelvin cooler.

The Berkeley/LBNL group has been developing detectors using the lens-coupled planar antenna. Figure 7.9 shows an array and closeup of a single pixel. The current status is that single pixels including lens, antenna, band-defining filters, and TES detectors have been tested and prototype arrays are currently under test.

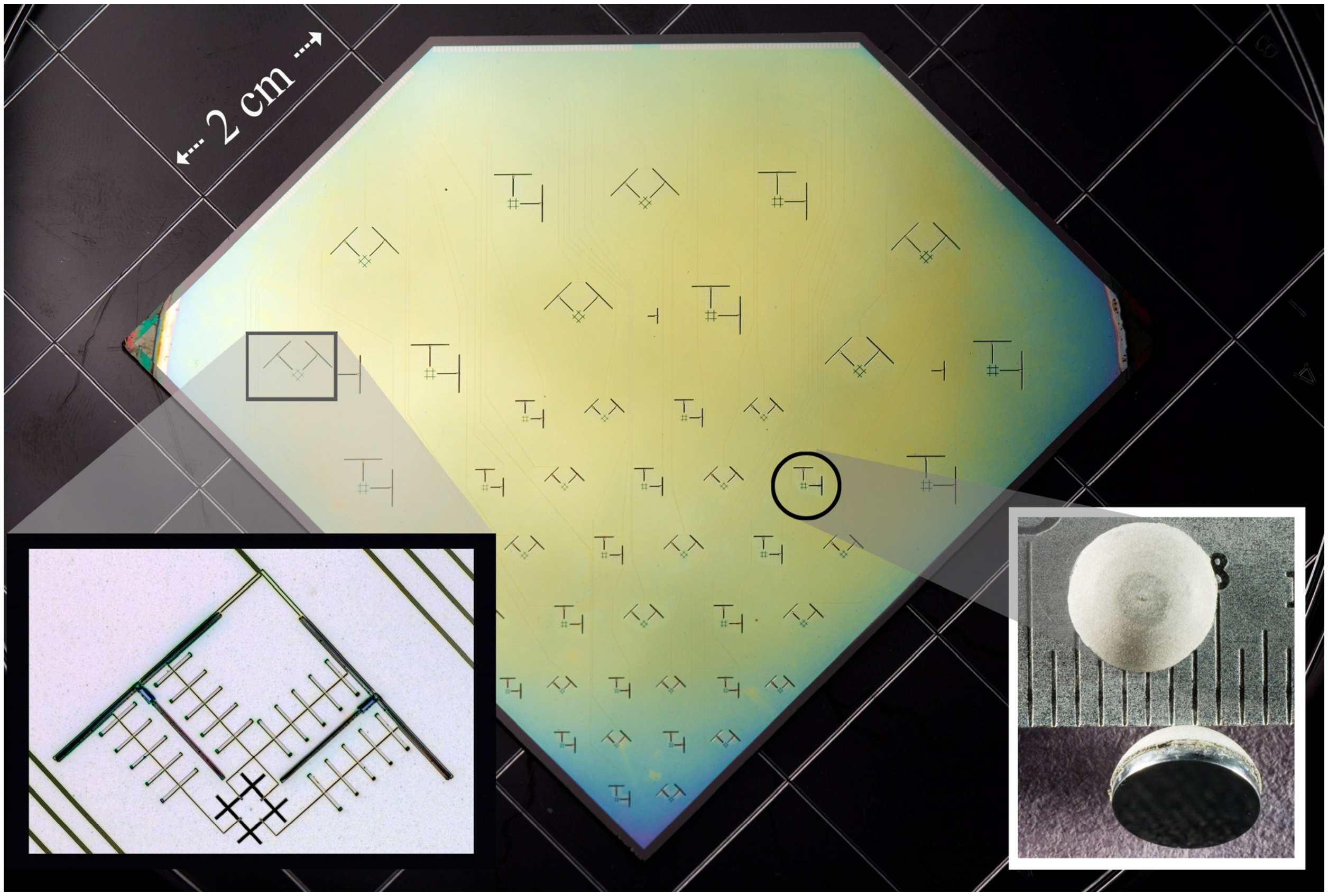

Figure 7.9. Array of TES bolometers optically coupled by a combination of planar antenna and contacting hemispherical lens. Left inset shows a close of of a single pixel where the antenna is at bottom and the RF filters connect the antenna to the “T” shaped TES bolometer. Right inset shows and hemispherical lens with antireflection coating. This array has 90 bolometers distributed between 90, 150 and 220 GHz. The bandpass filters use a distributed design with ¼ wavelength stubs. The anti-reflection coating is made from stycast and have been demonstrated to work optically and to withstand thermal cycling. Optical testing of single pixels from this wafer has been done.

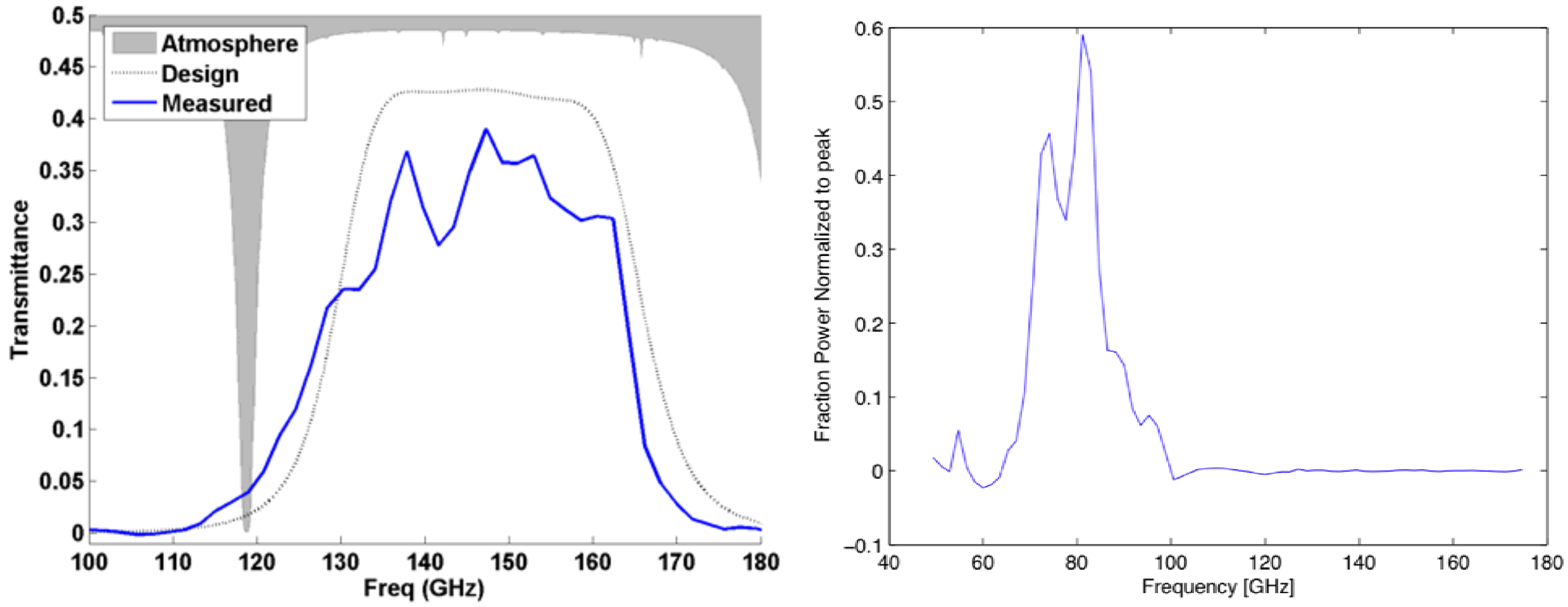

Figure 7.10. (Left) measured spectral response of a planar-antenna coupled pixel similar to that shown in Fig. 7.9. The efficiency is end-to-end for the entire receiver and would be improved by a factor of 1.3 by adding antireflection coatings to the optics. (Right) spectrum of multichroic antenna with a 90 GHz planar filter. End-to-end efficiency for the receiver is plotted in the y-axis. From this data the peak efficiency of the pixel is > 80%.

## 7.7 Low-Frequency Stability

The low-frequency stability requirement for EPIC-IM is set by the 0.5 RPM spin rate, which translates to a 1/f knee < 8 mHz, similar to the 16 mHz 1/f knee requirement for Planck. Current TES/SQUID systems are close to this level of performance (see Fig. 7.3), and single TES devices show 1/f noise down to 40 mHz (see Fig. 7.11). Techniques similar to those for stabilized NTD bolometers can be applied to TES bolometer systems. The bias can be modulated to remove any 1/f noise contribution in the readout. The signal carrier can be removed to reduce the requirement on system gain stability. Dark channels can be used to monitor the readout electronics, EMI/EMC pickup, and residual temperature drifts.

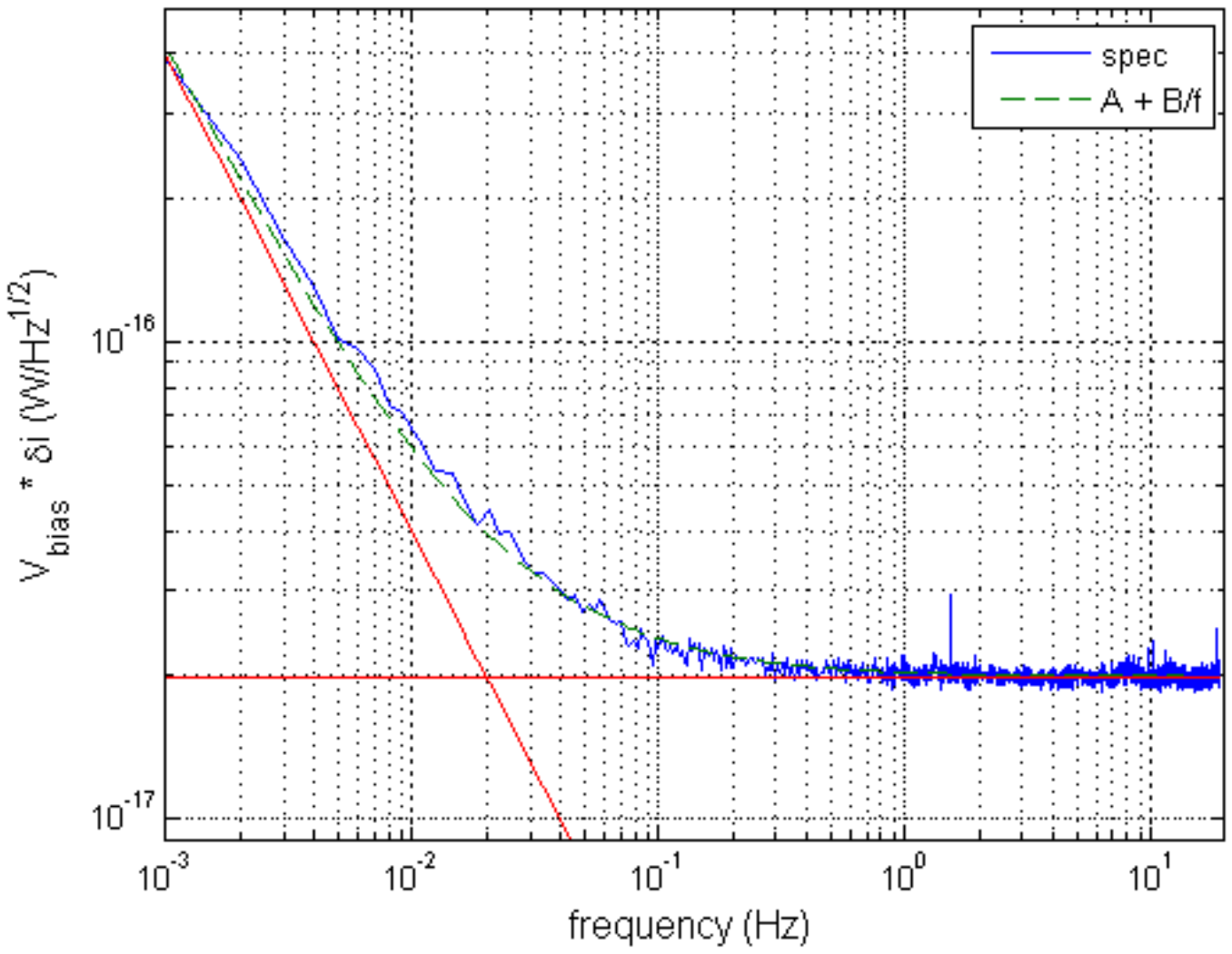

Fig. 7.11. Measured noise stability for a TES bolometer with NEP = 2 x $10^{-17}$ W/√Hz developed for a ground-based CMB polarimeter. The noise spectrum, obtained with a single SQUID amplifier, shows a 1/f knee at ~40 mHz. Based on tests of other devices, it appears the remaining 1/f noise is dominated by the test environment and not the device.

EPIC-IM entirely uses pair differencing, providing a level of immunity to common-mode 1/f noise sources. Differencing detector pairs reduces common-mode 1/f noise in the readout chain. Magnetic field susceptibility will also be largely common-mode. Temperature fluctuations in the optics and focal plane temperature are also common-mode, and can be removed by pair differencing. The required level of control for EPIC is already within that demonstrated by Planck using a combination of active and passive thermal stabilization.

# 8. Cooling

## 8.1 Passive Cooling

Passive radiative cooling makes use of deep space as a low temperature reservoir to sink heat from a spacecraft. When solar system dust and star lights are taken into account, the effective temperature of the sky is approximately 7 K [1]. This provides a large temperature difference from ambient (~300 K) for heat rejection and cooling. Radiative cooling has been successfully applied to solve cooling needs of many previous missions, the latest of which is the Spitzer Space Telescope. The V-Groove radiator design of EPIC follows the footsteps of the soon-to-be-launched Planck mission. The V-Groove radiator design consists of multiple stages of radiators, each having a V-shape cross section. The angle of the "V" is progressively narrower from the outer sun-facing shield toward the inner and colder shields. Thermal radiation between the two shields is guided into space by successive reflections on the shiny surfaces of the shields.

Two design options were explored - a four-shield design ('4 K telescope option) and a three shield design ('30 K telescope option). Fig. 8.1.1 shows EPIC with the four-shield design. The first shield refers to the shield facing the sun. The three-shield version is the same except that the fourth shield is removed. Each shield in Fig. 8.1.1 represents a doubled layered shield, which is needed to mitigate the risk of micro-meteorites puncturing the shield. The thermal effect of double-layering is included in the thermal design.

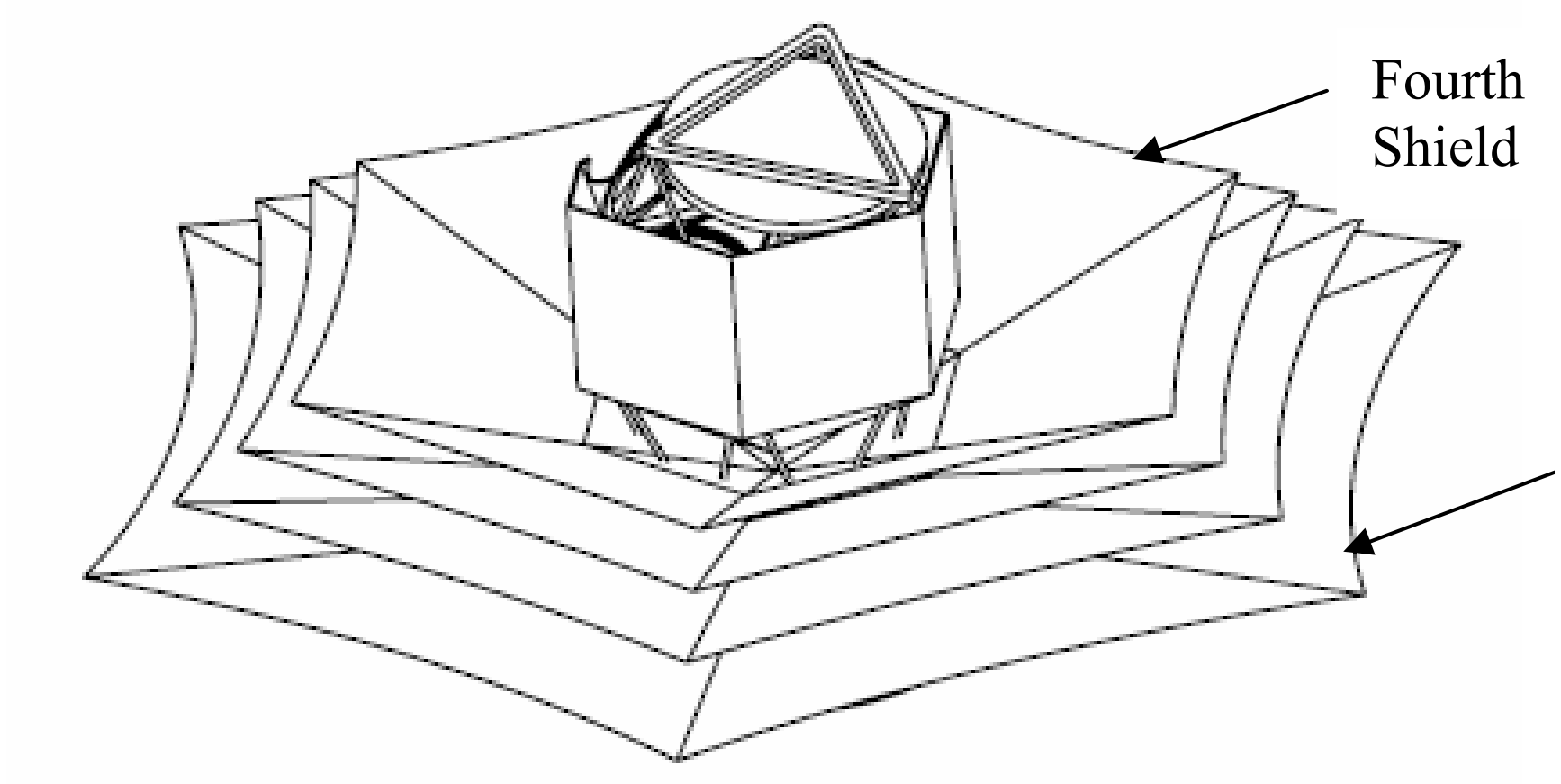

Fig. 8.1.1: EPIC with the four shields deployed.

### *8.1.1 Model Input*

The overall geometry is based on a ParaSolid drawing supplied by the mechanical design team. A solar incident flux of 1327 W/m$^2$ is applied at an angle of 45 degrees relative to the symmetry axis of the spacecraft. A power of 1.3 kW is applied to the spacecraft to simulate the total electronic dissipation there. A power of 8 mW is also applied to the proper cryogenic stage to simulate the total power of the Adiabatic Demagnetization Refrigerator (ADR). Table 8.1.1 summarizes the thermal properties of materials used for the various components. The diameter and the thickness of the gamma-alumina main struts were derived by the mechanical design team using structural analysis. Table 8.1.2 and Fig. 8.1.2 summarize the thermo-optical coating applied on various surfaces. The spacecraft surface and the sun-seeing side of the first shield is covered with silver Teflon for its low solar absorptivity and high infra-red emissivity, which helps reject solar heat input. Table 8.1.3 summarizes the optical properties of these coatings. The temperature dependent emissivity of aluminized Kapton was inherited from a previous study for the SAFIR proposal [2]. The temperature dependent thermal conductivities of the materials used are plotted in Fig. 8.1.3.

**Table 8.1.1.** Summary of Thermal Properties at 300 K.

| **Component** | **Material** | **Thermal conductivity** (W/m-K) | **Specific Heat** (J/kg-K) | **Density** (Kg/m$^3$) |
|---|---|---|---|---|
| Spacecraft | 5 mm thick Aluminum 6061-T6 | 155.8 * | 900 | 2702 |
| Center piece of all sunshields | 5 mm thick Aluminum 6061-T6 | 155.8 * | 900 | 2702 |
| Deployable piece of all sunshields | 0.127 mm thick Kapton | 0.12 | 1090 | 1420 |
| Main Struts | 5 cm diameter, 5 mm thick, 2.34 m long gamma-alumina | 1.8 * | 900 | 2000 |
| Mirrors and Focal Plane | 10 cm thick CFRP | 1 * | 837 | 2000 |
| Manganin Wires | 24.3 mm$^2$ cross section area, along strut. | 20 * | NA | NA |
| Gold shielding | 1.36 mm$^2$ cross section area, along strut. (99.9% purity, not annealed) | 200 * | NA | NA |
| HTS wires | 12 wires of 5 mm$^2$ each between the Optical Bench and the last shield. The length is 4 times the strut length along this section. | $3.7T^{0.62}$ | NA | NA |
| Brass leads connected to HTS Wires | 12 wires of 1.38 mm$^2$ and 12 wires of 2.78 mm$^2$ along strut from the last shield outward. | 90 * | NA | NA |
| Teflon Insulations | 1632 mm$^2$ cross section area, along the length of the strut. | 0.28 * | NA | NA |

*Temperature dependent properties are used in the model

**Table 8.1.2.** Optical coatings applied to components.

| | Center Solid Piece | | Deployed Kapton Piece | |
|---|---|---|---|---|
| | Top | Bot | Top/Out | Bot/In |
| Space Craft | S | S | | |
| 1st Shield | A | MLI/S | A | MLI/S |
| 2nd Shield | A | MLI/ | A | MLI/ |
| 3rd Shield | B | MLI/ | A | MLI/ |
| 4th Shield | B | MLI/ | A | MLI/ |
| Optical Box | A | MLI/ | | |
| Optical Box Side | | | A | MLI/ |

S = Silver Teflon; A = Aluminized Kepton; B = Black Paint; MLI = Multilayer Insulation, also used to substitute for a double layered shield of aluminized Kapton.

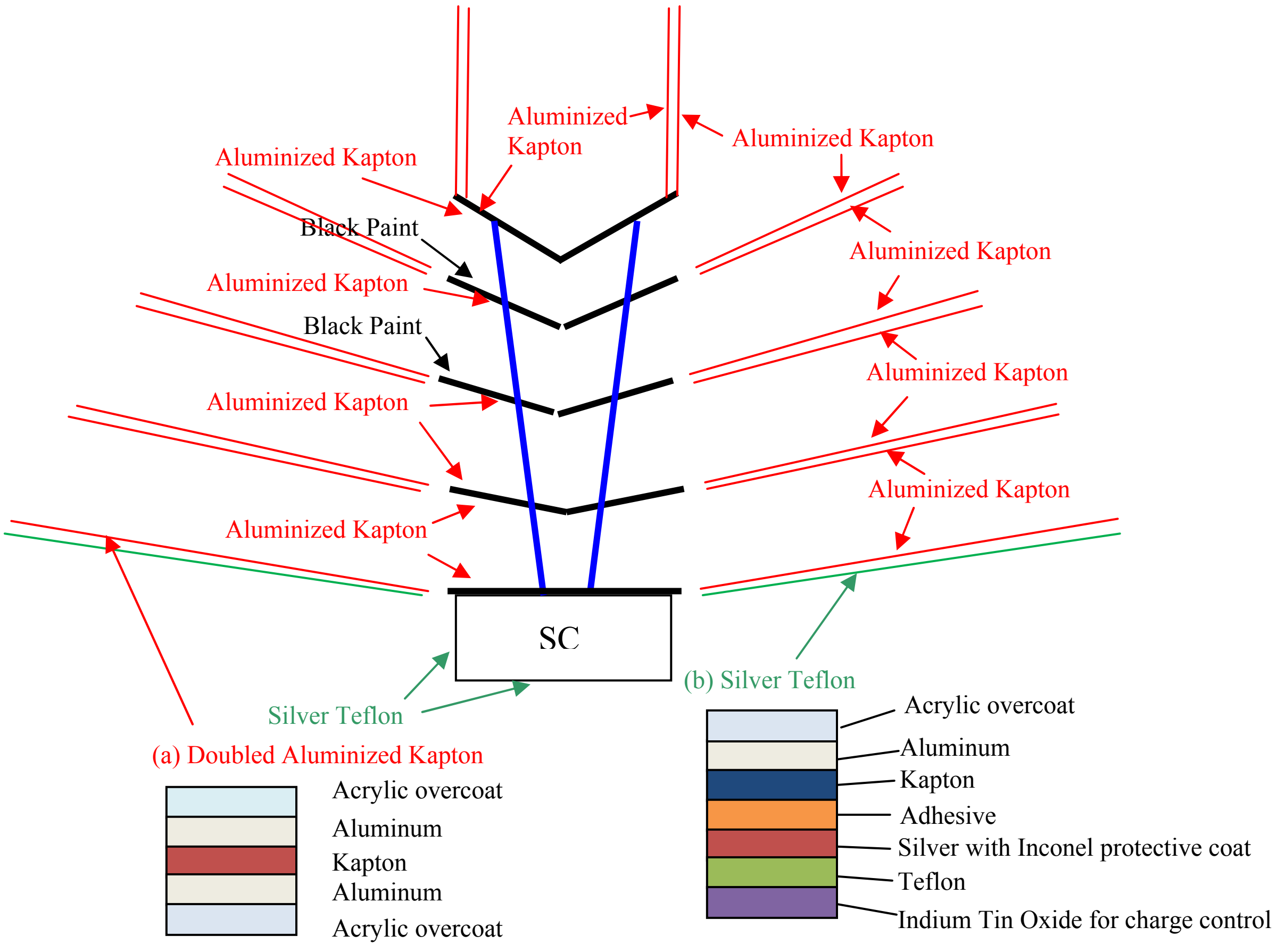

Fig. 8.1.2: Optical coatings applied to components of EPIC. a) Red color line represents a double aluminized Kapton film. b) Green line represents a silver Teflon film.

**Table 8.1.3.** Thermo-optical properties of coatings at 300 K from www.sheldahl.com.

| Coating | Solar Absorptivity α | Infrared Emissivity ε | Specularity | Thermal Conductivity (W/m-K) |
|---|---|---|---|---|
| Silver Teflon | 0.14 | 0.75 | 95% | NA |
| Aluminized Kapton | 0.14 | 0.056 * | 95% | NA |
| Black Paint | 0.94 | 0.9 | 100% | NA |
| MLI | NA | Effective ε = 0.05 | NA | $1.2 \times 10^{-6}$ |

*Temperature dependent properties are used in the model.

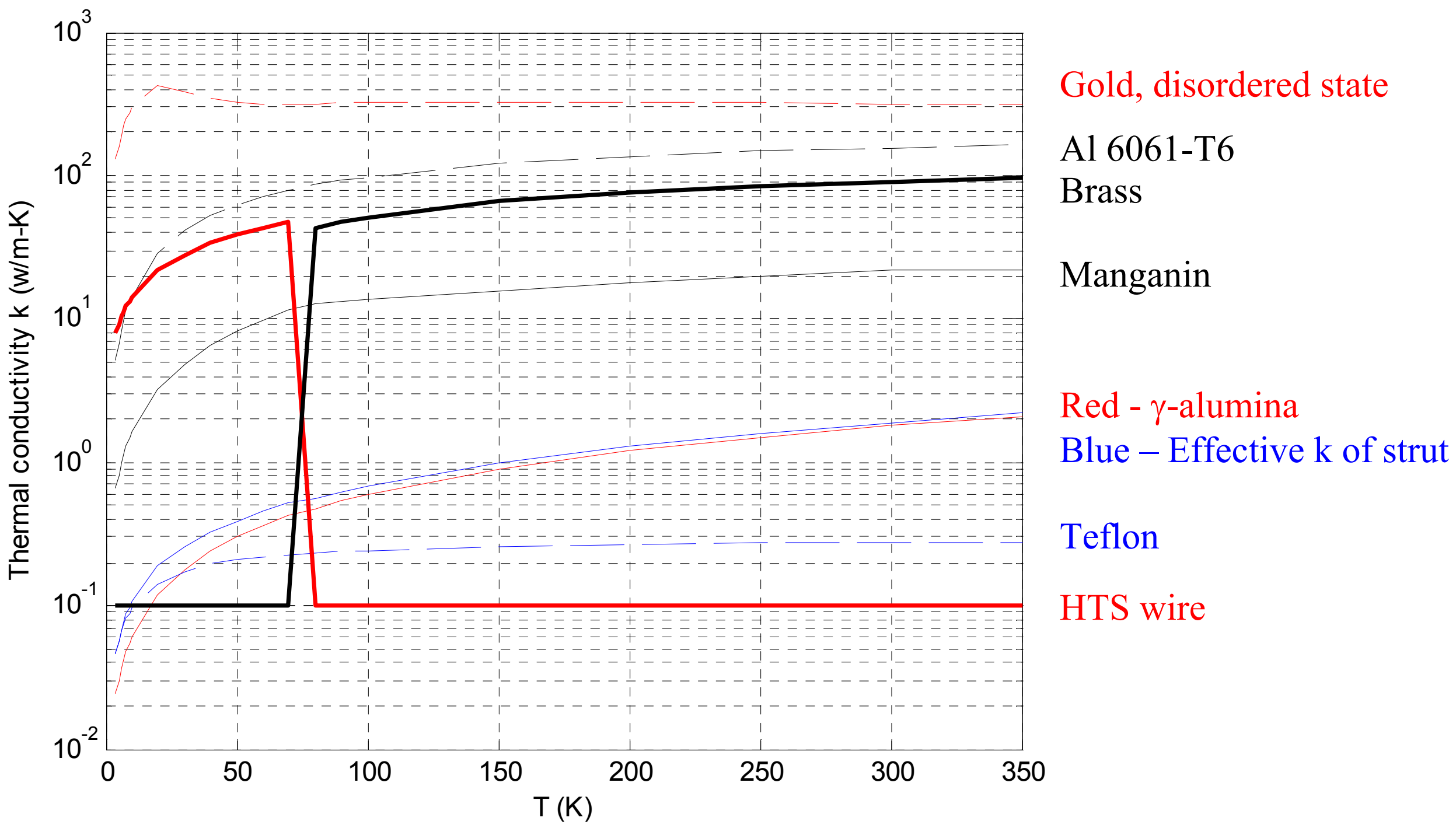

Fig. 8.1.3. Thermal conductivity of material used.

*8.1.2 Technique*

8.1.2a Tools: Thermal Desktop 5.1 Patch 4 is used for the model. Thermal Desktop is a graphical user interface that is built on top of a SINDA engine. It employs finite difference method to solve the heat equation in a sheet (2D geometry). 3D geometries are built by putting sheets together, and allowing thermal conduction to occur at the boundary of the sheets. Although true 3D heat flow can be treated by finite element method in Thermal Desktop, it is not used in the model because the problem being treated is primarily a radiative heat transfer problem, not a 3D heat flow problem. For radiative heat transfer, we use a module of Thermal Desktop called RADCAD which employs Monte Carlo ray trace technique to calculate the coupling between surfaces. The model ran on a PC with a 3.6 GHz Pentium-4 CPU and 4 GB of RAM.

8.1.2b Model Statistics: The model has 2900 nodes. For each node an average of 50,000 rays were shot for the Monte Carlo simulation. It took approximately 8 minutes to run.

8.1.2c Heat Conduction by Wires: It would be too laborious to input the geometry of each wire into the model. The approach taken is to increase the thermal conductivity of the struts by including the effect of heat conduction by all the wires. The effective conductivity is

$$k_{eff} = \Sigma\, k_i(A_i/L_i)(L_\gamma/A_\gamma),$$

where $k_i$, $A_i$ and $L_i$ are the thermal conductivity, the cross-section area and length of the wire and struts denoted by the subscript i, and $L_\gamma$ and $A_\gamma$ are the length and total cross-section areas of the 12 gamma-alumina struts. The results are shown by the blue lines in Fig. 8.1.4, which also shows that roughly 10% of conductive heat is carried by wires.

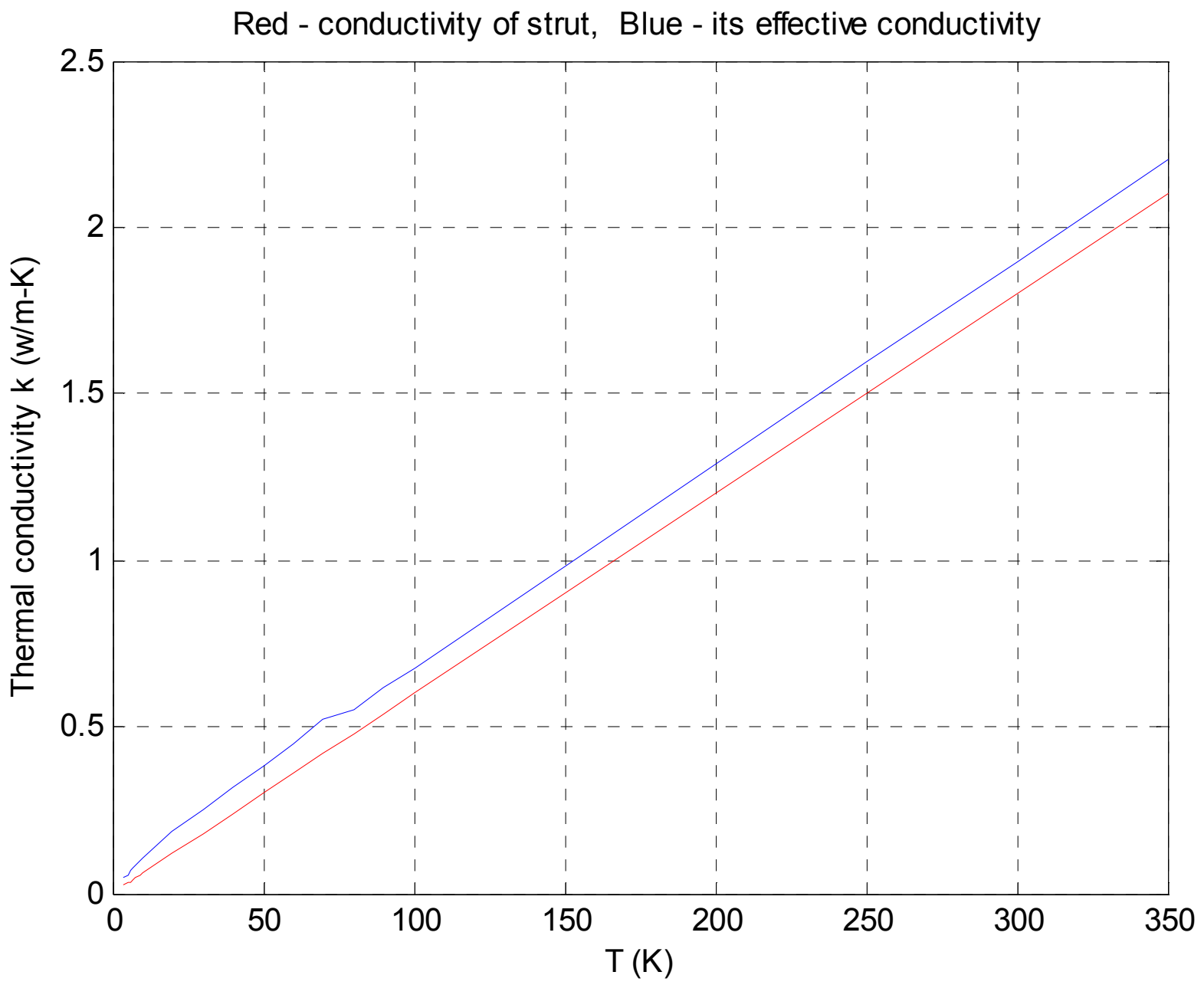

Fig. 8.1.4. The thermal conductivity of gamma-alumina (red) and $k_{eff}$ (blue).

8.1.2d Double-Layer Shield: Double-layer deployable shields are used for mitigation against the risk of micro-meteorites puncturing the shields. It also allows a temperature difference to develop from one layer to the other. Thermal Desktop allows the addition of a layer of MLI to simulate the effect of a double-layer shield. Mathematically, MLI is treated as a two-layer surface with an effective emissivity for the interior facing surfaces. We used an effective emissivity of 5 % for modeling double-layer shields. The effect of radiation escaping between the layers is not treated, resulting in a conservative calculation.

8.1.2e Lunar radiation: We do not include radiation from the moon in this model, because it appears to be a negligible effect. As shown in Fig. 8.1.5, the moon at its most extreme angle 8.1.2e Lunar radiation: We do not include radiation from the moon in this model, because it appears to be a negligible effect. As shown in Fig. 8.1.5, the moon at its most extreme angle only illuminates the top backing structure of the telescope. The sunshield prevents the moon from falling directly on any optical surface or inside the aperture stop. The radiation from the moon provides a heat input of 1.3 $mW/m^2$ on a black surface. If this small input is a problem, the lightweight 18 K optical shield can easily be extended to cover the top of the telescope.

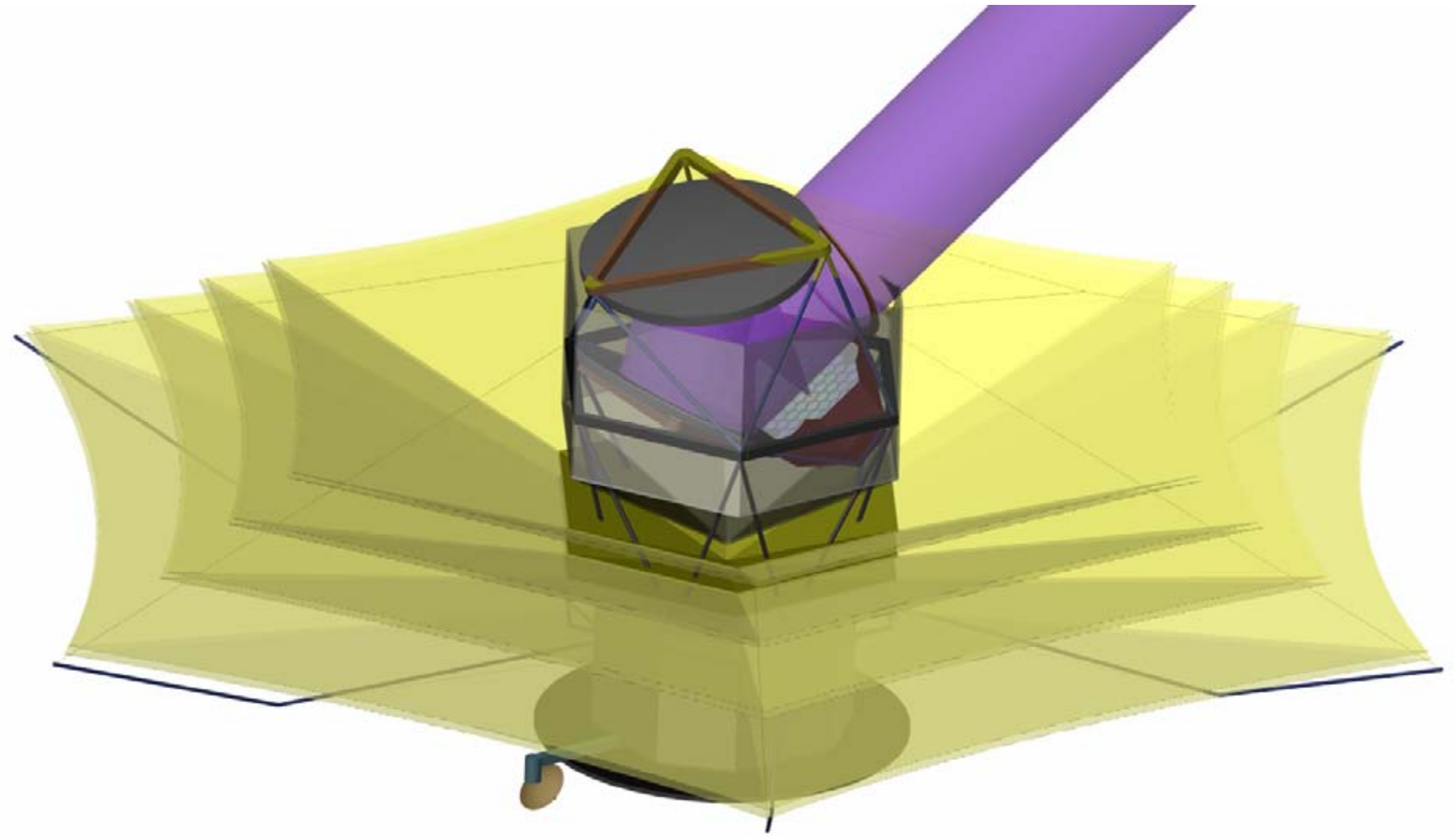

Fig. 8.1.5. Illumination from the worst-case moon angle, assumed to be when the moon is at apogee, at right angles from the sun-earth line, and in conjunction with the worst position in EPIC's L2 halo orbit. The sunshield prevents the moon from ever illuminating the entrance aperture or any optical surface. The moon however can illuminate the top of the telescope backing structure (shown as the mustard color). This thermal radiation is generally negligible, and could easily be shielded by extending the 18 K optical shield.

### *8.1.3 Model Results*

8.1.3a Four- Shield Option, Cryocooler Off: The color map of temperature is shown in the following figures. A dissipation of 8 mW is applied to the optical bench to simulate the power of the ADR. There is no cooling power from the cryocooler in this model. The heat transfer between the shields is summarized in Table IV. A detailed heat transfer map between components is also shown in Fig. 8.1.14.

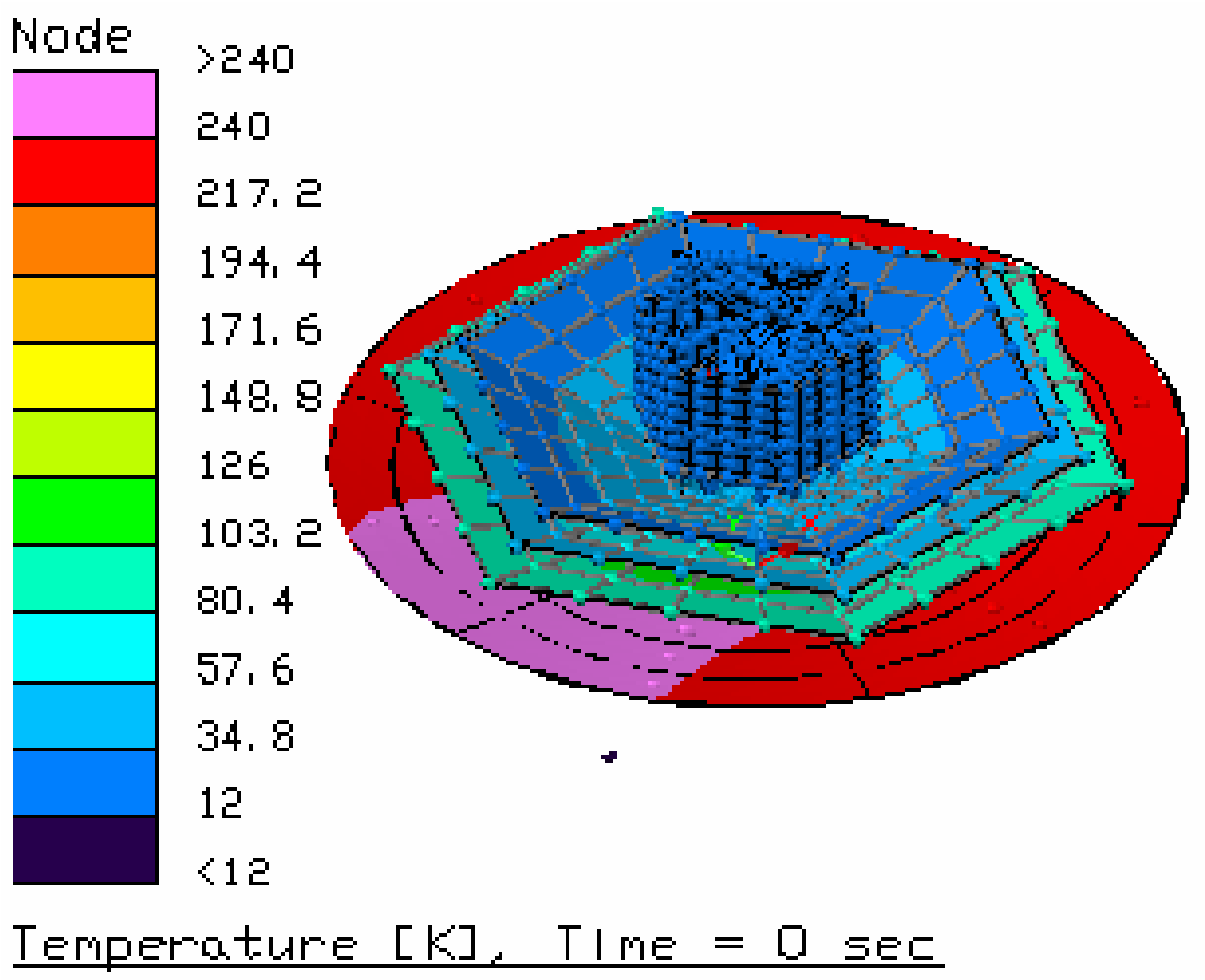

Fig. 8.1.6. 3-D color map of temperature of EPIC.

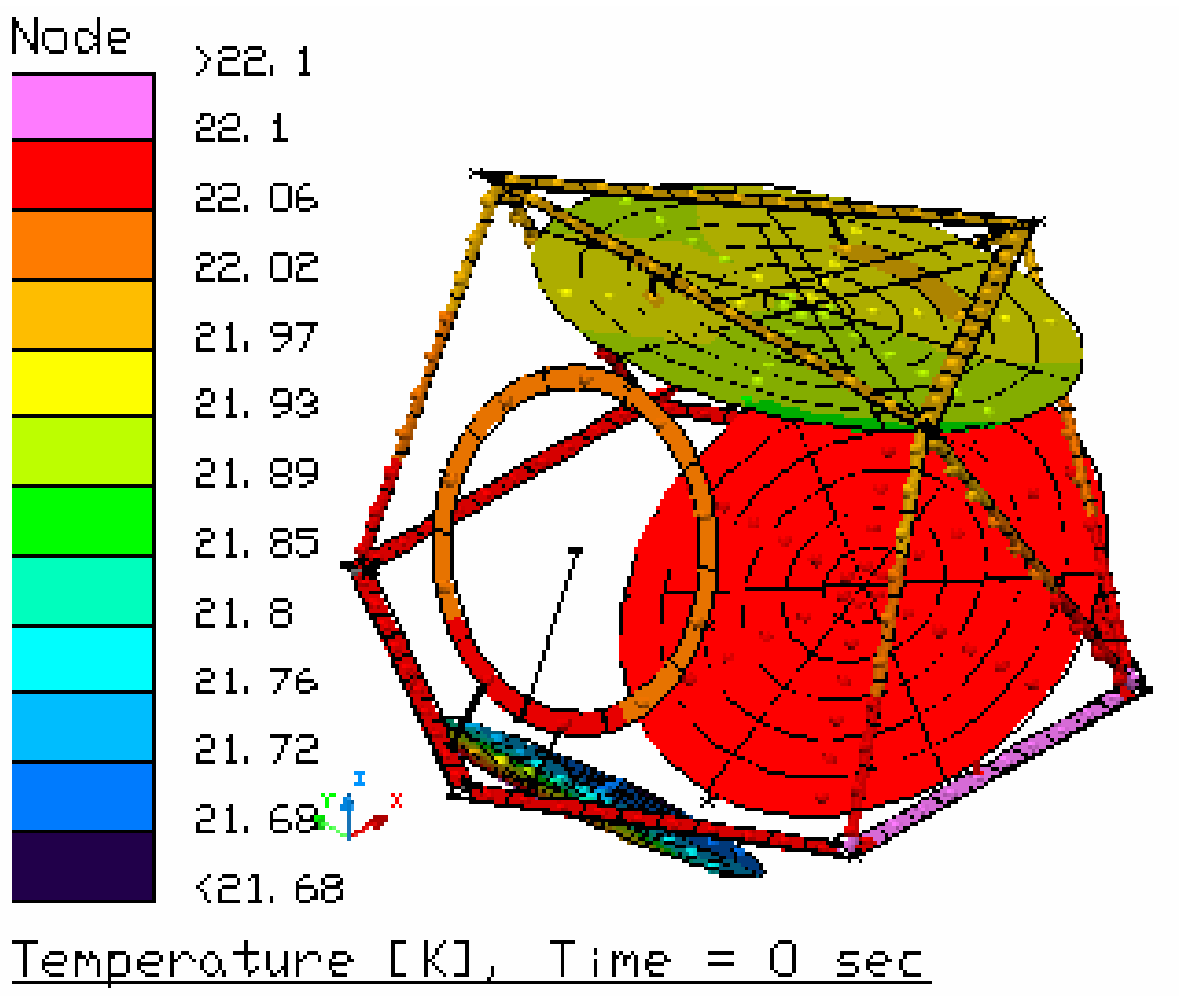

Fig. 8.1.7. Temperature of the telescope.

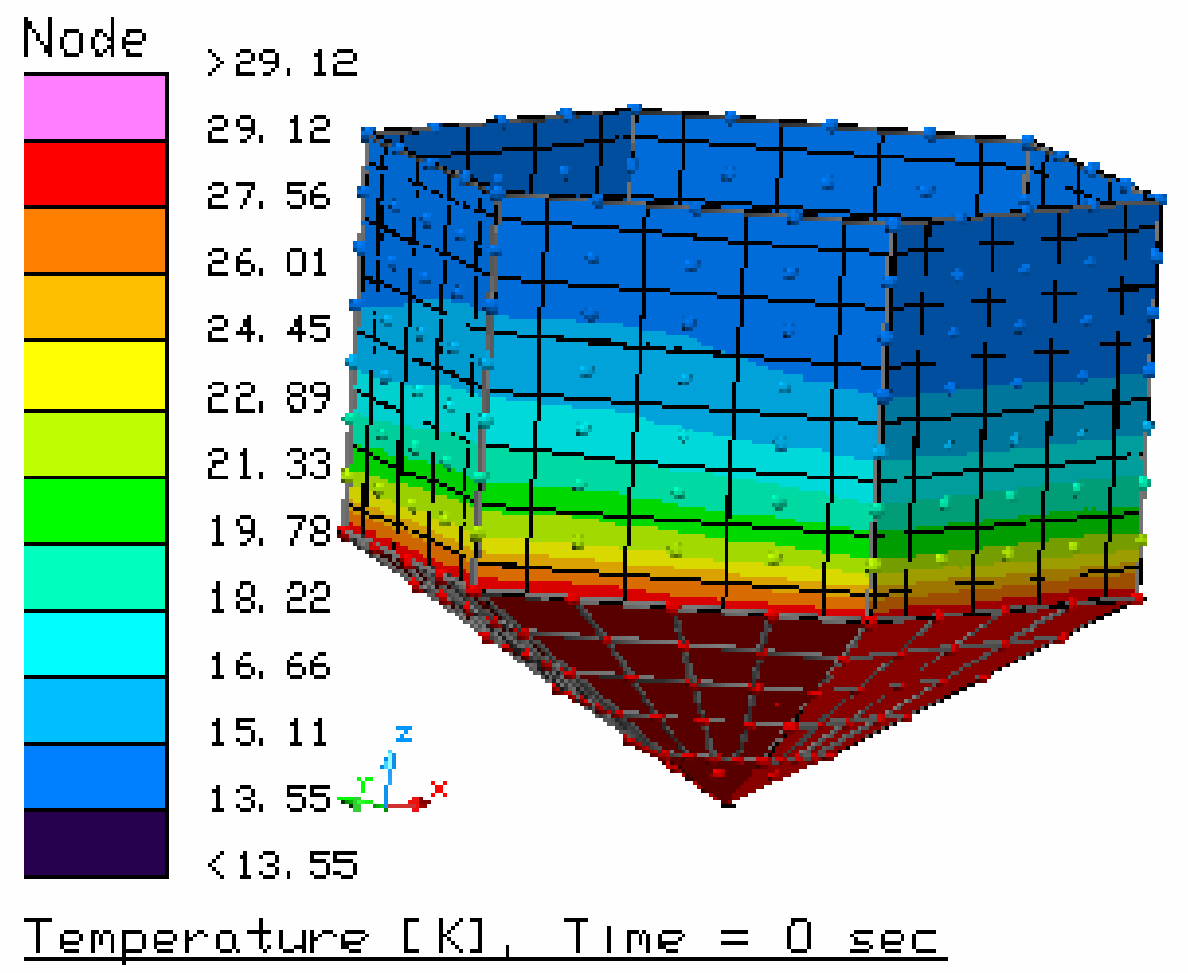

Fig. 8.1.8. Optical Box temperature.

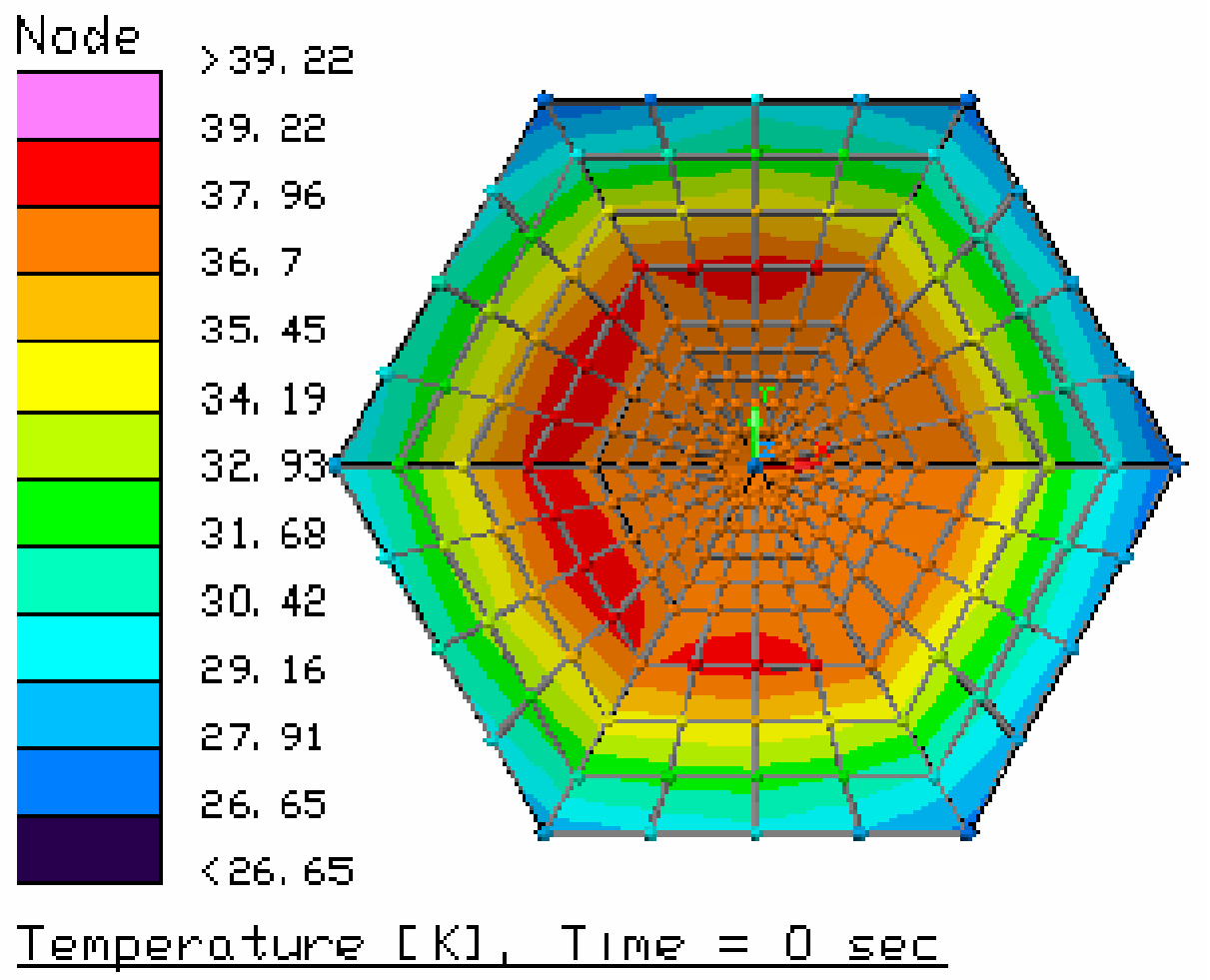

Fig. 8.1.9. 4th Shield temperature

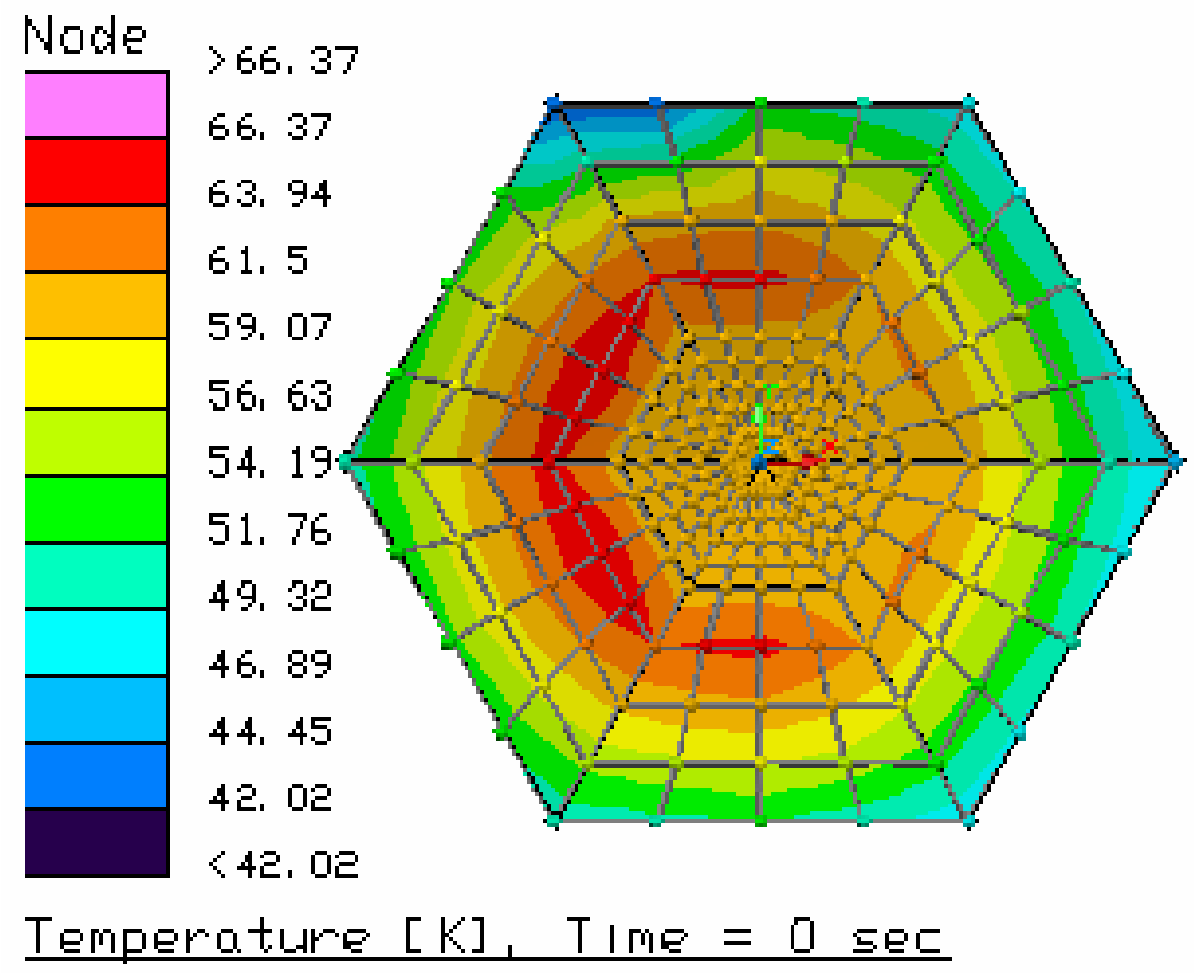

Fig. 8.1.10. 3rd Shield temperature

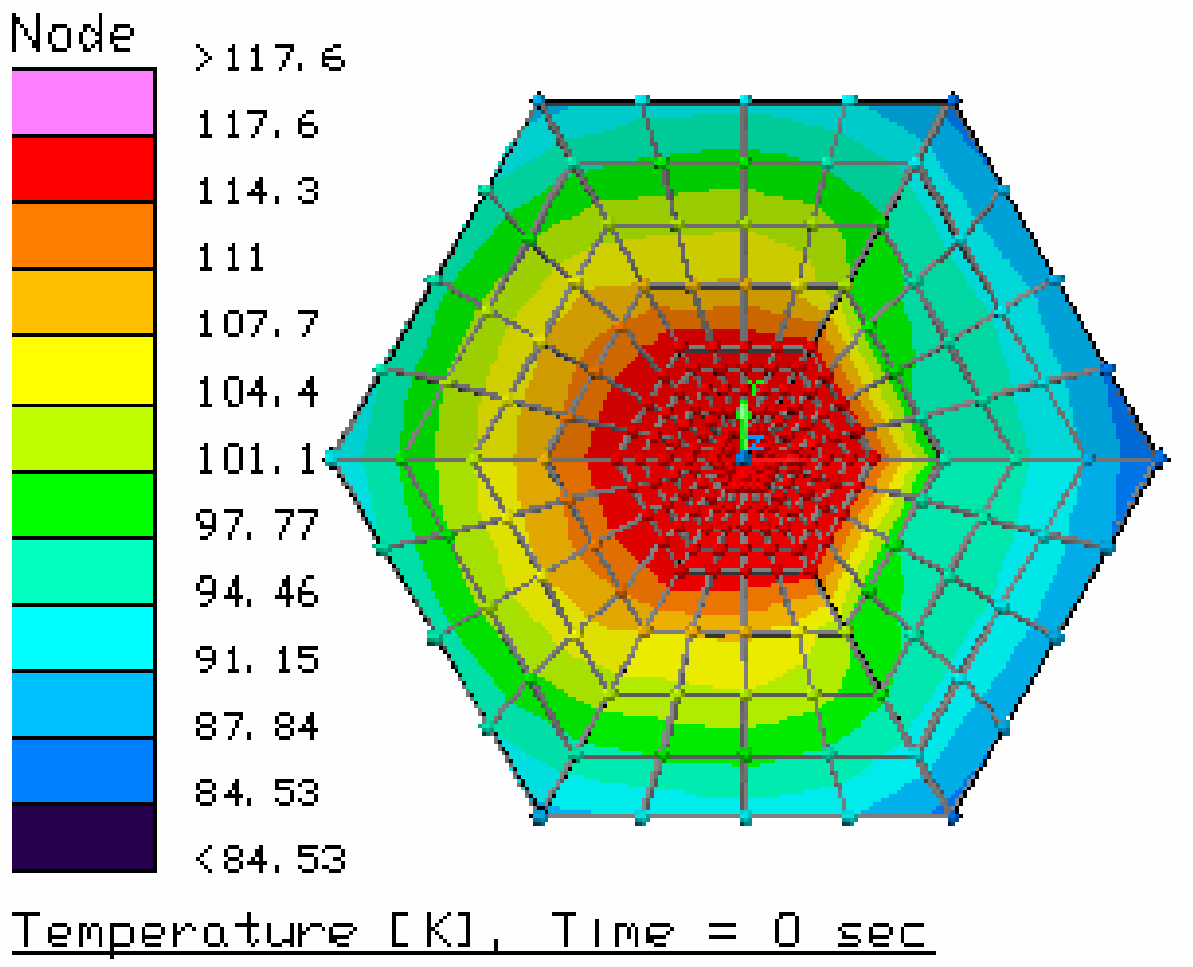

Fig. 8.1.11. 2nd Shield temperature

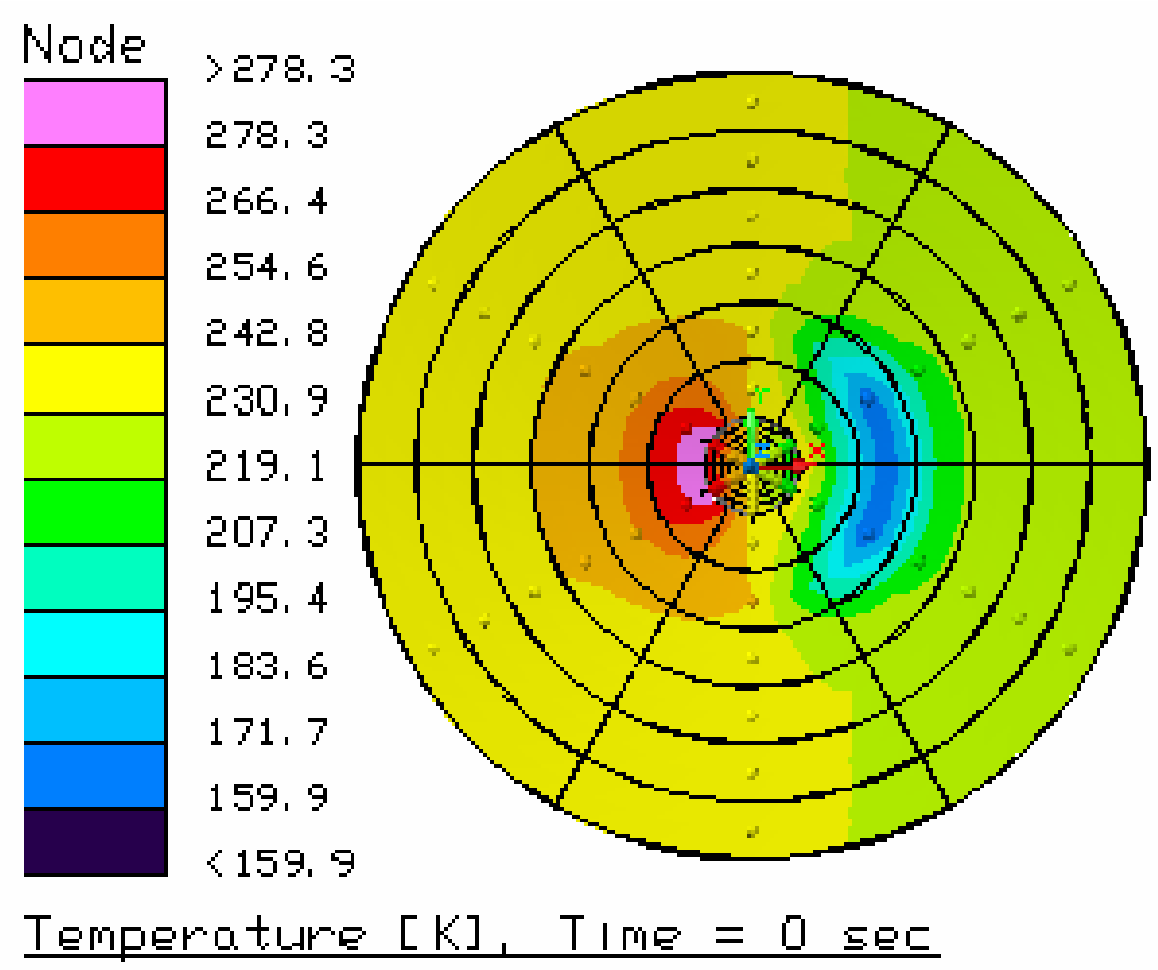

Fig. 8.1.12. 1st Shield temperature

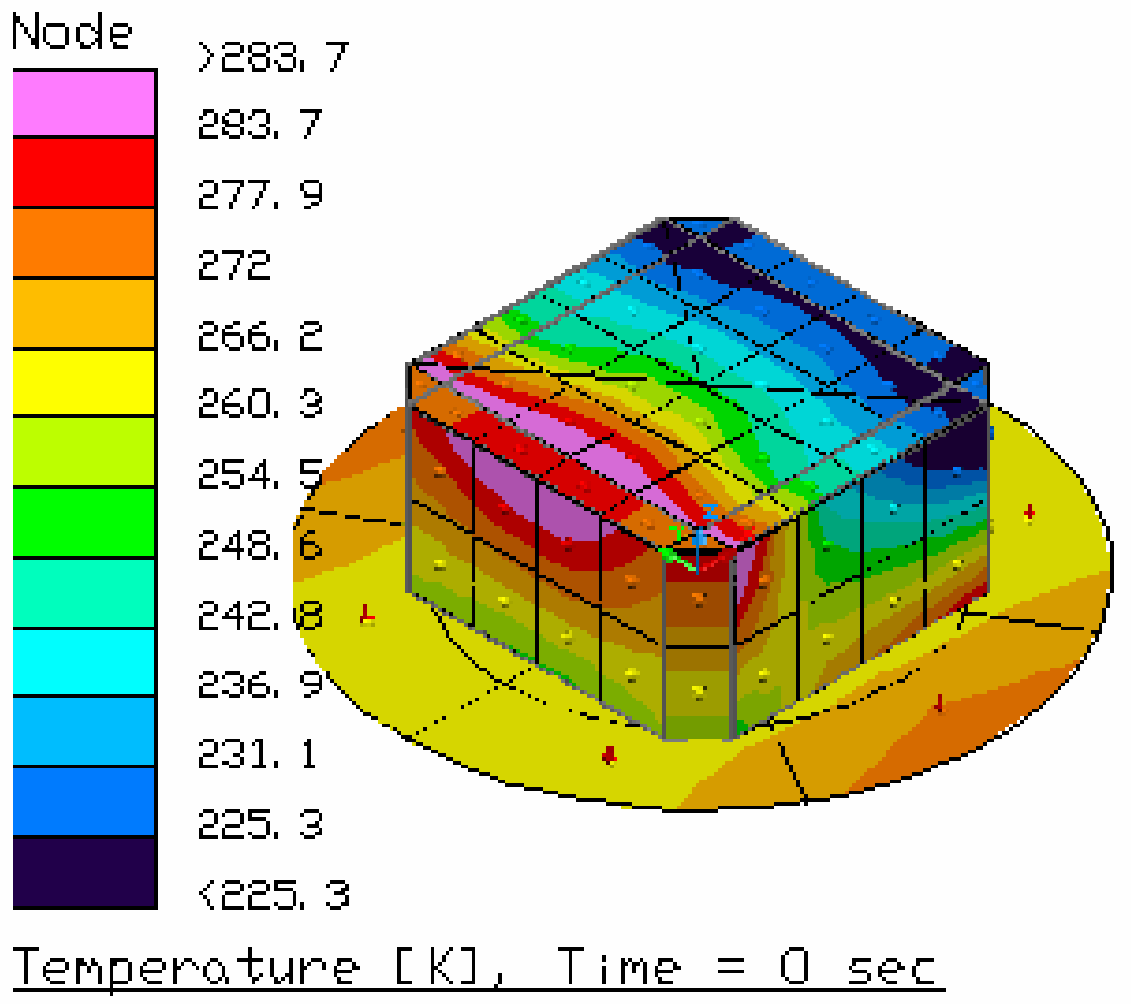

Fig. 8.1.13. Spacecraft temperature

**Table 8.1.4.** Heat transfer between different thermal stages.

| | T(K) | Radiative Heat Transfer to Next Stage (W) | Conductive Heat Transfer to Next Stage (W) | Radiative Heat Transfer to Space (W) | Thermal Resistance to next stage (K/W) |
|---|---|---|---|---|---|
| 1st Shield | 231 | 69.5 | 3.91 | 16,300 | 29.3 |
| 2nd Shield | 116.5 | 5.89 | 1.07 | 75.44 | 52.4 |
| 3rd Shield | 60.39 | 0.685 | 0.264 | 6.00 | 85.8 |
| 4th Shield | 37.75 | 0.0299 | 0.0266 | 0.892 | 323 |
| Optical Box | 29.15 | 0.00282 | 0.01213 | 0.0294 | 580 |
| Telescope | 22.12 | NA | NA | 0.0232 | NA |

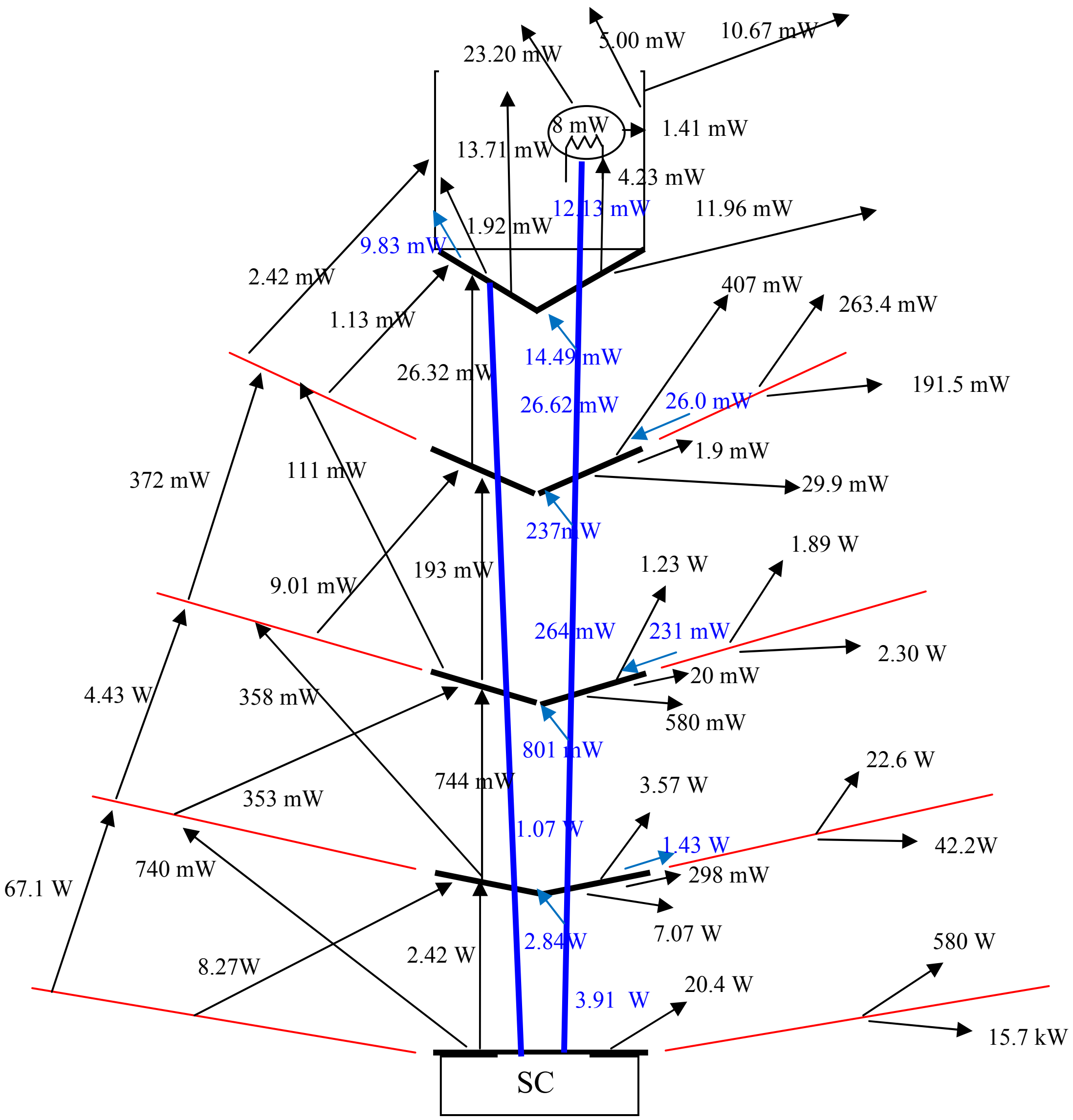

Fig. 8.1.14. Detailed heat transfer map between components without cooling power from cryocooler: Blue for conductive heat; Black for radiative heat. Two numbers are quoted for conducted heat for the wiring and the bipod.

8.1.3b Four- Shield Option, Cryocooler On: When the cryocooler is turned on, the Optical Box Bottom is thermally connected to a thermal stage of the cryocooler at 18 K temperature, and the cold stage of the cryocooler is connected to the Optical Bench. The temperature of the Optical Bench as a function of cooling power is given in Table 8.1.5.

**Table 8.1.5.** Cooling power at the Optical Bench versus temperature

| Cooling power (mW) | 0 | 10 | 15 | 18.5 | 20 | 20.25 | 20.5 | 21 |
|---|---|---|---|---|---|---|---|---|
| Optical Bench T (K) | 17.35 | 13.86 | 11.02 | 7.62 | 4.83 | 4.02 | 3.23 | 1.43 |
| Optics Box Bottom P (mW) | 78.2 | 74.0 | 71.4 | 69.1 | 68.0 | 67.8 | 67.5 | 67.0 |

The Optics Box Bottom is kept at 18 K.
The shaded column is the requirement on the cryocooler.

The following color maps are for the case in Table 8.1.5 with 20.25 mW cooling power at the Inner FP Box.

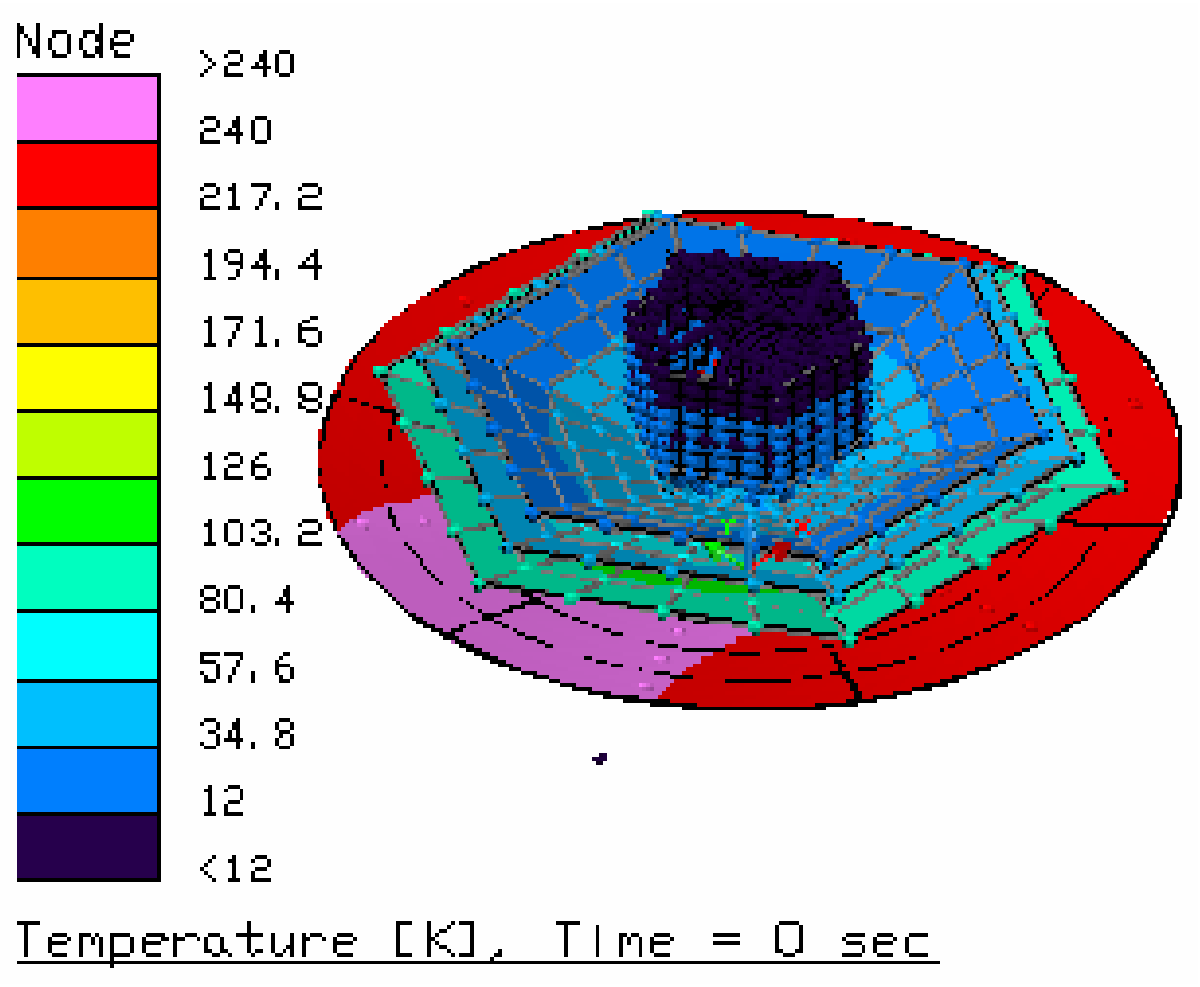

Fig. 8.1.15: 3D color map with cryocooler on.

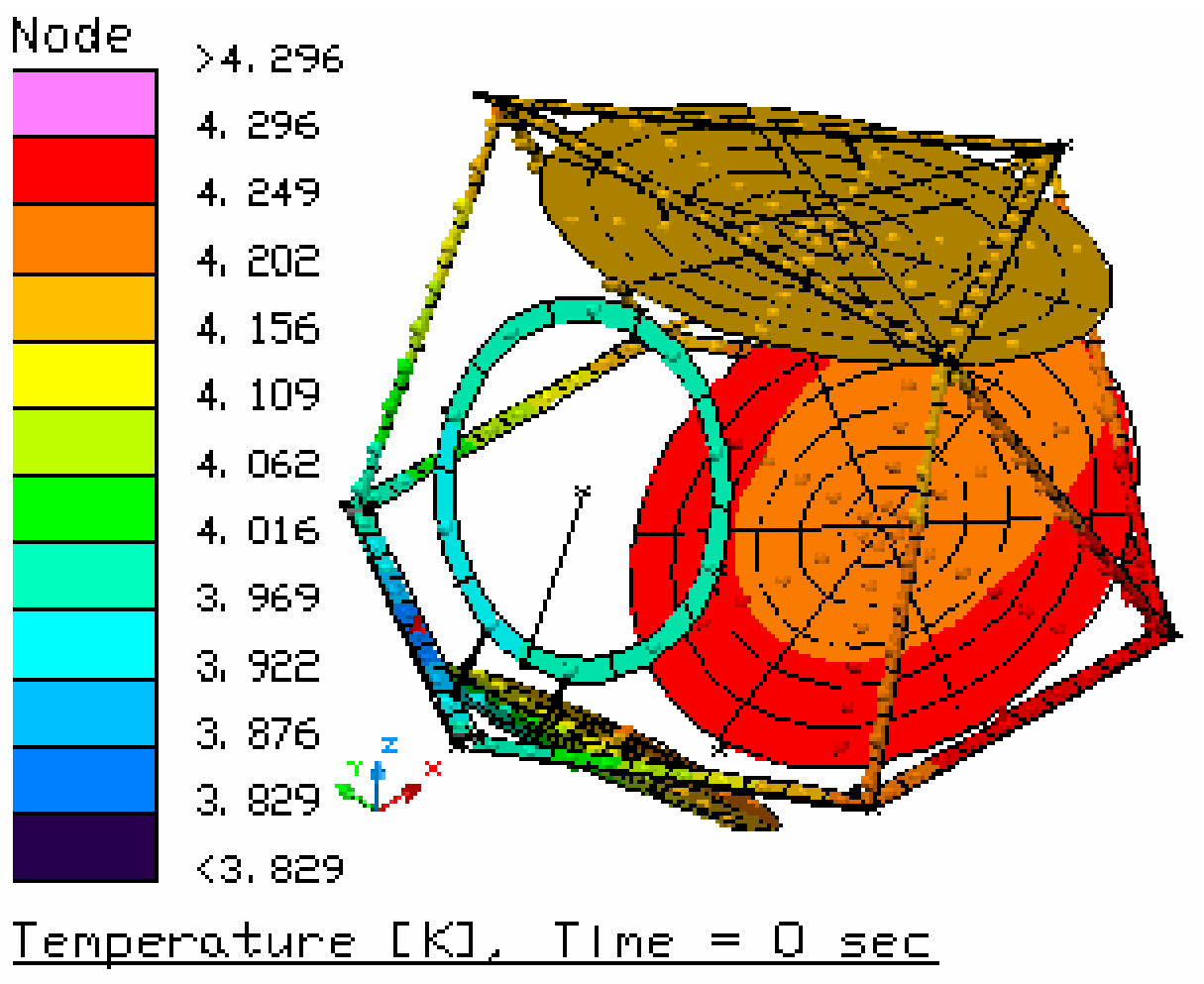

Fig. 8.1.16: Telescope with cryocooler on.

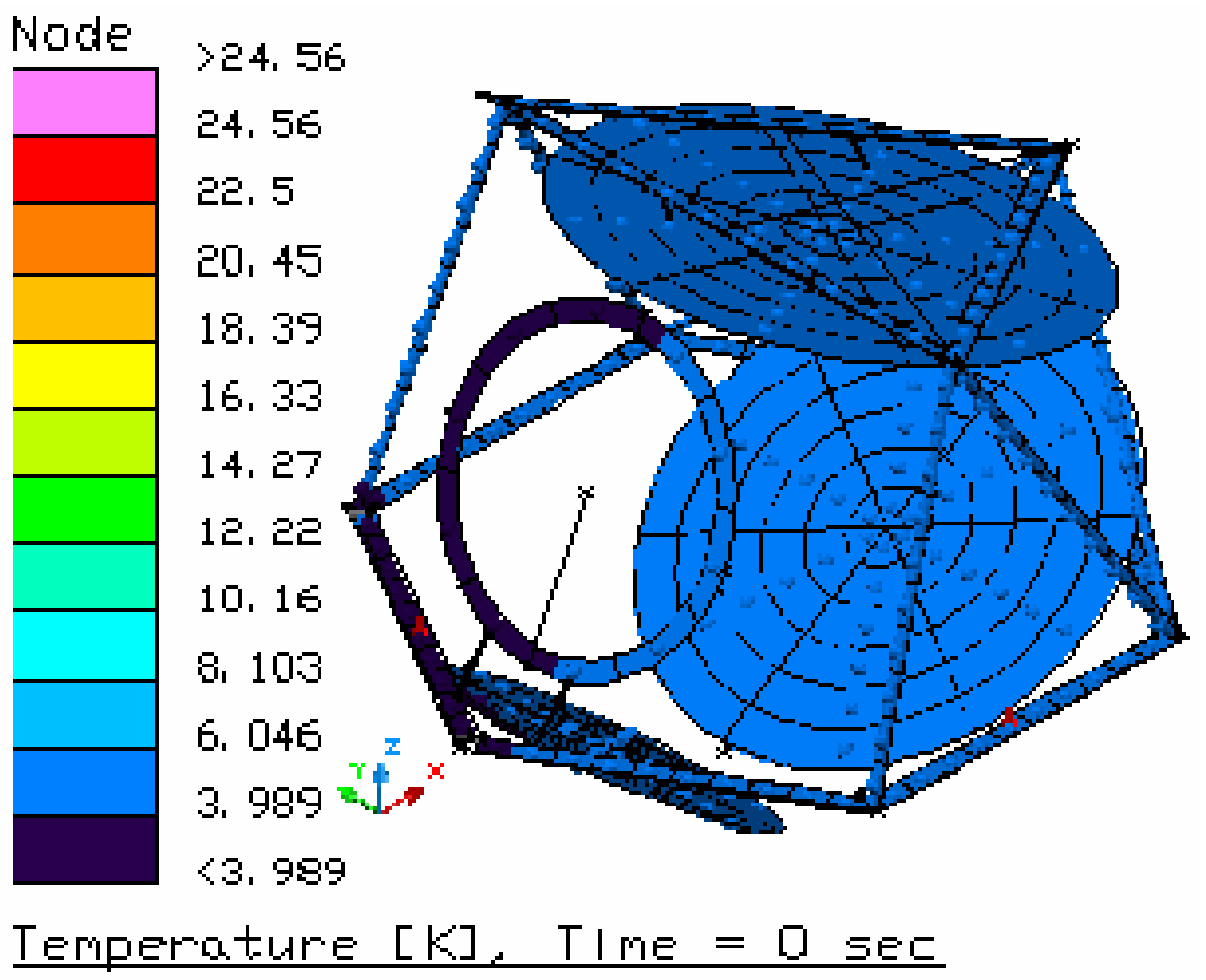

Fig. 8.1.17. Telescope with cryocooler on, plotted on the sample color scale as Fig. 8.1.25 .

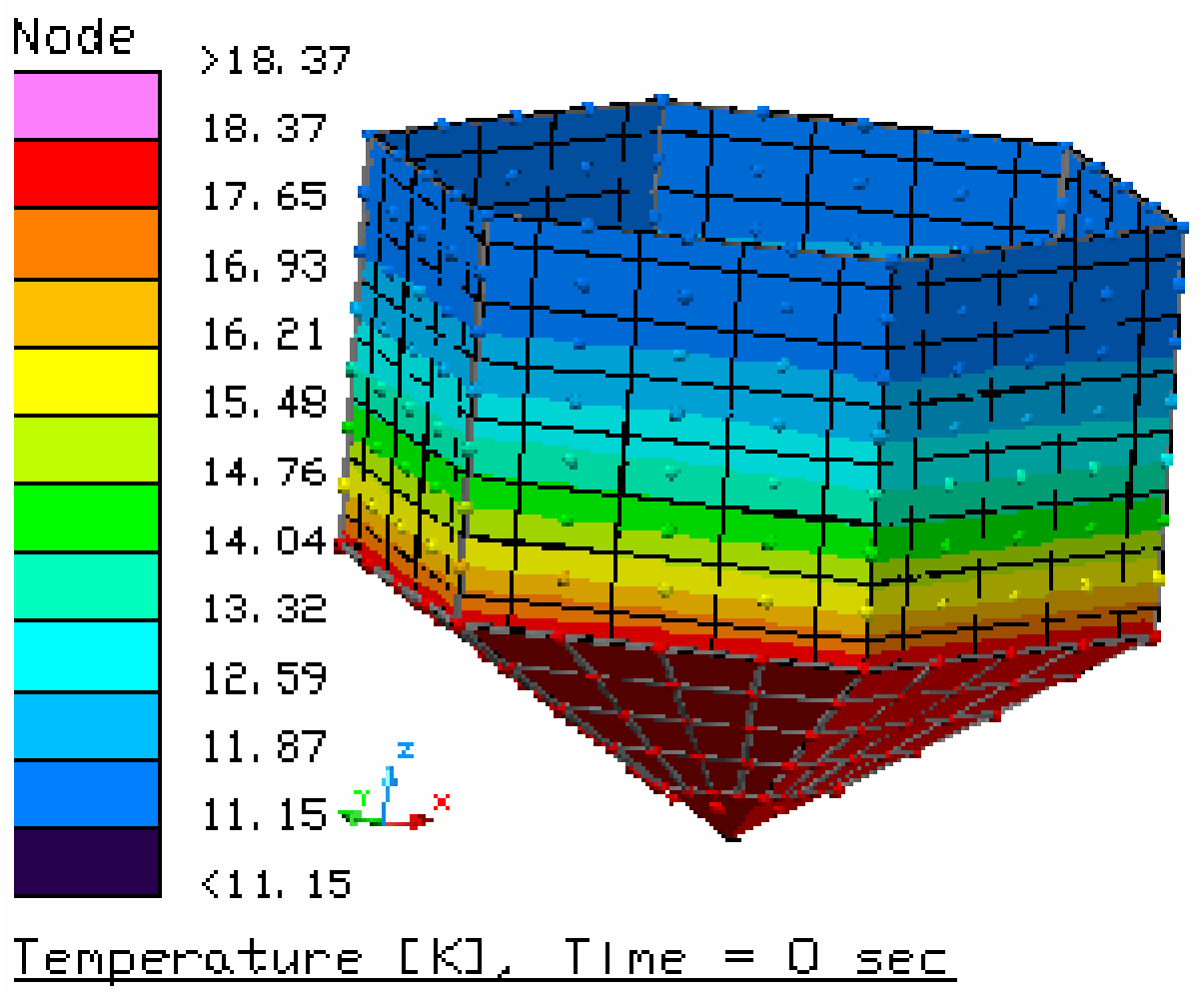

Fig. 8.1.18. Optical Box with cryocooler on.

Fig. 8.1.19 is a detailed heat transfer map between all the components with 20 mW cooling at the Optical Bench and 68 mW cooling at the Optical Box Bottom. Table 8.1.6 is derived from Fig. 8.1.19. It gives the heat balance at each of the components. The heat-in agrees with the heat-out to within 0.5%, which serves as a check of the results of the model.

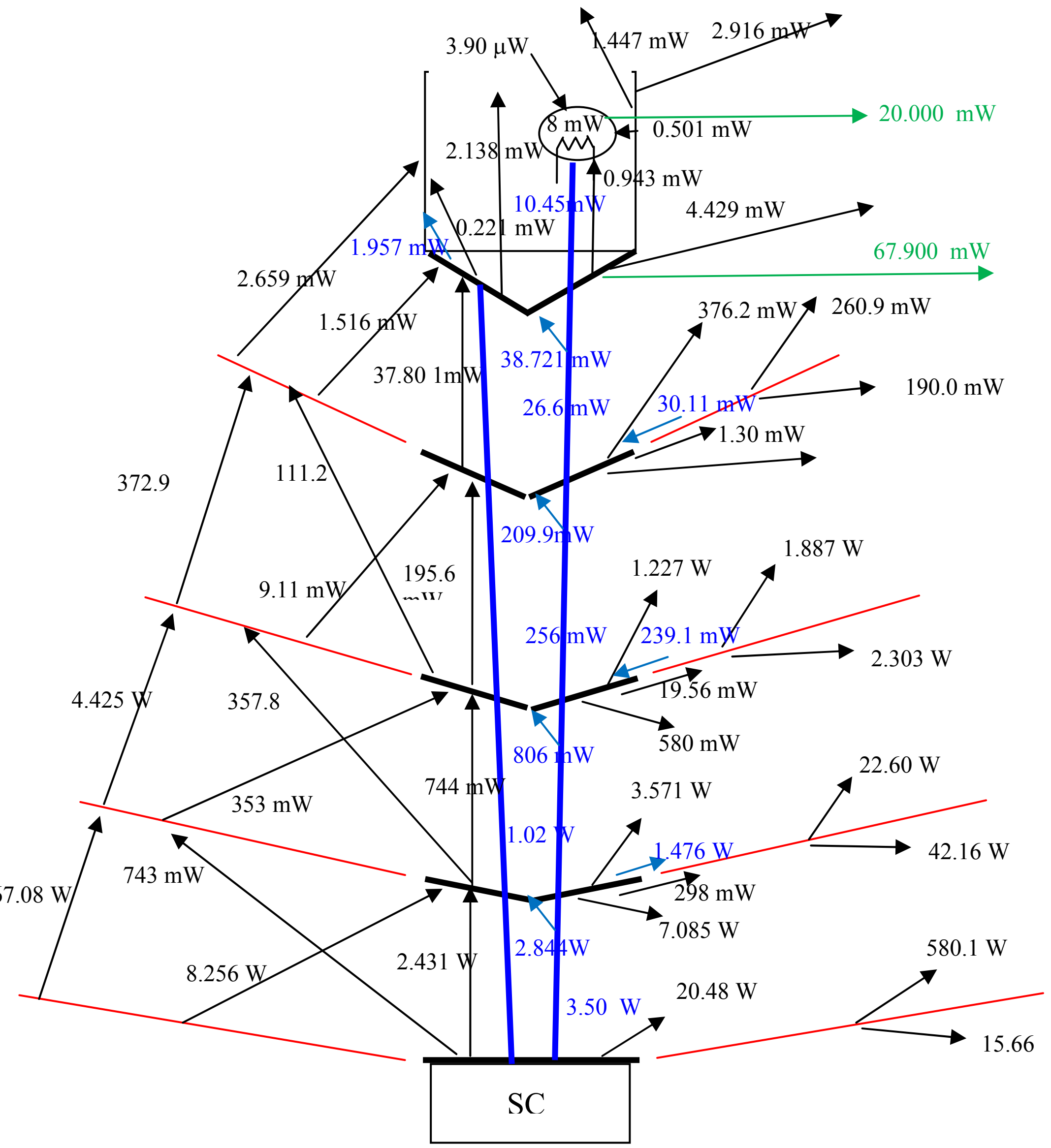

Fig. 8.1.19. Detailed heat transfer map between components with cryocooler on: Blue for conductive heat; Black for radiative heat; Green is for cryocooler cooling power. Two numbers are quoted for conducted heat for the wiring and the bipod.

**Table 8.1.6.** Heat balance at each component

| **Component Name** | **Heat Input** | **Heat Output** | **Error (%)** |
|---|---|---|---|
| Telescope | 19.90 mW | 20.00 mW | 0.5 |
| Optical Box Side | 4.847 mW | 4.864 mW | 0.35 |
| Optical Box Bottom | 78.038 mW | 77.588 mW | -0.58 |
| $4^{th}$ Shield Deployed | 444.7 mW | 444.7 mW | 0.00 |
| $4^{th}$ Shield V-groove | 465 mW | 465 mW | 0.00 |
| $3^{rd}$ Shield Deployed | 4.803 W | 4.811 W | 0.17 |
| $3^{rd}$ Shield V-groove | 2.142 W | 2.133 W | -0.42 |
| $2^{nd}$ Shield Deployed | 69.60 | 69.55 | -0.07 |
| $2^{nd}$ Shield V-groove | 13.53 | 13.53 | 0.00 |

8.1.2c Spinning the Spacecraft (Four- Shield Option, Cryocooler Off): We have also explored the effect of rotating the spacecraft at 0.5 rpm. The primary result is that the temperature becomes more uniform as expected. However a 1.2 K peak-to-peak sinusoidal temperature variation remains in the first shield (see Fig. 8.1.19). In the second shield, this 0.5 rpm time dependent signal drops to 0.3 mK (Fig. 8.1.21). Therefore the attenuation factor is ~4000 per shield. Beyond the second shield, the temperature variation in the model is limited by digitization noise, and the 0.5 rpm signal is not observable. However, using the attenuation factor of 4000 per stage, one can extrapolate to a temperature variation of 19 pico-K at the fourth shield.

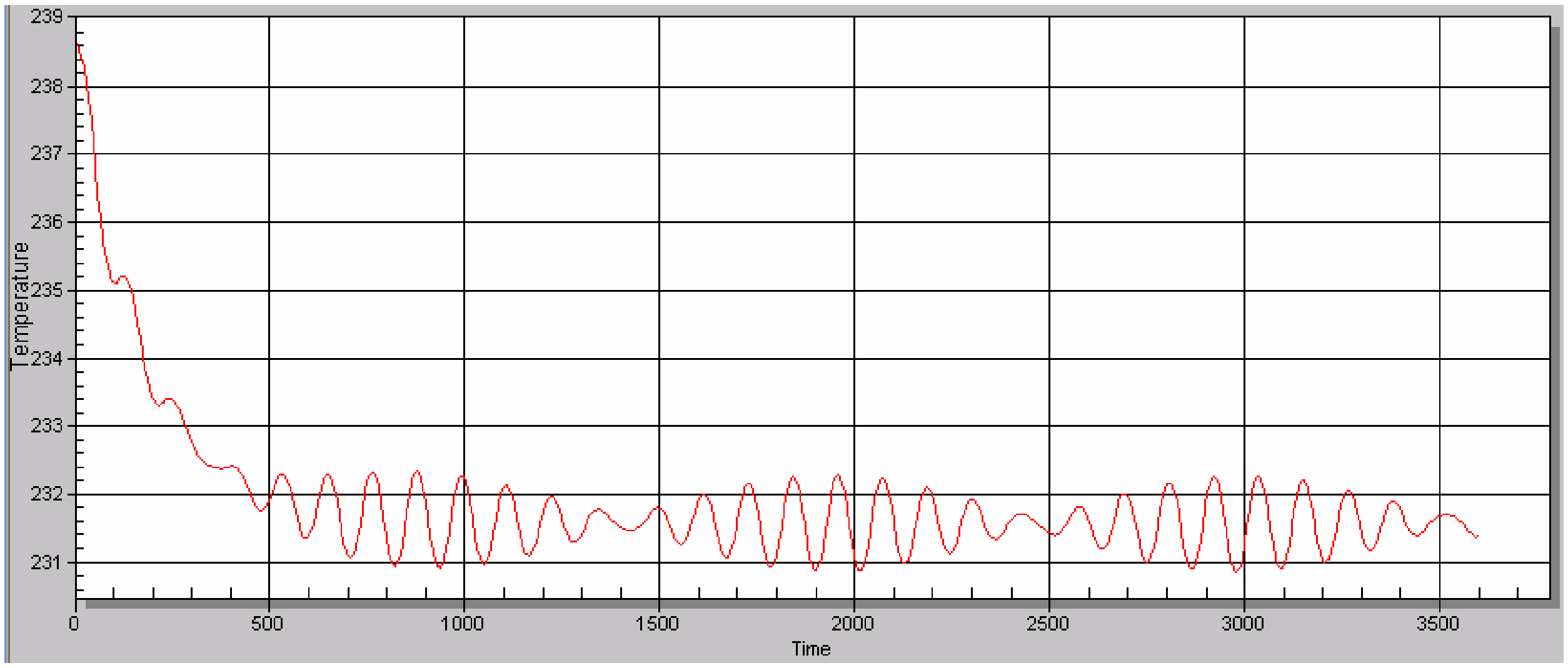

Fig. 8.1.20. $1^{st}$ Shield temperature versus time in units of second, showing the 1.2 K pp signal due to SC rotation.

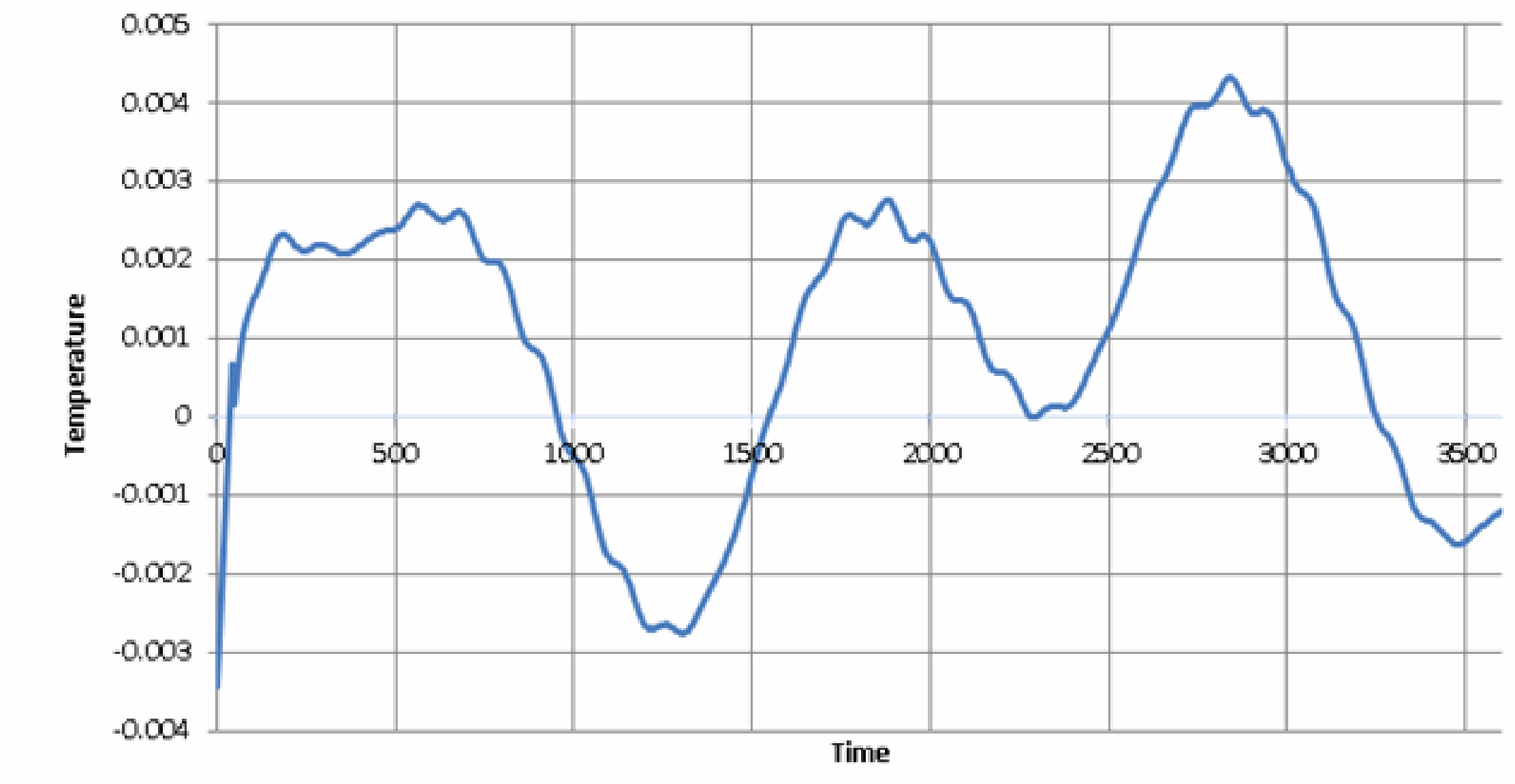

Fig. 8.1.21. Deviation plot of the Temperature the 2nd shield center panel. Rotation signals ~ 0.3 mK pp.

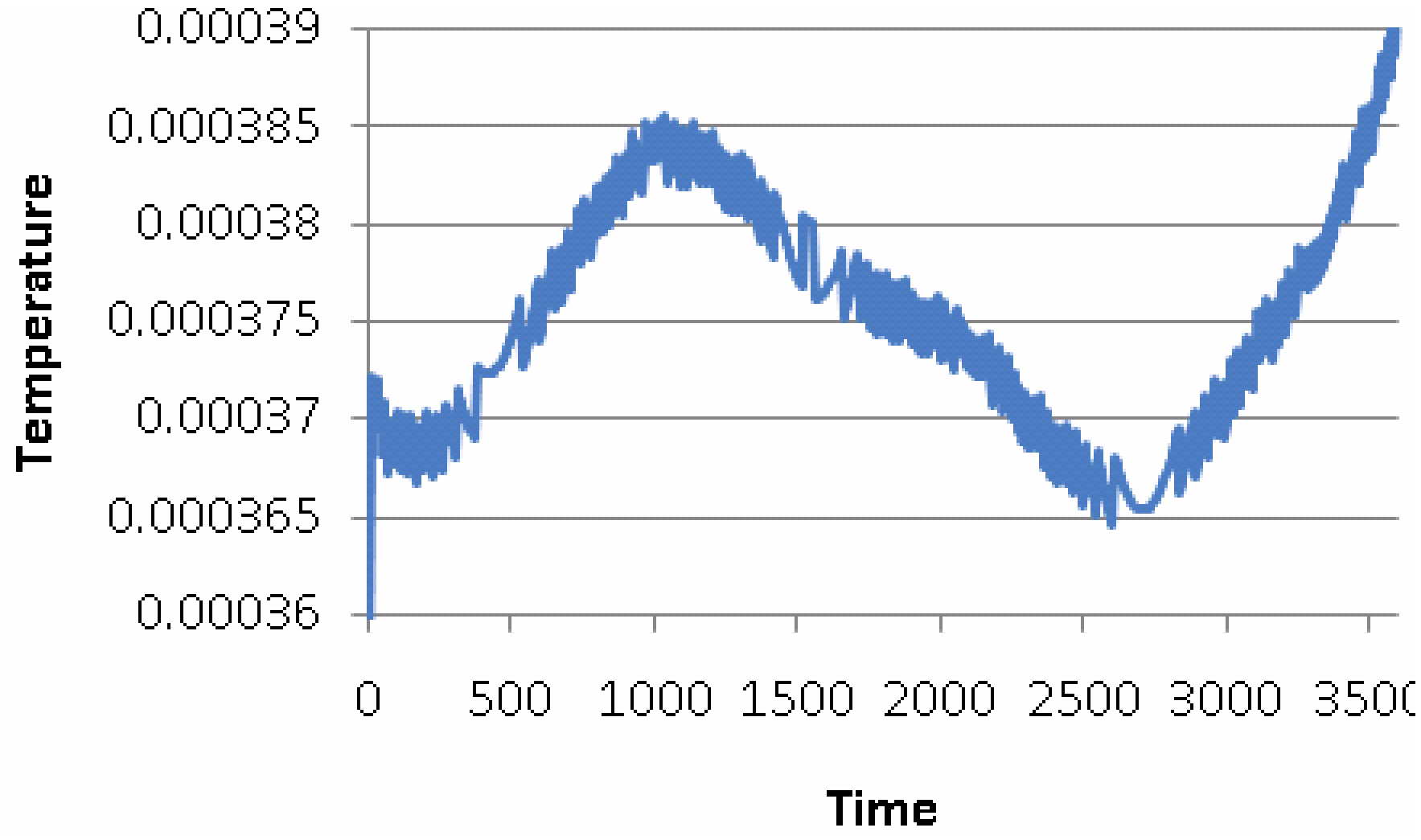

Fig. 8.1.22. Deviation plot of the Temperature of a point on the 3rd shield. The jitter is digitization noise at 3 micro-K level. Rotation signal is not observable.

8.1.3d Three- Shield Option, Cryocooler On: For this option, the area of the Focal Plane is reduced. It is only 27 % of that of the Four-Shield case. Additionally the focal plane is shielded by two radiation shields – the inner and the outer focal plane boxes. In operation, the outer focal plane box is cooled to 18 K, and the inner radiation box is cooled to 4 K by the cryocooler. The power dissipated by the ADR (8 mW) is applied to the inner radiation box. The temperature map for this case is shown in the following figures. The cooling power versus temperature at the inner focal plane box is given in Table 8.1.7.

**Table 8.1.7.** Cooling power at the Inner Focal Plane Box versus temperature

| | | | | | | | | |
|---|---|---|---|---|---|---|---|---|
| Inner FP box cooling power (mW) | 0 | 4 | 8 | 10 | 10.5 | 10.55 | 10.58 | 10.6 |
| Inner FP Box T (K) | 24.96 | 21.53 | 17.49 | 11.41 | 5.28 | 4.81 | 4.24 | 3.12 |
| Outer FP Box P (mW)<br>And T(K) | 7.09<br>18 | 4.13<br>18 | 0.88<br>18 | 0<br>13.96 | 0<br>8.30 | 0<br>7.89 | 0<br>7.28 | 0<br>6.16 |

The Outer FP Box is cooled to 18 K if it is above 18K. If it is below, no heat is applied. The shaded column is the requirement on the cryocooler.

The following color maps are for the case in Table 8.1.7 with 10.55 mW cooling power at the inner focal plane box.

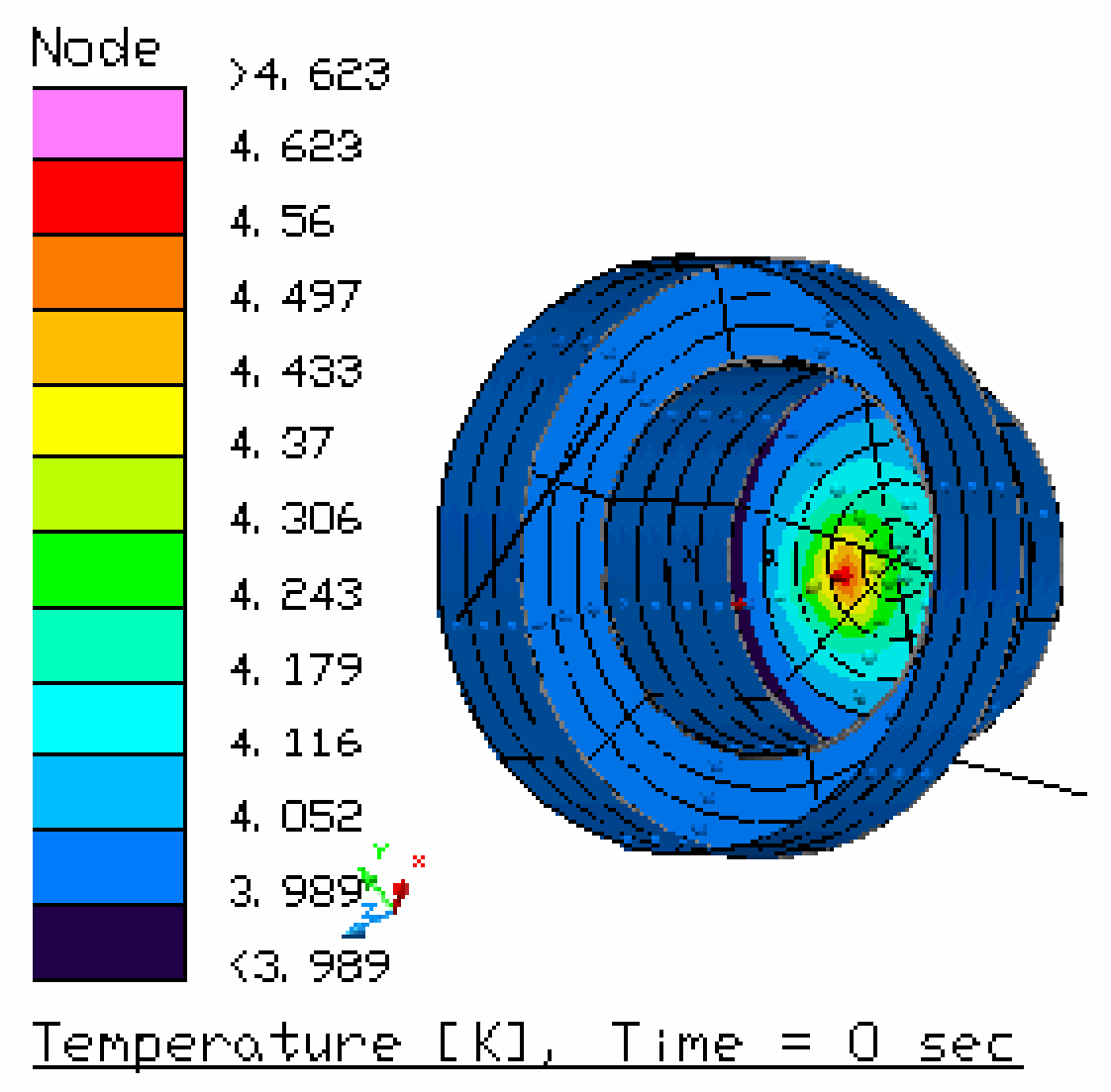

Fig. 8.1.23. Inner focal plane box.

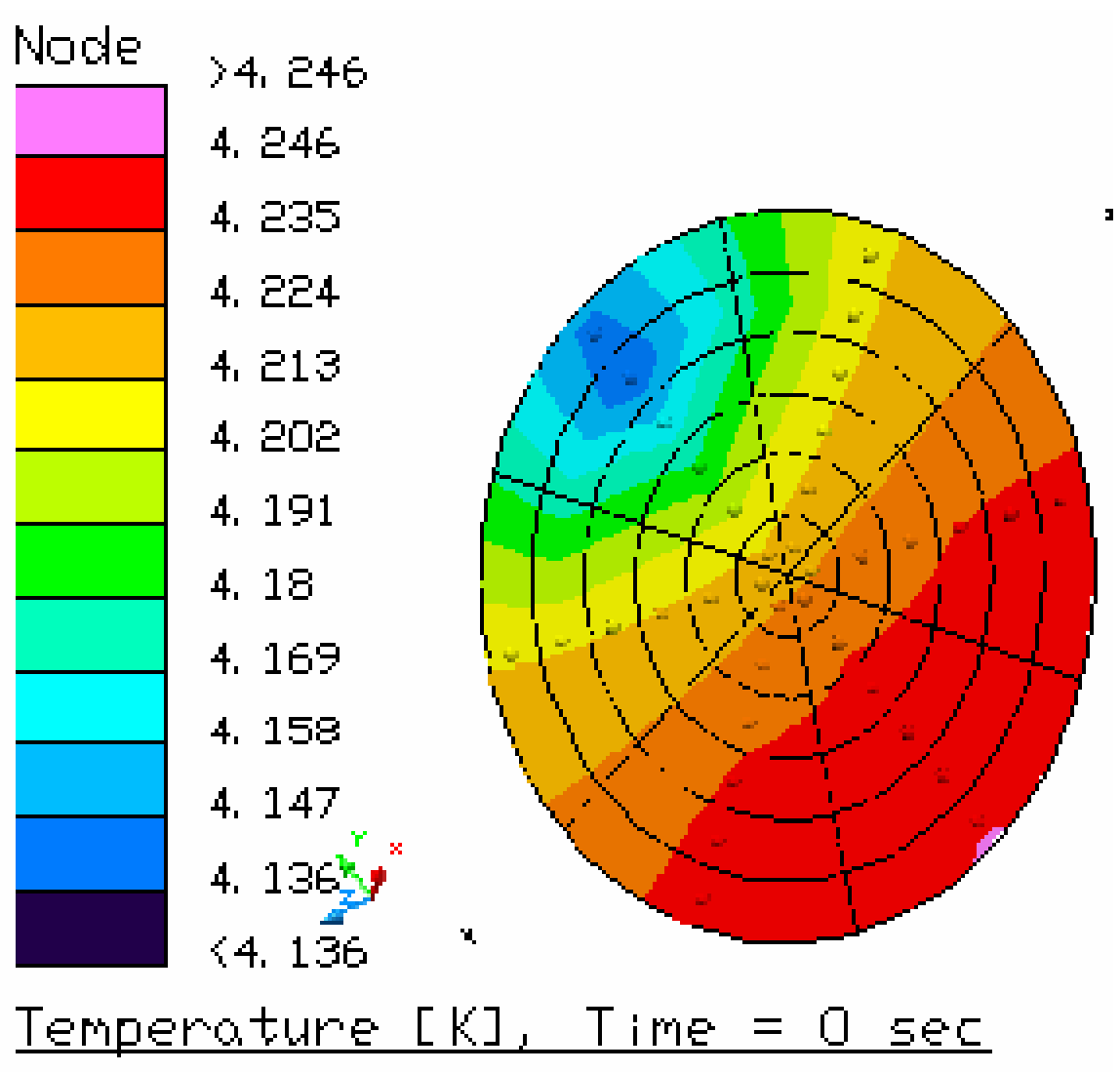

Fig. 8.1.24. Focal plane.

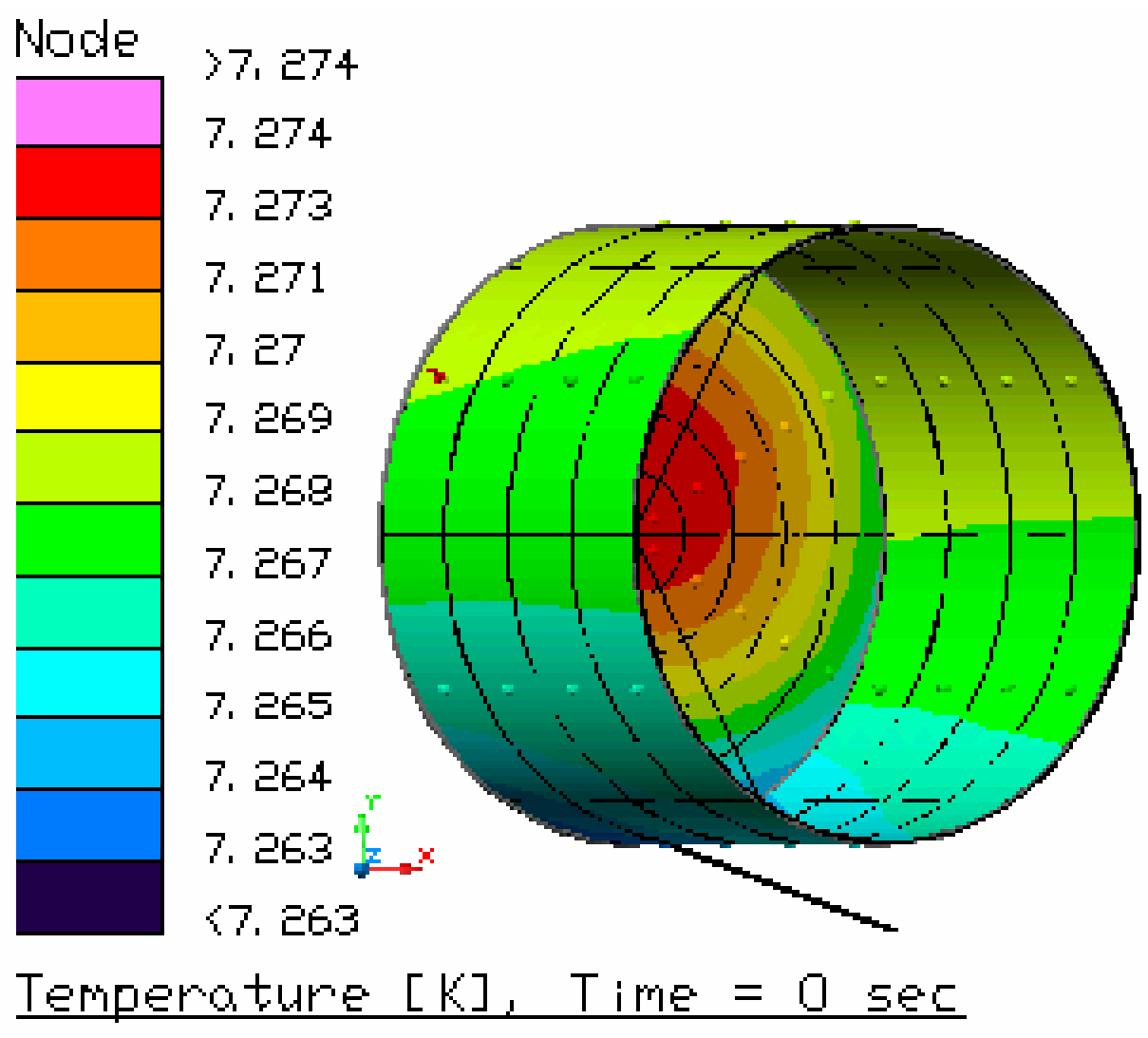

Fig. 8.1.25. Outer Focal Plane Box

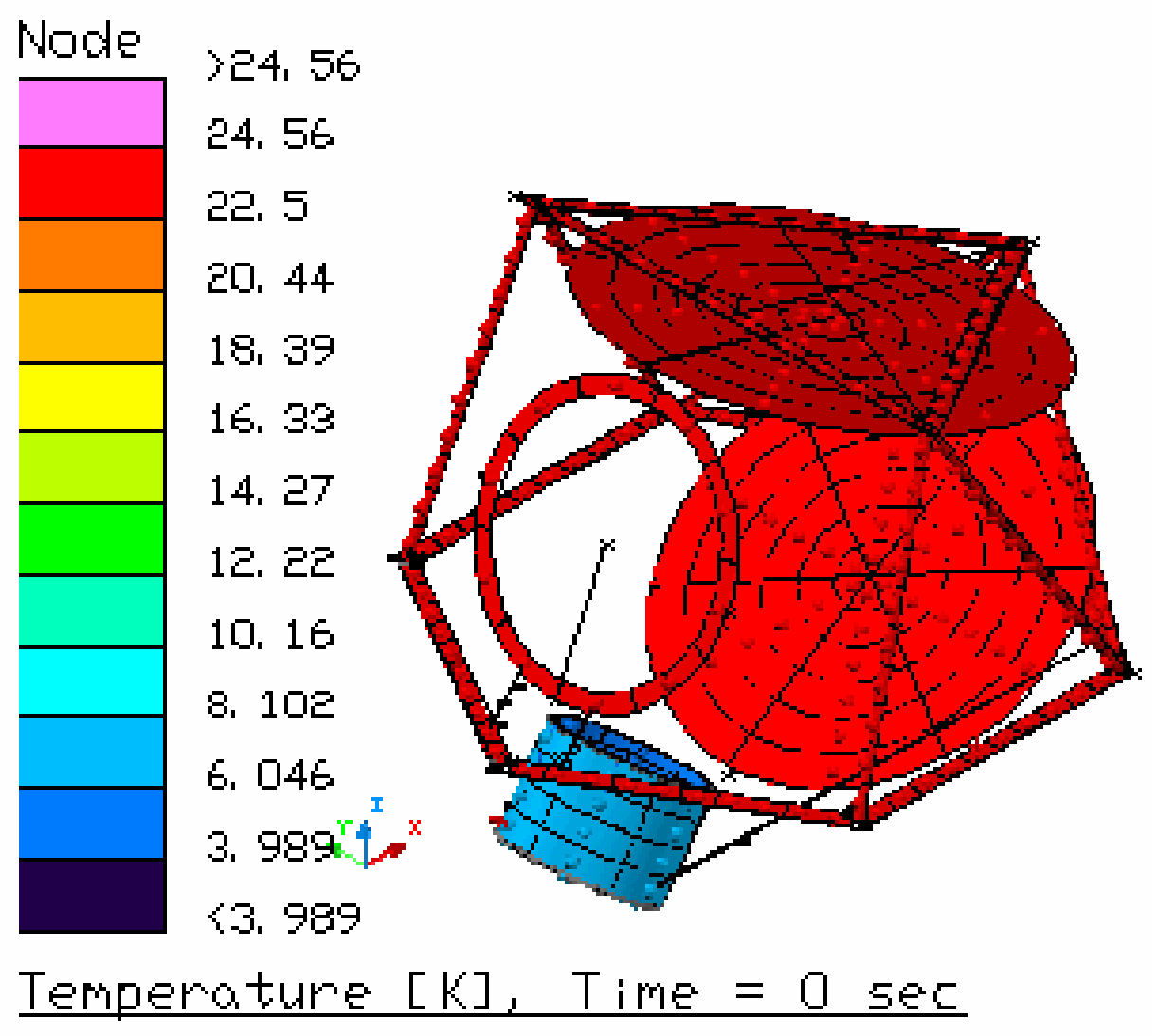

Fig. 8.1.26. Telescope

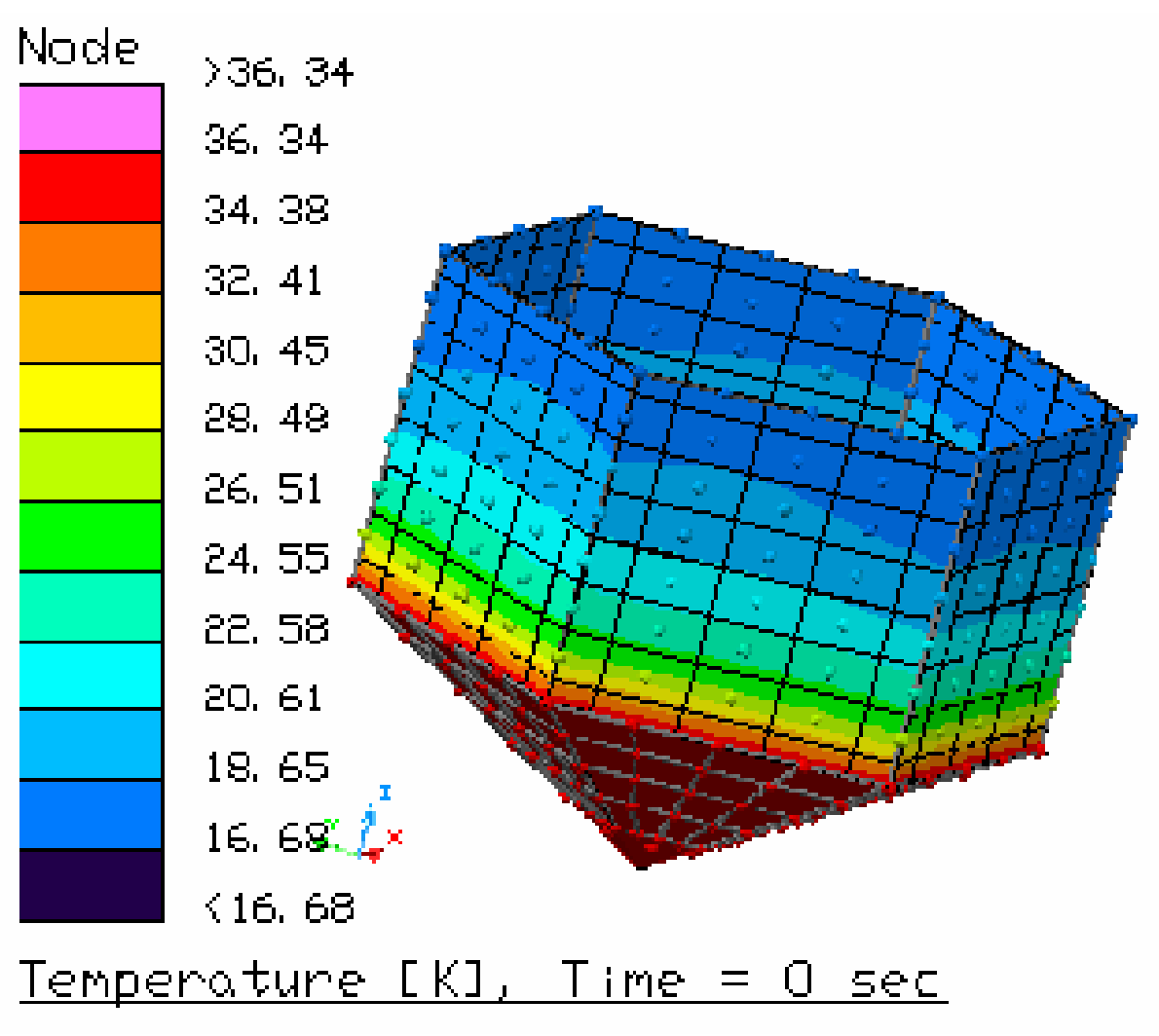

Fig. 8.1.27. Optical Box.

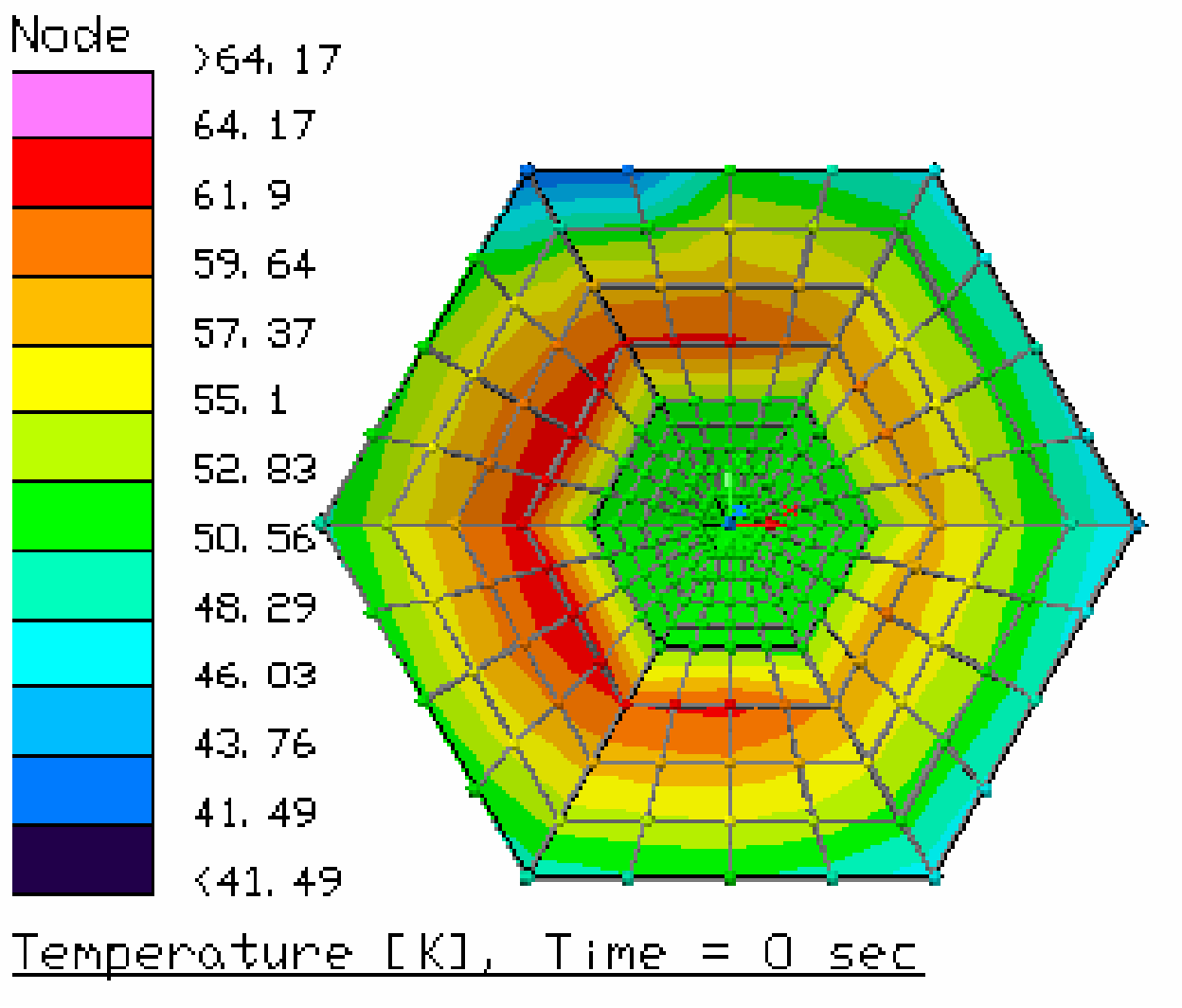

Fig. 8.1.28. 3rd Shield.

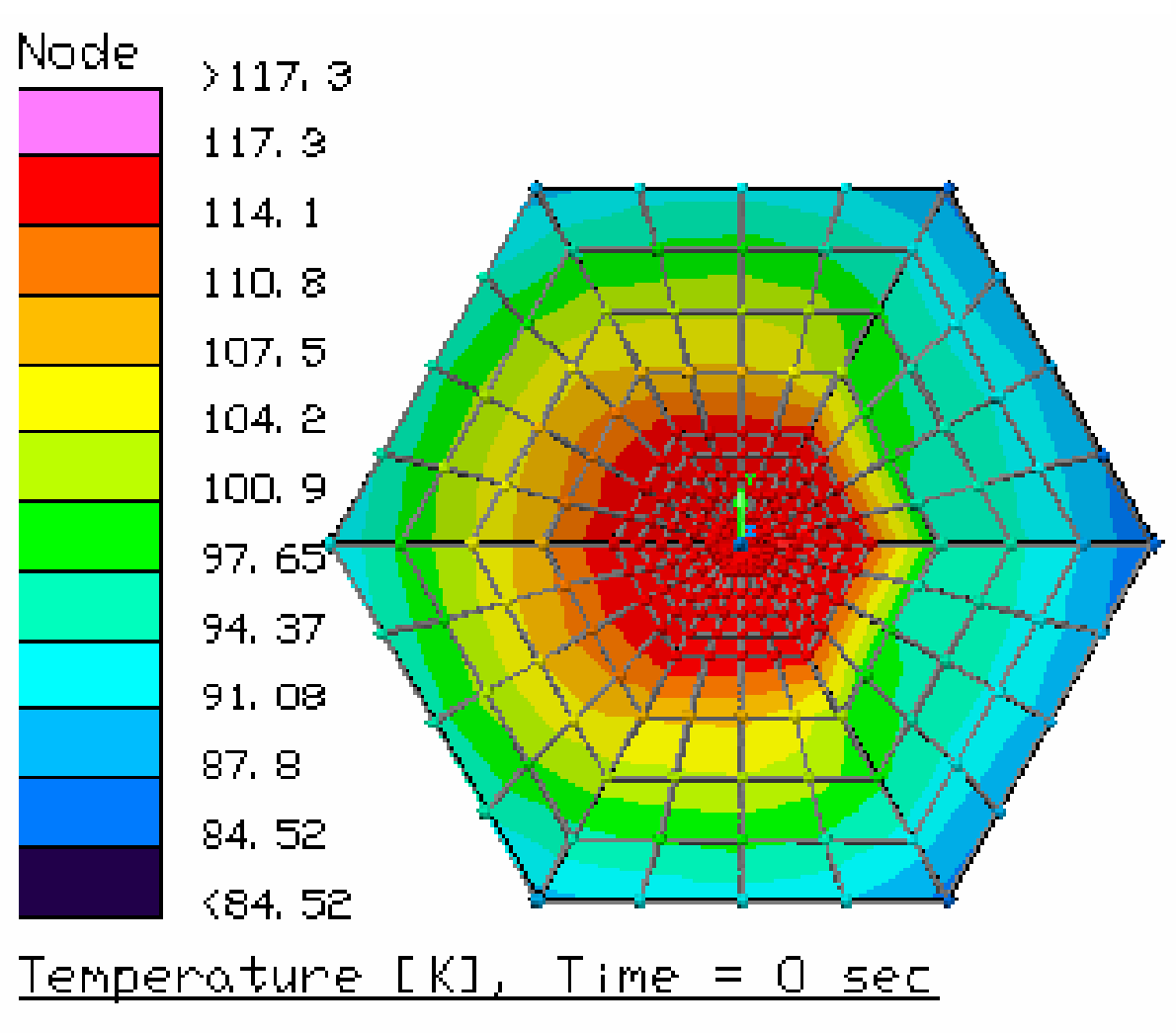

Fig. 8.1.29. 2nd Shield.

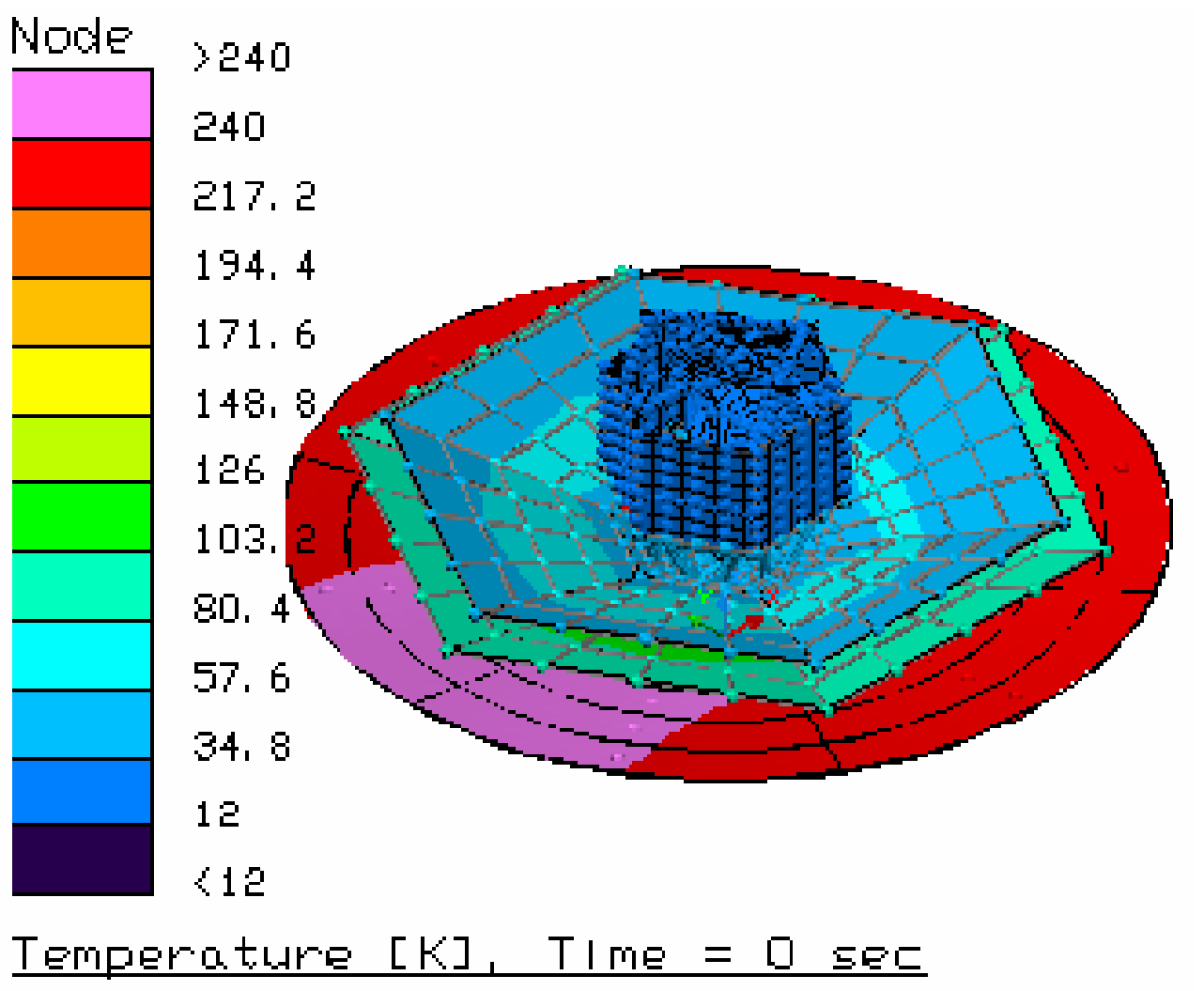

Fig. 8.1.30. Three-shield '30 K telescope' option.

## 8.2 4.4K Mechanical Cooling system

The EPIC cryocooler requirements are very similar to the requirements of the MIRI instrument cooler being developed by NGST for NASA's JWST. Fig. 8.2 is a block diagram of the proposed EPIC cooler and its differences from the MIRI cooler, containing photos of key components indicating their maturity. The chief hardware differences between the two designs is the addition of a second compressor stage to the JT cooler, removal of the JWST configuration specific CTA components required for the JWST mission and changes to thermal and mechanical interfaces. The second compressor stage is required to attain a higher pressure ratio ~10:1 vs ~3:1 for MIRI single stage. The higher pressure ratio provides a lower gas inlet pressure allowing 4.4K operation while maintaining the outlet pressure that establishes the mass flow rate for the required level of cooling. With these changes low risk modifications to areas such as the recuperator tubing diameter and JT restriction L/D will be required to achieve the 4.4K cooling without impacting overall cooler system efficiency relative to the MIRI cooler. With these few changes and modification of the cooler operating conditions the EPIC cooling requirements are readily accommodated.

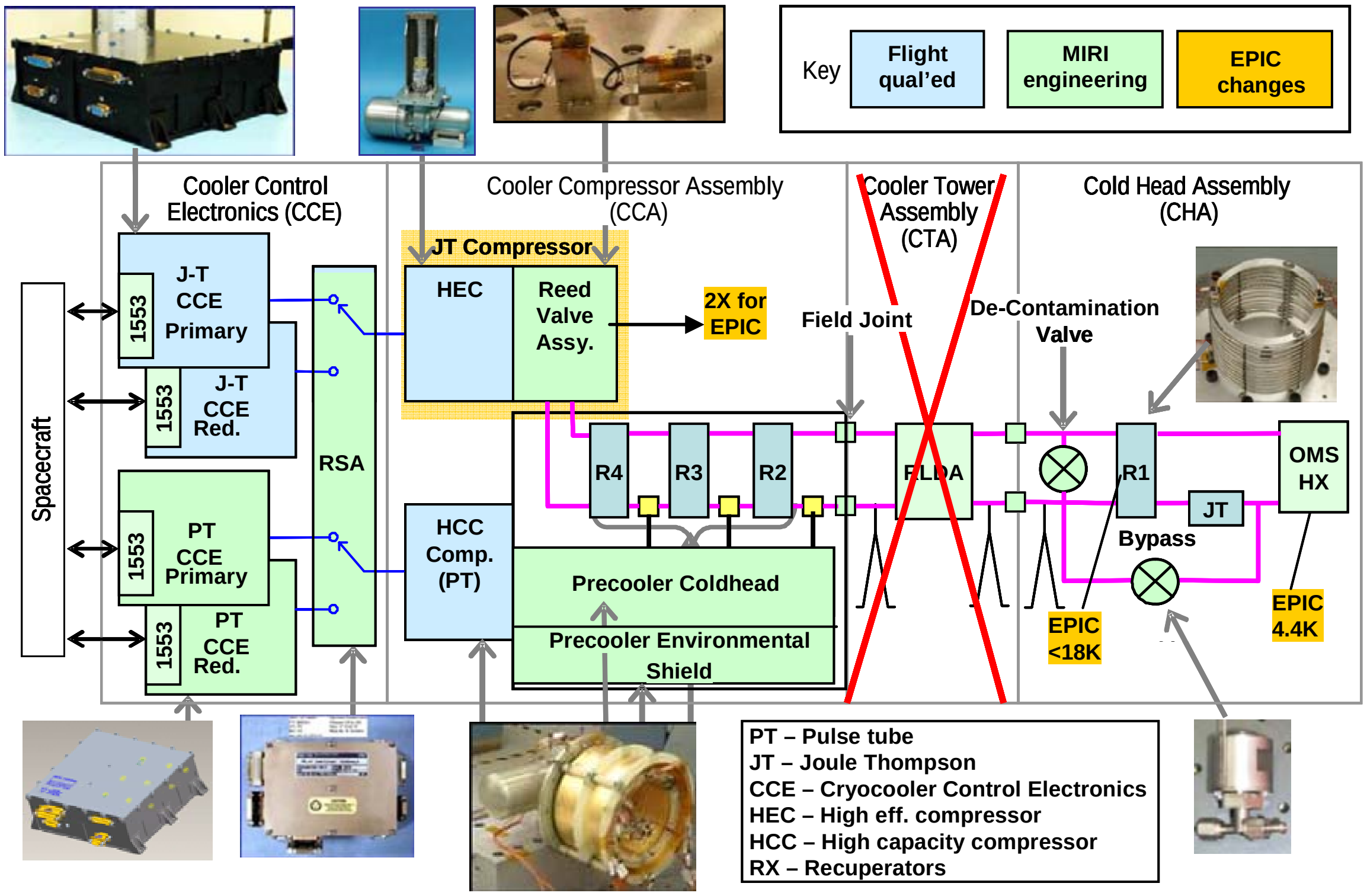

**Figure 8.2.1. EPIC Cooler Block Diagram Differences from MIRI Cooler**

*8.2.1 Cryocooler System Requirements*

The EPIC cryocooler will be used to precool a multistage adiabatic demagnetization refrigerator (ADR) cooling a sensor to 50 mK. The ADR requires cooling at a low temperature (4.4 K for EPIC) in order to cool the ADR's superconducting magnets, shields, heat switches, etc. and to pump the heat from the 50 mK sensor to the EPIC pre-cooler. In order to minimize the heat load (and therefore input power) the ADR and sensor assembly are thermally shielded at 4.4 K and the optical bench/cavity is actively cooled to less than 18 K. The other upper temperature stages, nominally 30 K and above are passively cooled by radiation at an L2 orbit. The EPIC cooler must provide all the optical bench/cavity and ADR/sensor assembly cooling assuming heat rejection to ~300 K. In the baseline configuration the 4.4 K stage is attached at a single point and conductively tied to cool the optics and baffle. If required, the conductive cooling path could be replaced by additional 4.4 K capillary cooling loops and heat exchangers to provide cooling at multiple points.

The EPIC pre-cooler thermal requirements were flowed down from the thermal and passive cooling analysis of the overall payload conceptual design. The cooler performance was modeled based on a MIRI cooler configuration, modified as described earlier. The precooler loads used in this analysis included 100 % cooling load margin to accommodate the early design stage. The cooler performance model is anchored against extensive test data taken on the MIRI program. The results in Table 8.2.1 confirm that the loads and temperatures required could be met with the minor modifications to the MIRI cooler design. The coolers incorporate flight proven active vibration control and we have assumed that this capability will suffice for EPIC. Alternatively, we can mount the cooler on a passive isolator similar to that used on the MIRI cooler or if the requirement is particularly stringent use an added 6 degree of freedom active vibration control system that we have previously demonstrated on another cooler was able to provide control to <5 mN in all axes. To incorporate an active system for this cooler would require some development and additional electronics.

**Table 8.2.1.** Cryocooler loads comparing the MIRI and EPIC-IM applications

| | MIRI | | EPIC (4 K telescope) | |
|---|---|---|---|---|
| | **Temperature (K)** | **Heat Load (mW)** | **Temperature (K)** | **Heat Load (mW)** |
| **Stage 4** | **6.2** | **65** | **4.4** | **42** |
| **Stage 3** | **17-18** | **78** | **<18** | **134** |
| **Reject Temperature** | **313 K** | | **300 K** | |
| **Bus Power (steady state)** | **400 W** | | **270 W** | |
| **Bus Power (cooldown)** | **475 W** | | **TBD** | |

The electronics and flight interfaces are very mature with 1553, RS422 and LVDS available for the various versions of the delivered flight electronics. The MIRI cooler electronics for the pulse tube and JT coolers are based on the delivered high TRL electronics design and their predecessor TRL 9 larger mass version (20 boxes). The JT cooler electronics for MIRI modified the heritage electronics to incorporate a 1553 communication interface and change the temperature measurement sensor to Cernox to measure 4.4K. The PT cooler electronics modified the heritage electronics to increase the power output capability, incorporate a 1553 communication interface and change the temperature measurement sensor to Cernox to measure 4.4 K

*8.2.2 Cooler System Description*

As shown in Fig. 8.2.1, the EPIC pre-cooler consists of a 3 stage pulse tube cooler for the 3 upper temperature stages acting as a precooler for the circulating $He^4$ lowest temperature (4.4 K) Joule Thomson stage. Small capillary circulation loops can be added to provide cooling for each of the shields. If the loops are configured similar to their configuration for MIRI, then both 4.4 K and ~18 K cooling is available for EPIC. The fully autonomous pulse tube and JT drive Cryocooler Control Electronics (CCE) are software driven and can be reprogrammed by upload in orbit, if necessary. The CCE includes a number of safety protect features to autonomously shut down the cooler if a fault is detected and in addition make available a large diagnostic database for download. Functions include autonomous flight proven single axis self-induced vibration control and very stable temperature control. It also includes active primary ripple current control back to the power bus that provides a considerable cost saving to the spacecraft. The electronic components are procurable to the highest reliability such as Class S and NASA

Grade I as implemented on GSFC's ABI program.

Fig. 8.2.1 shows the JT recuperators (R2, R3, R4) through which the circulating lower JT stage's pressurized $He^4$ is flowing and being precooled at each of the three pulse tube stages. This technology was taken to TRL 6 on the MIRI program. On MIRI the 18 K to 6 K recuperator R1 (very similar to the higher temperature recuperators), the JT expander and the bypass valve are located remotely from the precooler on the ISIM. These components will be integrated with the ADR/Sensor and optical bench/cavity on EPIC. The same design can also extend performance to 2.5 K if $He^3$ is used with minor redesign of the recuperators and JT expander. Extension to <2K requires additional JT compression stages.

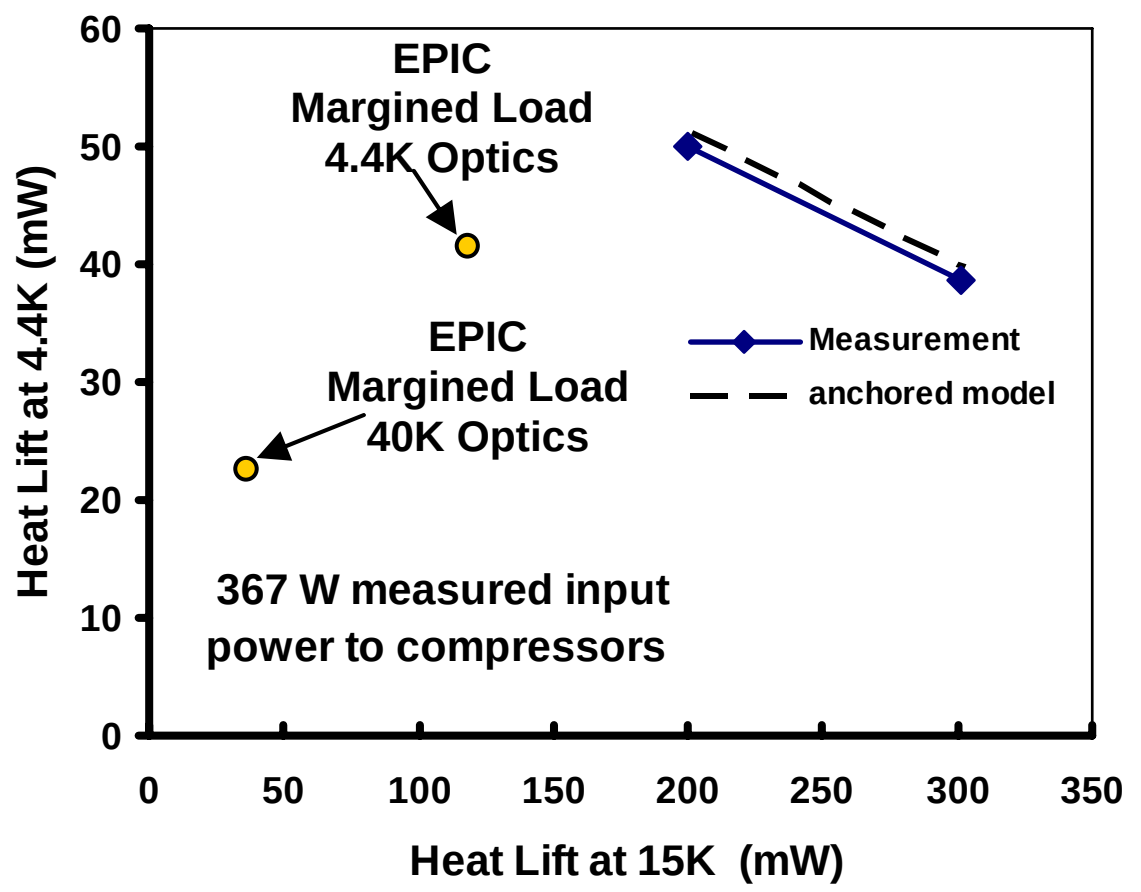

Fig. 8.2.2. Measured JT Cooling at 4.4K using MIRI EM Cooler

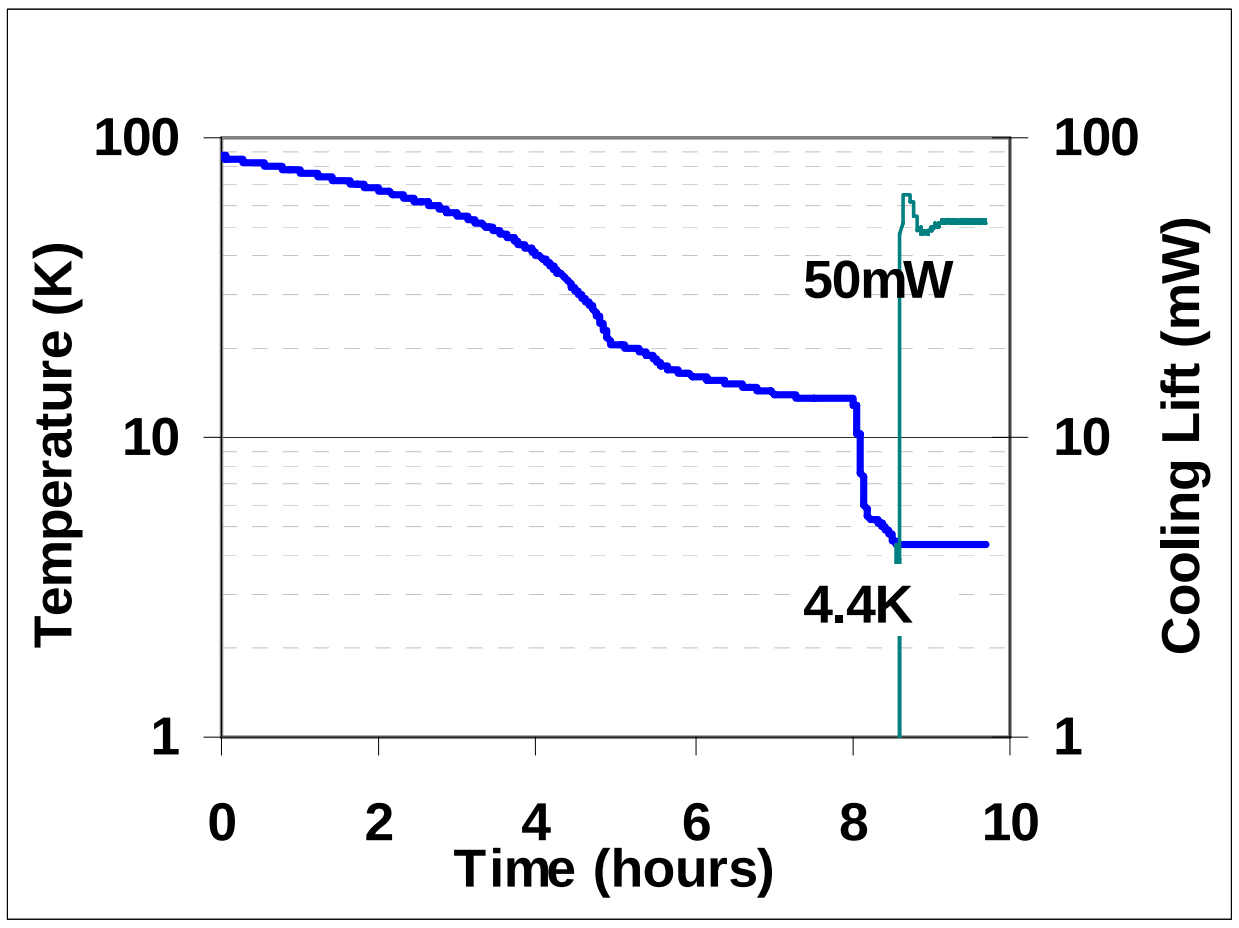

Fig. 8.2.3. Typical Cooldown using Bypass valve

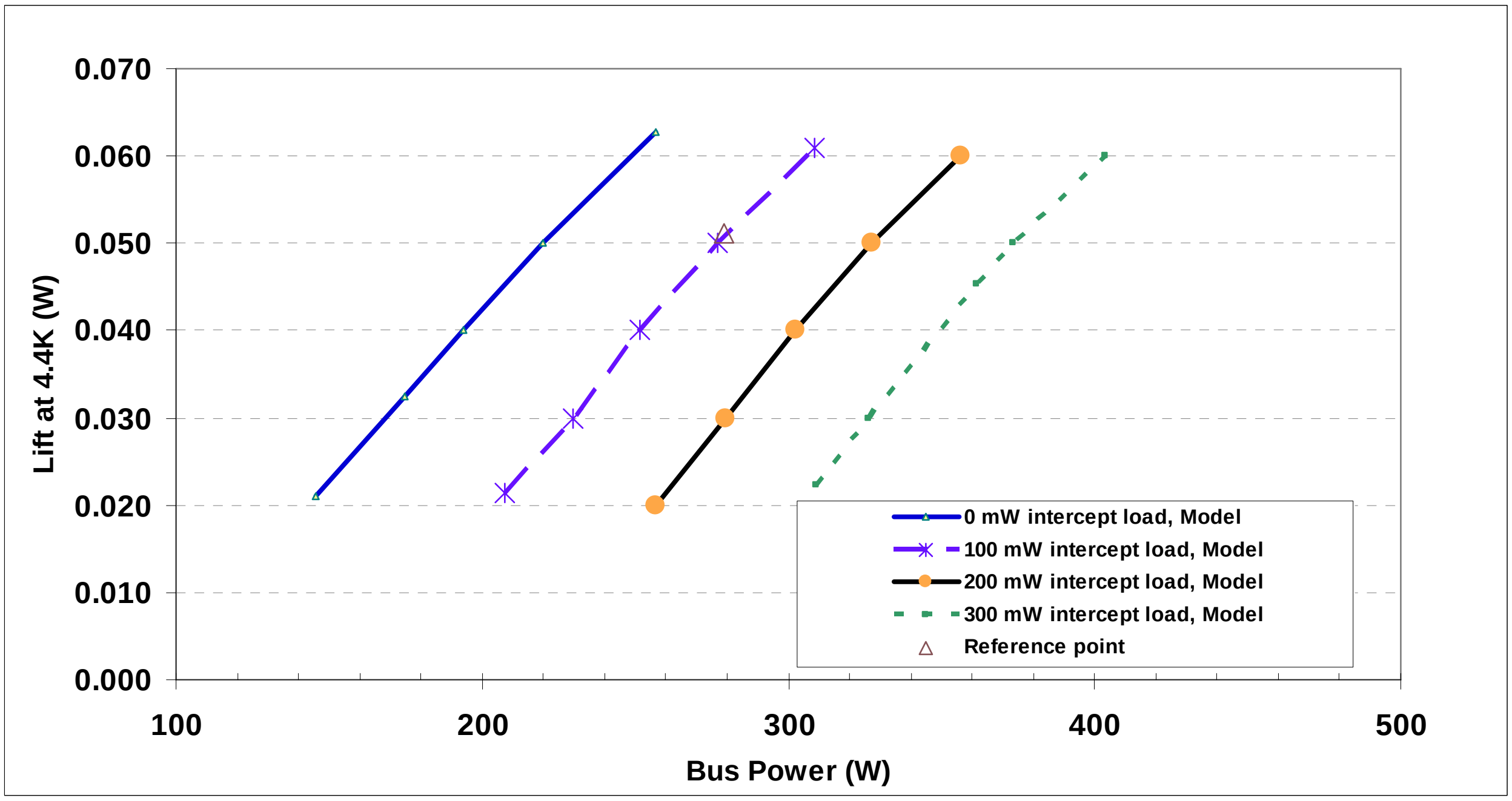

Fig. 8.2.4. Bus Power Estimates for Different Operating Points

**Table 8.2.2. Summary of EPIC 4 K Cooler Specifications**

| Instrument | Capabilities 4.4 K Optics | Capabilities 30 K Optics |
|---|---|---|
| **Mass (Best Estimate)** | **(Kg)** | **(Kg)** |
| Cooler Assembly (JT/ PT Pre-cooler) | 49.2 | 49.2 |
| Electronics (JT/Pre-cooler/Switch box) | 30.2 | 17.8 |
| **Total** | **79.4** | **67.0** |
| **Nominal Operating Condition** | | |
| Cooling Load @ 4.4K | 42 mW | 22 mW |
| Heat Reject Temperature | 300 K | 300 K |
| Bus Power at steady state | 270 W | 165 W |
| Peak cool down power | TBD W | TBD W |
| Operating Temperature Range (PT and JT coolers) | -20 to 50°C | |
| Non-operating Temperature Range (PT and JT coolers) | -40 to 70°C | |
| Operating Temperature Range (CCE) | -20 to 60°C | |
| Non-operating Temperature Range (CCE) | -35 to 75°C | |
| Launch Vibration (PT and JT coolers) | 14.2 Grms, 1 min | |
| Launch Vibration (CCE) | 14.2 Grms, 1 min | |
| Launch Vibration JT cooler 18K to 4.4K component | 25.8 Grms, 1 min | |
| Bus Voltage Range | 21V to 42V | |
| Ripple Current | 100 dB micro amps | |
| Communication Protocol | RS422/1553B | |
| Lifetime | >10 years | |

The TRL 9 HEC compressor with the addition of rectifying reed valves will be used to pressurize and circulate the $He^4$ JT gas. Both our models and test results indicate that we can meet the cooling powers, temperatures and input power using the same $He^4$ working fluid that we use on the MIRI program rather than resorting to a redesign of the recuperators and use of $He^3$. $He^3$ only becomes an attractive option if lower temperatures than 4.4 K are required. Temperatures and loads appropriate for EPIC were demonstrated with the JT loop using $He^4$ gas and two stages of compression for the JT loop with high and low pressures of ~0.7 bar and ~7 bar, respectively. The results of this testing shown in Fig 8.2.2 with the margined EPIC load point shows that the cooler itself has considerable additional margin.

Fig. 8.2.3 shows a typical cooldown from 100 K to 4.4 K (below the MIRI required 6 K) for the cooler system tested in the lab. The data shows the need to use a bypass valve for precooling. A similar procedure is anticipated for EPIC. Fig. 8.2.4 parametrically shows the bus power required for various combinations of 4.4 K and 15 K loads to illustrate the power sensitivity to different loads for the optical bench/cavity (15 K) and ADR/sensor assembly (4.4 K). Table 8.2.2 gives the cooler specifications and some key interface parameters.

## 8.3 Cooling to 100 mK

The sub-Kelvin architecture shown in Figs. 8.3.1 and 8.3.2, used to cool the bolometric detectors for EPIC, consists of a detector stage, a thermal intercept stage, and an outer shield cooled from a mechanical cryocooler to 4.4 K. The thermal intercept stage serves to intercept parasitic heat from supports, to cool infrared blocking filters and to buffer variations in the thermal environment. When the intercept stage is allowed to cool passively to steady state between the detector stage (100 mK) and the cryocooler (4.4 K) temperatures, the heat load to the detector stage is dominated by the parasitic heat. The cooling power required to lift this heat at the detector stage is $P_d = \Delta S(T_h, T_d)\ T_d/t_c$. Here $\Delta S(T_d,T_h)$ is the entropy lifted by a cooler to maintain the detectors at $T_d$ and dissipated at $T_h$ at the cryocooler every cycle period $t_c$. The heat load (or waste heat) from the cooler on the cryocooler is given by

$$P_h = \Delta S(T_h, T_d)\ T_h/t_c = P_d\ \Delta S(T_h, T_d)\ T_h/T_d.$$

The cycle period $t_c$, can be increased and the waste heat $P_h$ at $T_h$ reduced, both significantly as shown in Fig. 8.3.3, by using a part $\Delta S_i$ of the available coolant entropy $\Delta S$, to cool the intercept stage temperature $T_i$ below the passive steady state temperature. The reduced heat load on the detector stage is lifted by the remaining entropy $\Delta S_d = \Delta S - \Delta S_i$. The first step is to the amount of cooolant used for $\Delta S_d$ relative to $\Delta S_i$, given a fixed amount of coolant $\Delta S$, to float. Then second, for a given $\Delta S_d$, $\Delta S_i$ at fixed $\Delta S$, the temperature of the intercept stage, $T_i$, is tuned to maximize the cycle time of the ADR cooler system or to minimize heat load at 4.4 K. The two optimizations (maximum cycle time or minimum heat load at 4.4 K), yield slightly different values for $\Delta S_d$, $\Delta S_i$ and $T_i$. Once the cooler is built, $\Delta S_d$ and $\Delta S_i$ are fixed. The temperature of the intercept stage, $T_i$, can be tuned, to compensate for differences in the computed heat loads used for the design and the actual heat loads in the as built system. This architecture and optimization will result in a more efficient cooling system compared to a single stage cooling directly to the desired detector temperature and does not depend, in general, on the specific method used to cool the stages.

We considered 3 high technology readiness level (TRL) cooler types to provide cooling at the detector and intercept stages, an adiabatic demagnetization refrigerator [1] (ADR), a pumped $^3He$ evaporative cooler [2] and an open cycle $^3He/^4He$ dilution cooler [3]. Single shot designs of the ADR and pumped $^3He$ coolers have flown in space. Pumped $^3He$ and the open cycle dilution cooler have been built and will fly in space on *Herschel* and *Planck*.

### *8.3.1 Continuous Adiabatic Demagnetization Refrigerator*

For purposes of scoping the instrument, the baseline design for EPIC are paired ADR units. Adiabatic demagnetization of a paramagnetic salt was the first method used to achieve sub-Kelvin temperatures [7]. ADRs with superconducting solenoid magnets are commercially available for laboratory instruments and have been flown on balloons, rockets [8] and spacecraft [1]. For the ADR cooling cycle, shown in Fig. 8.3.4, the paramagnet is magnetized isothermally at $T_{DA}$ through path DA, where the heat of magnetization is conducted to a heat sink using a heat switch. Once at peak field, the heat switch is opened and the paramagnet is demagnetized to temperature $T_{AB}$ through path AB. At temperature $T_{BC}$, the stage absorbs heat isothermally at a much slower demagnetization rate through path BC. The steady state temperature of each stage is chosen by the magnetic field at which isothermal demagnetization begins. The ability to easily choose stage temperature with an ADR suits the two stage design for EPIC.

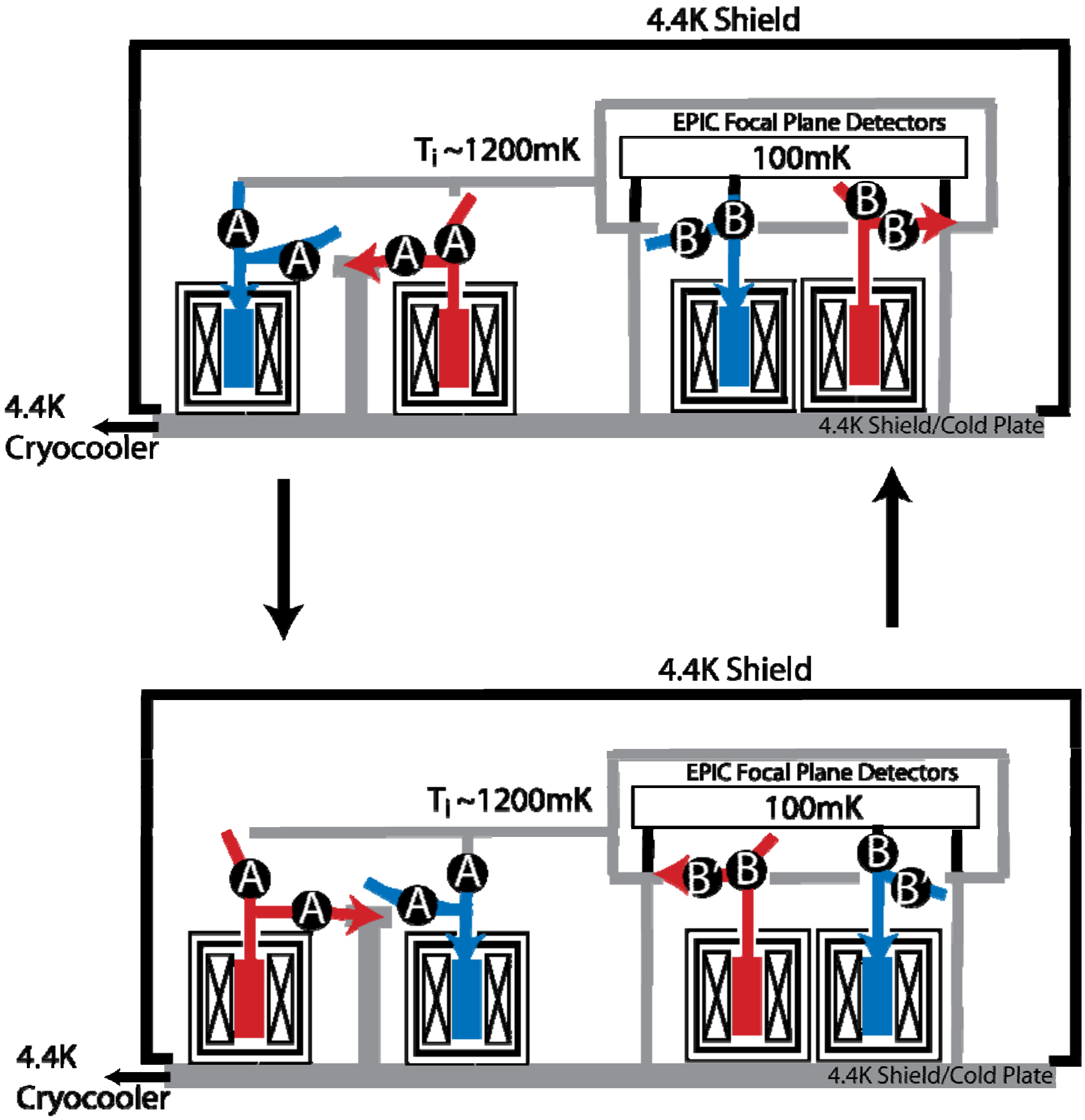

Fig. 8.3.1. Basic system layout of the sub Kelvin cooler for EPIC. The 'parallel' mode of continuous operation is shown here. Heat switches are labeled with A,B or B'. The arrow indicates the direction of heat flow. The EPIC focal plane detectors are cooled to 100mK and surrounded by a radiation shield (grey) held at the intercept temperature $T_i$. The blue and red units are paramagnets being magnetized (red) to dump heat to a heat sink or iso-thermally demagnetized (blue) to cool the detector or intercept stage.

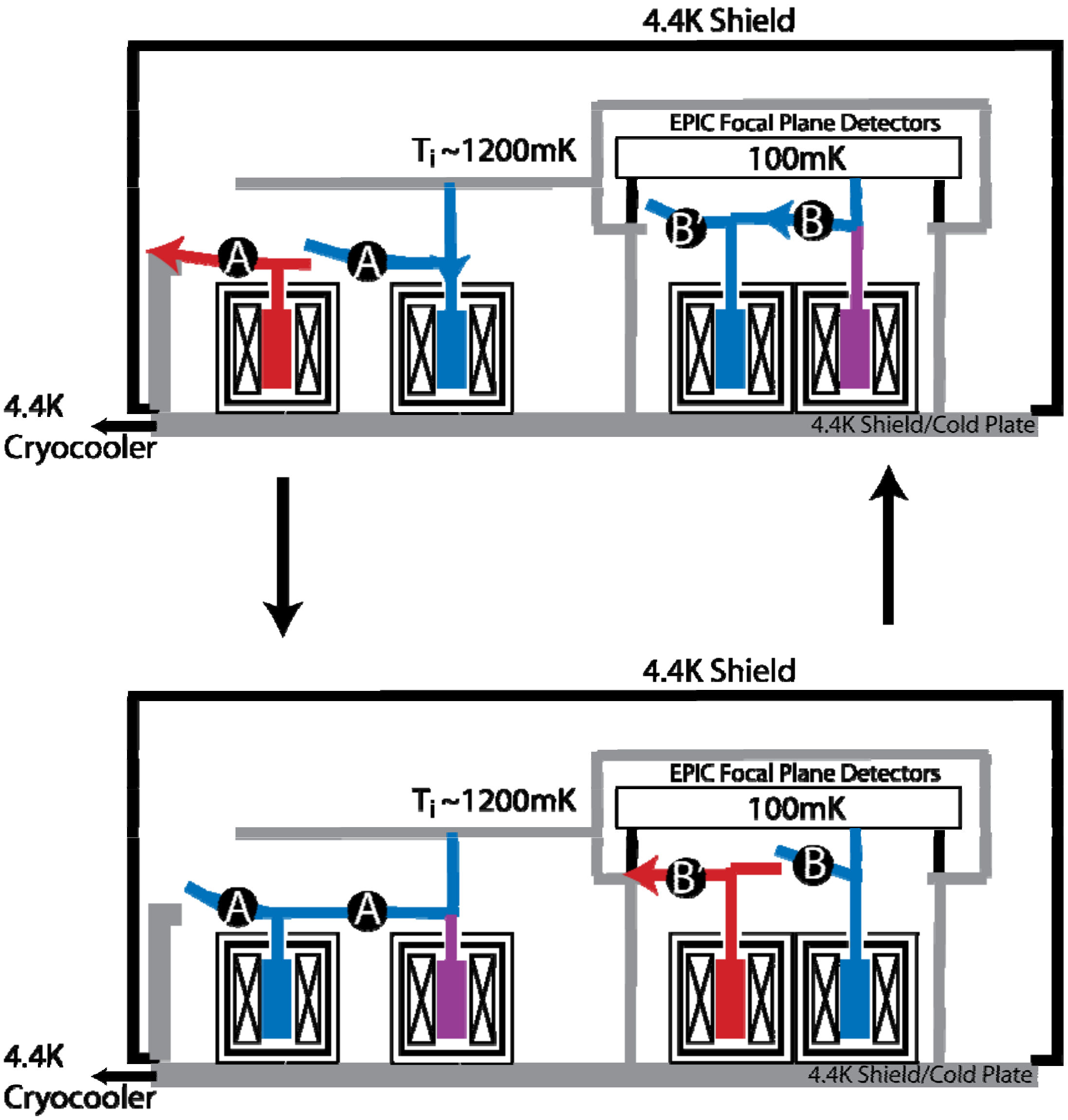

Fig. 8.3.2. Basic system layout of the sub Kelvin cooler for EPIC. The 'serial' mode of continuous operation is shown here. Heat switches are labeled with A,B or B'. The arrow indicates the direction of heat flow. The blue and red units are paramagnets being magnetized (red) to dump heat to a heat sink or isothermally demagnetized (blue) to cool the detector or intercept stage. The paramagnet during isothermal magnetization is colored purple.

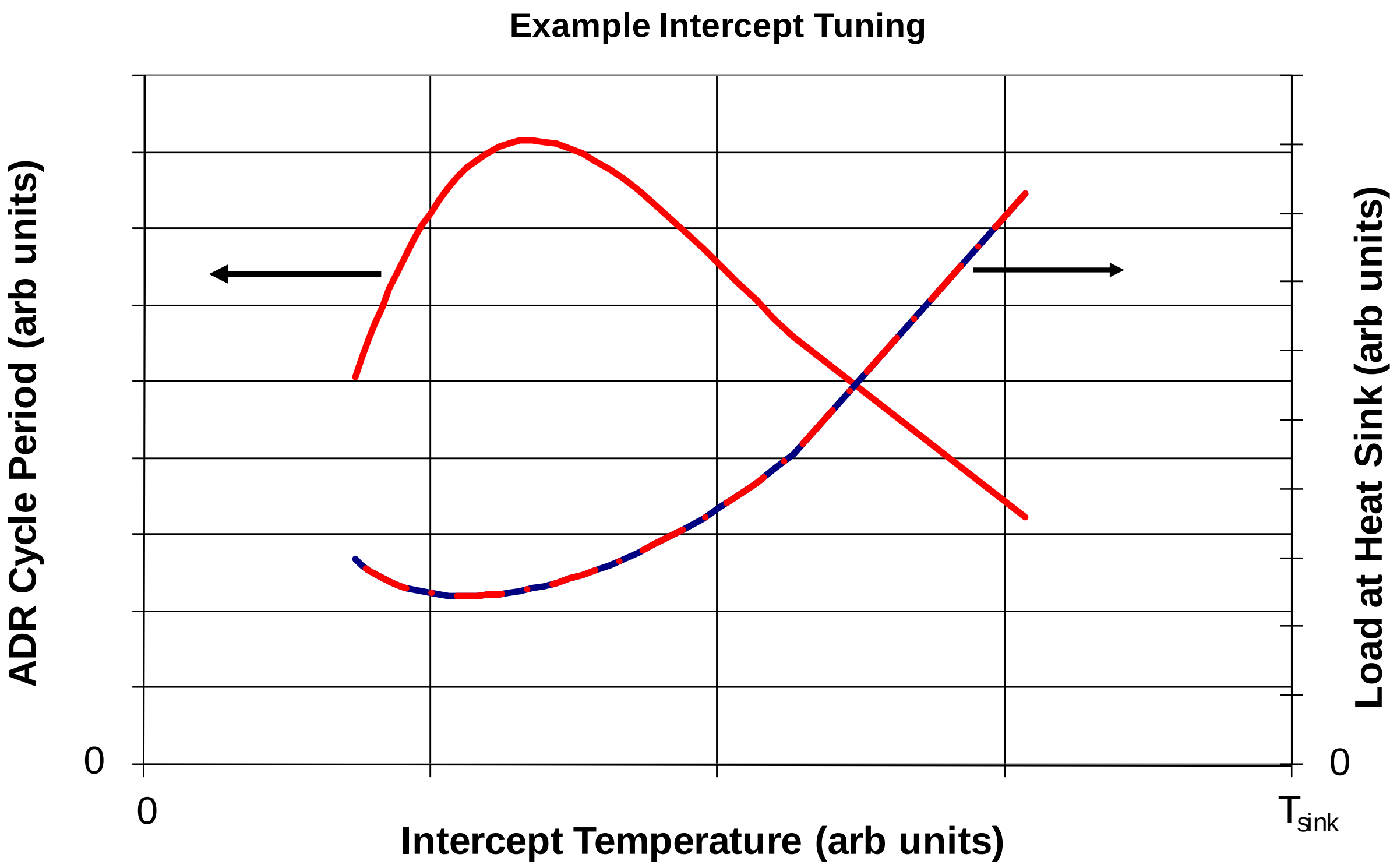

Fig. 8.3.3. Single shot hold time or cycle period and heat strap mass on a linear scale as a function of intercept stage temperature for an adiabatic cooler. Note the cycle period more than doubles and the power load into the heat sink is reduced by more than a factor 3. Also note that the intercept temperature that gives the maximum cycle period does not give the minimum power into the heat sink.

Paramagnets used for space flight ADRs that cool detectors to 100 mK are hydrated salts containing paramagnetic ions, either Chrome Potassium Alum (CPA) or Chrome Cesium Alum (CCA) grown onto gold or copper skeletons [1,9]. The gold or copper skeletons are connected to a gold plated copper bolt cold finger and the salt is encapsulated in a hermetic container to prevent long term dehydration. This assembly is commonly called a salt pill. The salt pill is supported within a 2-5 Tesla superconducting magnet using low thermal conductivity support such as tensioned kevlar. For the intercept stage at ~500 mK, Gadolinium Gallium Garnet (GGG) is attractive since it is more chemically stable, has a larger ion density and spin quantum number J than the chrome alums and does not order magnetically until <900 mK.

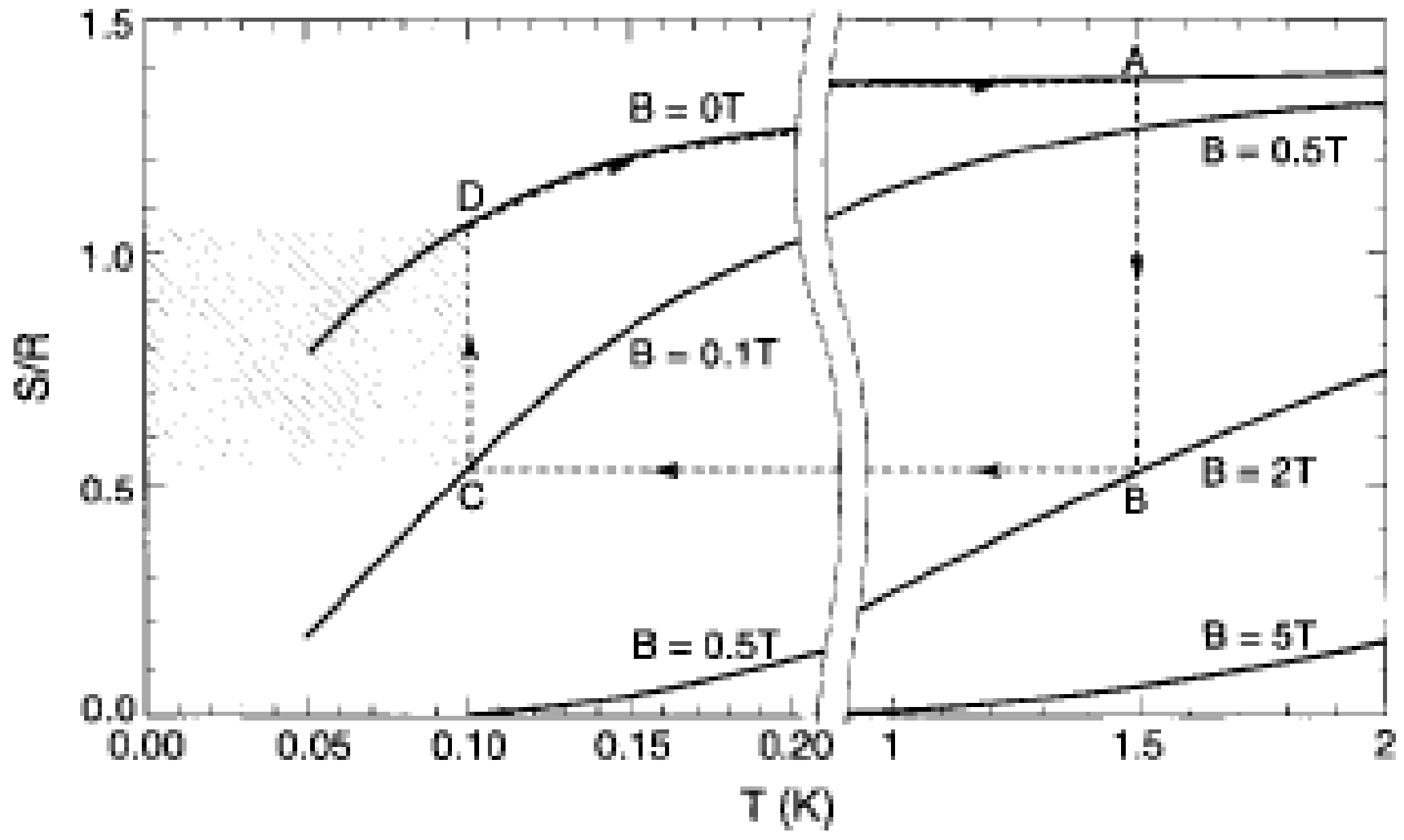

Fig. 8.3.4 Generic thermodynamic cycle for an ADR taken from Hagmann [10].

For EPIC, the ADR can be continuously cycled by either the parallel [4] Fig. 8.3.1 or serial [5, 6], Fig. 8.3.2, techniques to maintain constant temperature at the intercept and detector stages. In the case of parallel cycling, one ADR is connected to the stage via a heat switch and isolated from the high temperature stage by a second heat switch. A second ADR is identical to the first, but is dumping heat at the high temperature stage. These paired ADRs switch off cooling and recycling by using the heat switches as described in Fig. 8.3.1. In the case of serial cycling, the first ADR is either isothermally demagnetizing while a second ADR is dumping heat at the higher temperature stage. After the second ADR is fully magnetized, its heat switch opens and it cools down slightly below the temperature of the detector stage. Once at the lower temperature, the heat switch between the first ADR and second opens. Now the first ADR isothermally *magnetizes* dumping its heat of magnetization to the second ADR.

For the purposes of sizing the cooling system we consider a focal plane tiled with antenna-coupled voltage-biased superconducting Transition Edge Sensors (TES) bolometers cooled to 100 mK as described previously. The detector load is dominated by the power dissipated into the load resistance, 4 % of the bolometer operating resistance, used to maintain fixed bolometer voltage. The detectors are readout with an N x M array of multiplexed first stage DC SQUID ammeters mounted on the detector cold stage. Each set of N first stage SQUIDs are readout, via stripline cables, to M second stage high bandwidth series array SQUIDs mounted on the cold stage and cooled to 4.4 K by the cryocooler. We base the cooler design assuming the time-domain SQUID multiplexer readout of the TES. This places more demanding resource requirements on the cooler than the frequency-domain SQUID multiplexer, thus provides a worst case for the system performance. The detector stage is supported with low thermal conductance supports from an intercept stage. A radiation baffle with infrared blocking filters mounted from intercept stage surrounds the entire detector stage assembly. The intercept and detector stage assembly is supported from the 4.4 K stage. The supports between each stage are sized to support launch load with standard factors of safety to yield (1.2), ultimate (1.4), and buckling (2.0). The baseline design is made from high strength titanium which has very high yield and ultimate strength, 115 and 140 ksi respectively and low thermal conductivity, 150 $T^{2.7}$ μW/cm K. Supports made from tensioned Kevlar would have about a factor ~2 lower heat load on the stages but with additional design elements including the support posts, alignment fixtures and preloading springs to compensate for long term creep. The detector readout cables are kapton/constantin striplines similar to those used for SPIRE and heat sunk at the intercept stage, the 4K stage and at additional points in the cooling chain. The steady state temperature of the intercept stage with no active cooling is ~2.5 K. The mass accounts for a serial operation continuous ADR with a cycle time of ~1 hour. The power load from the 8 high temperature superconductor leads, of ~1.5mW, has been added to the total ADR cooler load.

**Table 8.3.1.** ADR Design Parameters for the 4 K and 30 K telescope configurations

| | Units | 4 K Telescope | 30 K Telescope |
|---|---|---|---|
| Detector System Power* | μW | 1.76 | 0.34 |
| IR Loading (Detector/Intercept)† | μW | 2.8 / 3.0 | 2.2 / 6.4 |
| Detector Stage Heat Lift | μW | 8.1 | 5.0 |
| Intercept Temperature | K | 1.03 | 1.00 |
| Intercept Stage Heat Lift | μW | 205 | 142 |
| Heat Strap Mass | kg | 1.1 | 0.7 |
| ADR System Mass | kg | 7.2 | 6.0 |
| ADR Cooler Load at 4.4K | mW | 5.5 | 4.3 |

*Detector system power taken for the worst case of time-domain SQUID multiplexing. Lower power would be achieved with FDM-SQUIDs, RF-multiplexed MKIDS, or TDM with improved lower power SQUIDs.

†Infrared loading presented in table 7.4. The first number is the power on 100 mK; second number is power on the ~1 K intercept stage.

Note: The steady state temperature of the intercept stage with no active cooling is ~2.5 K. The mass accounts for a serial operation continuous ADR with a cycle time of ~1 hour. The power load from the 8 high temperature superconductor leads of 1.5mW has been added to the total ADR cooler load.

8.3.1a Heat Switches: For the baseline cooler, 2 (4) gas gap heat switches (GGHS) [2,12], labeled (A) in Fig. 8.3.1 and 2-4 superconducting (SCHS) or magneto-resistive heat switches (MRHS), labeled B in Fig. 8.3.1, are required for the serial (parallel) cycling continuous ADR, for a total of 4 (8) heat switches for the entire cooler system. The detector stage heat switch is thermally anchored to the intercept stage to reduce the heat load to the detector stage when the heat switch is in the off state. The intercept stage heat switch is anchored to the liquid helium bath. The heat switches are nominally identical. The thermal design is driven such that the on state conductance of the intercept heat switch is high enough to conduct the heat of magnetization of both the detector and the intercept stage salt pills to the cryocooler at 4.4 K in half the cycle time. A typical measured on state is in the range 50 mW/K at 4.4 K. The main body of the heat switch is a pair of gold plated copper plugs each with a set of copper fins that, when assembled, are interleaved and encased in low thermal conductance thin walled low thermal conductivity alloy tubing. At operating temperature the switch is normally off. It is activated by heating a small charcoal capsule connected to the main body of the switch with a thin walled capillary. When heated, the charcoal desorbs $^3$He gas which thermally shorts the copper fins.

These GGHS cannot be adapted to work at 100 mK due to the low vapor pressure of $^3$He and hence low on state conduction at these temperatures. For this lower temperature range, a magnetically driven heat switch is the technology of choice. The heat switch design consists of a pure, annealed metallic foil that is switched on and off by using the magnetic field perpendicular to the foil and heat flow directions. The magnetic field works as a switch by turning ‘off’ heat flow by conduction electrons. In the off state, heat is conducted by the acoustic, or phonon, modes of the material. This leads to a switching ratio that scales ($k_e/k_p$) * $A/T^2$, where $k_e/k_p$ ~ 200-600 depends on the electronic and phonon specific heats and particle speeds and A is the ratio of scattering lengths, which depends on the material purity and state of annealing. Typical values of A are in the range $0.1 < A < 10$. A large switching ratio implies the desired small off state conductance. The electronic heat conduction is ‘turned off’ in superconductors (SC), such as Al (Tc ~ 1.1 K), Nb(Tc ~ 9.2 K), or V(Tc ~ 5.4 K), at temperatures well below the resistive transition temperature ($<T_c/5$). Switching to the on state requires fields < 2 kG (even for Nb which has a $H_{c2}$ ~ 2 kG). Al, Nb and V have high $k_e/k_p$ ~ 300-600. Al based heat switches with usable on state and switching ratios >1000 have been demonstrated at mK temperature as required to make the cADR work and should produce a usable (>1000) switching ratio at temperatures as high as 1 K. The SC material is clamped at the ends with gold plate copper end fittings and held, via low thermal conductance supports, within the high field region of a magnetic coil and shield assembly. The design of the coils and the shielding of the SCHS will incorporate ferromagnetic and superconducting layers to prevent leakage (in or out) at the < 1 G level at the surface of the shields. Using Nb or V as a SCHS will require more research since these materials are type II and can contain trapped flux in zero field which leads to a poor switching ratio. Temperature and field gradients have applied type II SC to push and dissipate trapped flux and can be incorporated in the HS design.

An alternative to using V or Nb to increase the usuable temperature range of the SCHS is a novel magnetically driven heat switch (a magnetoresistive heat switch (MRHS)) that was demonstrated recently at University of Wisconsin (UW) and Goddard. The MRHS is identical in form to the SCHS except that tungsten (W), is used at a temperature above its transition in place of the superconducting metal. The MRHS is normally on and is switched into an *off* state at magnetic fields of 10-30 kG. The UW tests indicate a low switching ratio of ~100 even though switching ratios >1000, comparable to those obtained with Al, are feasible based on the material properties of W. The low switching ratio in the UW test is surely limited by the method of attaching the W to copper end clamps. The MRHS concept is attractive since it can be used at temperatures >4 K where the SCHS cannot.

8.3.1b Paramagnet Sizing: Before the mass of the ADR system can be estimated, the paramagnet salt pill must be determined. For single shot systems, the paramagnet size was chosen to match a fixed observation time, usually 24 hours. The rest of the system, mainly the heat switches, were designed to support a high duty cycle. For a continuous ADR, the paramagnet volume is designed to be as small as possible, but is constrained by 3 factors. The minimum cycle time of the ADR, minimum heat capacity needed for adequate temperature regulation, and minimum ballast needed for the initial cool down. The primary constraint is to ensure a cycle time that is longer than the switching time of the heat switches. The magnet system is then sized around the salt pill. This magnet system, which consists of a magnet with low field to current ratio and a flux return shield, dominates the mass of the ADR system. For GGHS, the switching time is constrained by two factors, the “switch-off” time required for the charcoal to re-adsorb the helium gas and the “on-time” required to reject heat to the heat sink (cryocooler tip or the intercept stage cooler). The switch-off is typically 0.25 - 0.5 hours. To minimize the transient on the cryocooler, the on-time for the heat switch should be roughly the switch-off time which gives a minimum cycle time of 0.5 - 1 hour. For typical gas gap switches, the required on state conductance can be achieved easily. The Goddard group has operated a cADR with a minimum cycle time of 0.2 - 0.3 hour [5]. The magnetic heat switches tend to be faster that GGHS and are limited by the L/R time of the magnet and drive circuit or potentially, eddy current heating. The continuous ADR tested by Heer et al [6], using only SCHS, had a cycle time of 2 minutes. Regardless of the mechanism, cycling too fast results in extra heat load and the continuous cycle becomes unstable. There is also a choice to optimize the intercept temperature for minimum heat lift at the cryocooler heat sink into the heat or maximum hold time. As shown Fig. 8.3.5, optimizing for maximum hold time for a serially operated ADR yields the lowest cooler mass. However, the dispersion in mass between all options for a fixed cycle time is not large (~30%). For cost and system mass estimates, we use the mass for 1 hour cycle time and intercept optimized for maximum hold time.

The largest contributor to the cooler mass is that of the magnet and magnet shield. For our mass estimate studies, we considered a solenoid magnet wound on an aluminum spool with ferromagnetic shield to complete the magnetic flux circuit. The mass of each magnet and shield assembly depends primarily on the spacing of the wires and the desired field to current ratio. Stock commercial superconducting magnets are made with Nb-Ti wire spaced ~250 μm diameter Nb-Ti wire. Reducing the wire spacing increases the field to current ratio but increases the cost and risk of magnet failure to thermal and field cycling. With this wire winding, the magnet and shield assembly mass increases dramatically for field to current ratios in the range 0.3 - 0.5 T/A. For our mass estimate we use a current to field ratio of 0.4. The magnets are all mounted to the cryostat cold stage at < 2 K and shielded with Hiperco 50, a ferromagnetic material, sized to return all magnetic flux at peak field [10].

<u>8.3.1c Heat Straps</u>: The detectors in the EPIC focal plane are distributed over a large area ~100 cm in diameter. At sub-Kelvin temperatures, there are very few options to conduct heat from the focal planes to the point of cooling at the ADRs. The only liquids available are $^3$He or $^4$He, but no high TRL heat pipe using these fluids exists. Thus, the heat strap must be a highly conducting normal metal. The gradient, $dT = Q_i l/wt$, along a metallic strap of width w, thickness t, and conductivity $\kappa_0 T$, should be kept small $dT/T << 1$. Here $\kappa_0$ is constant and $Q_i$ is the power needed to be lifted by the cooler at the intercept (i = int) and detector (i = det) stages. These criteria fix the values of w and t. By substituting the $wt = m\rho l_j$, into the expression for the gradient and solving for m, we obtain an expression for the heat strap mass, m, in terms of fixed system parameters, $m = P_j\rho\ l_j^2/\kappa_0 TdT$. Here ρ is the density of the strap material. The mechanical design sets lengths of 30 cm for the heat straps to each stage each with a 1 % temperature gradient. At 100 mK, nearly all usable metals are superconducting or magnetic, so they have a low thermal conductivity or other undesirable properties such as permanent magnetic fields. This leaves copper (or beryllium) as the best available strap materials. We use $\kappa_0 = 1$ W/cm K for the strap material which is typical for moderately annealed, pure copper. The resulting total strap mass is shown in Table 8.3.1. Without the intercept stage, the power load at 100 mK would be ~10X higher. This would lead to heat straps to the detector stage approaching 10 kg.

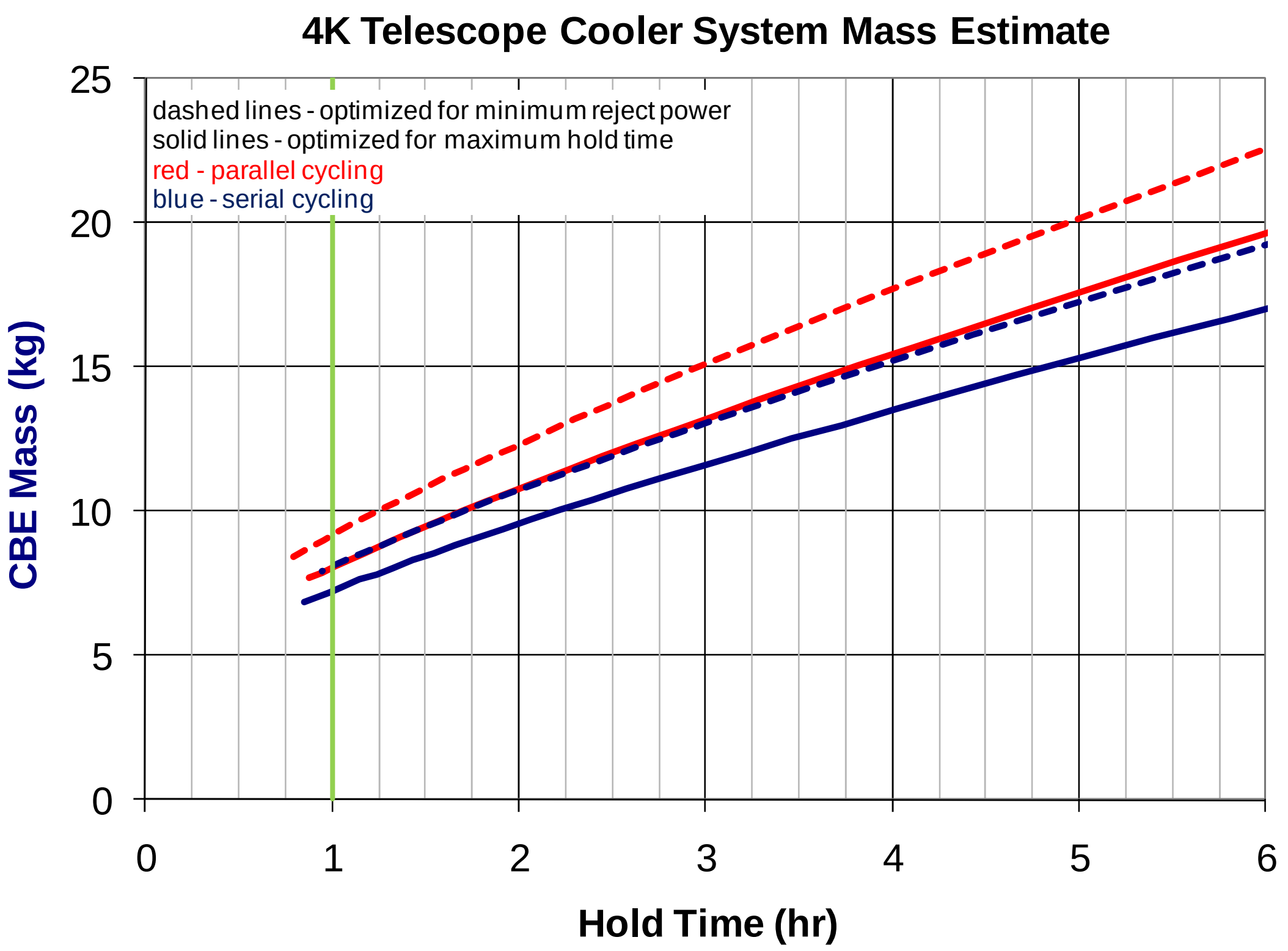

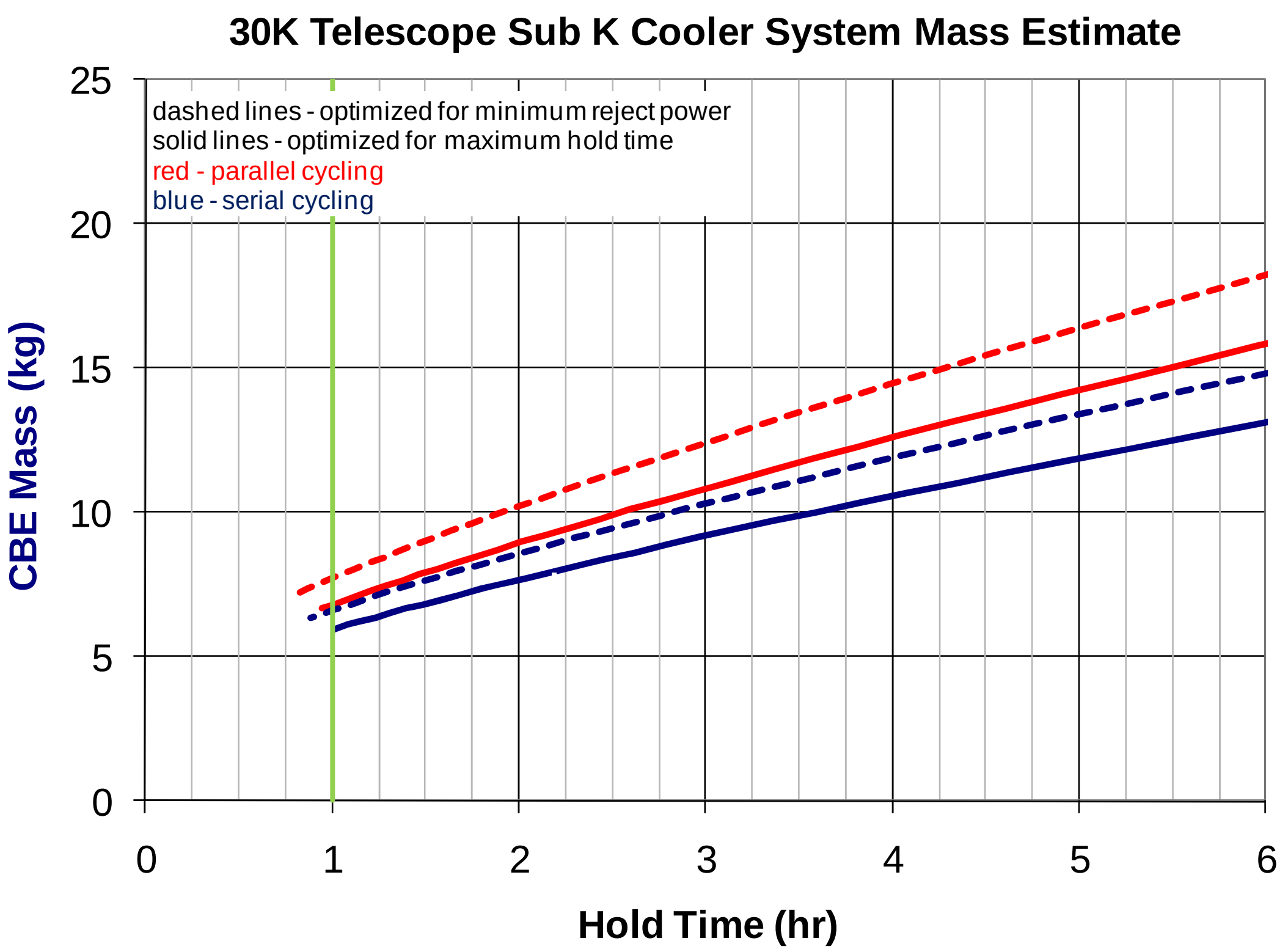

Figure 8.3.5 Mass as a function of hold time for the two stage cooler optimized for minimum heat lift at the cryocooler (dashed line) and maximum cycle time for parallel (red) and serial (blue) cycling strategies. For a cycle time of 1 hour, the minimum mass is for a serially cycled ADR optimized for maximum cycle time.

8.3.1d Magnet Leads: The magnet leads on the < 2 K cold stage are all superconducting Nb-Ti wired in parallel with high purity copper wire. AC losses in the magnet system are expected to be negligibly small [10]. Magnetic wiring along the cooling chain consists of high temperature superconductor (hiTc) wire such as the HTS Cryoblock wire available from American Superconductor [11] for stages ~80 K and normal resistive wire between high temperature stages and the spacecraft bus. The are used between the cold stage, to the first and second rigid V-groove. The second rigid V-groove is always below 100 K where the HiTc wire is well within the current rating rated at 10A with no resistive losses so there is little risk of the wire becoming resistively unstable. We use a power law fit to the thermal conductivity data of the HiTc wire, $\kappa = \kappa_0 T^\beta$ where $\kappa_0$ = 34 mW/cm $K^\beta$ and $\beta = 0.62$, a wire cross section of A = 0.0044 $cm^2$, and wire length ℓ ~ 100 cm along support struts in the thermal models of the cryostat and radiators. To facilitate ground testing with a 300 K vacuum shell and warm V-groove radiators, each HiTc wire is paired with a normal resistance wire, such as brass, that is sized to have the same conductive heat leak.

Normal resistive wiring for the magnet current leads is required from the second V-groove up to the spacecraft bus at ~300 K. The normal resistance wires are heat sunk at each thermal shield. The length of the wires is fixed by the spacecraft layout. The diameter of these wires is computed so that the average Joule heating per magnet cycle is equal to the parasitic heat leak. The Joule heating per cycle in each wire is, $P = \langle I^2\rangle_c R$, where $\langle I^2\rangle_c$ is the square of the magnet current averaged over one cycle, and R is the resistance of one of the two current leads driving a magnet. For typical low thermal conductance wires such as constantan or brass which have a resistivity nearly constant with temperature, the thermal conductance, $G = \kappa A/\ell$ can be approximated using the Wiedemann-Franz law G = LT/R, where L = 24.5 nW Ohm/$K^2$ is the Lorentz number. The total power into thermal stage j for N wires from thermal stage j+1 is $P_j = (NL/R)\,(T_{j+1}^2 - T_j^2)/2 + 1/2\, N\langle I^2\rangle_c R$, where 1/2 of the Joule heating power is conducted to each stage. The minimum $P_{ij} = 2\,(L\langle I^2\rangle_c(T_{j+1}^2 - T_j^2))^{1/2}$ for a pair of wires N = 2 to a single magnet is achieved for the optimum resistance $R = (L(T_{j+1}^2 - T_j^2)/\langle I^2\rangle_c)^{1/2}$ for each wire.

### *8.3.2 Closed-Cycle Planck Dilution Cooler*

At temperatures below the tri-critical mixing temperature < 0.86 K, mixtures of $^3He$ and $^4He$ phase separate into $^3He$ rich and $^4He$ rich ($^3He$ dilute) phases. At T = 0 K, the $^4He$ rich phase has a $^3He$ content of ~6.3%. At the interface between the two phases, driving flow of $^3He$, $dn_3/dt$, from the $^3He$ rich phase to the $^4He$ rich phase gives $dQ/dt = 84\, T^2\, dn_3/dt$ of cooling power dQ/dt at temperature T. For ground based dilution refrigerators, the $^3He$ rich phase floats on the $^4He$ rich phase which defines the interface and thus only works in gravity (or some acceleration of the fluids). A novel design of a dilution refrigerator, the open cycle dilution refrigerator (or Planck dilution cooler) [3] shown in Fig. 8.3.6 and described in Table 8.3.2, will be flown on the Planck spacecraft and was used for on the balloon borne Archeops telescope to cool bolometric detectors. It works by pumping precooled streams of pure $^3He$ and pure $^4He$ into a tubular mixing chamber heat exchanger. On combination, the $^3He$ fraction forms bubbles in a stream of $^4He$ which rapidly saturates with $^3He$ causing cooling at the bubble interfaces. The interface is stabilized by the laminar flow of the fluid through the tube and thus works without gravity [3] so it is suitable for cooling in space borne instruments. The measured cooling power, $dQ/dt = 33.6\, T^2\, dn_3/dt$, for the Planck dilution cooler is less than the 'ideal' case realized on the ground. The Planck dilution cooler flows 4X the necessary rate of $^3He$ (>6.9%) than needed to provide the cooling at $T_d$. This excess of $^3He$ defines bubbles and hence the interface to stabilize the flow and the cooling. The excess of $^3He$ flow produces an additional benefit. As the flowing helium mixture warms, the excess $^3He$ dissolves into the $^4He/^3He$ mixture yielding additional cooling $dQ_i$ at the intermediate stage(s) at temperature $T_i$ . For intercepts at temperature below ~350 mK, the flow of $^3He$ into the pure $^4He$, in fact, gives a cooling power that is larger than evaporative cooling of pure $^3He$. [15]

As shown in Table 8.3.2, the amount of helium required to run the cooler open cycle is large for Planck and is prohibitive for missions such as EPIC. A modification of the Planck dilution cooler to form a closed cycle system was recently demonstrated in the lab by Benoit and co-workers. Closing the cycle of the cooler allows much larger cooling power and can be used indefinitely. It requires a superfluid pump at < 2 K that purifies $^4He$ from the mixture and provides a small ~0.3 bar pressure head to force flow. The unpurified, $^3He$, rich stream is compressed using a pump very similar to those used for standard Joule-Thompson cryocoolers. The measured performance at 100 mK, $dQ/dT \sim 5\ \mu W$, and ultimate temperature of 39 mK, with ~5 mW of lift required at 1.7 K nearly meets the required cooling needs for EPIC as listed in Table 8.3.2. The heat load at the detector stage is lower than the ADR system since no heat switches are required.

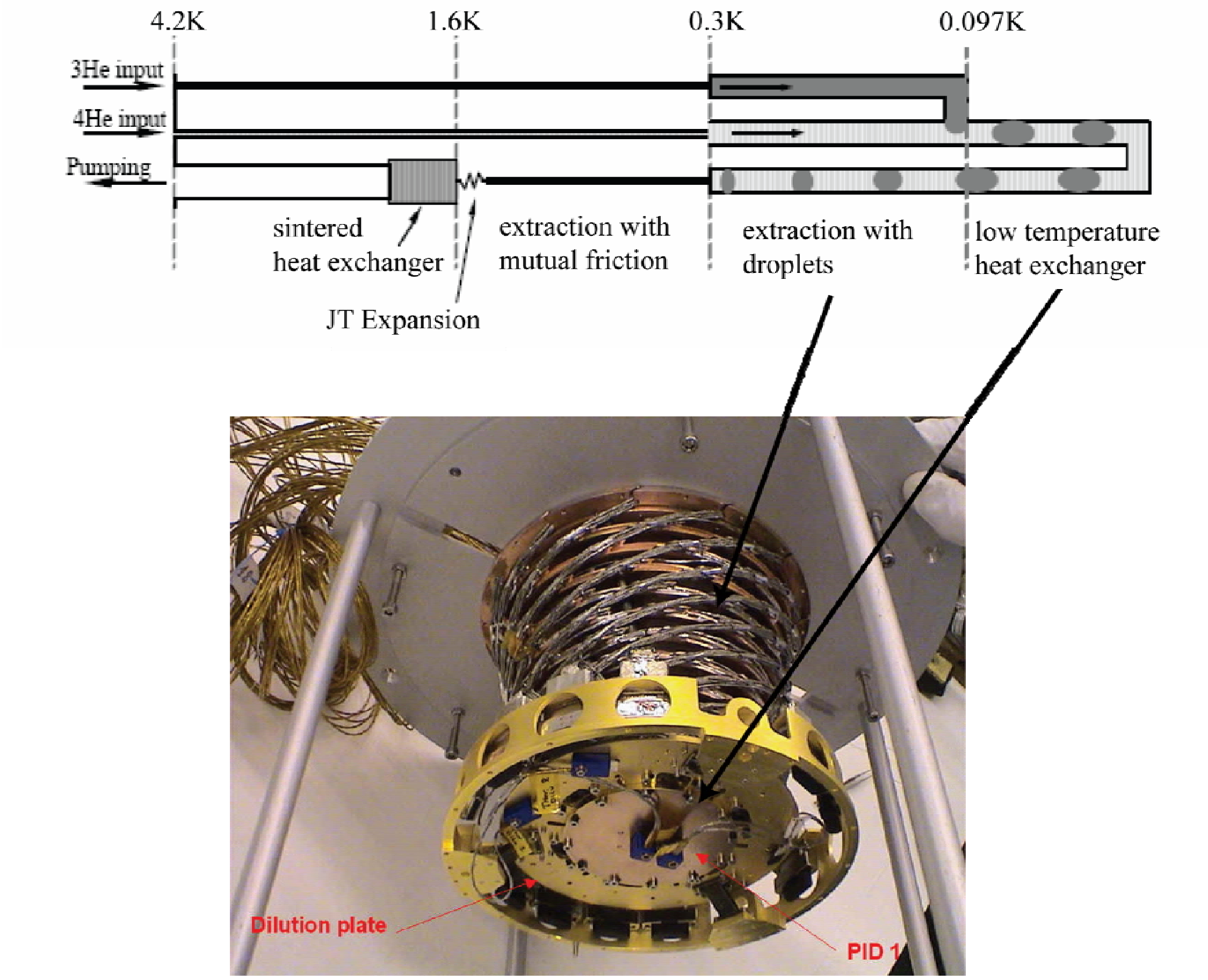

Fig. 8.3.6 Schematic of the open cycle dilution refrigerator (top) built for cooling the bolometric detectors on Planck (bottom). The bolometers are cooled to <100 mK at the low temperature heat exchanger plate. The coiled tube and copper structure is a multistage heat exchanger that extracts heat as the excess $^3He$ is diluted at the helium stream warms.

The Planck cooler, shown in Fig. 8.3.6, and closed cycle lab cooler, have many intercept stages to precool the wiring and struts. We have made a simpler model for EPIC using only one intercept stage. We have calculate the $dn_3/dt$ and $dn_4/dt$, needed to cool the EPIC detector arrays to $T_d$ = 100 mK and maintain the intercept temperature and then scaled from the lab tests of the closed cycle Planck dilution cooler, the values of heat lift needed at 1.7 K needed to accommodate the total flow of $^4$He. The condensation of the circulating $^3$He was accomplished, in the lab test, using JT expansion. Operational parameters for a single system that cools all detectors for the 4 K and 30 K telescope options are given in Table 8.3.2. For the open cycle Planck cooler, *excess* cooling power is provided by a Joule Thompson expansion valve (JT valve) on the output stream of the Planck 100 mK cooler, so cooling at 1.7 K is listed as a negative number. The required heat lift at 1.7 K is within the measured capability of the Japanese coolers[16] and expected for extension of coolers made in the US[17].

**Table 8.3.2.** Continuous Flow Dilution Cooler Parameters

| | **Units** | **4 K Telescope** | **30 K Telescope** | **Planck** |
|---|---|---|---|---|
| Intercept Temperature | mK | 145 | 180 | ~300 |
| Detector Stage Dissipation | μW | 4.6 | 2.5 | <0.1 |
| $dn_3/dt$ | μmole/s | 15 | 10 | 6.7 |
| $dn_4/dt$ | μmole/s | 200 | 110 | 20 |
| Cooling at 1.7 K | mW | 4.7 | 2.6 | -0.2 |
| $^3$He per year if open cycle | ℓ(STP) | 10550 | 6870 | 4730 |

Comparison of a continuous flow dilution cooler estimated for EPIC 4 K telescope and 30 K telescope at 100 mK to the baseline operation of the flight cooler installed in Planck.

### *8.3.3 Temperature Stabilization*

Launched in May 2009, Planck will be the first space-borne instrument utilizing only cryocoolers to achieve sub-Kelvin temperatures. The cooling chain, Fig. 8.3.7 consists of V-groove radiators cooling to ~50 K, the JPL-built hydrogen sorption cooler[18] cooling to <19K, a Rutherford Appleton Laboratory (RAL) stirling cycle cooler cooling to ~4.5 K and the open cycle dilution cooler [3] made by Air Liquide in France cooling to 100 mK. Planck will be launched 'warm', cools radiatively and then the active coolers will be started sequentially, warmest first, during cruise to the L2 lagrange point. Temperature stability achieved at 4 K and 100 mK in the full spacecraft system tests at the Centre Spatial de Liège (CSL) facility in Liege, Belgium, and shown in Fig. 8.3.8, are remarkable. Furthermore, the measured bolometer noise during the CSL tests is at the design value and is free from interference and low frequency $1/f^{\beta}$ noise over the signal bandwidth 0.016 – 40 Hz.

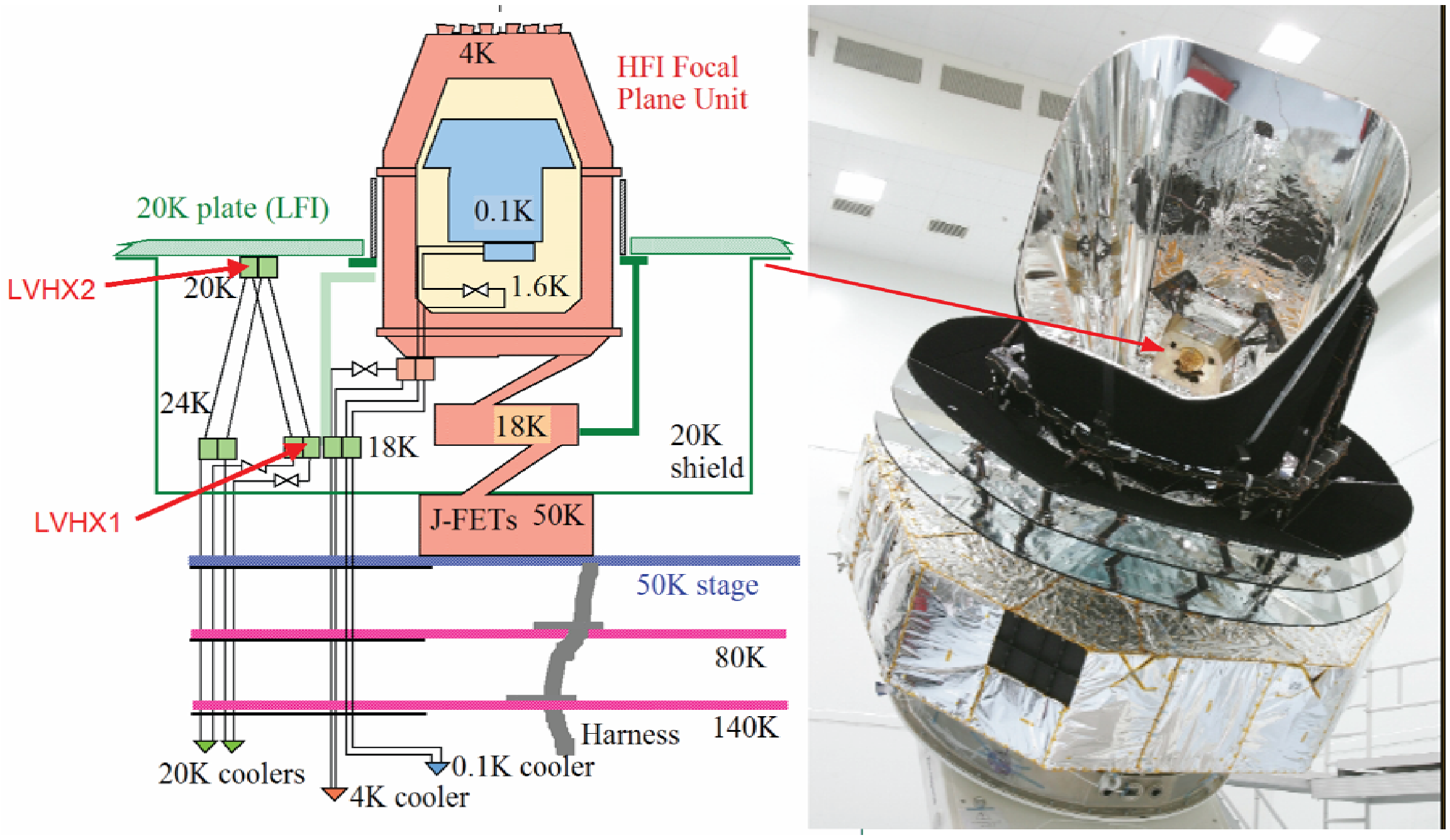

Fig. 8.3.7 Planck cooling chain (left) and picture of the completed Planck spacecraft (right).

The thermal control at the bolometer plate at 100 mK, Fig. 8.3.9, was achieved using both active control, passive thermal filtering, and surrounding the 100 mK plate with a thermally controlled shield. Mitigation of electromagnetic interference was carefully designed and tested to prevent RF heating [19]. For EPIC, all of these techniques will need to be used. The ADR salt pills serve as one source of 'high heat capacity material' at the point of cooling. The finite G of the heat straps with the heat capacity of the focal plane form a thermal filter with a time constant of 10's of seconds for the detector stage and 100's of seconds for the intercept stage thus active control will be needed on these stages. Temperature control of the intercept stage provides indirect single point control similar to the control of the dilution plate in HFI. Finally, the large extent of the focal plane will require control of gradients. When PID2, which was mounted near the center of the focal plane as indicated in Fig. 8.3.9, was used in Planck, it controlled temperature only at the thermometer. In fact, even with thick copper on the focal plane, time varying temperature gradients were induced, as measured by other thermometers on the focal plane, which caused significant drifts in the bolometer output. Control of these gradients by more careful design, including more detailed modeling of the heat straps and use of distributed heaters (or a heater "tape") around the exterior of the focal plane, and distributed high heat capacity materials will be needed. An option worth further investigation is the use of NIS junctions on the not as the main cooler[20,21] but for temperature control to source (as a heater) and sink (as a cooler) heat in zones where it is needed on individual focal plane detector tiles.

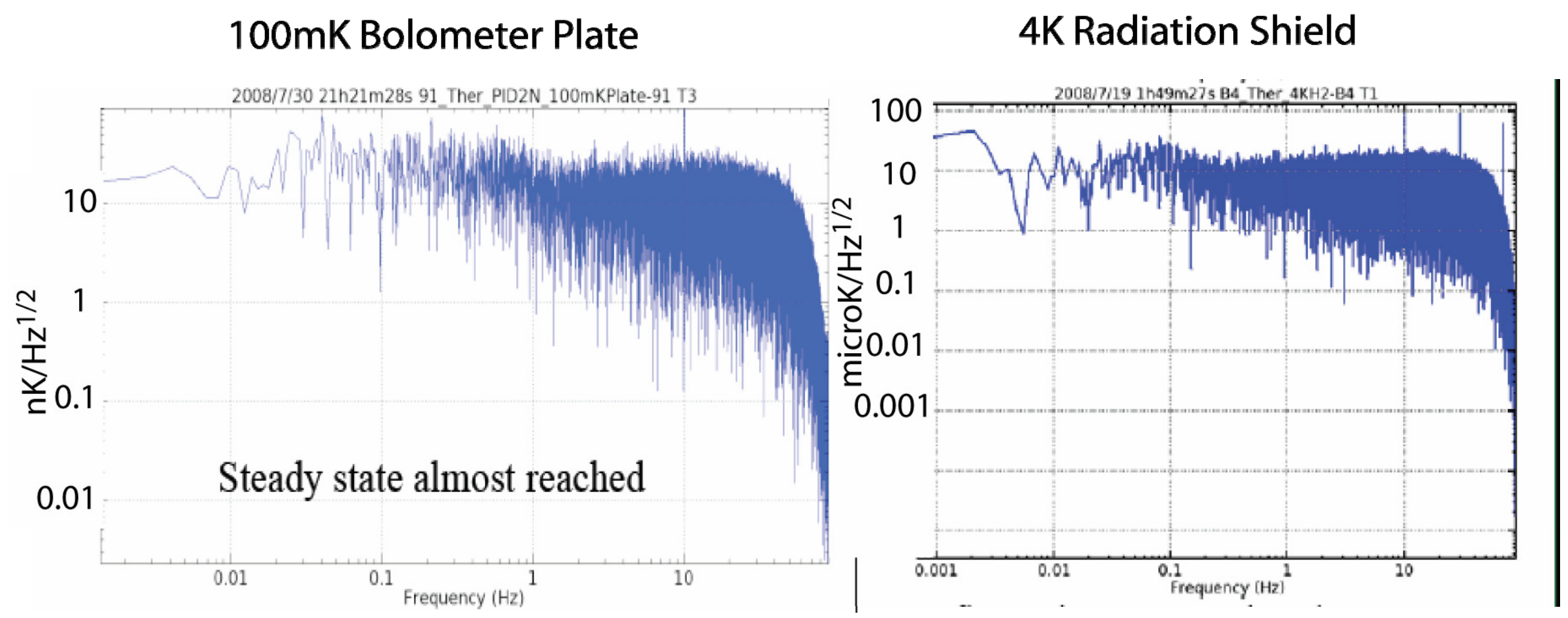

Fig. 8.3.8 Temperature stability achieved at 4 K and 100 mK the Planck spacecraft system tests at CSL. Graphs are courtesy of M. Piat, C. Leroy and J.L. Puget.

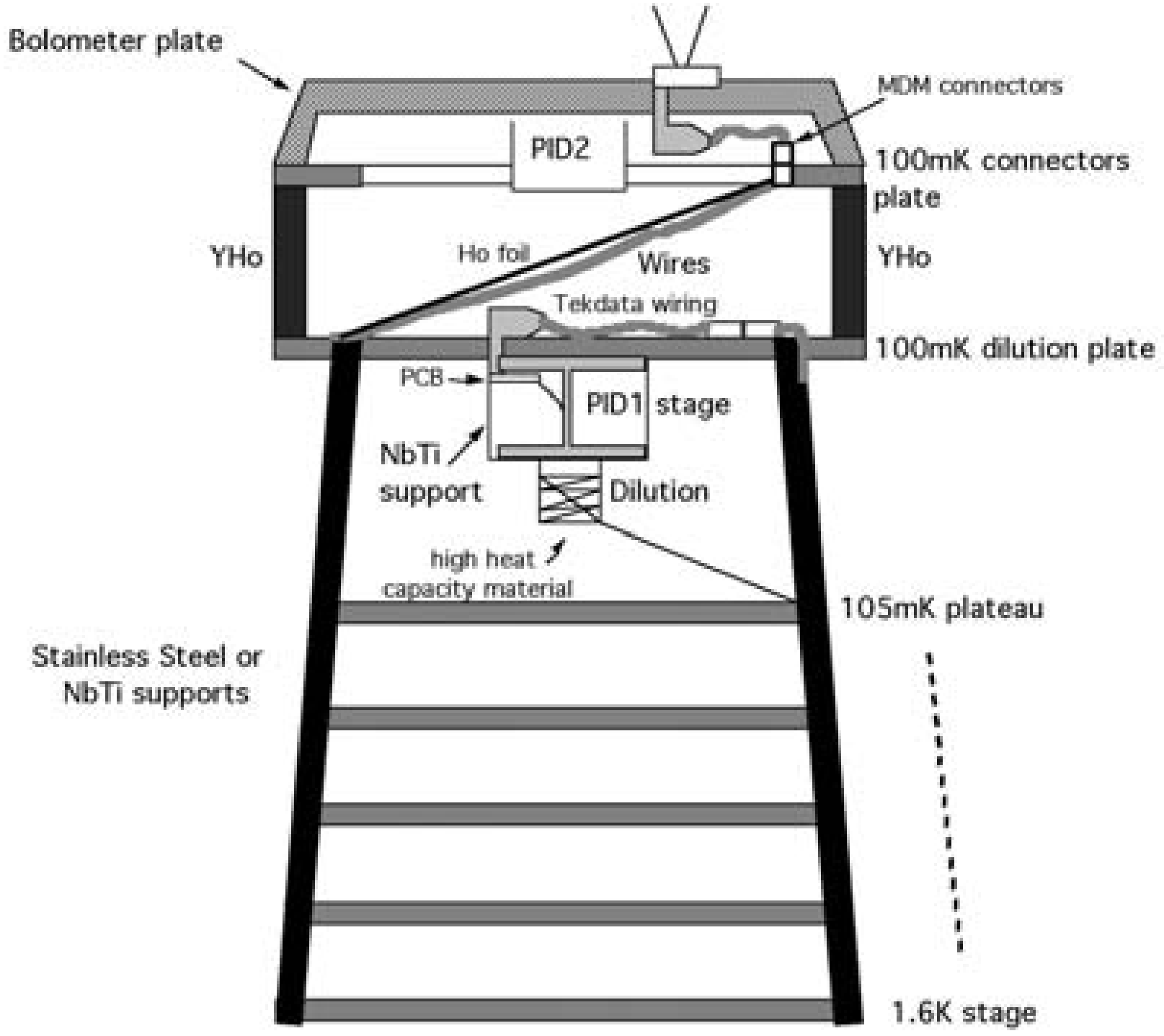

Fig. 8.3.9 Thermal architecture for the Planck HFI bolometers. The Planck HFI plate is isolated from the dilution plate by a high heat capacity alloy, HoY. The bolometer plate could also be controlled directly using PID2. In practice PID2 is not used since, the bolometer plate temperature is sufficiently stabilized by the combination of PID1, the thermal filter response of the HoY and control of ambient thermal radiation by temperature controlled 4K box that completely surrounds the components at 100 mK. (from C. Leroy, Ph.D Thesis, Toulouse, France).

# 9. Deployed Sunshade

This section builds upon the previous EPIC study [1] that focused on the development of a deployable, three-level sunshade. In that report, a sunshade was designed for a EPIC-LC refractor telescope design. Because the sunshade required for the low-cost refractor concept was much smaller, a folding-strut design was adopted in lieu of the original wrapped-rip deployment concept. The folding-strut concept is much simpler, but has a limit to the diameter of shade it can accommodate inside a launch vehicle. This current effort deals with a telescope configuration that requires a medium size sunshade diameter that can be accommodated in a launch fairing, and thus its deployment system is based on the folding-strut concept. In this section we describe a new deployment concept for the sunshade and present a mass estimate for this structural approach. In addition, this report includes a description and mass estimate for a non-deployed optics shield that serves as a radiation shield for the telescope optics.

## 9.1 Sunshade Requirements

The EPIC-IM crossed Dragone telescope is sized for launch in an Atlas rocket with a 4-m Extended Payload Fairing (EPF). While in a halo orbit at L2, the spacecraft will rotate at 0.5 rpm about its central axis. To prevent mechanical disturbances resulting from this rotation, the lowest natural frequency of the sunshade in the plane of rotation should be at least 10 times the rotation rate. Thus, the first mode should occur above 0.083 Hz. In addition to this frequency requirement, the load carrying members of the sunshade must have an acceptable factor of safety against buckling and material strength failure. In order to fit the sunshade support struts inside the launch vehicle, they are bent inwards towards the telescope when stowed. Because of this bend and the three layers of tensioned sunshade membrane material, the struts are subjected to bending stresses that dominate their design compared to the compressive loading. The layers of the sunshade are separated using a catenary system instead of rigid spreader bars. Also, there are two sunshade configurations under consideration in this study: the '4 K telescope' option requiring a 4-layer sunshade and the '30 K telescope' option requiring a 3-layer sunshade.

The optics shield (aka the 'optics tent') is a fixed structure that must have a shape to enclose the telescope optics with aluminized Kapton, survive launch loads, and not obscure the view of the telescope.

## 9.2 Sunshade Description

Deployed Configuration**:** The deployed 4 K telescope sunshade is shown in Fig. 9.1, along with the optics tent. The sunshade for the 4 K telescope is comprised of four dual layers of aluminized Kapton thin film that extends out from a central, rigid, four-layer aluminum honeycomb V-groove radiator. The bottom layer of the radiator is only a flat interface plate for connecting the sunshade assembly to the top deck of the spacecraft. The optics tent consists of an aluminum circular cross-section tube frame that is bolted together in a manner similar to the Aquarius sunshade. The side walls and six bottom portions are made of aluminized Kapton film that is attached to the frame using traditional lacing techniques. The aluminum frame is robust yet light, and can easily support the kevlar spreader cables, as discussed below. The 30 K telescope sunshade design is the same, except the inner-most dual-layer sunshade and its V-groove radiator have been removed.

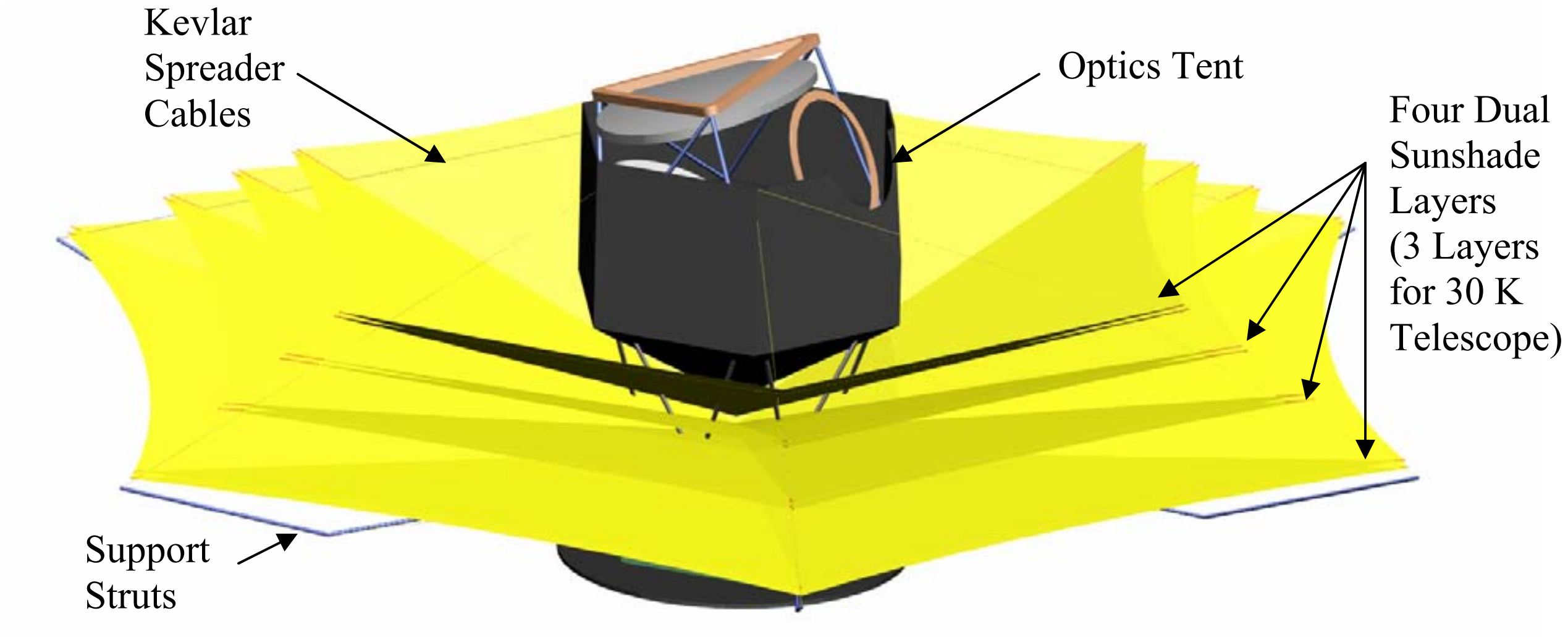

Fig. 9.1. Deployed 4 K telescope sunshade with optics tent (central V-groove radiators not visible).

From Fig. 9.1, the deployed sunshade is supported from below by six deployable struts whose tips are bent upwards. A Kevlar cable extends from the tip of each of these struts to connection points on the optics tent. These cables pass through and are connected to the sunshade membrane material as shown in detail in Fig. 9.2. The Kevlar cables must pass through the sunshade material as such for several reasons. First, the corner of the sunshade in line with the telescope aperture requires a "Y" shaped Kevlar cable in order to not obscure the aperture. This angle is kept constant for the other five cable locations for simplicity and symmetry. Also, the height of the support points on the optics tent (and support struts) is limited by the telescope optics and the fairing height. During the deployment process, described in detail below, this Kevlar cable becomes taught and establishes the correct separation distance between the sunshade layers. The short compression members shown in Fig. 9.2 ensure that the tension in the cable adequately stretches out each membrane layer from its tips.

Stowage and Deployment: The on-orbit, petal-like deployment of the EPIC sunshade shown in Fig. 9.1 is based on six circular cross-section struts hinged at their base and stowed vertically inside the launch vehicle fairing as shown in Fig. 9.3. Locating the struts on the outside of the entire mass of sunshade membrane material makes for easier venting of the Kapton during launch and control during deployment on-orbit.

For this design, base-hinged, thin-walled round tube struts are used as a baseline for their simplicity and structural efficiency. However, it may make sense to use an elliptical or other directionally-preferred cross-section due to the large bending loads carried by the struts. The sunshade will be delivered to I & T as a completely assembled structure. The bottom V-groove is a flat interface plate to the top deck of the spacecraft, and the support ring legs secure the deployable struts to the spacecraft. To ensure stability during launch, the struts and layers of membrane films are secured in place during launch with a series of cords that can be automatically cut after launch when the sunshade is ready to deploy.

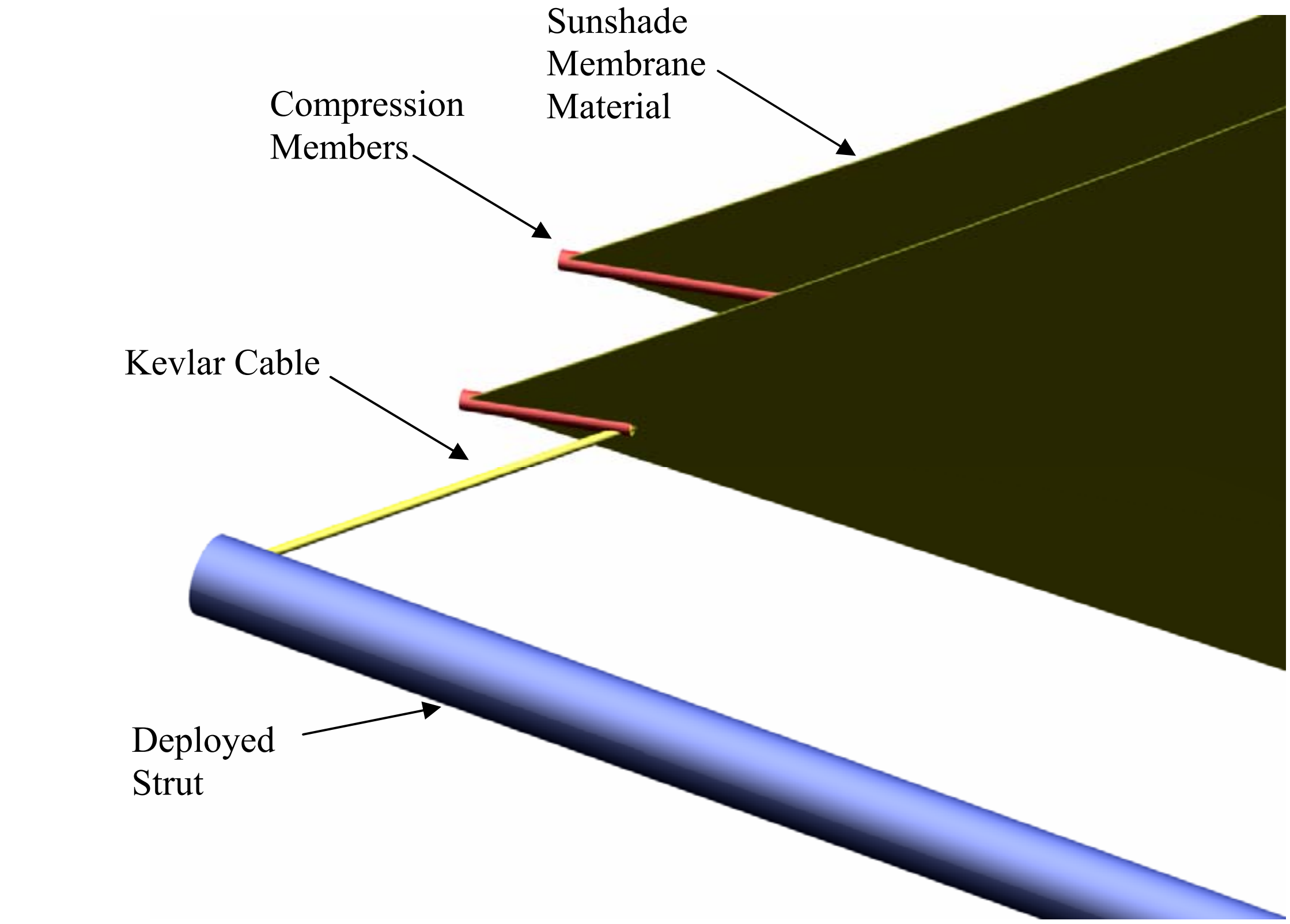

Fig 9.2. Detail view showing the Kevlar cord passing through the sunshade membrane layers towards the support points on the optics tent.

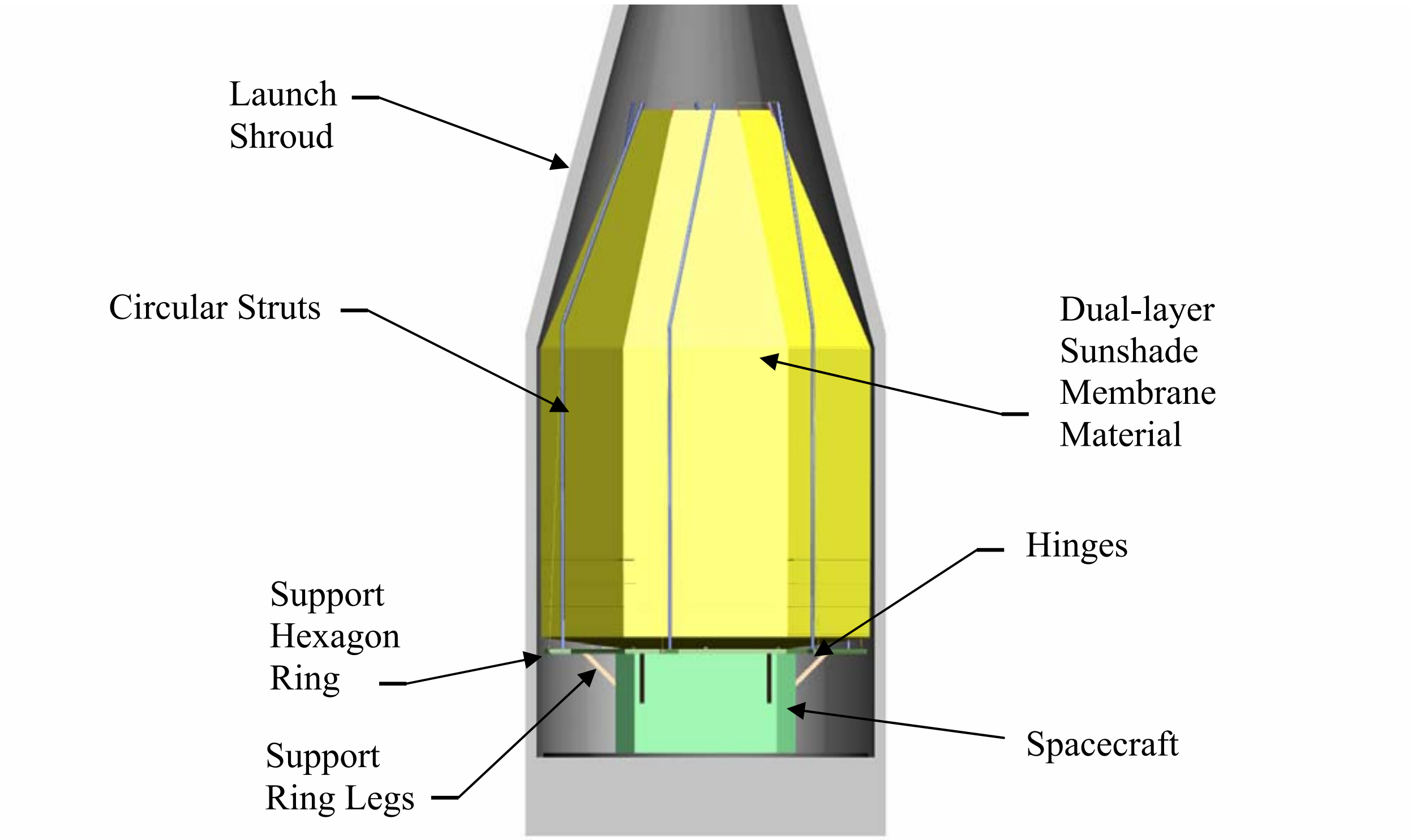

Fig. 9.3. Stowed configuration of the sunshade inside the Atlas V 401 launch shroud.

The deployment of the sunshade on orbit is controlled by a single electric motor-driven winch system mounted on the spacecraft interface. When activated, this winch will retract six cables that run along the length of each of the six deployable struts to connection points near their elbow. These cables will cause the struts to rotate about their hinged ends. To gain a bit of mechanical advantage and thus use a smaller motor, a small stand-off is included near the hinged base of the strut as shown in Fig. 9.4. The hinges will have a latching mechanism to lock the struts in their deployed configuration and thus not rely on the motor/winch to remain operational to keep the sunshade deployed during the entire mission.

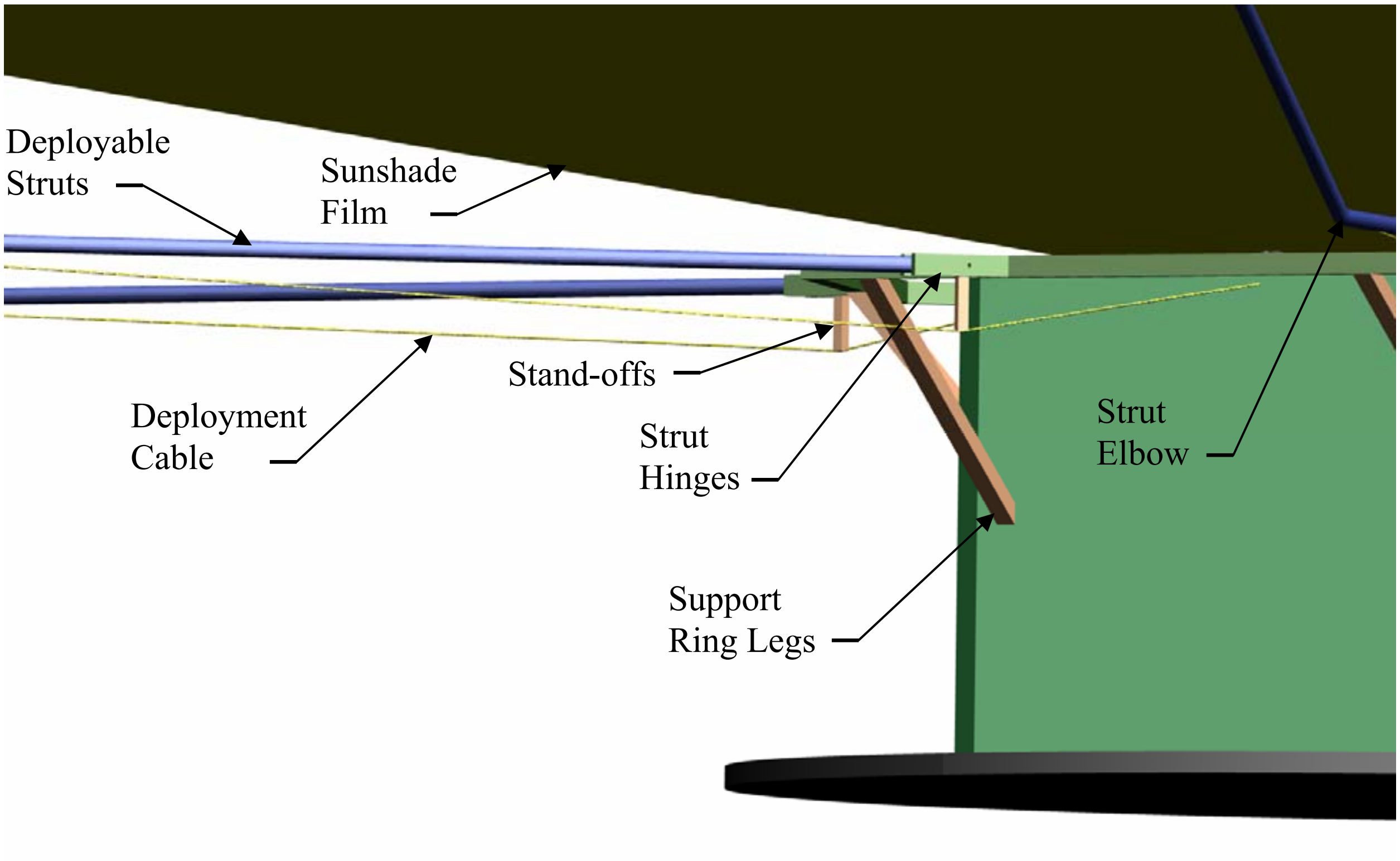

Fig. 9.4. Deployment cables along the struts, over stand-offs, and anchoring at the spacecraft interface.

The series of pictures in Fig. 9.5 illustrates the planned deployment motion of the sunshade. Fig. 9.5(a) depicts the sunshade fully stowed. When it is time to deploy the sunshade on-orbit, the cables restraining the deployable struts will be cut. In Figs. 9.5(b) and 9.5(c), the motor/winch system is engaged, causing the struts to deploy in a synchronized, controlled manner. In Fig. 9.5(d), the Kevlar spreader cable is beginning to tighten and thus the four layers of the sunshade are starting to separate and assume their intended shape. In Fig. 9.5(e), the struts are fully deployed in their straight configuration, and each of the six base hinges have locked into place. At this time the motor/winch system can be permanently powered off. Also, the Kevlar cables are fully taught, causing the Kapton membrane material of the sunshade to be properly tensioned to ensure a reasonably flat, smooth surface.

### 9.3 Technical Specifications

We provide a brief description of the analyses used to determine the materials, sizes, and masses of key components of the deployable sunshade.

**Fig 9.5. EPIC INTERMEDIATE MISSION: SUNSHIELD DEPLOYMENT**

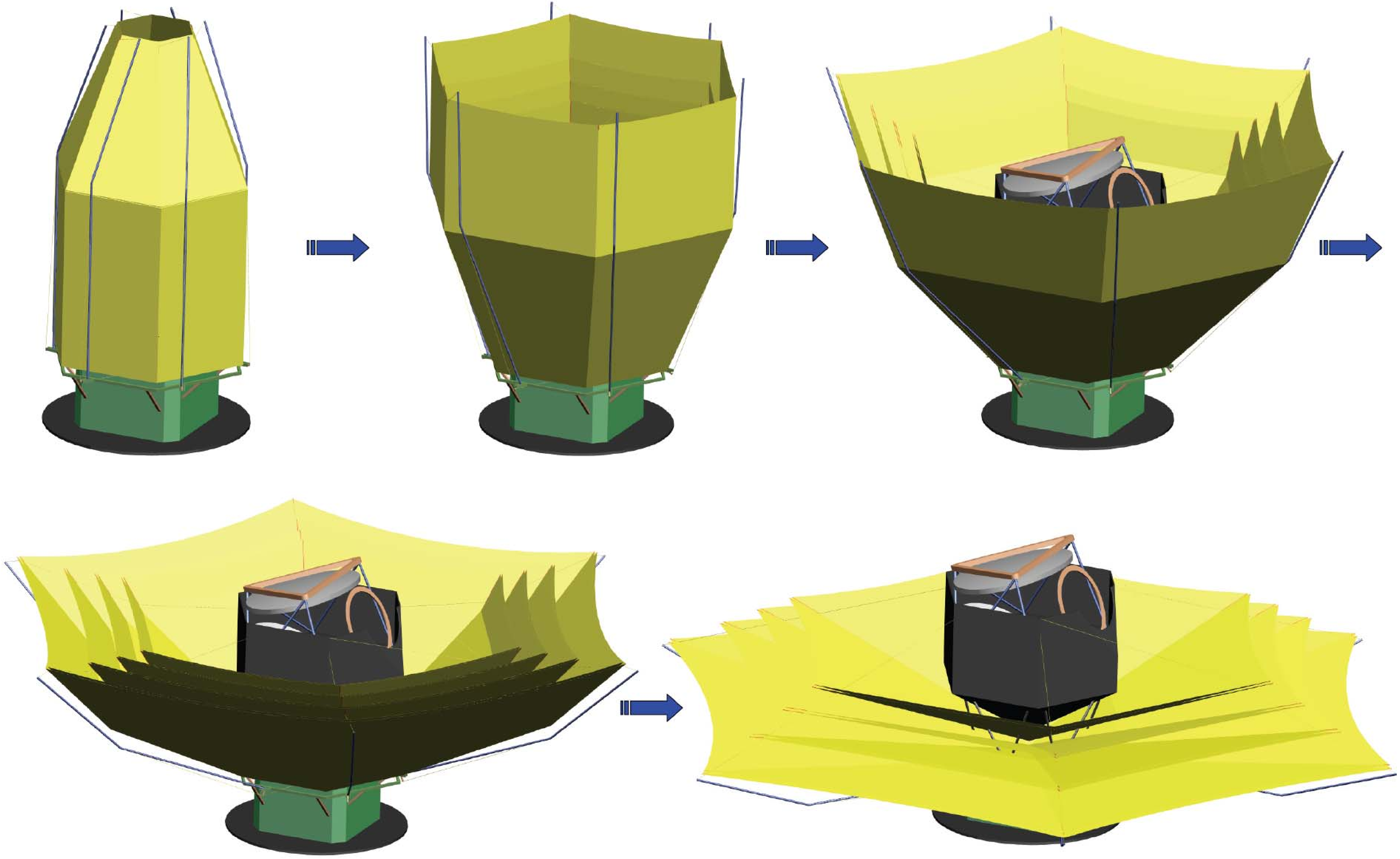

Structural Sizing – Sunshade**:** The hexagonal ring was sized in the same scaling manner as above for the optics tent horizontal members. This exercise indicates that a 38.1 mm square tube with 3.175 mm thick walls will adequately resist launch and on-orbit deployment loads. The same cross-section is assumed for the support ring legs.

In order to save mass, it is presumed that the deployable struts will be made of space-rated, low CTE, carbon fiber-reinforced composite material. Once deployed, the struts are loaded in compression and bending due to the tension developed in the Kapton membrane layers. This tension loading is transferred to the tip of the deployable struts through flexible Kapton cables, rather than through a rigid spreader bar, thus a new catenary-type of analysis was performed. It becomes quickly apparent that the bending stresses in the struts because of the elbow bend and lengthy outer portion dominate the design, over buckling and pure compressive stresses. A conservative design requires the struts be circular tubes with a 40 mm OD, 34 mm ID cross-section. Such a cross section ensures a natural frequency well in excess of the required 0.083 Hz based on inspection and simple calculation.

Structural Sizing – Optics Tent: The optics tent was scaled from the Aquarius sunshade, which was constructed from aluminum thin-walled tubes with custom end fittings that are used to bolt the structure together. The aluminum tubes in the optics tent in a plane perpendicular to the launch axis are about twice as long as those on the Aquarius sunshade, thus the launch loads will be about 4 times greater (assuming a similar MAC curve for the launch vehicle for EPIC). In order to maintain similar structural margins for the optics tent members, they should have approximately 4 times the moment of inertia as those on Aquarius. Assuming the thickness does not change, this scaling exercise gives a 50 mm OD, 48.57 mm ID circular tube cross-section. As mentioned earlier, the walls of the optics tent are aluminized Kapton attached using lacing techniques used for thermal blanketing.

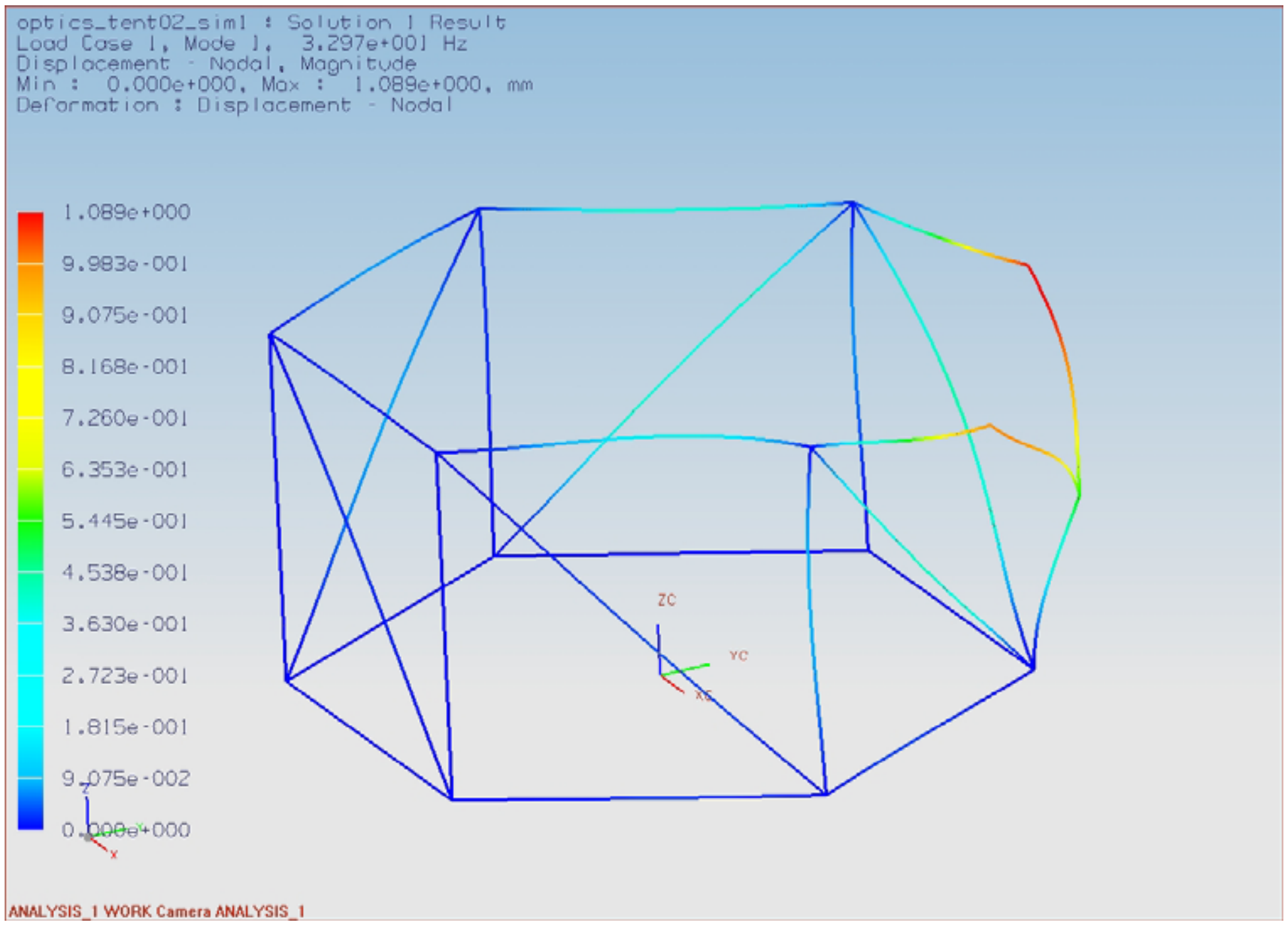

Fig. 9.6. First mode shape of the optics tent frame.

Once the structural members were sized for strength based on launch loads, a simple FEM of the entire optics tent structure was created to determine the first mode and ensure it was above the required 0.083 Hz. The first mode shape, which occurs at a frequency of 33 Hz is shown in Fig. 9.6. The addition of the diagonal elements on the sides of the frame greatly increases its stiffness, raising the frequency to 33 Hz from 10.8 Hz. Clearly, the structural design of the optics tent is driven by strength and not dynamics. The addition of non-structural film mass will have a very minor effect on the dynamics of the structure and will continue to be neglected for dynamic analysis.

**Table 9.1.** Mass estimate for the Optics Tent and Sunshade for 4 K Telescope Option

| Optics Tent | | | | | |
|---|---|---|---|---|---|
| **Item** | **Qty** | **area, m$^2$** | **Areal Density, kg/m$^2$** | **Total Mass, kg** | **Notes** |
| Sides, complete | 4 | 3.24 | 0.09 | 1.17 | Dual Layer Aluminized Kapton |
| Sides, w/hole | 2 | 2.71 | 0.09 | 0.49 | Dual Layer Aluminized Kapton |
| Side, angled | 1 | 0.37 | 0.09 | 0.03 | Dual Layer Aluminized Kapton |
| Bottom Panels | 6 | 1.73 | 0.09 | 0.93 | Dual Layer Aluminized Kapton |
| Frame joints/fasteners | 15 | - | - | 7.50 | 0.5kg each, Alum. scaled from AQ sunshade |
| **Item** | **Total Length, m** | **Cross section area, m$^2$** | **Density, kg/m$^3$** | **Total Mass, kg** | |
| Frame | 60 | 0.0001107 | 2710 | 18.00 | Alum, 33 Hz 1st mode (with diagonals), 1.35 Hz 1st local mode, scaled from AQ with same thickness |
| | | Optics Tent Total Mass | | 28.12 | kg |

| Sunshade | | | | | |
|---|---|---|---|---|---|
| **Item** | **Qty** | **area, m$^2$** | **Areal Density, kg/m$^2$** | **Total Mass, kg** | **Notes** |
| Sunshade 1 (Bottom) | 1 | 87.80 | 0.09 | 7.90 | Dual Layer Aluminized Kapton |
| Sunshade 2 | 1 | 63.05 | 0.09 | 5.67 | Dual Layer Aluminized Kapton |
| Sunshade 3 | 1 | 51.34 | 0.09 | 4.62 | Dual Layer Aluminized Kapton |
| Sunshade 4 (top) | 1 | 41.26 | 0.09 | 3.71 | Dual Layer Aluminized Kapton |
| V-Groove Radiator | 3 | 9.12 | 4.5 | 123.13 | Aluminum honeycomb |
| Bottom Interface Plate | 1 | 2.20 | 4.5 | 9.92 | Aluminum honeycomb |
| Hinges | 6 | - | - | 24.00 | 4kg each, wag from AQ hinges |
| Round Struts | 6 | - | - | 18.00 | 3 kg each (gr), 40mmOD, 3mm wall |
| **Item** | **Total Length, m** | **Cross section area, m$^2$** | **Density, kg/m$^3$** | **Total Mass, kg** | |
| Hexagon Frame | 10.74 | 4.44E-04 | 2710 | 12.91 | Alum. 1.5" sq tube with 1/8" wall |
| Frame Legs | 5.212 | 4.44E-04 | 2710 | 6.26 | Alum. 1.5" sq tube with 1/8" wall |
| Kevlar Cord | 30 | 7.92E-06 | 1440 | 0.34 | 1/8" dia cable |
| Compression Struts | 18.834 | 5.03E-05 | 1522 | 1.44 | graphite, 10mm OD, 2mm thick |
| Deployment motor/winch | 1 | - | - | 10.00 | wag |
| | Sunshade and V-Groove Total Mass | | | 227.91 | kg |

Mass Estimates: Based on the designed deployed geometry and the properties of the selected materials, mass estimates are given in Table 1 for the Optics Tent and the 14.8 m diameter sunshade for the 4 K Telescope option. Table 2 provides the same information for the 30 K telescope option, which removes the top layer of sunshade and its V-groove radiator panel, thus reducing the mass a bit. The Optics Tent remains unchanged conceptually for either option.

**Table 9.2.** Mass estimate for the Optics Tent and Sunshade for 30 K Telescope Option

| Optics Tent | | | | | |
|---|---|---|---|---|---|
| **Item** | **Qty** | **area, $m^2$** | **Areal Density, $kg/m^2$** | **Total Mass, kg** | **Notes** |
| Sides, complete | 4 | 3.240 | 0.09 | 1.17 | Dual Layer Aluminized Kapton |
| Sides, w/hole | 2 | 2.711 | 0.09 | 0.49 | Dual Layer Aluminized Kapton |
| Side, angled | 1 | 0.365 | 0.09 | 0.03 | Dual Layer Aluminized Kapton |
| Bottom Panels | 6 | 1.728 | 0.09 | 0.93 | Dual Layer Aluminized Kapton |
| Frame joints/fasteners | 15 | - | - | 7.50 | 0.5kg each, Alum. scaled from AQ sunshade |
| **Item** | **Total Length, m** | **Cross section area, $m^2$** | **Density, $kg/m^3$** | **Total Mass, kg** | |
| Frame | 60 | 1.11E-04 | 2710 | 18.00 | Aluminum, 33 Hz 1st mode (with diagonals), 1.35 Hz 1st local mode, scaled from AQ with same thickness |
| | | | Optics Tent Total Mass | 28.12 | kg |

| Sunshade | | | | | |
|---|---|---|---|---|---|
| **Item** | **Qty** | **area, $m^2$** | **Areal Density, $kg/m^2$** | **Total Mass, kg** | **Notes** |
| Sunshade 1 (Bottom) | 1 | 87.80 | 0.09 | 7.90 | Dual Layer Aluminized Kapton |
| Sunshade 2 | 1 | 63.05 | 0.09 | 5.67 | Dual Layer Aluminized Kapton |
| Sunshade 3 | 1 | 51.34 | 0.09 | 4.62 | Dual Layer Aluminized Kapton |
| V-Groove Radiator | 2 | 9.12 | 4.5 | 82.08 | Aluminum honeycomb |
| Bottom Interface Plate | 1 | 2.20 | 4.5 | 9.92 | Aluminum honeycomb |
| Hinges | 6 | - | - | 24.00 | 4kg each, wag from AQ hinges |
| Round Struts | 6 | - | - | 18.00 | 3 kg each (gr), 40mmOD, 3mm thic |
| **Item** | **Total Length, m** | **Cross section area, $m^2$** | **Density, $kg/m^3$** | **Total Mass, kg** | |
| Hexagon Frame | 10.74 | 4.44E-04 | 2710 | 12.91 | Alum, 1.5" sq tube with 1/8" wall |
| Frame Legs | 5.212 | 4.44E-04 | 2710 | 6.26 | Alum, 1.5" sq tube with 1/8" wall |
| Kevlar Cord | 30 | 7.92E-06 | 1440 | 0.34 | 1/8" dia cable |
| Compression Struts | 18.834 | 5.03E-05 | 1522 | 1.44 | graphite, 10mm OD, 2mm thick |
| Deployment motor system | 1 | - | - | 10.00 | wag |
| | | | Sunshade and V-Groove Total Mass | 183.15 | kg |

## 10. Instrument Cost and Schedule

The EPIC instrument cost estimate presented here was developed using a grassroots methodology which requires a detailed Work Breakdown Structure (WBS), a detailed development schedule, and a well developed integration and test flow for the instrument itself and with the spacecraft. For purposes of cost and schedule estimation, we assumed the focal plane consisted of antenna-coupled TES bolometers read out with time-domain multiplexed SQUID amplifiers. We also assumed the cooler is a continuous Adiabatic Demagnetization Refrigerator (ADR). No actual technology selection is being undertaken at this time however.

Note that the cost estimates in this document do not constitute an implementation-cost commitment on the part of JPL or Caltech. The accuracy of the cost estimate is commensurate with the level of understanding of the mission concept, typically Pre-Phase A, and should be viewed as indicative rather than predictive.

**Table 10.1** EPIC Instrument Detailed WBS and Total Cost in FY09 Dollars.

| EPIC Instrument WBS | 4K Option (FY09 $M) | 30K Option (FY09 $M) |
|---|---|---|
| **05 EPIC Instrument Current Total** | **286** | **242** |
| **05.01 Payload Management** | **2** | **2** |
| **05.02 Payload System Engineering** | **2** | **2** |
| **05.03 Instrument System** | **282** | **238** |
| **05.03.1 Tested Focal Plane Assy with Electronics** | **128** | **96** |
| 05.03.1.1 Focal Plane Assy with Electronics | 94 | 62 |
| *Detectors* | | |
| *Focal Plane Structure* | | |
| *Readout Electronics* | | |
| 05.03.1.2 ADR | 25 | 25 |
| *ADR build and Test* | | |
| *ADR Electronics* | | |
| *ADR I&T Support* | | |
| 05.03.1.3 Inst Ctrl Electronics & FPA Test | 9 | 9 |
| *Instrument Control Electronics* | | |
| *Focal Plane Assembly Test* | | |
| **05.03.2 Telescope** | **29.0** | **29.0** |
| **05.03.3 Tested 4K Cooling Sys & Structure** | **62** | **61** |
| 05.03.3.1 4K Cooling | 39.0 | 39.0 |
| 05.03.3.2 Sunshade Thermal System | 23 | 22 |
| *Scale Model Test*<br>*Thermal design of Sunshield & Radiator* | | |
| *V-Groove* | | |
| *BiPods* | | |
| *Optical Tent* | | |
| **05.03.4 Instrument I&T** | **25.0** | **21.0** |
| **05.03.5 Deployable Sunshield** | **38** | **31** |

The proposed total instrument development cost is $286M in FY2009 dollars (FY09$) for the baseline 4 K option. Since the EPIC instrument is relatively large and complex we have included in this cost estimate WBS items equivalent for "payload management" and "payload system engineering". These WBS elements are traditionally used in a project with a suite of instruments that need to be integrated into a payload system. On EPIC although there is a single instrument the complexity of its components warrants the addition of an overall instrument manager and system engineer. Table 10.1 shows the major instrument WBS items and corresponding cost. Section 10.4 provides detail descriptions of the WBS cost elements.

**Table 10.2** Grassroots Cost Estimation Process

| Step | Leader | Activity |
|---|---|---|
| 1 | PM | Develop EPIC Instrument Work Breakdown Structure |
| 2 | PM | Define key EPIC Instrument project milestones / schedule |
| 3 | PM | Issue Initial Costing Guideline with the following information:<br>a. WBS<br>b. Initial project schedule<br>c. Interface Assumptions<br>d. Instrument Functional Block Diagram<br>e. Costing guidelines and assumptions |
| 4 | CogE | Each WBS Cognizant Engineer perform initial costing based on the following work:<br>a. Define tasks in outline/bullet form<br>b. Develop preliminary WBS element schedule<br>c. Define receivables and deliverables<br>d. Validate or revise interface assumptions<br>e. Produce a list of WBS element costing assumptions<br>f. Establish an initial cost estimate |
| 5 | PM | The PM collects the initial costing information from each lead and conducts the following:<br>a. Integrate each individual schedule into a network project schedule.<br>b. Input in a pricing system and review<br>c. Review the list of assumptions from leads |
| 6 | PM/ CogEs | Cognizant Engineers review the initial cost, schedule and resource together to eliminate overlapping cost, wrong assumptions, etc. |
| 7 | PM | Issue a refined Costing Guideline with clear constraints in:<br>a. Scope of work/deliverables<br>b. Delivery schedule and flow of schedule |
| 8 | CogE | Each WBS Cognizant Engineer performs a revised cost estimate considering factors and constraints in items 7a – 7b and obtains organizational approval. |

**PM => Proposal Manager, CogE=> Cognizant Engineer*

## 10.1 Cost Estimation Methodology

Each WBS element is assigned to a WBS element cognizant engineer who is an expert in the element technical area to develop the element schedule, plan the integration and test flow, and define interfaces with other elements. The cognizant engineers have a recent, direct and hands-on flight development experience. This method requires a thorough understanding of the subsystem design and the plan and schedule for the development flow, knowledge of the interrelationships of tasks and their associated deliverables and receivables, an understanding of available facilities and equipment, and a detailed staffing plan and an agreement on make/buy strategy. The grass roots method, illustrated in Table 10.2, forms the basis of our cost estimates for the instrument module. These costs were then confirmed with Team X for the mission study.

Fig. 10.1. EPIC Instrument Work Breakdown Structure with Costs in FY09$

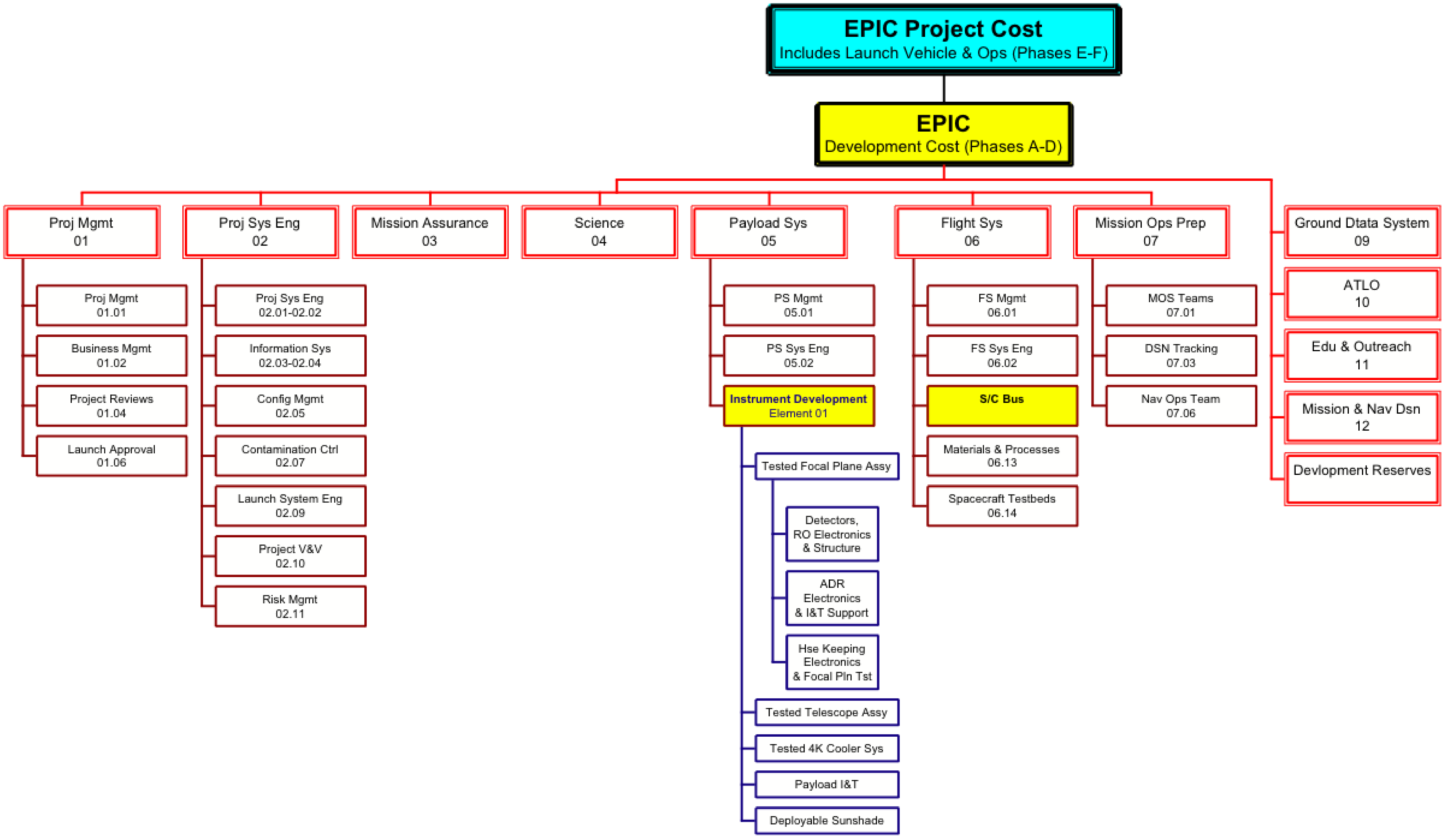

Our cost estimate was based upon the Work Breakdown Structure (WBS) shown in Fig. 10.1 (with further details in Table 10.1), the schedule shown in Fig. 10.2 and the instrument and integration and test flow shown in Fig. 10.3. Team members prepared and developed the estimates at the lowest possible levels of the WBS. We draw heavily on recent actual cost collected from like WBS elements for the design, analysis, fabrication and test, and for the procurements of cryogenic and optic components. The estimate includes current burden rates, inflation factors, travel rates and salary rates. The EPIC team reviewed the rollup grassroots cost estimate to ensure that the estimate is current, accurate and complete.

**10.2 Schedule**

Fig. 10.2 shows the development schedule for the EPIC project. The schedule is driven by the major instrument component development time (42 months), the significant integration and test time required for each component (12 months), the integration of the instrument (12 months) and for the instrument integration with the spacecraft (18 months). The resulting total development time Phases A through D is 7 years. In contrast the spacecraft development schedule is 3.5 years. We based instrument hardware development durations from recent experiences with Herschel, Planck, and MIRI.

The project master schedule shows all mission phases, major milestones, and the development critical path. This schedule is driven by the detector fabrication and test time, which requires 42 months prior to integration with the focal plane structure as shown in Fig. 10.2. The other critical instrument components shown in this schedule are the telescope, the 4 K cooler systems, the ADR, the warm electronics, the sunshield and the structures for the focal plane and the V-grooves.

Phase A is twelve months and is for the preliminary instrument design and includes detail design of long lead items such as the detectors, the mirrors, the 4 K cooler system and the sunshield. Phase B is 18 months, in which the detail design and engineering models are built. Phase C is 36 months. The start of Phase C marks the beginning of the flight build for the detectors, the ADR, the electronics, the sunshield and instrument structures. Phase C also includes all component test bed development and component integration and test. The last twelve months of phase C are for instrument integration. Phase D is 18 months and is for instrument integration with the spacecraft.

Conforming to JPL Flight Project Practices the schedule has the equivalent of one month funded reserve per year during development and two months funded reserve per year for assembly and test shown in green. For the detector development these funded reserves are doubled.

**10.3 Integration and Test Flow**

A key element in developing accurate and complete cost estimates is a detailed understanding of the integration and test flow of the instrument and the accompanying testbeds and facilities needed. Fig 10.3 shows the I&T flow and the planned tests and type, warm or cold.
The first panel lists the major components their development time including component integration and test in months. The next panel, subsystem I&T, as stated before is allocated 12 months. Here the detectors, the ADR, the focal plane structure and electronics are integrated allowing the focal plane assembly to be fully tested. The telescope assembly is also integrated and tested. The 4 K cooler system is integrated with the V-groove structure and also tested. The sunshield assembly is integrated and deployment tests are conducted. The next panel is

Fig. 10.2. Payload Development Schedule

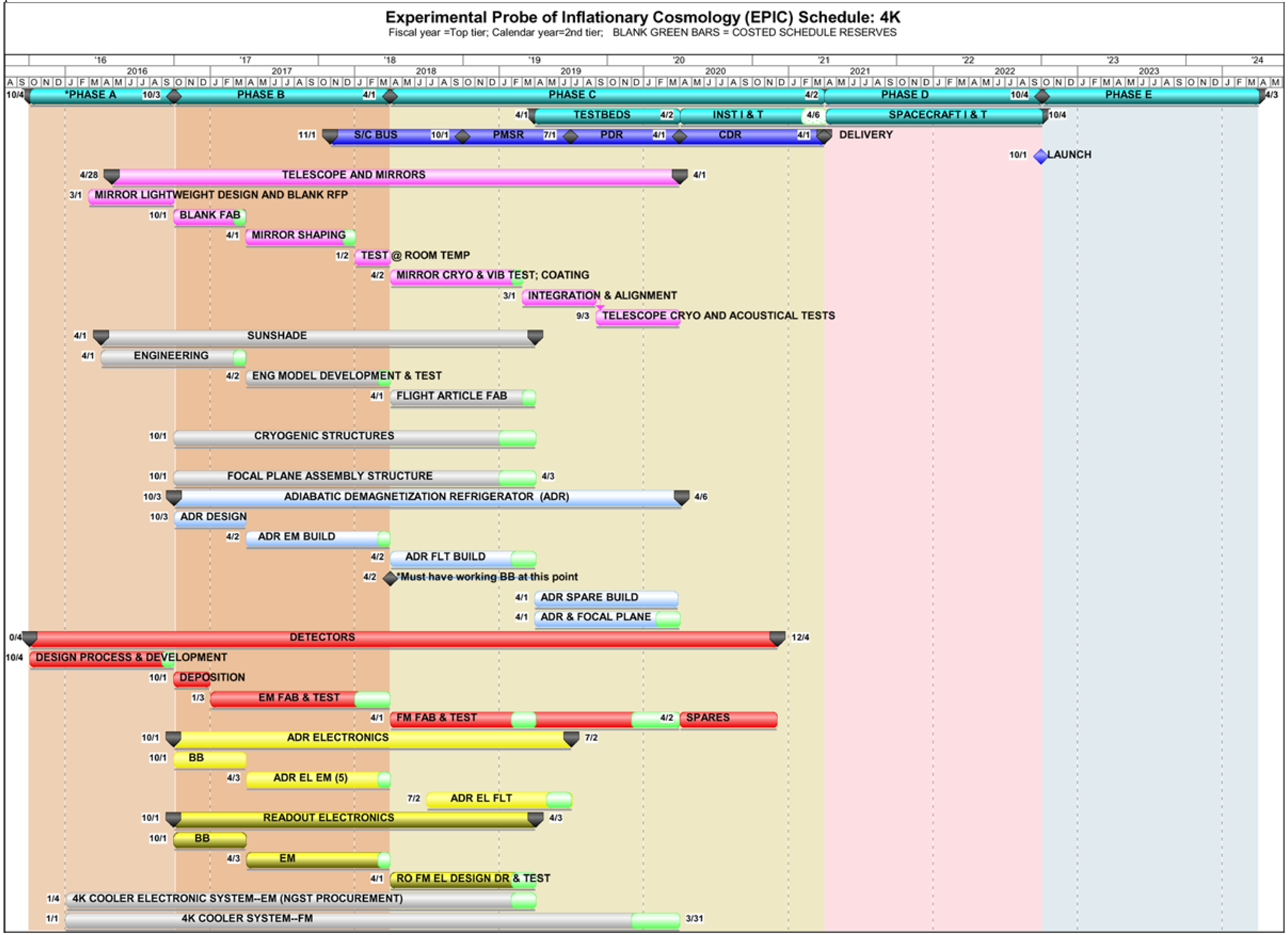

As instructed by the decadal review, we assume a start in October 2010 for purposes of assessing the cost of mission phases A-F

Fig. 10.3. I&T Flow and the Planned Tests and Type.

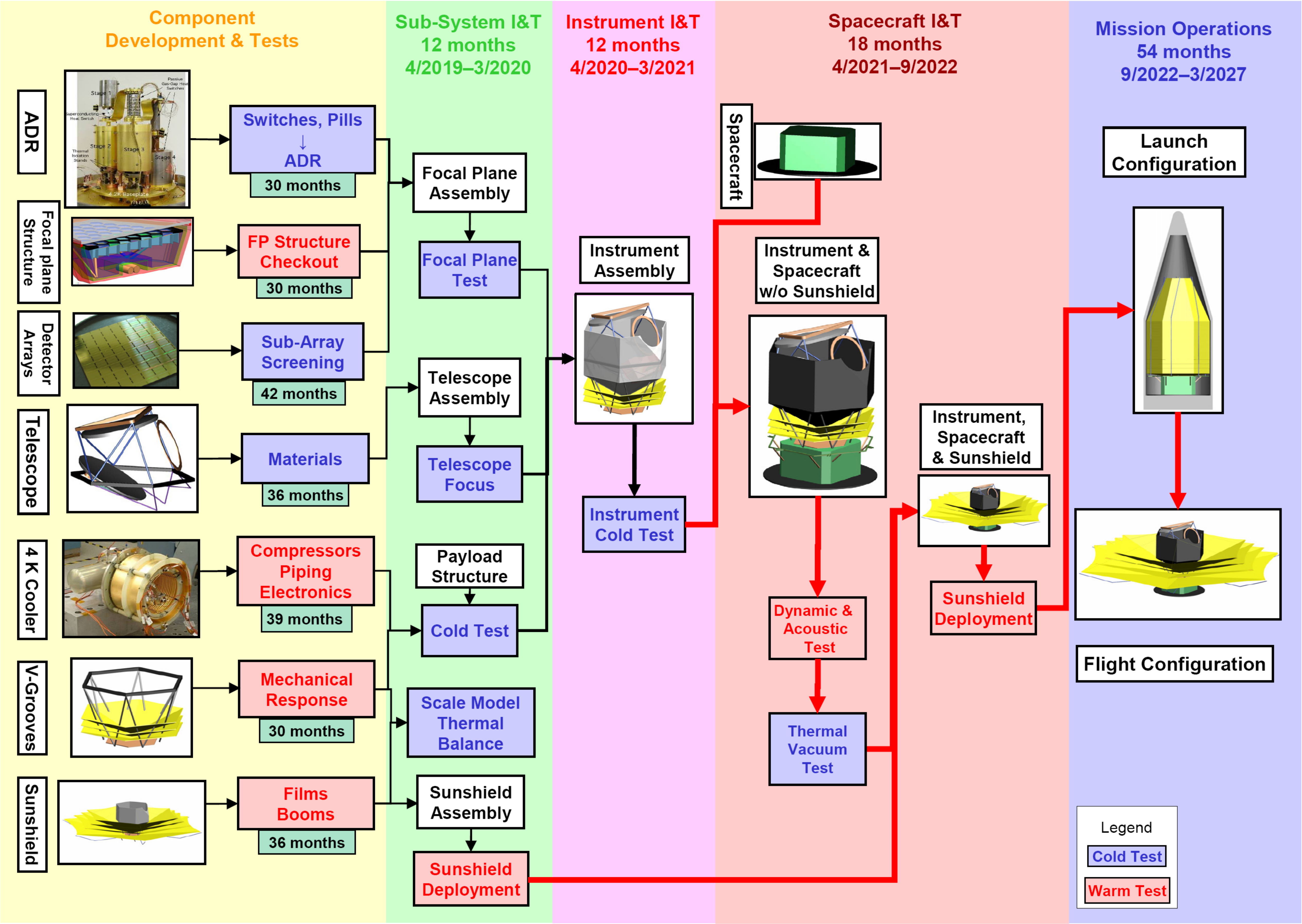

instrument integration and test, also allocated twelve months. Here the assemblies from the previous panels are now integrated into a complete instrument and functional tests are conducted.

The next panel first shows the spacecraft being integrated to the instrument without the sunshield and undergoing functional and environmental tests. Next the sunshield is integrated and deployment tests are repeated for the complete system. This is the traditional ATLO phase and is allocated 18 months. The last panel shows the launch and flight configurations.

## 10.4 Work Breakdown Structure Cost Element Descriptions

Below is a detailed description of each element in the instrument WBS.

### 05.01 Payload Management

Lead and manage the overall EPIC instrument for the project for delivery to the spacecraft. Includes the business and administrative planning, organizing, directing, coordinating, controlling, and approval processes used to deliver EPIC instrument goals. It includes coordinating the instrument risk management processes, creation and maintenance of project master risk list, interfacing with the instrument business personnel for development, and maintenance of the instrument soft lien list. Document products include project risk management plan, significant risk list and metrics on risk items. This element was costed based on past experience with similar tasks and modified to account for the long EPIC schedule. Note that for EPIC the payload and instrument are the same since there is only one instrument.

### 05.02 Payload System Engineering

The PSE defines and implements the instrument's overall system architecture. Includes requirements structure, flow-down, definition, and management; defining intersystem interfaces, project external interfaces; defining fault protection and system design margins and guidelines; conducting trade studies; managing project technical resources; manage the project action-item list. Documentation products include project review plan, review agendas, and review reports Also includes project engineering-document development tasks, such as system engineering reports, project requirements documents, system description documents, ICDs, V&V requirements, and instrument test plans & test/verification matrix. This element was costed based on past experience with similar tasks and modified to account for the long EPIC schedule.

### 05.03 Instrument System

### 05.03.1 Tested Focal Plane Assembly with Electronics

*05.03.1.1 Focal Plane assembly with Electronics*

As shown in Table 10.1, the focal plane assembly cost includes the detectors, the structure that hosts the detectors and the read out electronics for the detectors.

Detectors: It is assumed that that the detectors and readouts are at TRL 6 and that one deposition system and one 4-tile cryogenic testbed are in use and operational at start of the project. The flight spares consist of 50 % of the tiles needed and a 50 % fabrication yield as shown in Table 10.3. It is also assumed that the cryogenic test beds use cryogen-free dilution refrigerators. As stated before two months funded schedule reserve per year were allocated in the detector schedule. The estimate also includes purchase of an e-beam evaporation deposition system, a deep trench reactive ion etcher, and a dielectric deposition system, which are currently process bottlenecks.

**Table 10.3** Detector Tiles: Total Needed, Spares and Expected Yield.

| | 4 K Telescope | | | | | 30 K Telescope | | | | |
|---|---|---|---|---|---|---|---|---|---|---|
| **Freq (GHz)** | **#Flight tiles** | **#Spare tiles** | **Total[1] tiles x2** | **#Tiles for testing** | **Total #tiles needed** | **#Flight tiles** | **#Spare tiles** | **Total[1] tiles x2** | **#Tiles for testing** | **Total #tiles needed** |
| 30 | 6 | 3 | 18 | 2 | 20 | 4 | 2 | 12 | 2 | 14 |
| 45 | 14 | 7 | 42 | 2 | 44 | 6 | 3 | 18 | 2 | 20 |
| 70 | 18 | 9 | 54 | 2 | 56 | 8 | 4 | 24 | 2 | 26 |
| 100 | 18 | 9 | 54 | 2 | 56 | 6 | 3 | 18 | 2 | 20 |
| 150 | 12 | 6 | 36 | 2 | 38 | | | | | |
| 220 & 350 | 4 | 2 | 12 | 2 | 14 | 6 | 3 | 18 | 2 | 20 |
| 500 & 850 | 1 | 1 | 4 | 2 | 6 | 1 | 1 | 4 | 2 | 6 |
| **Totals** | **73** | **37** | **220** | **14** | **234** | **31** | **16** | **94** | **12** | **106** |

[1]Fabrication planning assumes 50 % yield. However these additional tiles would be stopped after testing and prior to delivery.

Structure: The structure cost estimate is based on a detailed estimate done for a similar focal plane for a proposed instrument for the SPICA far-infrared telescope.

Readout Electronics: For the electronic boards the cost includes a bread-board (BB), engineering model (EM), a qualification unit, and the flight units (FM) with 50 % sparing policy on flight boards and the EGSE is included. For the 30 K option the design is based on existing functioning ground hardware architecture and upgrade to space flight qualified parts. For the 4 K option a re-architecture effort is costed due to significant interface increase in number of focal plane detectors.

*05.03.1.2. The Adiabatic Demagnetization Refrigerator (ADR)*

ADR Build and Test: The ADR consists of the magnet subsystem; (magnets, harness and shield), the paramagnetic coolant; (Cesium Chrome Alum in hermetic container with thermal bus and garnet in holder with thermal bus), Gas Gap heat switch; Superconducting heat switch; thermal straps and mechanical structure. The cost includes BB, EM and flight unit with 50 % flight spares. The cost assumes the magnet lead harness will be at TRL 6. There is a concern with excess parasitic thermal loads and testbed failures may cause schedule slip and schedule reserves has been added to mitigate this concern.

ADR Electronics: This consists of four Current Controller boards (each similar in design) and one housekeeping unit with eight low power on/off switches. As with other electronics, included is BB, EM, Qual unit, flight units with 50% spares and EGSE.

ADR I&T Support: The cost of the ADR team to support the instrument I&T is book kept here.

*05.03.1.3. Instrument Control Electronics and Focal Plane Assembly Test*

Instrument Control Electronics (ICE): The ICE is based on the Microwave Radiometer instrument electronics developed for the Juno mission. The control electronics consists of three parts the instrument Command and Data Unit (CDU) that provides the interface with the spacecraft (includes processor for onboard compression), the House Keeping Unit (HKU) that provides instrument low-level control and the Power Distribution Unit (PDU) that provides power conversion and distribution to the instrument subsystem elements.

Focal Plane Assembly Test: The full testing of the integrated focal plane assembly is included in this cost element.

**05.03.02 Telescope**

At the end of Phase C a fully tested telescope will be delivered for integration and system testing. In the schedule we have allocated, four months for the mirror definition and acquisition process, six months for blank fabrication and nine months for figuring the mirror. In the telescope development it takes 15 months to fabricate each mirror and as a result both mirrors (primary and secondary) will be fabricated in parallel. Once the mirror meets the specs at room temperature it requires 11 months for cryo-testing, any re-work, thermal cycling, vibration and acoustical testing, and finally coating. The telescope I&T is a 13-month activity and includes alignment, cryo-testing, shimming, and further cryo-testing. We checked the grassroots cost estimate of $29M with the costs of the manufacture and testing of the Herschel and Planck telescopes, provided to us by European partners. After scaling for differences in diameter and operating wavelength, we find that our grassroots estimate is in reasonable accord with these as-built costs. The major cost components of the telescope are the design, fabrication of M1 and M2, cryo-testing, fabrication and assembly of telescope support ring, and overall integration and testing.

**05.03.3 Tested 4K Cooling System and Structure**

This WBS item cost estimate was based on an analogous system built for the MIRI instrument, provided by the MIRI project team at JPL. The MIRI cooler is a close analog of this system, the EPIC cooler requiring one additional compressor to reach 4 K. The design of the EPIC cooler assumes 100 % design margin on the estimated heat loads. The cost estimate included a scaled cost for the manufacture of the cooler, and includes contract burden and JPL design oversight during the development period.

**05.03.04 Instrument I&T**

The cost was based on two approaches, an analogous cost of other equally complex instrument I&T costs as a fraction of instrument development cost and a grassroots cost estimate for the workforce and facility costs to complete I&T.

**05.03.5 Deployable Sunshield**

A grassroots cost estimate was used to cost this element. The estimated cost includes JPL contracts burden on procured parts as well as JPL design oversight during development. The total cost is estimated at $39M, with the major component being contracts for deployable booms, reflector membranes, stowage system, and deployment control electronics, electrical and mechanical ground support equipment.

# 11. Spacecraft

We have carried out a definition of the EPIC-IM spacecraft in a JPL TeamX study. The spacecraft requirements are generally conventional, and may be realized with commercial options with existing space-qualified components. We describe the spacecraft functions, and component specifications detailed by TeamX. Available commercial options are assessed and costed in chapter 12.

## 11.1 Scientific Operations

EPIC carries out scientific observations from an L2 halo orbit. We reach L2 approximately 170 days after launch by means of a transfer orbit using lunar assist (see [1] for details of the orbit). The delta-V budget of 170 m/s includes 50 m/s for clean-up of launch injection errors, a conservative trajectory correction strategy, and 4 years of orbit correction at L2. Note that launch injection errors have been reduced due to the choice of vehicle from the earlier study. We take 95 % probability on all maneuver errors add then include an additional 10 % overall margin. The sunshield is deployed early in the mission in order to begin passive telescope and instrument cooling en route to L2.

**Table 11.1** Mission Design Summary

| | |
|---|---|
| **Orbit** | L2 Halo |
| **Mission Life** | 1 year at L2 (minimum mission)<br>4 years at L2 (design life) |
| **Maximum Eclipse Period** | 0 (the Halo orbit at L2 is designed to avoid eclipses) |
| **Spacecraft dry bus mass** | 792 kg, includes 43 % contingency |
| **Spacecraft propellant mass** | 295 kg |
| **Launch vehicle** | Atlas V 401 |
| **Launch vehicle mass margin** | 1329 kg (37 %) in the 4 K telescope option<br>1622 kg (45 %) in the 30 K telescope option |

Once at L2, the instrument executes a single observing mode which consists of a spinning/precessing scan strategy (see Fig. 3.5). This strategy provides uniform and redundant coverage of the sky and efficiently rotates the telescope direction on all regions of the sky. These maneuvers can be accomplished with a zero-momentum spacecraft, similar to the design used on WMAP. Data are transmitted to earth once per day via a two-axis gimballed High-Gain antenna (HGA). With this repeated sequence of events and continuous observing, operations are rather simple. The sequence of operations is summarized in Tables 11.2 and 11.3.

**Table 11.2** Science Observations Operations

| **Mission Operation** | **Rate** |
|---|---|
| Spin Spacecraft | Continuous, 0.5 rpm |
| Precess Spin Axis | Continuous, 1 rph |
| ADR Operations | Continuous |
| Downlink | Once every 24 hours |
| Maintain Orbit | Small maneuvers every few weeks |

**Table 11.3** Mission Operations and Ground Data Systems

| Down link Information | Value, units |
|---|---|
| Number of Data Dumps per Day | One 8-hour pass per day (baseline) |
| Downlink Frequency Band | Near-Earth Ka-Band for 4 K option<br>Near-Earth X-Band for 30 K option |
| Telemetry Data Rate | 28 Mbps via Ka-Band w/High-Gain Antenna for 4 K<br>3 Mbps via X-band w/toroidal beam antenna for 30 K |
| S/C Transmitting Antenna Type(s) and Gain(s) | 0.5 m HGA, 44 dB @X-Band for 4 K<br>Toroidal-beam antenna, 9.0 dB @ X-Band for 30 K |
| Spacecraft transmitter peak power | 97 W for 4 K option<br>192 W for 40 K option (total power) |
| Downlink Receive Antenna Gain (34-m DSN) | 76 dB @ Ka-Band<br>68.3 dB @ X-Band |
| Transmitting Power Amplifier Output | 3.5 W for 4 K option<br>100 W for 40K option (RF power) |
| **Uplink Information** | **Value, units** |
| Number of Uplinks per Day | 1 |
| Uplink Frequency Band | 7.17 GHz (Near-Earth X-Band) |
| Telecommand Data Rate | 100 kbps |
| S/C Receiving Antenna Type(s) and Gain(s) | 0.5 m X/Ka&X HGA<br>30.3 dB; X-band LGAs, 7.7 dBi |

## 11.2 Instrument Data Generation

We calculate the data generation rate of the instrument in Table 11.4. This estimation requires that the detectors are fast enough to avoid beam smearing, $\tau < (1/2\pi)\ \theta_{FWHM}/d\theta/dt$, and the data streams are sampled at the Nyquist frequency for this time constant. The effective compression is assumed to be 4 bits per sample, essentially the same compression realized for Planck HFI. Finally we include a factor of 2 safety margin by doubling the sampling rate. While this sets the approximate data requirements, a final requirement should be derived from a full simulation of sampling and compression.

**Table 11.4** Data Rate Generation

| Freq [GHz] | Beam [arcmin] | $\tau_{req}$ [ms][1] | Sample Rate [Hz][2] | 4 K Telescope | | 30 K Telescope | |
|---|---|---|---|---|---|---|---|
| | | | | Ndet [#] | Data Rate [kbps][3] | Ndet [#] | Data Rate [kbps][3] |
| 30 | 28 | 30 | 21 | 84 | 7 | 24 | 2 |
| 45 | 19 | 20 | 32 | 364 | 46 | 84 | 9 |
| 70 | 12 | 13 | 49 | 1332 | 260 | 208 | 35 |
| 100 | 8.4 | 9 | 71 | 2196 | 620 | 444 | 110 |
| 150 | 5.6 | 6 | 106 | 3048 | 1300 | 516 | 190 |
| 220 | 3.8 | 4 | 155 | 1296 | 800 | 408 | 220 |
| 340 | 2.5 | 2.7 | 240 | 744 | 710 | 120 | 100 |
| 500 | 1.7 | 1.8 | 350 | 1092 | 1500 | 108 | 130 |
| 850 | 1.0 | 1.1 | 600 | 938 | 2200 | 110 | 230 |
| **Total** | | | | **11094** | **7500** | **2022** | **1000** |

[1]Required speed of response $\tau_{req} < (1/2\pi)\ \theta_{FWMH}/d\theta/dt$ for 0.5 rpm spin rate at 55°

[2]Sample rate at twice Nyquist, $\nu_{samp} = 2/\pi\tau_{req}$

[3]Data generation rate at 4 bits per sample = $4\ N_{det}\ \nu_{samp}$

## 11.3 Spacecraft Requirements

The EPIC Intermediate mission option is sized for an Atlas V 401 launch vehicle with a 3.65 m fairing, as shown in Fig. 3.4. The payload includes the focal plane assembly, electronics, telescope, 4 K cooling system, sub-K cooling, deployable sunshield, V-groove radiators, and support struts between the instrument and spacecraft. A summary of the payload and spacecraft mass and power is listed in Tables 3.6 and 3.7.

**Table 11.5** Spacecraft Requirements

| | | Units | EPIC-IM 4K Requirement | EPIC-IM 40K Requirement |
|---|---|---|---|---|
| Compatibility | Payload Power (OAV) | W | ≥ 643 (incl. 43 % cont.) | ≥ 432 (incl. 43 % cont.) |
| | Payload Mass Capability of Bus | kg | ≥ 1163 (incl. 43 % cont.) | ≥ 948 (incl. 43 % cont.) |
| | Bus Dry Mass (w/o Payload) | kg | < 2000, incl. contingency | < 2200, incl. contingency |
| | Science Data Downlink Rate | kbps | 28,000 | 3,000 |
| | Science Data Storage Capability | Gbit | 1,400 | 170 |
| | Pointing Knowledge | arcsec | < 36 (3σ) | < 36 (3σ) |
| | Pointing Control | arcsec | < 3600 (3σ) | < 3600 (3σ) |
| | Pointing stability (jitter) | | 45"/50msec (3σ) | 45"/50msec (3σ) |
| | Spin rate | deg/min | 180 | 180 |
| | Mission Design Life | years | 4.5 | 4.5 |
| | Compatible LVs | | Atlas V 401 | Atlas V 401 |
| | Orbit | | Earth-Sun L2 | Earth-Sun L2 |
| | Internal Volume Available for Payload | | Sufficient for warm electronics | Sufficient for warm electronics |
| Description | Attitude Control System | | 3-axis momentum compensated | 3-axis momentum compensated |
| | Batteries | type/Ah | Two at 24 Ah each | Two at 24 Ah each |
| | Arrays | Type/ area | Triple junction GaAs 6.8 $m^2$ body mounted | Triple junction GaAs 6.2 $m^2$ body mounted |
| | Nominal Voltage | V | 28 | 28 |
| | C&DH Bus Architecture | | 422 or LVDS | 422 or LVDS |
| | Downlink Formats | | CCSDS | CCSDS |
| | Downlink Band | | Near-Earth Ka- and X-bands | Near-Earth X-band |
| | Structure | | Al or composite | Al or composite |
| | Propulsion | | Monoprop Hydrazine | Monoprop Hydrazine |
| | Propellant Capacity | kg | 295 | 295 |
| | Mass Delta-V | m/s | 170 | 170 |
| Programmatic | Nominal Schedule | months | 38 | 38 |
| | Contract Options | | Enhance C&DH for high data rates and add Ka-Band telecom | Replace S/C telecom with toroidal antenna |
| | | | Body mounted solar panels | Body mounted solar panels |
| | | | Modify propulsion tanks | Modify propulsion tanks |
| | | | Modify bus structure | Modify bus structure |

NOTE: the values supplied in this table are the EPIC requirements -- not the specifications for any particular implementation. The vendor for the spacecraft bus for this mission has not yet been selected.

We assume EPIC will operate with a spacecraft bus with high heritage from an existing design. The spacecraft itself requires no new technology (the one exception here is a custom-designed X-band downlink antenna producing a toroidal beam for the 30 K option, which would be provided equipment to the spacecraft vendor). EPIC requires a bus-mounted solar panel on the sun-facing side of the bus. The deployable sunshield would be a provided payload element and is not part of the spacecraft.

**Table 11.6** Spacecraft Bus Design Characteristics from TeamX Study

| | **Spacecraft bus** | **4 K Telescope Option** | **30 K Telescope Option** |
|---|---|---|---|
| **Structure** | Structures material | Aluminum or composite | Aluminum or composite |
| | Number of articulated structures | Two-axis gimbaled HGA | None |
| | Number of deployed structures | None (Deployed sunshade in payload) | None (Deployed sunshade in payload) |
| **T/C** | Type of thermal control used | Passive | Passive |
| **Propulsion** | Estimated delta-V budget | 170 m/s | 170 m/s |
| | Propulsion type | Hydrazine | Hydrazine |
| | Number of thrusters and tanks | Eight 1-lbf thrusters<br>One tank | Eight 1-lbf thrusters<br>One tank |
| | Specific impulse of each propulsion mode | 220 s | 220 s |
| **Attitude Control** | Control method | 3-axis, momentum compensated | 3-axis, momentum compensated |
| | Control reference | Inertial | Inertial |
| | Attitude control capability | 1.0 deg | 1.0 deg |
| | Attitude knowledge limit | 36 arcsec (3σ) | 36 arcsec (3σ) |
| | Agility requirements | None | None |
| | Articulation/#–axes | None | None |
| | <u>SENSORS</u>:<br>Sun Sensors (14)<br>Star Trackers (2)<br>IMU (2)<br><br><u>ACTUATORS</u>:<br>Reaction Wheels (4) | <br><br>20 arcsec accuracy<br>0.005 arcsec/sec stability<br><br><br>150 Nms momentum, 0.1 Nm torque | <br><br>20 arcsec accuracy<br>0.005 arcsec/sec stability<br><br><br>150 Nms momentum, 0.1 Nm torque |
| **C & DH** | Spacecraft housekeeping data rate | 10 kbps | 10 kbps |
| | Data storage capacity | 1,400 Gbits | 768 Gbits |
| | Maximum storage record rate | 300 Mbps | 8 Mbps |
| | Maximum storage playback rate | 300 Mbps | 8 Mbps |
| **Power** | Solar array configuration | Fixed, body-mounted solar panels | Fixed, body-mounted solar panels |
| | Array size | 6.8 $m^2$ | 6.2 $m^2$ |
| | Solar cell type | Triple-junction Ga-As | Triple-junction Ga-As |
| | Expected power generation | 1430 W BOL | 1206 W BOL |
| | On-orbit average power consumption | 1292 W (incl. 43% contingency) | 920 W (incl. 43% contingency) |
| | Battery type | Li-Ion (two) | Li-Ion (two) |
| | Battery storage capacity | 40 Ah | 40 Ah |

# 12. Mission Cost

The EPIC team generated a payload development schedule, and this was merged with the spacecraft development schedule to produce the overall EPIC Project schedule shown in Fig. 10.2. Team X utilized these schedules, together with EPIC Team grassroots and analogy cost estimates for the payload elements to estimate the total Project Cost.

## 12.1 Project Schedule

Our development schedule for the payload is 94 months. The spacecraft development is assumed to be 38 months, and is decoupled from the payload development schedule. Flight System ATLO (where the payload is integrated with the spacecraft, and then mated with the launch vehicle, through launch + 30 days) is 18 months. This schedule was adopted in analogy with ATLO durations for similar cryogenic missions.

EPIC-IM's approach is parallel development and testing early in the hardware implementation, in order to minimize the risk and cost of tests and system level later in the schedule. The telescope, focal plane, cooling, and deployed sunshield subsystems are all developed and tested independently prior to payload integration and test.

The thermal system is verified by a combination of testing and modeling, following the design philosophy of Spitzer, which demonstrated a successful radiatively cooled cryogenic system built with adequate margins combined with analysis and a limited test program. The passive cooling system is first tested independently of the spacecraft or instrument, in order to simplify tests at final integration. To verify the performance of the passive cooling system, we will measure the infrared and thermal properties of the materials and carry out a thermal balance test on a scale model of the radiators and sunshield.

As the spacecraft and payload are integrated and tested separately, the cryogenic payload does not impose unusual or demanding requirements on the spacecraft bus during development and testing. After the payload and spacecraft have been integrated, a thermal balance test of the spacecraft will be conducted, since the payload and sunshield thermal performance will be already verified separately.

<u>Sunshade</u>: The EPIC-IM sunshield uses four nested double-layer deployed sunshields made from aluminized kapton membranes. The 30 K option uses three nested double-layer deployed shields. The tip-to-tip diameter of the deployed sunshield is 14.8 m in both options. We will use a gravitational offload system to comprehensively test the deployment of the structures and kapton membranes. Gravitational offload techniques have been used to carry out testing of numerous structures of this size, mostly deployable antennas, with a high rate of success. Several aerospace companies currently have this test capability. Kapton membranes of the flight design and folding arrangement are integral to the deployment test. Offloading the weight of the membranes is not a significant concern. We will also test venting of the folded membranes in stowed configuration.

It is important to note that the sunshield only plays a minor role in passive cooling. The sunshield's function is to block radiation from the sun and warmer sunshield layers from viewing the cryostat and optics, and to simply reflect radiated thermal power from the internal V-groove coolers to space. Thus thermal tests of the sunshield can be limited in scope, a scale model to test thermal properties.

Spacecraft: The spacecraft bus requirements are not demanding, and can be accommodated using a modified commercial bus. Modifications may include upgraded C&DH and Telecom subsystems to accommodate the high data rates for the 4 K option, a strengthened bus structure, body-mounted solar panels, and (for the 30 K option) the JPL-provided toroidal antenna. All spacecraft components (except the toroidal antenna) are off-the-shelf, flight-proven commercial hardware. The 38-month spacecraft development schedule was provided by JPL's TeamX, which assumed a custom built bus using off-the-shelf commercial components for all subsystems.

### 12.2 Cost Estimate

Costs were generated by JPL's Advanced Concurrent Engineering Design Team (Team X), which includes experts in science, mission design, instruments, programmatics, ground systems, and every spacecraft subsystem. Team members synthesize their own expertise and discipline-specific models to generate complete mission studies including cost details. JPL has used Team X to generate well over 600 project studies. The Team X cost estimates summarized in this document were generated as part of a Pre-Phase-A preliminary concept study, are model-based, were prepared without consideration of potential industry participation, and do not constitute an implementation-cost commitment on the part of JPL or Caltech. The accuracy of the cost estimate is commensurate with the level of understanding of the mission concept, typically Pre-Phase A, and should be viewed as indicative rather than predictive.

Prior to the Team X session, the grassroots payload cost and schedule was developed as described in chapter 10. This assessment included all components including the deployable sunshade, telescope, focal plane detector arrays, cryogenic cooling systems, and warm and cold readout electronics, and including steps of payload integration and test. These costs were scaled from actual costs on similar hardware delivered for JWST, Planck, and Herschel where applicable. The grassroots cost estimate (CBE, without reserves) for the payload was $286M (FY09) for the 4 K option and $242M (FY09) for the 30 K option. A breakdown of the payload element costs is shown in Table 10.1.

The spacecraft subsystems are generally within the range of existing flight-proven technologies. Based on a request for information (RFI) to industry to determine whether a modified commercial bus was capable of meeting the requirements of the 4 K option (specified in Table 11.5), we used a quasi-grassroots input to the Team X project cost model for the flight system at a cost of $114M. The Flight System WBS 6.0 includes Flight System Management, Flight System Systems Engineering, Materials and Processes, and Testbeds, including spacecraft bus I&T and I&T of the full flight system (spacecraft bus + payload).

The resulting project costs are summarized in Table 12.1, assuming a 4 K telescope case operating for 4 years with a commercial spacecraft.

We investigated the potential cost savings of 2 descope options. One option is to reduce to 1 year of science operations (compared to the baseline 4 years). The implications of shorter observations time at L2 is 2x reduced sensitivity, and less redundancy for checking systematic errors. The reduced operations phase reduces costs by about $55M, including reserves, primarily in the Science and Mission Operations areas.

**Table 12.1.** Summary of EPIC Mission Costs: 4K Telescope, 4 year science operations at L2

| **WBS Elements** | $M FY'09 Dollars* |
|---|---|
| **Project Cost (including Launch Vehicle)** | **920** |
| **Development Costs (Phases A-D)** | **684** |
| 1.0, 2.0, and 3.0: Management, Systems Engr. and Mission Assurance | 51 |
| 4.0 Science | 15 |
| 5.0 Payload System | 286 |
| 6.0 Flight System | 114 |
| 7.0 Mission Operations Preparation | 13 |
| 8.0 Launch Vehicle | 136 |
| 9.0 Ground Data Systems | 13 |
| 10.0 ATLO | 28 |
| 11.0 Education and Public Outreach | 2 |
| 12.0 Mission and Navigation Design | 6 |
| Development Reserves (30%) | 156 |
| **Operations Costs (Phases E-F)** | **100** |
| Operations Reserves (10%)** | 11 |

Notes
 * Individual WBS elements have been rounded to 2 significant digits.
** On all elements except DSN tracking costs

**Table 12.2.** EPIC Mission Descope Cost Reductions

| | **1 Year Science Operations** | **30 K Telescope** |
|---|---|---|
| **WBS Elements** | | |
| **Project Cost (including Launch Vehicle)** | **-55** | **-100** |
| **Development Costs (Phases A-D)** | **0** | **-100** |
| 1.0, 2.0, and 3.0: Mgmt, Sys. Engr. and MA | 0 | -4 |
| 4.0 Science | 0 | 0 |
| 5.0 Payload System | 0 | -44 |
| 6.0 Flight System | 0 | -28 |
| 7.0 Mission Operations Preparation | 0 | 0 |
| 8.0 Launch Vehicle | 0 | 0 |
| 9.0 Ground Data Systems | 0 | 0 |
| 10.0 ATLO | 0 | 0 |
| 11.0 Education and Public Outreach | 0 | 0 |
| 12.0 Mission and Navigation Design | 0 | 0 |
| Development Reserves (30%) | 0 | -24 |
| **Operations Costs (Phases E-F)** | **-55** | **0** |
| Operations Reserves (10%) | -5 | 0 |

We also investigated the 30 K telescope option. This option has reduced sensitivity, especially in the highest frequency bands, due to a combination of increased photon background and reduced focal plane density. For the EPIC mission design using a 30K telescope, the overall Team X mission cost is almost exactly $100M less than the 4 K telescope version, and this cost

difference is entirely in the Development Phases. There is a difference of about $45M in the payload costs, primarily due to the difference in cost of the Focal Planes and readout electronics as shown in Table 10.1. However, there is a modest cost savings of approximately $8M due to the simpler (three-layer) sunshade in the 30 K option as compared to the four-layer sunshade in the 4 K option. The other major difference is in the Spacecraft cost. Due to the much higher data rates and volumes in the 4 K options, the spacecraft C&DH, Telecom, and Software costs are lower in the 30K version, resulting in an overall WBS 6.0 cost savings of approximately $28M. These lower costs correspondingly reduce the cost of development reserves. As noted, most of the cost savings derive from reductions in the number of detectors and associated electronics and data handling, and only a small fraction is associated with the reduced thermal requirements. The majority of these savings can be realized simply by scaling back the focal plane without changing the cooling system.

## Appendix A. Systematics Propagation to Parameter Estimation

Since the ultimate goal of CMB observations is to confirm the standard cosmological model, or alternatively find departures from this model, a matter of primary importance is cosmological parameter estimation – this is the ultimate distilled yield (a dozen parameters) of every CMB experiment. Beam systematics, as well as other systematics, have to be controlled to a sufficiently high precision that their impact on parameter estimation will be negligible or, at least, lie within the (arbitrarily set) tolerance range. The systematics requirements discussed in section 5 are a step towards this goal but a requirement on the systematics level at a single multipole (while well motivated) may be too conservative since, in reality, data analysis will always make use of all accessible multipoles, and as can be seen from Fig. 5.2, the requirements/goals on beam systematics at $\ell = 100$ typically result in a much larger signal-to-noise ratio at lower $\ell$. For this reason it is advantageous to propagate beam systematics through Fisher-matrix and Monte Carlo Markov Chain (MCMC) parameter estimation and assess the allowed beam mismatch (i.e. differential ellipticity, gain, beamwidth, and pointing, as well as beam rotation) that meets a certain requirement on the allowed cosmological parameter bias. Clearly, this task is feasible only with analytic approximations for beam systematics since a full simulation of the sky with HEALPIX (and for a range of beam ellipticities, gains, etc.) is too demanding. O'Dea, Challinor & Johnson (2007) [12] applied a Fisher matrix estimation to cosmological parameters in the presence of beam systematics. The Fisher matrix elements read

$$F_{ij} = \frac{1}{2}\sum_{l}(2l+1)f_{sky}Trace\left[C^{-1}\frac{\partial C}{\partial \lambda_i}C^{-1}\frac{\partial C}{\partial \lambda_j}\right]$$

where $f_{\rm sky}$ is the fraction of observed sky, $C$ is a generalized matrix that contains all temperature, polarization and cross-correlation power spectra, and $\lambda_i$ are the cosmological parameters of the cosmological model. The uncertainty in the cosmological parameter is then obtained from the inverse of the Fisher matrix $\sigma(\lambda_i) = \sqrt{(F^{-1})_{ii}}$.

In the presence of systematics the power spectra that appear in the Fisher matrix elements are modified by adding the systematic spectra to the true underlying speactra. This always increases the error on the estimated parameter. In addition to the increase in statistical error beam systematics can also significantly *bias* of the estimated parameter (for example, if the induced B-mode power spectra are larger or comparable to the primordial B-mode). This latter effect turns out to be the more significant. The conservative requirement of section 5, if satisfied, guarantees that this is not the case with EPIC since the signal-to-noise ratio of the primordial B-mode to the systematic-induced B-mode is kept larger than 3 times the cosmic variance at $\ell \sim 100$. However, as pointed out above, this might be too restrictive a requirement.

The effect of bias can be estimated from products of first derivatives of the power spectra and the systematics power spectra [1,12] and we do not specify the exact expressions here. We introduce two parameters $\delta_i$ and $\beta_i$ which measure the fractional bias and statistical uncertainty, respectively, in the parameter $\lambda_i$ (given in units of the statistical uncertainty) $\delta_i = \Delta\lambda_i/\sigma_{\lambda i}$ and $\beta_i = \Delta\lambda_i/\sigma_{\lambda i}$, where the statistical error is obtained from the Fisher matrix whose elements (which include derivatives of the CMB power spectra with respect to the cosmological parameter in question calculated at its fiducial value). As mentioned above, it is always the bias $\delta_{\lambda i}$, which sets the most restrictive requirements on beam systematics. We note that both [1,12] adopted the tolerance level $\delta_\lambda$, $\beta_{\lambda i} \leq 0.1$). Incorporating CMB lensing into the Fisher matrix procedure is straightforward to do by including $C_\ell^{dd}$ and $C_\ell^{Td}$, the power spectrum of the lensing deflection angle and its correlation with temperature anisotropy, respectively. $C_\ell^{dd}$ is very sensitive to neutrino masses via its effect on structure formation due of their free streaming and the suppression of LSS on scales smaller than their free streaming scale. Lensing extraction of the CMB can tighten constraints on neutrino masses by a factor of a few in the absence of beam systematics. Therefore, it is essential to control B-mode systematics on sub-degree scales, where the lensing-induced B-mode peaks ([1]). MCMC is a standard tool for CMB parameter estimation and was employed by [1] to scrutinize the procedure of bias estimation described above since it uses only the information from the peak of the likelihood function and, in general, large biases (larger than the statistical uncertainty) are underestimated by this method. A full MCMC calculation demonstrates that the Fisher-matrix-based calculation of parameter bias is an adequate approximation for biases smaller than the original statistical error (i.e. $\delta_{\lambda i} \leq 1$, where the most sensitive parameters to B-mode polarization -- and therefore to beam-systematics-- are: the tensor-to-scalar ratio, neutrino mass, dark energy equation of state, and any other parameter in the cosmological model which impacts structure formation on scales of a few tens of Mpc). Since it was found in [1] that the inflationary science is more demanding than the lensing science as far as beam systematics are concerned (and as is evident from comparing Figures 5.2 and 5.3) we employed a Fisher matrix estimate to the allowed systematics under the condition $\delta_r \leq 0.1$. Table A.1 shows the results for EPIC-IM (4 K).

**Table A.1** Tolerance levels on beam systematics

| | **$\Delta g$ [$10^{-5}$]** | **$\Delta\mu$ [$10^{-3}$]** | **$\Delta\rho/\sigma$ [$10^{-3}$]** | **$\Delta e$ [$10^{-3}$]** | **$\varepsilon$ [']** |
|---|---|---|---|---|---|
| **EPIC-IM (4K)** | 9.2 | 9.2 | 0.7 | 2.7 | 0.9 |

Note: Tolerances obtained from the requirement that beam systematics do not bias a detection of r = 0.01 by more than 10% of the uncertainty with which this inflationary B-mode signal would have been detected in an ideal (systematics-free) case. Note that differential gain, beamwidth and pointing depend on the scanning strategy and we approximated a uniform, but not ideal, scanning strategy (see [1 & 3]); an assumption that very-well approximates EPIC's scanning strategy. These constraints, obtained from propagating beam systematics into parameter estimation, are comparable to the threshold goal values and requirements (at 100 GHz) shown in Table 5.4. We recall here that the 100 and 150 GHz bands carry most of the sensitivity of EPIC. We also note that the requirement that the tensor-to-scalar ratio *bias* is an order of magnitude smaller than the statistical uncertainty is *very* conservative.

## Appendix B. Cosmological Birefringence

`Cosmic birefringence' is an ancillary science target that CMB polarization probes via the TB and EB correlations it induces (caused, in theory, by rotation of the polarization plane by parity-violating terms in the electromagnetic Lagrangian (e.g. axionic fields). This effect has been constrained but not detected by existing CMB data from Boomerang, WMAP and QUAD by searching for small inconsistencies between the observed $C_\ell^{TE}$, $C_\ell^{E}$, and $C_\ell^{T}$. In the standard cosmological model $C_\ell^{TB} = 0 = C_\ell^{EB}$ so any rotation of the polarization plane will result in $C_\ell^{TB} \neq 0$ and $C_\ell^{EB} \neq 0$ which are a result of power 'leakage' primarily from $C_\ell^{TE}$ and $C_\ell^{EE}$, respectively. Fig. B.1 illustrates how the TB correlation induced by a $\approx 1^{o}$ cosmological rotation compares with the TB systematics we derived based on the inflationary requirements. As shown in [3], neither differential gain nor beamwidth induce TB and EB correlations, so we only show the contributions of differential pointing, ellipticity and rotation. In this case, EPIC will not be able to set upper bounds on possible cosmological rotation of the polarization plane better than 1 degree. While the requirement of <2 arcminute rotation will certainly allow tighter constraints, it is the effect of differential ellipticity which challenges a measurement of cosmological birefringence the most, as can be seen from Fig. B.1. The likely reason, as discussed in Miller, Shimon & Keating 2009b [13], is that both $C_\ell^{TB}$ and $C_\ell^{EB}$ are linear in B while $C_\ell^{B}$ is quadratic in B. Setting constraints on the small beam parameters to avoid systematics larger than 10% in $C_\ell^{B}$ typically results in 1/ε and 1/e larger fractional errors in $C_\ell^{TB}$ and $C_\ell^{EB}$. This argument demonstrates that CMB polarimeters optimized for the inflation/lensing science will not yield a systematics-free measurement of $C_\ell^{TB}$ or $C_\ell^{EB}$ (which would be 'smoking guns' for cosmic birefringence). Rather, the constraining such non-standard physics via the TT, TE, EE and possibly BB should be used. Unless ellipticity is somehow sufficiently suppressed, this single systematic will limit any detection of cosmological rotation below 1 degree.

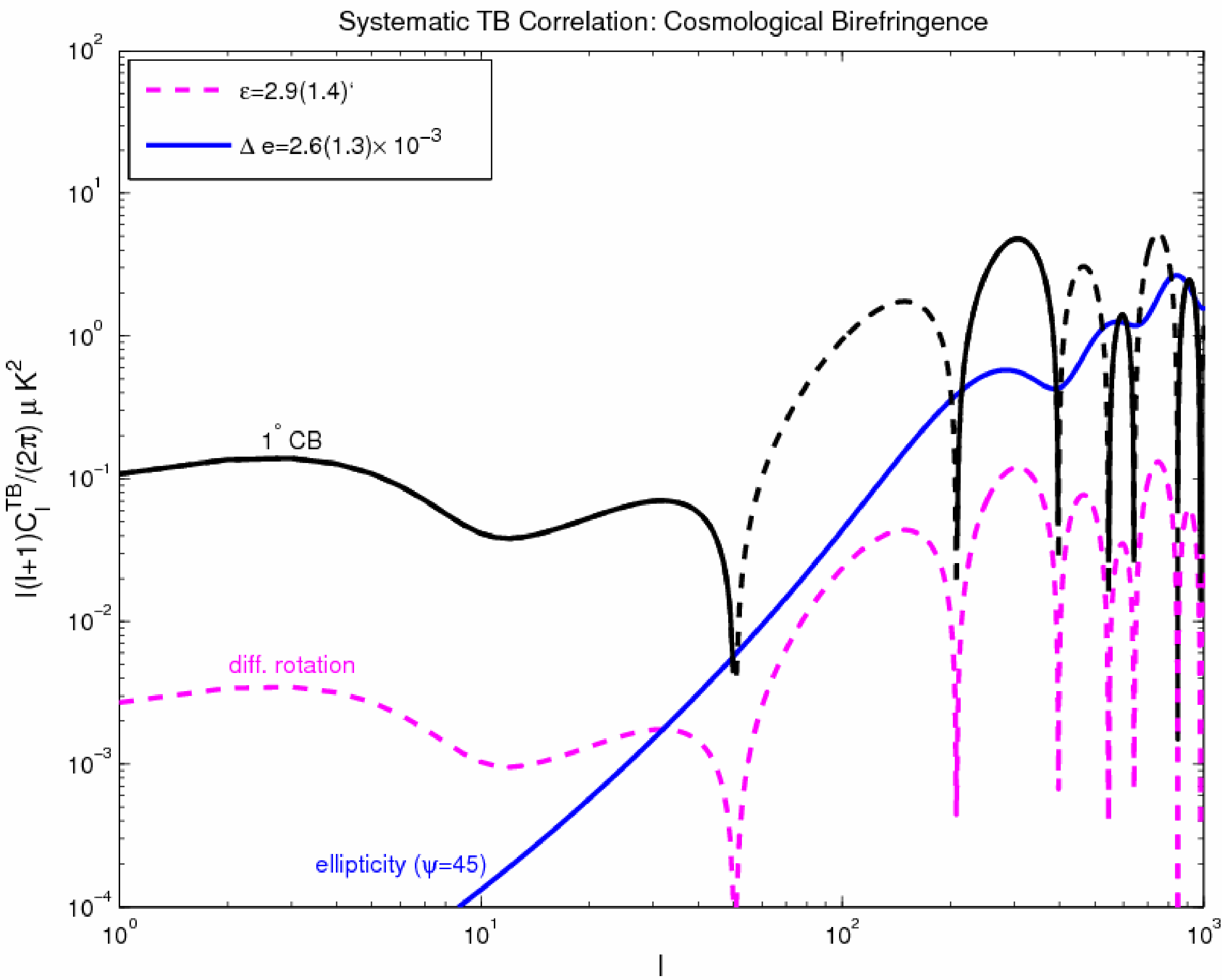

Fig. B.1. The impact of spurious TB correlations on the CB science for the same beam systematics levels as in Figure 1.2. Shown are absolute values; dashed parts of the curves correspond to negative values. Black curve corresponds to CB of 1 degree. The effect of pointing error is very small and is not shown. Differential gain and beamwidth induce no TB cross-correlations. In both cases the beam systematics (optimized for and calibrated by the inflationary signal) are at least an order of magnitude smaller than the lensing signal in all multipoles of interest. However, this imples that a CB at the few arcminute level, for example, will be overwhelmed by beam systematics. This is consistent with the findings of Miller, Shimon, & Keating's [1] findings that optimizing beam systematics for inflationary and lensing science will *not necessarily* be sufficient for CB science.

# References

**Summary**

[1] Bock, J. et al. (DOE/NASA/NSF Interagency Task Force on CMB Research Chaired by Rai Weiss), available at arXiv.org:astro-ph/0604101 (2006).
[2] Bock, J. et al. The Experimental Probe of Inflationary Cosmology (EPIC): A Mission Concept Study for NASA's Einstein Inflation Probe, available at arXiv.org:0805.4207 (2008).

**Chapter 1. Science**

[1] Netterrfield, C.B. et al. Astrophys. J. 571, 604 (2002).
[3] Lee, A. T. et al. Astrophys. J. 561, 1 (2001).
[4] Halverson, N. W. et al. Astrophys. J. 568, 38 (2002).
[5] Pearson, T. J. et al. Astrophys. J. 591, 556 (2002).
[6] Benoit, A. et al. Astron. and Astroph. 399, 19 (2003).
[7] Kuo, C.L. et al. Astrophys. J. 600, 32 (2004).
[8] Kovac, J. M. et al. Astrophys. J. 561, 1 (2002).
[9] Bennett, C. L. et al. Astrophys. J. Supp. 148, 97 (2003).
[10] Baumann, D. et al. CMBpol Mission Concept Study : Probing Inflation with CMB Polarization, available at arXiv.org :0811.3919 (2008).
[11] Page, L. et al. arXiv.org:astro-ph/0603450 (2006).
[12] Haslam, C. G. T., Stoffel, H., Kearsey, S., Osborne, J. L. & Phillips, S. Nature 289, 470 (1981).
[13] Finkbeiner, D. P., Davis, M. & Schlegel, D. J. Astrophys. J. 524, 867 (1999).
[14] Spergel, D. et al. Astrophys. J. Supp. 170, 377 (2007).
[15] Komatsu, E. et al. Astrophys. J. Supp. 180, 330 (2009).
[16] Yadav, A. P. S. & Wandelt, B. D. Phys. Rev. Lett. 100, 181301 (2008).
[17] Smith, K. et al. arXiv.org : 0901.2572 (2009).
[18] Bock, J. et al. The Experimental Probe of Inflationary Cosmology (EPIC): A Mission Concept Study for NASA's Einstein Inflation Probe, available at arXiv.org:0805.4207 (2008).
[19] Guth, A., Phys. Rev. D 23, 347 (1981).
[20] Albrecht, A. & Steinhardt, P. J. Phys. Rev. Lett. 48, 1220 (1982).
[21] Linde, A. D. Phys. Lett. B. 108, 389 (1982).
[22] Abbott, L. & Wise, M., Astrophys. J. 282, L47 (1984).
[23] Starobinskii, A., JTEP Lett. 30, 682 (1979).
[24] Randall, L., Soljacic, M. & Guth, A. Nucl. Phys. B 472, 377 (1996).
[25] Rajagopal, M. & Romani, R. W. Astrophys. J. 446, 543 (1995).
[26] Baumann, D. & Zaldarriaga, M. arXiv.org:0901.0958 (2009).
[27] Zaldarriaga , M. & Seljak, U., Phys. Rev. D 55, 1830 (1997).
[28] Kamionkowski, M., Kosowsky, A. & Stebbins, A. Phys. Rev. Lett. 78, 2058 (1997).
[29] Lyth, D. Phys. Rev. Lett. 78, 1861 (1997).
[30] Buchbinder. E. I. et al. Phys. Rev. D. 76, 123503 (2007).
[31] Battefeld, T. & Watson, S. Rev. Mod. Phys. 78, 435 (2006).
[32] Lehners, J.-L. arXiv.org: 0806.1245 (2008).
[33] Steinhardt & Turok, Science 296, 1436 (2002).
[34] Dunkley, J. et al. CMBpol Mission Concept Study: Prospects for polarized foreground removal, arXiv.org:0811.3915 (2008).
[35] Planck Collaboration, The Scientific Programme of Planck, arXiv.org:astro-ph/0604069 (2006).
[36] Maldacena, J., J. High. Ener. Phys. 0305, 013 (2003).
[37] Bartolo, N. et al. Phys. Rept. 402, 103 (2004).
[38] Lehners, J.-L. & Steinhardt, P. J. Phys. Rev. D. 77, 063533 (2008).
[39] Zahn, O. & Zaldarriaga, M. Phys. Rev. D 67, 063002 (2003).
[40] Lue, A., Wang, L. & Kamionkowski, M. Phys. Rev. Lett. 83, 1506 (1999).
[41] Gubitosi, G., Pagano, L., Amelino-Camelia, G., Melchiorri, A. & Cooray, A. in preparation (2009).
[42] Yadav, A. P. S. et al. arXiv.org:0902.4466 (2009).
[43] Gluscevic, V., Kamionkowski, M. & Cooray, A. in preparation (2009).
[44] Smith, T. L., Pierpaoli, E. & Kamionkowski, M. Phys. Rev. Lett. 97, 021301 (2006).
[45] Kaplinghat, M et al. Astrophys. J. 583, 24 (2003).

[46] Mortonson, M. J. & Hu, W. Astrophys. J. 672, 737 (2008).
[47] Zaldarriaga, M. et al. CMBpol Mission Concept Study : Reionization Science with the Cosmic Microwave Background, arXiv.org :0811.3918 (2008).
[48] Fan, X. et al. Astron. J. 132, 117 (2006).
[49] Lewis, A. & Challinor, A. Phys. Rept. 429, 1 (2006).
[50] Smith, K. et al. CMBpol Mission Concept Study: Gravitational Lensing, arXiv.org:0811.3916 (2008).
[51] Hu, W. & Okamoto, T. Astrophys. J. 574, 566 (2002).
[52] Hirata, C. & Seljak, U. Phys. Rev. D 67, 043001 (2003).
[53] Kesden, M., Cooray, A. & Kamionkowski, M. Phys. Rev. D. 67, 123507 (2003).
[54] Kaplinghat, M., Knox, L. & Song, Y.-S. Phys. Rev. Lett. 91, 241301 (2003).
[55] Cooray, A. Astron. Astrophys. 348, 31 (1999).
[56] De Bernardis, F. et al. Phys. Rev. D. 78, 083535 (2008).
[57] Fukada, S. et al. (Super-Kamiokande Collaboration), Phys. Rev. L ett, 86, 5656 (2001).
[58] Ahmad, Q. R. et al. (SNO Collaboration), Phys. Rev. Lett., 87, 071301 (2001).
[59] Hannestad, S. Prog. Part. Nucl. Phys. 57, 309 (2006).
[60] Lesgourgues, J. & Pastor, S. Phys. Rept., 429, 307 (2006).
[61] Caldwell, R., Cooray, A. & Melchiorri, A. Phys. Rev. D. 76, 023507 (2007).
[62] Bertschinger, E. Astrophys. J. 648, 797 (2006).
[63] Daniel, S. F., Caldwell, R., Cooray, A., Serra, P. & Melchiorri, A. Phys. Rev. D, arXiv.org:0901.0919 (2009).
[64] Serra, P., Cooray, A., Daniel, S. F., Caldwell, R. & Melchiorri, A. Phys. Rev. Lett. arXiv.org:0901.0917 (2009).
[65] Doran, M., Robbers, G. & Wetterich, C. Phys. Rev. D. 75, 023003 (2007).
[66] Kamionkowski, M. & Loeb, A. Phys. Rev. D., 56, 4511 (1997).
[67] Cooray, A & Baumann, D. Phys. Rev. D., 67, 063505 (2003).
[68] Cooray, A. & Hu, W. Astrophys. J. 534, 533 (2000).
[69] Crutcher, R. M., presented at The Cosmic Agitator, Lexington, KY (2008).
[70] Lazarian, A., J., CMBPol workshop, arXiv.org:0811.1020 (2008).
[71] Hiltner, W. A., Science, 109, 165 (1949).
[72] Schleuning, D. A., ApJ, 493, 811 (1998).
[73] Ward-Thompson, D., et al., ApJ, 537, L135 (2000).
[74] Chandrasekhar, S, & Fermi, E., ApJ, 118, 113 (1953).
[75] Davis, L., Physical Review, 81, 890 (1951).
[76] Ostriker, E., Stone, J. M., & Gammie, C. F., ApJ, 546, 980 (2001).
[77] Padoan, P., et al., ApJ, 559, 1005 (2001).
[78] Heitsch, F., et al., ApJ, 561, 800 (2001).
[79] Falceta-Gonçalves, D., et al., ApJ, 679, 537 (2008).
[80] Gordon, K. D., et al., ApJ, 638, L87 (2006).
[81] Knox, L. & Song, Y.-S. Phys. Rev. Lett. 89, 011303 (2002).
[82] Kesden, M., Cooray, A. & Kamionkowski, M. Phys. Rev. Lett. 89, 011304 (2002).
[83] Li, H., Dowell, C. D., Goodman, A., Hildebrand, R., & Novak, G. ApJ, submitted (2009).

**Chapter 2. Foreground Removal**

[1] Dunkley, J. et al. CMBpol Mission Concept Study: Prospects for polarized foreground removal, arXiv.org:0811.3915 (2008).
[2] Finkbeiner, D. P., Davis, M. & Schlegel, D. J. Astrophys. J. 524, 867 (1999).
[3] Draine B. T. & Lazarian, A. Astroph. J. Lett. 494, L19 (1998).
[4] Hinshaw, G. et al. arXiv.org:astro-ph/0603451 (2006).
[5] Lazarian, A. & Draine, B. T. Astroph. J. Lett. 536, L15 (2000).
[6] Lazarian, A. & Finkbeiner, D. New Astron. Rev. 47, 1107 (2003).
[7] Draine B. T. & Lazarian, A. Astroph. J. 512, 740 (1999).
[8] Page, L. et al., arXiv.org:astro-ph/0603450 (2006).
[9] Tucci, M., Mart´ınez-Gonz´alez, E., Toffolatti, L., Gonz´alez-Nuevo, J., & De Zotti, G. Mon. Not. Roy. Astron. Soc. 349, 1267 (2004).
[10] Keating, B., Timbie, P., Polnarev, A. & Steinberger, J. Astroph. J. 495, 580 (1998).
[11] Eriksen,H. K. et al., Astrophys. J. 641, 665 (2006).
[12] Davies, R. D., Dickinson, C., Banday, A. J., Jaffe, T. R., Gorski, K. M. & Davis, R. J. Mon. Not. Roy. Astron. Soc. 370, 1125 (2006).

[13] Davies, R. D., Watson, R. A. & Gutierrez, C. M. Mon. Not. Roy. Astron. Soc. 278, 925 (1996).
[14] Tegmark, M., de Oliveira-Costa, A. & Hamilton, A. Phys. Rev. D 68, 123523 (2003).
[15] Jonas, J. L., Baart, E. E., & Nicolson, G. D. Mon. Not. Roy. Astron. Soc. 297, 977 (1998).
[16] Reich, W. Astron. & Astroph. 48, 219 (1982).
[17] Reich, P. & Reich, W. Astron. & Astroph. 63, 205 (1986).
[18] Haslam, C. G. T., Stoffel, H., Kearsey, S., Osborne, J. L. & Phillips, S. Nature 289, 470 (1981).
[19] Schlegel, D. J., Finkbeiner, D. P.,& Davis, M. Astrophys. J. 500, 525 (1998).
[20] Delabrouille, J. et al MNRAS 346, 1089 (2003)
[21] Betoule, M. et al arXiv:0901.1056 (2009).[22] Dobler, G., Draine, B.T., & Finkbeiner, D.P. 2008, ApJ, submitted, astro-ph/0811.1040.
[23] de Oliveira-Costa, A., et al. ApJ, 567, 363 (2002).
[24] Casassus, S., et al. ApJ, 639, 951 (2006).
[22] Amblard, A., Cooray, A., Kaplinghat, M. Physical Review D 75, 083508 (2007).

**Chapter 3. Mission Overview**

[1] J. Bock et al. 2008, "The Experimental Probe of Inflationary Cosmology (EPIC): A Mission Concept Study for NASA's Einstein Inflation Probe", arXiv 0805.4207.
[2] R. Weiss et al. 2006, "Task Force on Cosmic Microwave Background Research, astro-ph 0604101.
[3] S. Meyer et al. 2009, "A Program of Technology Development and of Sub-Orbital Observations of the Cosmic Microwave Background Polarization Leading to and Including a Satellite Mission", programmatic decadal white paper available at http://cmbpol.uchicago.edu/.

**Chapter 4. Technology Readiness**

[1] A. Sirbi, A. Benoit, M. Caussignac, and S. Pujol, "An open cycle dilution refrigerator for operation in zero-g," Czech J. of Phys. 46(S-5), pp. 2799–2800, 1996.
[2] Puget, J.-L. and Benoit, A. and Camus, P., "A closed cycle dilution refrigerator for space." Presentation at the Novel Approaches to Infrared Astrophysics Workshop, Keck Institute for Space Studies, 2009.
[3] C. Heer, C. Barnes, and J. Daunt, "The design and operation of a magnetic refrigerator for maintaining temperatures below 1K," *Rev. Sci. Inst.* **25**, pp. 1088–1098, 1954.
[4] P. Shirron, E. Canavan, M. Dipirro, J. Francis, M. Jackson, J. Tuttle, T. King, and M. Grabowski, "Development of a cryogen free continuous ADR for the Constellation-X mission," *Cryogenics* **44**, pp. 581–588, 2004.
[5] W. McRae, E. Smith, J. Beamish, and J. Parpia, "An electronic demagnetization stage for the 0.5K to 1.8K temperature range," *J.Low Temp. Phys.* **121**, pp. 809–814, 2000.
[6] http://cmbpol.uchicago.edu/papers.php

**Chapter 5. Systematic Errors**

[1] Miller, N.J, Shimon, M. & Keating, B.G., 2009a,"CMB Beam Systematics: Impact on Lensing Parameter Estimation", PRD, 79, 063008.
[2] Hu, W., Hedman, M.M., & Zaldarriaga, M. 2003, Ph Rev D, 67, 3004.
[3] Shimon, M., Keating, B.G., Ponthieu, N., Hivon, E. 2008 ,"CMB Polarization Systematics Due to Beam Asymmetry: Impact on Inflationary Science", PRD, 77, 3033.
[4] Piat, M. et al. 2002, "Cosmic background dipole measurements with the Planck High Frequency Instrument", A&A, 393, 359.
[5] The HFI consortium 1998, "High Frequency Instrument for the Planck Mission" a proposal to the European Space Agency.
[6] Leroy, C. et al. 2006, "Performances of the Planck-HFI cryogenic thermal control system", SPIE 6265, 13.
[7] MacTavish, C.J. et al. 2008, "Spider Optimization: Probing the Systematics of a B-Mode Experiment", ApJ, 689, 655
[8] Takahashi et al. 2008, "CMB Polarimetry with BICEP: Instrument Characterization, Calibration and Performance", Proc. SPIE, 7020, 70201D
[9] Bock et al. 2008, "The Experimental Probe of Inflationary Cosmology: A Mission Concept Study for NASA's Einstein Inflation Probe", arXiv 0805.4207.
[10] Smith, K.M, et al., 2008,"CMBPol Mission Concept Study: Gravitational Lensing", arXiv: 0811.3916.
[11] Rosset, C. et al. 2007, "Beam mismatch effects in cosmic microwave background polarization measurements", A&A, 464, 405.

**Appendix A & B**

[12] O'Dea, D., Challinor, A. & Johnson, B.R., 2007, MNRAS, 376, 1767.
[13] Miller, N.J, Shimon, M. & Keating, B.G., 2009b,"CMB Polarization Systematics Due to Beam Asymmetry: Impact on Cosmological Birefringence", PRD accepted, arXiv: 0903.1116.

**Chapter 6. Crossed Dragone Telescope**

[1] CMBpol Mission Concept Study: Optical Elements for a CMBPol Mission,(2008) available at http://cmbpol.uchicago.edu/.
[2] Christophe Granet. Designing classical dragonian offset dual-reflector antennas from combinations of prescribed geometric parameters. *Antennas and Propagation Magazine,,IEEE*, 43(6):1045-9243, Dec 2001.

**Chapter 7. Focal Plane**

[1] M.J. Griffin, J.J. Bock, and W.J. Gear 2002, "The Relative Performance of Filled and Feedhorn-Coupled Focal-Plane Architectures," *Applied Optics*, 41, 6543.
[2] K.D. Irwin & G.C. Hilton 2005, "Transition-Edge Sensors", in Cryogenic Particle Detection, Springer Berlin: Heidelberg, v. 99.
[3] J.C. Mather 1982, "Bolometer noise: nonequilibrium theory," Applied Optics, 21, 1125.
[4] J.C. Mather 1984, "Bolometers: ultimate sensitivity, optimization, and amplifier coupling," Applied Optics, 23, 584.

**Chapter 8. Cooling**

*8.1 Passive Cooling*

[1] Kelsall T, et al. 1998, "The COBE Diffuse Infrared Background Experiment Search for the Cosmic Infrared Background. II. Model of the Interplanetary Dust Cloud", ApJ, 508, 44.
[2] H. W. Yorke et al. 2004, "Thermal Design Trades for SAFIR Architecture Concepts", SPIE, 5487, 1617.

*8.3 Cooling to 100 mK*

[1] S. Breon, et al. 1999, Cryogenics **39,** 677.
[2] L. Duband, L Hui, and A Lange 1990, Cryogenics **30,** 263.
[3] A. Sirbi, A. Benoit, M. Caussignax, and S. Pujol 1996, Czech. J. Phys. **46,** 2799.
[4] W.J. McRae, E.N.Smith, J. Parpia, and J. Beamish 2000, J. Low Temp. Phys. **121**, 809.
[5] P. Shirron et al. 20002, Cryogenics **41,** 789.
[6] C. Heer, C. Barnes, and J. Daunt 1954, Rev. Sci. Inst. **25,** 1088.
[7] W.F. Giaque, et al 1936, J. Am. Chem. Soc. **58,** 1114.
[8] D. McCammon et al 1997, Nucl. Instrum. Methods.
[9] C. Hagmann, D.J.Benford, and P.L. Richards 1994, Cryogenics **34,** 213.
[10] C. Hagmann and P.L. Richards 1995, Cryogenics **35,** 303.
[11] American Superconductor, HTS wire, http://www.amsuper.com/htsWire/103668710751.cf
[12] J. Torre and G. Chanin 1984, Rev. Sci. Inst. **55**, 213.
[13] E. Smith, J. Parpia, and J. Beamish 2000, J. Low Temp. Phys. **119**, 507.
[14] P. Shirron, et al. 1999, Adv. Cryo. Eng **25**, 256.
[15] R. Srinivasan 1997,
http://www.iisc.ernet.in/~academy/resonance/Jun1997/pdf/June1997p06-14.pdf
[16] H. Sugita et al. 2008, "Cryogenic system for the infrared space telescope SPICA," in SPIE 7010.
[17] M. Petach et al. 2008, "Mechanical cooler for IXO and other space based sensors," *American Astronmical Society Meeting Poster*
[18] P. Bhandari et al. 2004, Cryogenics **44**, 395.
[19] D. Yvon et al. 2008, J. Low Temp. Phys. **151**, 448.
[20] A. M. Clark et al. 2005, Appl. Phys. Lett. **86**, 173508.
[21] G. C. O'Neil et al. 2008, Appl. Phys. Lett. **92**, 163501.

**Chapter 9. Deployed Sunshade**

[1] J. Bock et al. 2008, "The Experimental Probe of Inflationary Cosmology (EPIC): A Mission Concept Study for NASA's Einstein Inflation Probe", arXiv 0805.4207.

**Chapter 11. Spacecraft**

[1] J. Bock et al. 2008, “The Experimental Probe of Inflationary Cosmology (EPIC): A Mission Concept Study for NASA's Einstein Inflation Probe”, arXiv 0805.4207.

Acknowledgment:
This research was carried out at the Jet Propulsion Laboratory, California Institute of Technology, under a contract with the National Aeronautics and Space Administration